\documentclass[twoside,11pt]{Latex/Classes/PhDthesisPSnPDF}
\pdfoutput=1
\usepackage{graphics,epsfig}
\usepackage{amsmath}
\usepackage{amsfonts}
\usepackage{amssymb}
\usepackage{caption}

\ifpdf  
    \pdfcatalog { /PageMode (/UseOutlines)
                  /OpenAction (fitbh)  }
\fi

\title{Searching for $P-$ and $CP-$ odd effects in heavy ion collisions}

  \author{\href{mailto:xumeu@icc.ub.edu}{Xumeu Planells Noguera}}
  \collegeordept{\href{http://www.ecm.ub.edu}{Departament d'Estructura i Constituents de la Mat\`eria
}\\
  \&\\
  \href{http://icc.ub.edu}{Institut de Ci\`encies del Cosmos}
}
  \university{\href{http://www.ub.edu}{Universitat de Barcelona}\\
  \vspace{2em}
  Advisors:\\
    \vspace{0.5em}
  \href{mailto:espriu@icc.ub.edu}{Dom\`enec Espriu Climent}\\
    \vspace{1em}
  \href{mailto:andrianov@icc.ub.edu}{Alexander A. Andrianov}}
  \crest{\vspace{0em}\hspace{-4em}\includegraphics[width=15cm]{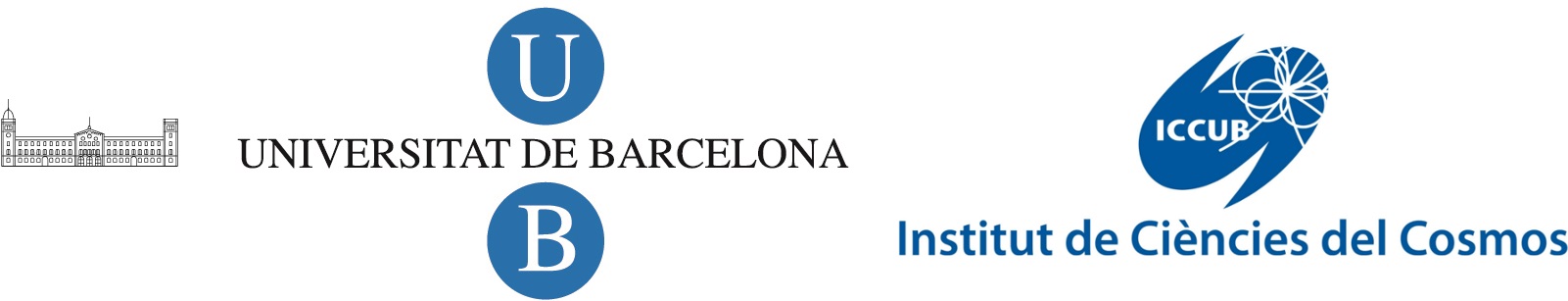}}

\degree{Philosophi\ae Doctor (PhD) in Physics\\
(Programa de Doctorado en F\' isica)}
\degreedate{November 2014}

\begin{document}


\maketitle

\begin{abstracts}        

In this thesis we study the possibility that QCD breaks parity at high temperatures and densities, a scenario that may be tested in heavy ion collisions. We present two different approaches to this problem. On the one hand, analytical studies with effective models suggest that QCD may break parity in dense systems due to condensation of pseudoscalar mesons in the isotriplet channel. On the other hand, $P-$ and $CP-$ odd bubbles may appear in a finite volume due to local large topological fluctuations in a hot medium. The presence of a large magnetic field or angular momentum may lead to the so-called Chiral Magnetic Effect. We assume topological fluctuations as the source of parity violation in heavy ion collisions throughout. This effect may be treated in a quasi-equilibrium description by means of a non-trivial axial chemical potential.

\medskip

We consider the 'two flavour' Nambu--Jona-Lasinio model in the presence of a vector and an axial external chemical potentials and study the phase structure of the model at zero temperature. The Nambu--Jona-Lasinio model is often used as a toy replica of QCD and it is therefore interesting to explore the consequences of adding external vector and axial chemical potentials in this model, mostly motivated by claims that such external drivers could trigger a phase where parity could be broken in QCD. We are also motivated by some lattice analysis that attempt to understand the nature of the so-called Aoki phase using this simplified model. Analogies and differences with the expected behaviour in QCD are discussed and the limitations of the model are pointed out.

\medskip

We also consider the low energy realization of QCD in terms of mesons when an axial chemical potential is present; a situation that may be relevant in heavy ion collisions. We shall demonstrate that the presence of an axial charge has profound consequences on scalar/pseudoscalar meson physics. The most notorious effect is the appearance of an explicit source of parity breaking. The eigenstates of strong interactions do not have a definite parity and interactions that would otherwise be forbidden compete with the familiar ones. We focus on scalars and pseudoscalars that are described by a generalized linear sigma model. We comment briefly on the screening role of axial vectors in formation of effective axial charge and on the possible experimental relevance of our results, whose consequences may have been already seen at RHIC.

\medskip

Finally, we investigate how local parity breaking may affect vector physics. A modified dispersion relation is derived for the lightest vector mesons $\rho$ and $\omega$. They are characterised by a mass splitting depending on their polarization. This effect predicts a natural overproduction of lepton pairs in the vicinity of the $\rho-\omega$ resonance peak as well as a polarization asymmetry around this peak. The dilepton excess seems relevant to explain the anomalous dielectron yield quoted by PHENIX/STAR. We present a detailed analysis of the angular distribution associated to the lepton pairs created from these mesons searching for polarization dependencies. Two angular variables are found to carry the main information related to the parity-breaking effect. Possible signatures for experimental detection of local parity breaking are discussed.

\end{abstracts}

\frontmatter

\begin{dedication} 

A Jennifer Mesa, por haber sido una parte esencial de mi vida durante esta tesis.

\medskip

Y a mis padres, por darme la oportunidad de aprender y formarme como persona.

\bigskip
\bigskip

To Jennifer Mesa, for being an essential part of my life during this thesis.

\medskip

And to my parents, for giving me the chance to learn and grow as a person.

\end{dedication}

\cleardoublepage
\cleardoublepage
\begin{agradecimientos} 

Quiero agradecer al Prof. Dom\`enec Espriu la oportunidad que me brind\' o para iniciar este viaje, en el que hemos colaborado en un ambiente muy ameno y distendido. Han sido unos a\~nos de trabajo constante pero agradable y enriquecedor tanto a nivel profesional como personal. Tambi\' en me gustar\' ia agradecer al Prof. Alexander Andrianov su hospitalidad y su eterna sonrisa. No s\' e qu\' e habr\' ia sido de m\' i en San Petersburgo sin \' el. Quiero expresar mi gratitud tambi\' en a David d'Enterria por su infinita paciencia conmigo. Nunca pens\' e que dos meses pudiesen ser tan productivos en tantos aspectos.

\medskip

No puedo olvidar a toda aquella gente que ha hecho que esta experiencia haya sido una aventura inolvidable. Gracias a mis compa\~neros de doctorado del departamento de ECM Albert Renau, Vicente Rives y Juan Gonz\' alez. Siempre nos quedar\' an esas partidas de mus. Y a mis colegas durante la carrera Jos\' e Daniel Madrigal, H\' ector Garc\' ia, y en especial Santi Roca y Enrique Pinto, por aquellos momentos en el buffet en que todo se ve\' ia mejor mientras nos hart\' abamos de pizza.

\medskip

Quiero agradecer tambi\' en la hospitalidad que recib\' i en Rusia por parte no s\' olo del Prof. Alexander Andrianov y su esposa, sino tambi\' en de su hermano Vladimir, su estudiante Sergey Kolevatov, y Vladimir Zherebchevsky y la gente del Ciclotr\' on de la Universidad Estatal de San Petersburgo.

\medskip

A pesar del tiempo que ha pasado, quiero agradecer el papel que han jugado en mi vida algunos profesores durante mi etapa formativa como Gabi Vidagany y Amparo Fuster, extraordinarios profesores y modelos de persona. Y tambi\' en a Conrad P\' erez, sin cuyos consejos mi experiencia en la universidad habr\' ia sido mucho m\' as tortuosa.

\medskip

Quiero dar las gracias a todos aquellos que han estado junto a m\' i desde hace a\~nos y me han apoyado para lo bueno y para lo malo. Gracias a toda mi familia y en especial a mi madre, porque su presencia y su apoyo incondicional no pueden pasarse por alto. Y a mi gente de Ibiza ya sea en Barcelona o en la 'quinta China'. Me refiero a Santi Mar\' i, Christian Mar\' in, Adri\' an Baena, Laura Tur, Mar\' ia Ram\' on y muchos m\' as.

\medskip

Por \' ultimo quiero agradecer el soporte econ\' omico que me ha facilitado la beca FPU AP2009-1855.

\end{agradecimientos}
\begin{acknowledgements}      

I want to thank Prof. Dom\`enec Espriu for the opportunity he provided me to start this journey, in which we collaborated in a very friendly and relaxed atmosphere. It has been a time of constant but pleasant work that was enriching for me both on the professional and personal level. I also want to thank Prof. Alexander Andrianov for his hospitality and his eternal smile. I don't know what I would have done without him in Saint Petersburg. I also want to express my gratitude to David d'Enterria for his infinite patience with me. I had never thought that two months could be so productive in so many aspects.

\medskip

I can't forget all the people who have made this experience an unforgettable adventure. Thanks to my fellow PhD students from the ECM department Albert Renau, Vicente Rives, Juan Gonz\' alez. We'll always have those 'mus' games. And to my degree mates Jos\' e Daniel Madrigal, H\' ector Garc\' ia, and specially Santi Roca and Enrique Pinto, for those moments at the buffet in which everything looked better while we were eating pizza to burst.

\medskip

I also want to thank the hospitality that I received in Russia, not only from Prof. Alexander Andrianov and his wife, but also from his brother Vladimir, his student Sergey Kolevatov, and Vladimir Zherebchevsky and the people from the Cyclotron at the Saint Petersburg State University.

\medskip

Despite the time that has passed, I want to thank some of the teachers I had during my education for the role they played in my life like Gabi Vidagany and Amparo Fuster, extraordinary teachers and role models. And also Conrad P\' erez, because my experience at the university would have been much more tortuous without his advice.

\medskip

I want to acknowledge all the people who have been with me for years and have supported me through thick and thin. Thanks to my family and specially to my mother, because her presence and unconditional support cannot be missed. And to my people from Ibiza either in Barcelona or thousands of miles away. I am talking about Santi Mar\' i, Christian Mar\' in, Adri\' an Baena, Laura Tur, Mar\' ia Ram\' on and many others.

\medskip

Last, I want to thank the financial support from Grant FPU AP2009-1855.

\end{acknowledgements}


\setcounter{secnumdepth}{3} 
\setcounter{tocdepth}{3}    
\tableofcontents            


\mainmatter

\chapter{Introduction}\label{ChIntro}
\graphicspath{{1-intro/figures/}{figures/}}

The nuclear strong force is known to be the responsible for binding the atomic nuclei despite the electromagnetic repulsion between protons. This interaction is one of the four fundamental forces of Nature, like gravity and the electromagnetic and weak forces. The last two interactions can be treated together by means of the unified theory of electro-weak interaction. Electro-weak and strong interactions, but not gravity, can be formulated in terms of a quantum field theory that is known as the Standard Model (SM) of elementary particles. The behaviour of elementary particles is ruled by this successful theory that has been contrasted many times with experiments.

\medskip

The firsts attempts to deal with strong interactions can be found in the middle of the last century. In 1961 Gell-Mann proposed a phenomenological model \cite{eightfold} to structure the large amount of new hadrons that were being found in accelerator experiments at that time. Over the next decade, the nuclear strong force was formulated at the quantum level by Quantum Chromodynamics (QCD) \cite{QCD}, a non-abelian $SU(3)$ Yang-Mills theory that describes the fundamental interactions between quarks and gluons \cite{QCDSU3}, called partons. Quarks and gluons were introduced as the basic components of hadrons and have an associated colour quantum number due to the symmetry group of QCD. The QCD Lagrangian reads
\begin{align}\label{QCDlag}
\nonumber \mathcal L_{\text{QCD}}&=-\frac14G^{\mu\nu,a}G_{\mu\nu}^a+\bar q(i\gamma^\mu D_\mu-m)q,\\
D_\mu&=\partial_\mu -igG_\mu^a\lambda^a, \quad G_{\mu\nu}^a=\partial_\mu G_\nu^a-\partial_\nu G_\mu^a+g f^{abc}G_\mu^bG_\nu^c,
\end{align}
where $q$ is the quark field with two implicit indices in colour and flavour spaces and the colour-dependent gluon field is $G_\mu^a$. The strong coupling constant $g$ and the Gell-Mann matrices $\lambda^a$ in the adjoint representation of $SU(3)$ are also used. There are 3 colours and 6 flavours of quarks. Quarks are massive fermionic states lying in the fundamental representation of the QCD symmetry group while gluons are the massless force carriers, equivalently to photons in electromagnetism, and they belong to the adjoint representation of $SU(3)$.

\medskip

The strong attraction between a quark-antiquark pair requires a large amount of energy to separate them. Such energy excites the vacuum state up to the point where the creation of a new quark pair becomes energetically favourable and the new particles quickly screen the original charge. This is the qualitative description of quark confinement, which is responsible for the fact that coloured states are never observed alone in Nature. This phenomenon is only known to occur in QCD and believed to be linked to its non-abelian structure. The lack of a rigorous proof of confinement still constitutes a remarkable challenge for theoretical physicists nowadays.

\medskip

It was experimentally observed that there are only two main structures formed by quarks. Hadrons are separated into baryons and mesons. Baryons are 3-quark states while mesons are composed by a quark-antiquark pair. In order to preserve the Pauli exclusion principle in the formation of baryons a new quantum number, the colour charge, was introduced. Assuming that every quark is associated to one colour, both structures can be arranged to be colourless. This is the reason why it is commonly stated that only colourless (or white) states are observable in Nature.

\medskip

In addition to the gauge symmetry, the QCD Lagrangian \eqref{QCDlag} shows an invariance under three global discrete symmetries \cite{PS,coleman}. Parity $P$ reverses the handedness of space that corresponds to reverting the momentum of a particle without flipping its spin. The parity transformation is given by
\begin{equation}
q(\vec x,t)\to \eta_P\gamma^0 q(-\vec x,t), \qquad G_0(\vec x,t)\to G_0(-\vec x,t), \qquad G_i(\vec x,t)\to -G_i(-\vec x,t),
\end{equation}
where $\eta_P$ is a possible generic phase that is normally taken to be 1 for quarks and leptons without any loss of generality. For antiparticles, $\eta_P$ is chosen to be -1. Charge conjugation $C$ interchanges particles and antiparticles without any modification of momentum or spin through the transformation
\begin{equation}
q(\vec x,t)\to -i\eta_C C \bar q^T(\vec x,t), \qquad G_\mu(\vec x,t)\to -G_\mu^T(\vec x,t),
\end{equation}
where $\eta_C$ is another phase and $C$ is the charge conjugation matrix that obeys the following relations
\begin{equation}
C\gamma_\mu C^{-1}=-\gamma_\mu^T, \qquad C\gamma_5 C^{-1}=\gamma_5^T, C^{\dagger}=C^{-1}=-C
\end{equation}
and in the Dirac representation is given by $C=i\gamma^0\gamma^2$. Finally, time reversal $T$ corresponds to an interchange of the forward and backward light-cones, corresponding to reversing the momentum and the spin of a particle via
\begin{equation}
q(\vec x,t)\to -i\eta_T\gamma_5C q(\vec x,-t), \qquad G_0(\vec x,t)\to -G_0(\vec x,-t), \qquad G_i(\vec x,t)\to G_i(\vec x,-t),
\end{equation}
where $\eta_T$ is again an arbitrary phase. QCD, as any other Lorentz-invariant local quantum field theory with a Hermitian Hamiltonian is also invariant under a simultaneous $CPT$ symmetry. 

\medskip

The quark spectrum is spread over a wide range of masses from $m_{u,d}\sim 5$ MeV up to $m_t\sim 175$ GeV, leading to a clear energy/mass hierarchy. At a given energy scale $\Lambda$, the quarks can be separated into light quarks with masses $m_q$ and heavy quarks with $m_Q$ in such way that the relation $m_q\lesssim \Lambda \ll m_Q$ holds. In this framework at low energies, the effects of heavy quarks amount to redefinitions of the light quark masses $m_q$ and the coupling constants involving the light quarks up to $\Lambda/m_Q$ corrections. This is known as the decoupling theorem \cite{AppelCarazz}. Low energy theories \cite{ChPT,ChPT2} are usually formulated to describe the physics of strong interactions below the chiral symmetry breaking scale $\Lambda\sim 1$ GeV and hence the number of quark flavours is normally taken to be $N_f=2,3$ always including $u$ and $d$ quarks, and sometimes the $s$ quark. For analyses at higher energy scales, other techniques like Heavy Quark Effective Theory are used  \cite{HQET}.

\medskip
  
In the limit of massless quarks, QCD exhibits an exact $U(N_f)_L\times U(N_f)_R$ global symmetry at the classical level, where $N_f$ is the number of quark flavours that are considered. Under such transformation the quark fields are modified in the following way:
\begin{equation}
q_{L,R}(x)\to\exp\left (\frac i2 \theta^a_{L,R}T^a\right )q_{L,R}(x), \qquad q_{L,R}(x)\to\exp\left (\frac i2 \tilde\theta_{L,R}\right )q_{L,R}(x),
\end{equation}
where $T^a$ belong to the adjoint representation of $SU(N_f)$ and $\theta^a_{L,R},\tilde\theta_{L,R}$ are some arbitrary constants that parametrize the transformation. Gluon fields are unaffected by this global symmetry. Any $U(N)$ transformation can be decomposed into $SU(N)\times U(1)$ and thus, the previous global symmetry can be rewritten as $SU(N_f)_L\times SU(N_f)_R\times U(1)_V\times U(1)_A$, where the vector part $V$ is the sum of the left- and right-handed quark numbers and the axial component $A$ corresponds to the subtraction of the previous quantities.

\medskip

Quantum corrections reveal that the singlet axial-vector current associated to the symmetry under $U(1)_A$, which is conserved at the classical level, develops an anomaly and thus, the true symmetry of the quantum theory reduces to $SU(N_f)_L\times SU(N_f)_R\times U(1)_V$. In Chapter \ref{Chfase} we will address a more detailed explanation of the axial anomaly. The $U(1)_V$ symmetry is related to the baryon number conservation while the rest is usually referred to as the chiral symmetry. If the chiral symmetry were a true symmetry of QCD, one would expect a spectrum of degenerate states with opposite parity. However, this is not observed in Nature implying that the symmetry is broken. In fact, the symmetry is broken spontaneously as the generators of the $SU(N_f)_L\times SU(N_f)_R/SU(N_f)_V$ quotient space do not annihilate the ground state and therefore each of them is associated to a massless Goldstone boson. The physical states corresponding to these Goldstone bosons are the lightest mesons $\pi$, $\eta$ and $K$, the latter being important only if $N_f=3$. The finite masses of these particles are understood as a consequence of the explicit symmetry breaking due to the finite quark masses in the QCD Lagrangian. Although exact chiral symmetry is not actually realized in QCD, this treatment may be used as an approximation to work in QCD at low energies developing the so-called Chiral Perturbation Theory ($\chi$PT), which is discussed below.

\medskip

The QCD coupling $g$ determines the strength of the interactions between quarks and gluons and the gluon self-interaction. Gluon self-interactions are allowed due to the non-abelian nature of QCD, unlike the case of electromagnetism where photons can only interact with other particles. The renormalization of a quantum theory leads to the running of its coupling constant, which is not actually a real constant but depends on the energy scale $\mu$ at which a process takes place. A $SU(N_c)$ gauge theory with $N_f$ quark flavours shows \cite{asymfreed} that the $\beta$ function at one loop
\begin{equation}
\beta(g)_{\text{1-loop}}=\frac{\partial g}{\partial\log\mu}=-\frac{g^3}{16\pi^2}\left (\frac{11}3N_c-\frac23N_f\right )
\end{equation}
is negative for $N_f<\frac{11}2N_c$. In particular, the fact that QCD with $N_c=3$ and $N_f=6$ quarks is characterised by a negative $\beta$ function leads to a coupling constant decreasing with rising energy. At high energies QCD is not a strongly interacting theory any more and approaches the free limit. This effect is called asymptotic freedom and in a way it corresponds to the opposite behaviour of the confinement described at low energies.

\medskip

This picture of QCD allows to use perturbative techniques for the calculations of processes at high energies. The perturbative short-distance parton cross sections and the non-perturbative long-distance quantities may be combined due to QCD factorization theorems \cite{factorizationQCD}. The long-distance contribution includes the parton distribution functions, fragmentation functions, different kinds of form factors, etc. Indeed this method becomes more accurate as energies are further increased due to QCD asymptotic freedom.

\medskip

At low energies, perturbative calculations are meaningless due to the large coupling among quarks and gluons. Non-perturbative effects cannot be avoided in this regime and therefore many difficulties appear making QCD hard to analytically or numerically solve. Several approaches have been addressed to overcome these problems such as numerical simulations on the lattice \cite{wilson}, effective field theories \cite{ChPT,ChPT2} or $1/N_c$ expansions \cite{1/N}.

\medskip

One possibility to rigorously define a quantum field theory consists in discretizing it. Lattice calculations are a non-perturbative mechanism where QCD is formulated on a grid of points in Euclidian space-time. A fixed and finite spacing between the points of the grid represents a UV regulator. If the lattice size is taken to be infinitely large and the separation of its sites infinitesimally small, continuum QCD is recovered \cite{wilson}. The Feynman path integral approach is used to compute correlation functions of hadronic operators and matrix elements of any operator between hadronic states in terms of the fundamental quark and gluon degrees of freedom. The pion mass $m_\pi$ is normally used to test the reliability of lattice calculations since it is the lightest hadron state and therefore its corresponding wave length is the largest one. If the lattice is not large enough, the pion wave function does not fit in the lattice. This is why large quark masses are normally used in this description leading to a non-physical value for the pion mass. Nevertheless, large lattices with small spacing correspond to a huge number of grid sites leading to computational simulations that need much more time to be performed. In addition, the corrections to finite-volume effects have to be calculated for a realistic comparison with experiments.

\medskip

A different approach for low energy QCD consists in using effective field theories (EFT). Since they are anyway not visible at large distances, quarks or gluons are integrated out of the model while hadrons become the relevant low-energy degrees of freedom. Imposing the symmetries of QCD to the theory, any observable can be calculated by means of a sort of perturbative expansion in which a power-counting scheme determines the importance of the quantum amplitudes. In the same way the increasing powers of a small coupling constant are used in perturbation theory, EFT usually take advantage of the ratio of a small parameter such as mass, momentum/energy. These theories correspond to the low-energy realization of QCD and therefore they are only valid up to some maximum cut-off scale, making them non-renormalizable. $\chi$PT has proven to be a highly successful effective theory where the interactions among the lightest meson states are described (for a review see \cite{ChPT,ChPT2}). Due to the large $s$ quark mass, the convergence of the expansion in the $SU(3)$ sector is somewhat slower as compared with the $SU(2)$ version. A generalisation may be done by accommodating baryon states to the theory leading to additional baryon-baryon and baryon-meson interactions. It is important to note that $\chi$PT uses as only ingredients the symmetries of QCD and it is therefore not a model but a low energy realization.

\medskip

In this thesis we shall use two low energy models: the Nambu--Jona-Lasinio model (NJL) and the sigma model. NJL is a microscopic model including four-fermion vertices that replace the interactions of gluons in QCD. Due to the presence of these terms, the model is not renormalizable but the absence of gluons allows the theory to be bosonised. After this process, one obtains a meson theory including scalar and pseudoscalar fields that can be associated to a sigma model with concrete values for its parameters. In turn the sigma model corresponds to a particular theory that derives from EFT based on $\chi$PT assuming scalar dominance of the interaction at low energies. In both NJL and the sigma model chiral symmetry is spontaneously broken due to the formation of a dynamical chiral quark condensate. This condensate provides a constituent mass for the quarks. In the $u,d$ sector, it is expected to be of the order of 300 MeV (based in the mass of nucleons) in contrast with the light mass of current quarks, of order 5 MeV. More details of these two models will be given in Chapters \ref{ChNJL} and \ref{ChSmodel}.

\medskip

When no other method works, the only way to investigate the behaviour of QCD at low energies consists in performing an expansion in terms of the small parameter $1/N_c$. The large $N_c$ limit of a $SU(N_c)$ gauge theory might lead to think that a larger gauge group complicates the theory since it involves an increase of the number of dynamical degrees of freedom. However, this limit can be shown to be a reasonable approximation even for the case of QCD where $N_c=3$ \cite{1/N}. In the large-$N_c$ limit the number of gluon species grows as $N_c^2$ while the number of quarks is of order $N_c$. QCD becomes a theory of an infinite number of non-interacting zero-width resonances \cite{1/N-resonanc}, which are mesons and glueballs since baryon masses grow with $N_c$ and can be decoupled and safely ignored. The $1/N_c$ expansion is normally formulated by assigning to each Feynman diagram a power of $N_c$ \cite{1/N-form}, which is determined by the topology of the diagram. The leading contributions are 'planar' and do not contain quark loops while subleading terms are suppressed by a factor $1/N_c$ due to quark loops and each non-planar gluon exchange is suppressed by $1/N_c^2$. The qualitatively important point of this approach is the $N_c$-scaling behaviour of the different diagrams that allows to classify them depending on their relevance. Since $1/N_c$ is not a very small parameter, a perturbative expansion in $1/N_c$ may require the inclusion of a large number of higher order terms if an accurate result is searched.

\section{The QCD phase diagram}

The fact that highly energetic partons could interact almost freely suggests the possibility that confined matter may eventually undergo a phase transition. The system is led to a deconfined regime where the main degrees of freedom of QCD are quarks and gluons instead of hadrons and therefore colourless states do not exist any more.

\medskip

The study of deconfinement has been addressed by many different collaborations and techniques. QCD lattice calculations can give reliable results on thermodynamic properties of strongly interacting matter only for very small baryon chemical potentials. In this case a crossover type of the deconfinement transition is predicted \cite{lattcrossover}. Lattice QCD also suggests a phase transition from hadronic phase to quark gluon plasma at $T_c\sim140-190$ MeV \cite{lattcrit} (for reviews, see \cite{revlattcrit}).

\medskip

This new phase, the so-called quark-gluon plasma (QGP), could be reached at high temperatures and densities, conditions that are reachable in heavy ion collisions (HIC). In fact, recent results from Relativistic Heavy Ion Collider (RHIC) show strong evidence that QGP has been created in the Au+Au collisions at $\sqrt{s_{NN}}=200$ GeV\cite{QGPatRHIC}.

\medskip

By modifying the experimental conditions such as the collision energy or impact parameter, it is possible to explore a broad range of temperatures and densities unravelling the different phases of QCD. While the system is heated and compressed, hadrons are expected to occupy more and more space leading to a ultimate overlapping that triggers a parton 'percolation' between hadrons thus being released. In addition, light quarks lose their constituent mass inside hadrons and acquire their genuine mass triggering a restoration of chiral symmetry. One of the scientific goals of the current high energy heavy ion program is to quantify properties of this QGP matter, such as the equation of state.

\medskip

In a HIC experiment, heavy nuclei are accelerated to ultra-relativistic speeds before they impact. Immediately after the collision the highly boosted nuclei create a strongly interacting media that reaches extremely high temperatures of the order of several hundreds of MeV, and the density of nuclear matter may be widely exceeded. Understanding the behaviour of the hot and dense fireball constitutes one of the most notorious challenges in QCD. These extreme thermodynamic conditions are believed to exist in two astrophysical contexts: a few microseconds after the 'Big Bang' a transitory stage of strongly interacting matter survived with a temperature $\sim 200$ MeV (around $10^{12}$ K) and a small net baryon excess; in the core of neutron stars densities are expected to be a few times the nuclear density. New regions of the QCD phase diagram can be explored by means of current accelerator experiments \cite{Andronic:2009gj,Blaizot:2009ws}. The first collisions were performed at the Alternating Grading Synchrotron (AGS) in Brookhaven (BNL) with center-of-mass energies around $\sqrt s\sim 5$ GeV per nucleon and later, the CERN Super-proton-Synchrotron (SPS) achieved $\sqrt s\sim$20 GeV per nucleon among many other accelerator experiments. Currently RHIC at BNL and the LHC at CERN reach energies as high as $\sqrt s\sim 200$ GeV and 5 TeV per nucleon, respectively.

\begin{figure}[h!]
\centering
\includegraphics[scale=1.8]{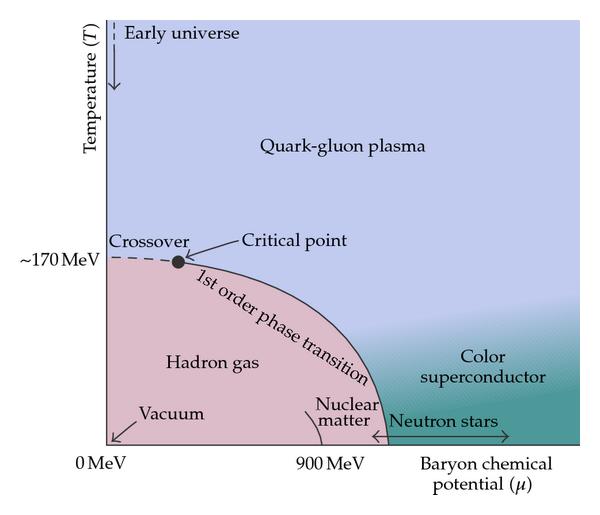}
\caption{QCD phase diagram depending on the temperature and density extracted from \cite{QCDphasediag}. Realistic and speculative phases are indicated.
}\label{QCDphasediag}
\end{figure}

The phenomenon of deconfinement has also been investigated in accelerator experiments. The crossover transition to the deconfined phase that predicts lattice QCD seems to be confirmed by recent RHIC experiments \cite{crossovRHIC}. In the regime of moderate densities, different approximations predict a possible first order phase transition to the deconfined phase suggesting the existence of a critical point that has been intensively searched at RHIC. One of the main goals of the RHIC low-energy program is to find this critical point by scanning different regions of the phase space colliding gold nuclei at various energies. In order to search for a possible critical point, higher moments of the net-proton distributions are measured in Au+Au collisions over a wide range of chemical potentials. Deviations of several moments with respect to theoretical expectations have been observed at beam energies below $\sqrt{s_{NN}}=27$ GeV per nucleon \cite{critpoint}. However, higher statistics are needed to obtain definite conclusions. From the theoretical point of view, many different approaches have posed and investigated this critical point.

\medskip

Dealing with strongly coupled matter complicates the theoretical understanding of the properties and dynamics of the medium. In the last years an increasing interest in strongly coupled techniques led to the study of the liquid plasmas that arise as hot deconfined phases of non-abelian gauge theories which have holographically dual descriptions as gravitational theories in 4+1 dimensions containing a black hole horizon. A simple example of this gauge/gravity duality has been applied in a strongly coupled $\mathcal N=4$ supersymmetric Yang-Mills theory in the limit of a large number of colours. These sort of analyses have provided plenty of qualitatives insights into the properties of the QGP \cite{casalderrey}. In particular, in the limit where the QCD coupling is large at all relevant energy scales the gauge/gravity duality allowed to use holographic calculations in order to obtain detailed dynamical information on the energy loss processes. 

\medskip

Theoretical studies of the QGP thermodynamics revealed that it behaves as a near-ideal Fermi liquid, although research on flow characteristics is ongoing. An important development in recent years has been the conjecture that for any thermal fluid, a fundamental bound on the value of the viscosity may exist \cite{viscbound}. The uncertainty principle requires a non-zero viscosity. More precisely, the conjecture states that the ratio of viscosity to entropy density must satisfy the relation $\eta/s\geq 1/4\pi$ and different investigations have been developed in order to experimentally measure this ratio \cite{viscos}. The combination of the different results seems to conclude that the produced matter is within a factor of 2-3 of the conjectured bound and therefore, it should correspond to the most ideal fluid that has been ever measured. However, there is evidence that the matter does not behave as a quasi-ideal state of free quarks and gluons, but rather as an almost perfect dense fluid with a nearly complete absorption of high momentum probes \cite{zajc}.

\medskip

In addition to the QGP and the hadronic phase, other thermodynamical phases have been under discussion together with their corresponding phase transitions. In Fig. \ref{QCDphasediag} a tentative phase diagram of QCD is shown depending on the temperature and density. For example, the liquid-gas transition from the nuclear matter to the hadron gas is believed to occur for temperatures of the order of $T\lesssim 10$ MeV and chemical potentials $\mu\simeq 900$ MeV. Another phase is the so-called colour-flavour locked phase (CFL) \cite{CFL}. At very high densities the 3 lightest quarks $u,s,d$ may form Cooper pairs that correlate their colour and flavour properties leading to a colour superconducting behaviour. The CFL would break the QCD symmetries and exhibit a very different spectrum and transport properties. No experimental evidences of this phase have been found so far but it is conjectured that this form of matter may exist in the core of neutron stars. All in all the structure of the phase diagram at high baryon densities remains rather uncertain.

\section{Exploring the phase diagram with heavy ion collisions}

\begin{figure}[h!]
\centering
\includegraphics[scale=0.4]{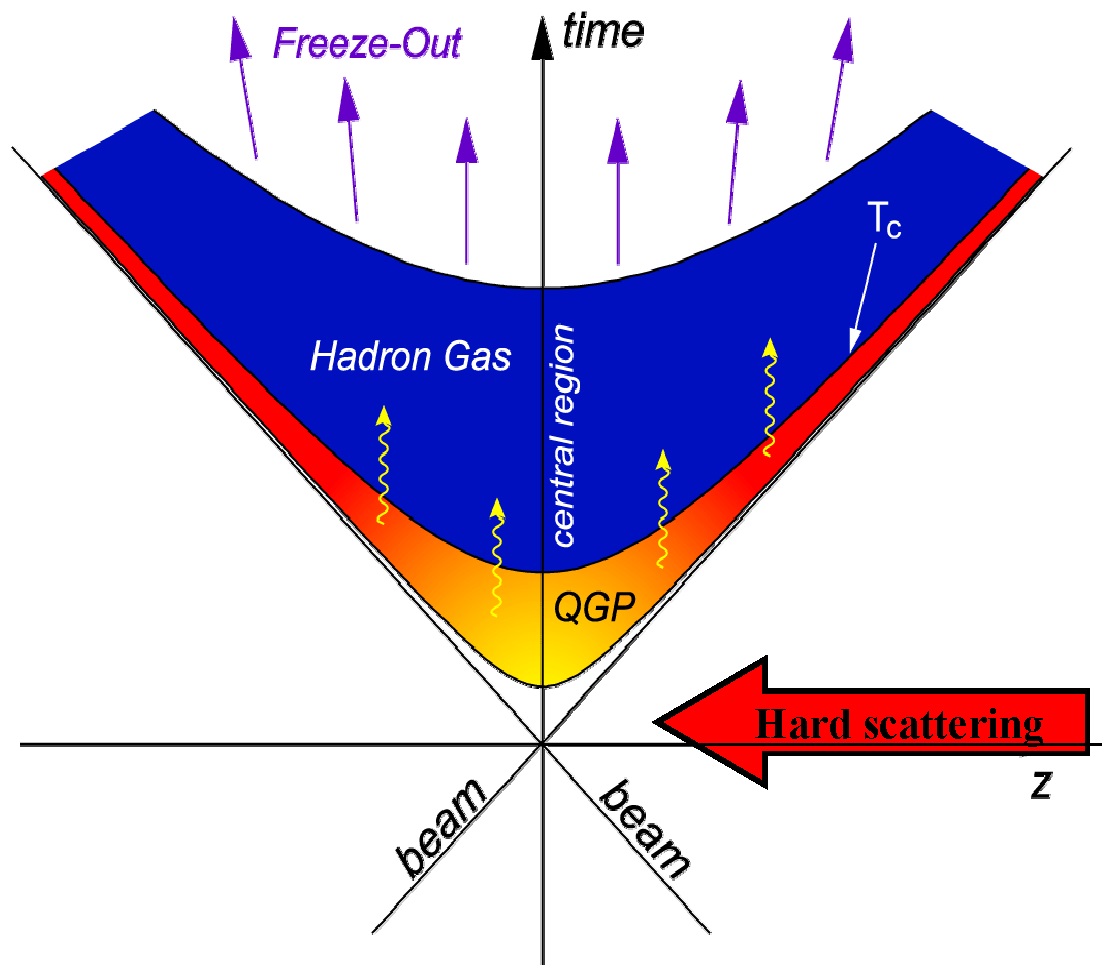}
\caption{Evolution of the nuclear medium created from a HIC depending on time. After the collision the system heats up and reaches the QGP where quarks and gluons become deconfined. Next, a slow cooling process takes place pushing the system to the hadron phase where partons are arranged into baryons and mesons. Finally, when the freezout temperature is met, hadrons do not interact any more.}\label{HIC}
\end{figure}

The dynamics of the nuclear fireball created from a HIC is not trivial. In Fig. \ref{HIC} we present a sketch of a HIC evolving in time and the corresponding phase where the system is found at each moment. At the initial stage, the fireball is compressed for a short period of time between 0.5 and 1 fm. During this time, the system is thought to reach temperatures as high as $\sim$600-700 MeV depending on the model and the experiment (see \cite{LHClastcall} and references therein) and enter the QGP. In the next stage, all the phenomenology associated to a deconfined phase takes place while the system starts to thermalize due to strong interactions among the fireball components. The temperature slowly drops down to $\sim 200$ MeV, a situation where it is still unclear if a mixed phase exists. After an expansion that lasts for 5 to 10 fm, conventional hydrodynamic models and lattice calculations suggest that around $T_c\sim 140-190$ MeV, a phase transition leads the system to the hadronic phase, where quarks and gluons regroup again to form hadrons. The experimental probes from hadron interactions may thus provide a valuable insight into the QGP and the deconfinement phase transition. Finally, when the system is cold enough, say $T\sim 100$ MeV, collisions among hadrons cease leading to the freezout of the fireball components, which survive as the final particle states.

\medskip

At high energies, or equivalently, at small values of Bjorken $x$, gluon densities are very large. Under these circumstances QCD may be described by a many-body theory of weakly coupled partons, the so-called colour Glass Condensate (CGC) \cite{CGC}. Two highly boosted nuclei in the first stages of a HIC can be considered as two pancakes of CGC. The gluons in the nuclear wavefunctions are disordered and evolve very slowly relative to the time scale of a HIC similar to the behaviour of glasses. Their high density is the responsible for the term condensate. This picture of nuclear matter consists in a coherent classical field which decays into nearly on-shell partons and eventually thermalize to form the QGP. This approach allows to study the initial conditions and equilibration in HIC. In fact, the CGC has to be considered as the non-equilibrium stage preceding the QGP. For this reason, this matter is often called Glasma \cite{lappi,Glasma}.

\medskip

\begin{figure}[h!]
\centering
\includegraphics[scale=0.3]{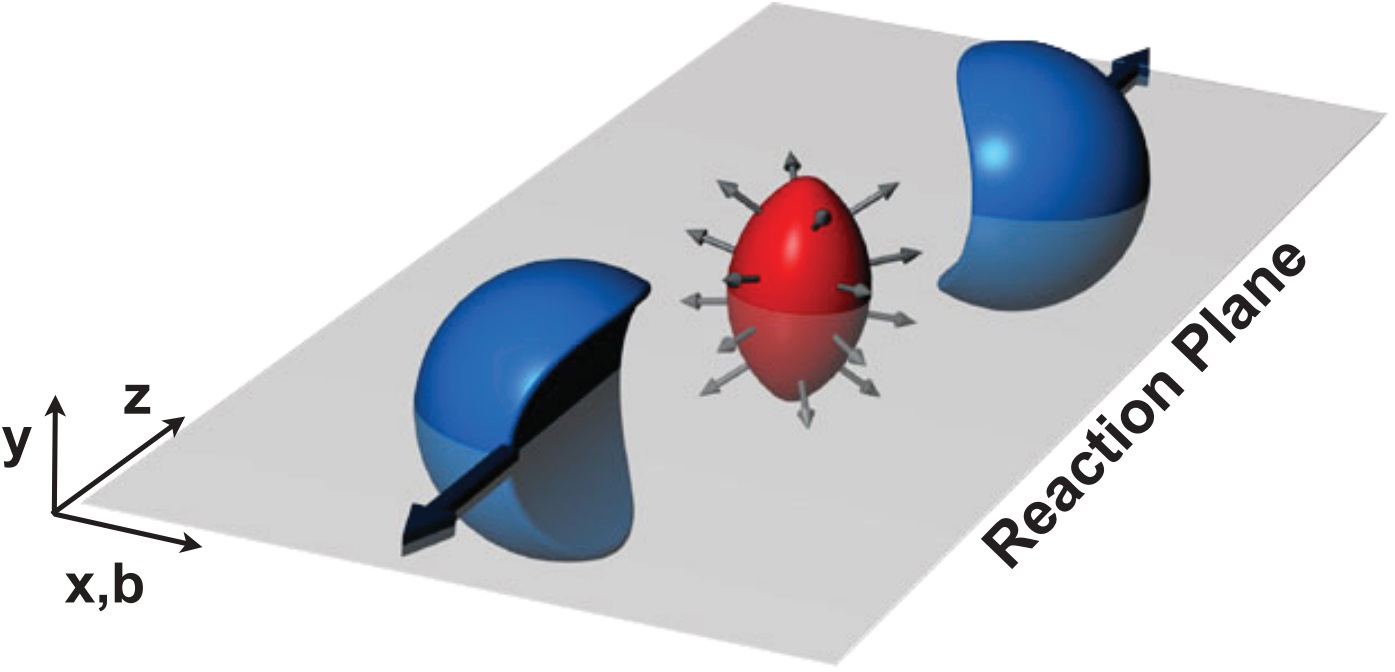}
\caption{In non-central collisions the central region where the two nuclei overlap shows an almond-like shape (graphic extracted from \cite{almond}) leading to an initial anisotropy with respect to the reaction plane. This is translated into anisotropies in particle momentum distributions.}\label{almond}
\end{figure}

One way to analyse the properties of QCD matter in HIC is via the study of the azimuthal anisotropy induced in non-central collisions (for reviews see \cite{flowreview,almond}). Anisotropic collective flow, i.e. anisotropies in particle momentum distributions with respect to the reaction plane, is highly sensitive to the system properties in the first stages of its evolution (see Fig. \ref{almond}). This is due to the initial asymmetries in the geometry of the system that rapidly wipe out while the system cools down. Early probes are normally characterised by weak interactions but the analysis of collective flow corresponds to a unique hadronic observable. The particle azimuthal distribution measured with respect to the reaction plane is conveniently expanded as a Fourier series \cite{flowFourier}. The different harmonics, generically referred to as $n$-th order flow, of this expansion depend on the rapidity, the transverse momentum and the impact parameter of the collision and are measured at different experiments \cite{flowexp}. Many different hydrodynamical calculations have been addressed to provide qualitative descriptions of collective flows (see \cite{casalderrey} and references therein).

\medskip

When the HIC fireball cools down and enters the hadronic phase, quarks and gluons rearrange to form hadrons. Matter becomes colourless again creating a gas of hadrons. Thermal equilibrium within the gas has been argued to be reached in RHIC and LHC unlike other previous machines like SPS. Many different particles may conform this medium but the partition function is dominated by the lightest degrees of freedom. Thus the fireball can be approximated by gas of pions. Other light meson states may be present in small amounts but baryons are certainly too heavy to play a significant role in this context.

\medskip

The dynamical evolution of the hadron gas needs to be well understood as it corresponds to the intermediate step between the QGP and the final states that are measured in experiment detectors. The $\rho$ meson is a key player of the hadron gas thermalization. First, $\rho$ mesons strongly interact with pions and their short mean free path allows for multiple scattering processes leading to a regeneration of these particles from pions, i.e., the effective number of $\rho$ mesons is enhanced with respect to the direct production from the hadronization process. And second, studies based on effective hadronic theories seem to agree that the $\rho$ meson spectral function is severely modified in-medium leading to a strong broadening of the resonance \cite{rapp,renkr,zahed} with only a little mass shift \cite{brown}. 

\medskip

On the one hand, a broadening of the $\rho$ meson spectral function may result from its scattering off the baryons in the dense hadronic medium \cite{rapp}. An extended many-body hadronic model of the $\rho$ in hot meson matter that includes resonance excitations and pion Bose enhancement in the $\rho$ self-energy leads to a broadening of around $\Delta\Gamma_\rho\sim 80$ MeV at a temperature $T=150$ MeV and a pion density $\varrho_\pi\simeq 0.75\varrho_0$ \cite{RappT}, where $\varrho_0$ is the nuclear density. In cold nuclear matter, other investigations have been developed in $\rho N$ resonance models including the in-medium pion cloud \cite{Rappmu} showing a stronger broadening of $\sim300$ MeV at $\varrho_N\simeq \varrho_0$ and a comparable mass shift of $\sim40$ MeV. On the other hand, the 'dropping' vector meson mass scenario in the dense fireball \cite{dropmass} establishes a link between hadron masses and the quark condensate, the order parameter of the chiral symmetry restoration. The $\rho$ mass scales \cite{brown} with the quark condensate and the latter drops due to the high baryon density rather than high temperature. Note that both approaches rely on having a high baryon density. The relative merits of these two different approaches have attracted much debate trying to explain the low invariant mass dilepton excess that has been measured in different experiments \cite{CERES,NA60,phenix}.

\section{Hard probes in heavy ion collisions}

A variety of signals of the strongly interacting nuclear medium and its dynamics have been proposed. Hard probes, i.e. processes involving large momentum exchanges, may explore small distance physics provided that constituents quarks are heavy enough, a regime that can be well described with perturbative QCD leading to a good theoretical control of the subject.

\medskip

An important hard probe is the different heavy quarkonia states and their in-medium modifications \cite{MatsuiSatz}. The presence of QGP screens the attractive force between a quark-antiquark pair. The maximum distance which allows the formation of quarkonium decreases with the medium temperature. The interaction of these states with a hot medium leads to the modification of their properties and to their complete dissociation at sufficiently high temperatures. Gauge/string duality has shown that heavy quarkonium mesons remain bound in the plasma and the fact that their dissociation temperature drops with increasing meson velocity \cite{casalderrey}. Such fracture entails a suppression of quarkonium production in nuclear collisions as compared to their proton-proton equivalent. Most studies of quarkonium have focused on charmonia states such as the $J/\psi$ due to the small $b$-quark production cross section at low energies. Nevertheless, charm quarks are not sufficiently heavy for a reliable theoretical description and thus, their interactions with the QGP and their production mechanism in HIC are somewhat uncertain \cite{quarkonium}. A strong suppression of $J/\psi$ mesons in HIC has been shown \cite{Jpsi,CMSquarkonium} but the suppression pattern is still puzzling and the interpretation of data is not totally clear yet since quarkonium is also suppressed in proton-nucleus collisions \cite{pA}. Recent data on bottomonium suppression through the study of $\Upsilon$ mesons have been published by LHC \cite{CMSquarkonium} and RHIC \cite{RHICUpsilon} allowing a better theoretical understanding of this channel.

\medskip

The high energy jets produced in a collision are severely affected by a strongly interacting medium. When a high-momentum parton is produced in the initial stage of a nucleus-nucleus collision, it will multiply interact with the medium constituents before its hadronization. The energy of such partons is reduced due to collisional energy loss \cite{perkins} and medium-induced gluon radiation \cite{gross}. The energy of these partons is stored in the QGP and propagates as a collective excitation or conical flow \cite{conicflow}. This effect leads to peaks in particle correlations that have been experimentally observed \cite{conicflowexp}. The effect of jet quenching in the QGP is the main motivation for studying jets as well as high-momentum particle spectra and particle correlations in HIC. The study of high energy jets is one of the best hard probes to investigate the nuclear medium as their production occurs at very high energy scales where perturbative QCD guarantees a good theoretical description. Similarly, many of the properties of jets in vacuum are also controlled by physics at high energy scales and are thus well understood theoretically. The observed deviations of different properties in HIC and in vacuum must be due to the interaction of the different jet components with the nuclear medium. Accurate jet reconstruction allows measurements of the jet fragmentation functions and provide insight on the properties of the hot and dense QGP. Evidences of parton energy loss were first observed at RHIC from the suppression of high $p_T$ particles studying the nuclear modification factor \cite{highptsup1} and the suppression of back-to-back correlations \cite{highptsup2}. LHC also observed jet quenching \cite{highptsup3} and SPS partially did \cite{SPSjet}.

\medskip

Besides hard probes, many other signatures of the medium created in HIC have been discussed. Large topological fluctuations in the QCD vacuum may induce the breaking of parity in nucleus-nucleus collisions leading to the Chiral Magnetic Effect \cite{kharzeev}, which predicts a charge separation across the reaction plane. In addition, photons and dileptons (same-flavour lepton and antilepton pairs) are considered powerful probes due to their negligible interactions with the final state hadrons. We will explain these two aspects in detail in the following sections.

\section{The Chiral Magnetic Effect}\label{SecCME}

\begin{figure}[h!]
\centering
\includegraphics[scale=0.3]{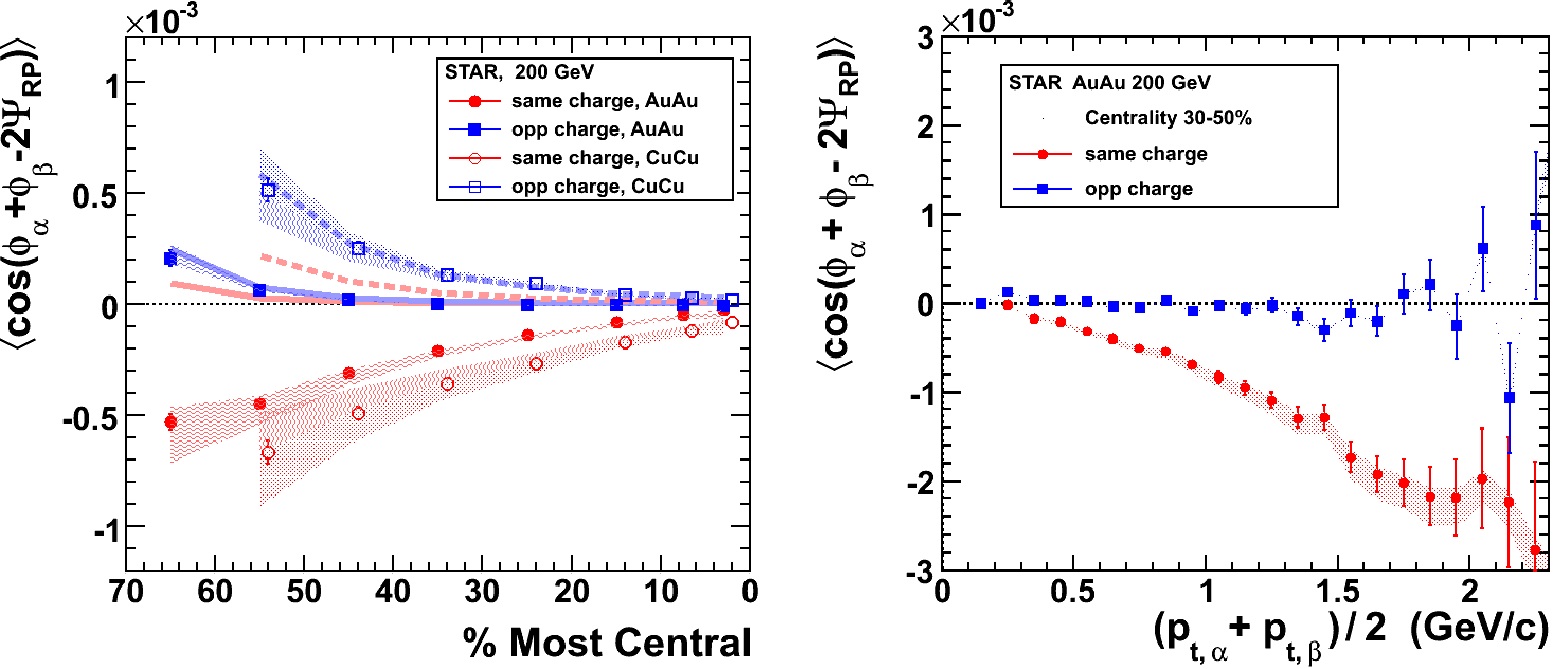}
\caption{The observable that reflects same- and opposite- charge azimuth correlations is displayed depending on the centrality of the collision for Au+Au collisions and Cu+Cu collisions (left) and the semi-sum of transverse momenta only for Au+Au collisions (right) extracted from \cite{star}. In the left panel, the thick solid (Au+Au) and dashed (Cu+Cu) lines represent theoretical calculations of 3-particle correlations.}\label{star}
\end{figure}

A different effect that has received a lot of attention is the charge separation detected in nucleus-nucleus collisions at STAR \cite{star}. The measurement of a non-trivial three-particle observable reflects that event-by-event same- and opposite- charge azimuth correlations lead to a small separation of charges across the reaction plane in a HIC. This observable has the advantage of suppressing the background correlations unrelated to the reaction plane. The measurement of this effect has been assumed to be an experimental confirmation of the Chiral Magnetic Effect (CME) \cite{kharzeev}, which assumes that local parity-odd domains may appear in HIC. We will discuss this effect in more detail in Chapter \ref{Chfase}. The pair correlations of same and opposite charge due to the CME are expected to be similar in magnitude and opposite in sign. This could be modified by the medium resulting in the dilution of the correlations between opposite sign particles \cite{kharzeev}. A clear difference in the correlation strength between the same and opposite charge particle combinations seems to be observed (see Fig. \ref{star}). The simulation of opposite-charge correlations of three particle correlations is similar to the measured signal but the same-charge signal is much larger and of opposite sign compared to the predicted one. Such phenomenon has also been observed at ALICE in LHC \cite{ALICECME}.

The centrality dependence shown in the left panel of Fig. \ref{star} agrees with the charge separation relative to the reaction plane. On the other hand, the magnitude of the effect for same charge particles increases with average pair $p_T$ both at RHIC and LHC. This observation is in contradiction with the initial expectation \cite{kharzeev}, which predicts the dominance of low $p_T$ particles. In any case, no definite conclusions can be stated due to the lack of realistic model calculations for the dependences on the centrality and pair momentum.

\medskip

Different magnetic field dynamics are expected to be induced in LHC and RHIC, a fact that should be quantified. The parity-breaking signal indicates the presence of a charge separation whose magnitude shrinks with diminishing beam energy, suggesting that the QGP volume shrinks with lower energy collisions, leading to less charge separation. It has been argued that at higher energies the strength of the CME should decrease since the magnetic field decays much faster \cite{CME-B}.

\medskip

The analysis of different charge azimuthal correlations in order to search for charge separations in a deconfined state of matter may hint to the possibility of probing new effects in hot and dense QCD. In particular, the appearance of local large topological fluctuations inducing metastable $P$- and $CP$-odd domains could be detected together with the electromagnetic current induced by large angular momentum or magnetic field in non-central HIC. In addition to the presented study, analyses of higher order harmonic correlations are planned and may help to provide a deeper insight into the charge dependent correlations observed at RHIC and LHC.

\section{The problem of low-mass dileptons}\label{dilepexcess}

Dileptons are considered the most useful electromagnetic probe to unravel the main features of the hot medium created in HIC. They are produced continuously carrying information of the entire evolution of the system and due to the existence of an additional variable, their invariant mass $M_{\ell\ell}$, dileptons possess a better signal to background ratio compared to photons. Depending on the stage of the collision, dileptons are emitted from different sources. Before the collision dileptons are produced through coherent bremsstrahlung due to the electromagnetic repulsion among the two nuclei \cite{Bremsstr-dilep} but this contribution is negligible compared to other processes that follow \cite{late-dilep}. During the heating of the nuclear system and before equilibrium is reached, Drell-Yan annihilation provides the main source of dileptons that may be dominant above $M_{\ell\ell}\simeq 3$ GeV \cite{DrellYan}. When the system enters the QGP and thermalizes, dileptons are mainly created from perturbative quark-antiquark scattering. Next, the plasma evolves to a hadron gas where dileptons are emitted from pion and kaon annihilation (among other channels) \cite{piK-annihil}. The presence of light vector meson resonances like $\rho$, $\omega$ and $\phi$ are responsible for an enhancement of such processes due to their connection between the initial two-meson states and the final dileptons \cite{vectmesons-dilep}. The dilepton invariant mass corresponds to the vector meson mass distribution at the precise moment of the decay allowing to obtain a deep insight into the hot medium. At the final stages of the collision when freezout is reached, only long-lived hadronic states (mainly mesons) like $\pi^0$, $\eta$ and $\omega$ are able to produce lepton pairs through their Dalitz decay as they do not interact among themselves \cite{KW}. These processes correspond to an initial state decaying into a pair of particles, one of them being a virtual photon that subsequently leads to a lepton pair. Due to the kinematics of these decays, all the mentioned meson sources are relevant for the production of dileptons with an invariant mass below 1 GeV.

\medskip

Due to the relevance of dileptons, many studies have been undertaken in order to unravel the main properties of the nuclear medium. Perhaps one of the most important and oldest analysis consists in measuring the dilepton invariant mass distribution, which has been analysed by different experimental collaborations during the last decades. In proton-proton collisions the shape of the detected distribution can be well reproduced by the so-called hadronic 'cocktail', the sum of several final-state hadron decays with known abundances. In fact, $pp$ data is normally used to fit the weight of the different hadrons in order to be used for other predictions. And while proton-nucleus data can be well understood from $pp$ results, central nucleus-nucleus collisions have presented severe discrepancies with respect to the expected theoretical contribution \cite{CERES,NA60,phenix}, in particular, a strong enhancement of dileptons with invariant mass below 1 GeV. This disagreement reflects that in-medium effects have to be considered to explain the abnormal $AA$ results.

\medskip

The anomalous dilepton yield from nucleus-nucleus collisions has been a long-standing problem that has been under investigation by many different groups, both theoretical and experimental (for a review see \cite{reviews}). No dilepton measurements were performed at the AGS accelerator, where high baryon densities were expected. The DLS collaboration at BEVALAC released the first dielectron results for fixed target Ca+Ca, C+C, He+Ca and Nb+Nb reactions at a beam energy of 1 GeV per nucleon. The first generation of DLS data \cite{DLSv1} were consistent with the results from transport model calculations \cite{transpmod} including the conventional dileptons sources as $pn$ bremsstrahlung, light meson Dalitz decays, direct decays of vector mesons and pion-pion annihilation, without incorporating any medium effects. However, an updated version of the DLS measurement on the full data sample and an improved analysis showed an enhancement of almost one order of magnitude in the cross section compared to the previous results \cite{DLSv2}. This discrepancy has persisted for years and many attempts have been addressed to disentangle the newest DLS data. In fact, all in-medium scenarios that successfully have explained the dilepton enhancement at higher energies in more powerful accelerators failed to describe the new DLS dilepton data, a problem that was soon known as the 'DLS puzzle'. 

\medskip

The DLS puzzle motivated the HADES experiment at SIS, a new generation experiment that was designed to study C+C, Ni+Ni and Au+Au collisions at beam energies up to 8A GeV with a fixed target. The dielectron excess observed at DLS was confirmed by HADES data \cite{HADES} running C+C collisions at 1A and 2A GeV. Recently, new approaches have been trying to tackle this problem. On the one hand, the effect was partially interpreted as being due to enhanced $\eta$ meson production in proton-neutron scattering \cite{HADES-eta}. On the other hand, an enhanced bremsstrahlung contribution in line with one boson exchange calculations, led to a rather successful agreement with the published DLS and HADES data on C+C collisions \cite{BratkCass}. However, the same calculations still underestimate the dilepton production in Ca+Ca reactions and thus, the issue still remains controversial.

\medskip

At SPS the low-mass dilepton excess in central nucleus-nucleus collisions was observed again. Two different beams were used: a sulphur beam at 200 GeV/nucleon and a lead one at 158 GeV/nucleon. CERES collaboration reported a dielectron excess \cite{CERES} where fixed target S+Au and Pb+Au reactions were analysed. In addition, they showed that the enhancement, particularly important in the mass region around 400-600 MeV, covered a wide range in transverse momentum, but was largest at low transverse momentum. Similar abnormal results for low-mass dimuons were observed at HELIOS/3 \cite{HELIOS} with S+W reactions. The results from S+Cu and S+U reactions studied by the NA38 collaboration together with the Pb+Pb collisions at NA50 revealed no significant excess of dimuons \cite{NA38/50}, a fact that could most probably related to a large $M_T$-cut applied in their analysis. These independent measurements of the low-mass dilepton excess at the SPS triggered a huge theoretical research also motivated by a possible connection to the breaking of chiral symmetry. The importance of baryon density suggested by in-medium effects on the $\rho$ meson encouraged the CERES collaboration to measure low-mass dielectron in Pb+Au collisions at the lower energy of 40A GeV where a higher baryon density and a lower temperature were expected \cite{CERES40AGEV}. Many different approaches were addressed to describe the dilepton excess at SPS, where it is not clear that thermal equilibrium was reached. Finally, the data could apparently be explained by means of in-medium modifications of the $\rho$ meson due to its strong hadronic coupling and other less conventional assumptions \cite{rapp,dropmass,HatsudaLee,brown,BratkCass}.

\medskip

The dilepton physics program of the NA38/NA50, HELIOS/3 and CERES collaboration was continued by the NA60 experiment also at SPS. Nucleus-nucleus collisions were investigated via dimuon production in In+In reactions at 158 GeV/nucleon with a fixed target. The experiment exhibited a clear excess of dimuons at low energies that increases with centrality and low pair $p_T$ \cite{scomparin-arnaldi2}. The most successful models at that time were compared with the high quality data of NA60, which could isolate the $\rho$ channel from the rest of hadronic sources. In particular, the dropping mass and broadening scenarios show similar results for masses below the $\omega$ resonance peak due to the limited precision of the data, but between the $\omega$ and the $\phi$ resonance peaks the dropping mass case seems to be ruled out \cite{NA60}. The broadening mass scenario \cite{rapp} is reinforced by the NA60 data, together with other recent calculations that are claimed to reproduce the dilepton excess \cite{renkr,zahed}. Yet, the still considerable experimental uncertainties may allow other effects to be accommodated.

\medskip

\begin{figure}[h!]
\centering
\includegraphics[scale=0.28]{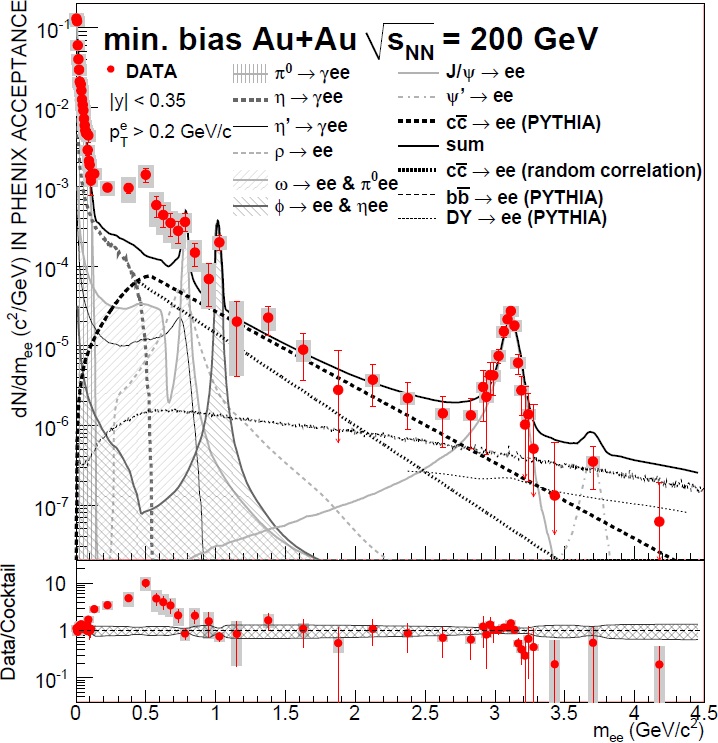}
\caption{Dielectron mass spectrum in minimum-bias Au+Au collisions at PHENIX compared to the cocktail of known hadronic sources extracted from \cite{phenix}. The ratio of data to the cocktail is shown in the bottom panel. A clear enhancement of almost one order of magnitude is observed at low invariant masses.}\label{phenix}
\end{figure}

RHIC is the first accelerator with two highly energetic beams running at $\sqrt{s_{NN}}=200$ GeV/nucleon, where Au+Au collisions are currently carried out and analysed by PHENIX and STAR experiments. PHENIX has measured a serious dielectron excess at low energies that is present at all $p_T$ but it is more pronounced at low pair $p_T$ \cite{phenix}. A very strong centrality dependence is observed for the low-mass continuum. The enhancement appears in central collisions where it reaches a factor of almost 8 (see Fig. \ref{phenix}). PHENIX results appear different from those observed at the SPS: while at SPS the excess was closer to the $\rho$ meson resonance peak, at PHENIX it spreads towards lower invariant masses. In addition, the integrated yield increases faster with the centrality of the collisions than the number of participating nucleons. The contribution of correlated pairs from the semi-leptonic decays of open charm mesons, negligible at SPS energies, is an important source that has to be included in the hadronic cocktail at RHIC energies. In contrast to previous experiments, PHENIX low-mass dilepton results have not been correctly reproduced by any of the successful models that apparently explained the excess before.

\medskip

The STAR collaboration also reported a dielectron enhancement below the $\omega$ peak \cite{STARdilept} but it was observed to be significantly lower than what was measured by PHENIX. Due to this moderate excess, theoretical calculations including an in-medium broadened $\rho$ meson can possibly explain the STAR data. The centrality and $p_T$ dependence is also comparable with expectations from model calculations.

\medskip

Finally, low invariant mass dilepton measurements have not yet been published by LHC collaborations.

\medskip

A similar dilepton excess with invariant mass between 1 and 3 GeV has been also measured at SPS and RHIC. All experiments at SPS reported an excess with respect to the expected yield from the two main contributions in this mass region, Drell-Yan and semi-leptonic charm decay \cite{HELIOS,NA38/50-IMR}. Several explanations were investigated based in the phenomenological hypothesis of an enhanced charm production, such as secondary Drell-Yan emission, final-state rescatterings that broaden the $m_T$ distribution of charmed mesons enhancing the yield, and all sources of lepton pairs from secondary meson interactions. NA60 concluded that the excess of dimuons in the intermediate mass region corresponds to the thermal radiation emitted in the early stages of the collision \cite{NA602}. However, PHENIX showed that the inclusive production seems to agree with the simulated open charm contribution \cite{phenix}.

\medskip

We are living an era with plenty of ongoing experiments that may obtain relevant information about QCD matter. Many models have tried to solve or explain different problems with the addition of new parameters or observables that need to be measured. Thanks to the current experimental work, the measurement of such parameters may help to confirm or rule out all these models. In addition, one cannot exclude the fact that new phenomena could be detected. This scenario constitutes an encouraging source of motivation for both theoretical and experimental research.

\section{Scope and overview}

In the context of HIC, new effects can be studied in QCD due to high temperatures and densities. In particular, in this thesis we will explore the possibility that the QCD vacuum breaks parity in a finite volume. The Vafa-Witten theorem \cite{vw} forbids the spontaneous breaking of discrete symmetries in gauge theories. However, this theorem does not apply when one deals with a dense system. The high temperatures and densities reached in the strongly interacting medium induced in HIC lead to a realistic scenario where the violation of parity in QCD could be perceptible.

\medskip

The violation of parity in QCD has been already studied by different collaborations and approaches \cite{kharzeev,anesp} that we will discuss. Here we are interested in describing the thermodynamic conditions that could lead to this phenomenon in HIC as well as the phenomenology associated to it. If this effect takes place, the properties of in-medium hadrons would be severely distorted leading to unexpected modifications of the distributions of final state particles. We will show in detail how scalar and vector mesons are influenced by a parity-odd medium and in addition, we will search for possible signatures that could help to detect this effect at RHIC or LHC.

\medskip

In Chapter \ref{Chfase} we will focus on the possible existence of a new phase characterised by the violation of parity in QCD. We will analyse two different approaches to explore this effect. The first of them corresponds to the study of QCD at finite density by means of effective lagrangian techniques coupled to a non-trivial baryon chemical potential. A generalised linear sigma model retaining the two lightest multiplets of scalar and pseudoscalar fields of $SU(2)$ and satisfying the QCD symmetries in vacuum shows that for a given range of $\mu\neq0$ a non-vanishing pseudoscalar condensate may arise leading to a new phase of QCD \cite{anesp}. The second possibility is related to the presence of metastable $P-$ and $CP-$ odd bubbles in hot matter due to large topological fluctuations in a finite volume. In this context a large magnetic field or angular momentum could induce an electric field perpendicular to the reaction plane and hence an electric dipole moment \cite{kharzeev}. This is the Chiral Magnetic Effect that we already presented in Section \ref{SecCME} and has been claimed to be detected in peripheral HIC at STAR \cite{star}. Following the lines of the latter approach we will investigate the violation of parity in QCD as a consequence of large topological fluctuations corresponding to a non-equilibrium effect that may be driven by an external axial chemical potential $\mu_5$. 

\medskip

In Chapter \ref{ChNJL} we will implement the Nambu--Jona-Lasinio model with a vector and an axial chemical potentials at zero temperature in order to investigate the phase diagram of the model searching for possible phases with pseudoscalar condensation leading to parity violation. We will see that a non-zero $\mu_5$ severely modifies the known properties of the NJL model while the presence of $\mu$ does not affect much our considerations. Provided that some numerical relations between the parameters of the model hold, spontaneous breaking of parity is found to be stable. In particular, we will find a phase with a non-trivial isospin singlet pseudoscalar condensate as well as a scalar condensate down to a critical value of the axial chemical potential. However, NJL with such characteristic parameters has nothing to do with QCD since the masses of the known states in vacuum are not compatible with the phenomenology. Despite being unable to draw definite conclusions for QCD, the values of $\mu_5$ leading to a thermodynamically stable phase with parity violation suggest the range of the axial chemical potential where a parity-breaking phase has to be searched in HIC experiments.

\medskip

After investigating the possible values of $\mu_5$ that could lead parity violation, we address a detailed study of the in-medium modifications of mesons. First of all, Chapter \ref{ChSmodel} is devoted to the study of scalar and pseudoscalar mesons in a medium characterised by a non-trivial axial chemical potential. We will implement a generalised sigma model including the lightest scalar and pseudoscalar mesons in the $SU(3)$ representation in order to investigate the mixing among states of different parity and same flavour induced by a parity-odd medium, where in fact, the distinction of scalar and pseudoscalar states is meaningless. The axial chemical potential will be introduced as an external spurious axial-vector field through a covariant derivative. In-medium mesons will exhibit energy-dependent properties as spectrum and widths. We will see that the lightest degrees of freedom in each isospin channel become tachyonic states at high energies. In addition, the pion gas that constitutes the heavy ion fireball in the hadronic phase could instead be a gas of distorted pions due to the breaking of parity. We will thus compute the widths of all these states into distorted pions so as to search for possible broad resonances that could suffer from a regeneration process as the $\rho$ meson does. In particular, the distorted $\eta$ meson seems to be in thermal equilibrium inside the 'pion' gas due to its remarkably large width, unlike in vacuum.

\medskip

Vector mesons are treated in Chapter \ref{ChVMD}. We will introduce the parity-breaking effect through the Chern-Simons term since the mixing of vector mesons with different parity is rather uncertain. By means of the Vector Meson Dominance model (VMD), we will introduce the lightest states in the $SU(3)$ representation. However, as we will argue in the next chapter, the $\phi$ meson decouples from the rest of the system and it can be simplified to $SU(2)$. A mixing will be found among photons and $\rho$ and $\omega$ mesons due to the VMD mechanism itself and the parity-violating effect. After resolving the mixed system we will find a distorted dispersion relation for vector mesons while photons are not affected by parity-odd effects. As in the case of spin-zero mesons, vector mesons will show a frame-dependent effective mass and a tachyonic behaviour for one of their polarizations. The in-medium modifications of vector mesons can be tested through their decays into dileptons. The distorted shape of the $\rho$ and $\omega$ spectral functions predicts an enhancement of the dilepton production around their resonance peak. We will show that this enhancement could help to partially explain the low-mass dilepton excess that has been presented in Section \ref{dilepexcess} for the NA60/PHENIX/STAR experiments. But this signal can be rather tricky to detect. The distorted shape of the $\rho$ and $\omega$ spectral functions reveal a polarization asymmetry around the vacuum peak. Hence, we will also describe how the parity-breaking effect may be detected in HIC by means of a restriction of the angular coverage of the final lepton states. Two suitable angular variables will be presented in order to search for polarization dependencies and we will explain how a combined analysis of one of these variables together with the dilepton invariant mass could be an important signature to detect the possible violation of parity in HIC.

\medskip

Finally, in Chapter \ref{Chconcl} we present the main conclusions and future perspectives of this work.
\chapter{A new phase of QCD with parity violation}\label{Chfase}
\graphicspath{{2-fase/figures/}{figures/}}

During the last years the behaviour of baryonic matter under extreme conditions has received a lot of attention \cite{Jacobs:2007dw,Blaizot:2011xf,nora}. Many different models and approximations have appeared trying to obtain a deeper insight into the dynamics of heavy ion collisions. A particular investigation line has been recently studied triggering a significant spark among the scientific community: the possibility that QCD breaks parity at finite temperature and/or density. Indeed, one of the very few non-perturbative analytic results of QCD is the Vafa-Witten theorem \cite{vw}. According to it, vector-like global symmetries such as parity, charge conjugation, isospin and baryon number in vector-like gauge theories like QCD cannot be spontaneously broken while the $\theta$ angle is zero (see Eq. \eqref{pbreakoper} and subsequent discussion). However this theorem does not apply to dense QCD matter where the partition function is not any more positive definite due to the presence of a highly non-trivial fermion determinant. In addition, out-of-equilibrium symmetry-breaking effects driven by finite temperatures are not forbidden by the Vafa-Witten theorem. Lorentz--non-invariant $P$-odd operators are allowed to have non-zero expectation values at finite density $\mu\neq0$ \cite{vwmu} and finite temperature if the system is out of equilibrium \cite{vwT}. The possibility that in extreme conditions of temperature and baryon density such discrete symmetries might be broken in QCD has been discussed recently by different collaborations \cite{kharzeev,anesp,zhitn,avep}, whose different approaches we will describe with some detail in this section.

\medskip

Invariance under parity or charge conjugation are some of the characteristic footprints of strong interactions. However, the QCD Lagrangian may exhibit some terms explicitly break the $CP$ symmetry. The corresponding operators with dimension 4 (or less) read
\begin{equation}\label{pbreakoper}
\mathcal{O}_g=\frac12\varepsilon^{\mu\nu\rho\sigma}\text{Tr}\left (G_{\mu\nu}G_{\rho\sigma}\right ) \equiv \text{Tr}\left (G^{\mu\nu}\widetilde G_{\mu\nu}\right ), \qquad \mathcal{O}_q=i\bar q \hat m_q\gamma_5 q.
\end{equation}
The dimension-4 operator $\mathcal{O}_g$ normally appears coupled to a constant \emph{angle} $\theta$, the so-called $\theta$-term. If one introduces the QCD Lagrangian with complex quark masses including the quark operator $\mathcal{O}_q$, a $U(1)_A$ transformation of the quark fields $q_f\to\exp(i\gamma_5\theta_f/2)q_f$ can remove the imaginary part of quark masses \cite{coleman} with an appropriate choice of the coefficients $\theta_f=\arctan(\hat m_q/m)$, following the notation introduced in Eq. \eqref{QCDlag}. However, eliminating the term $\mathcal{O}_q$ from the Lagrangian involves the replacement $\theta\to\theta+\sum_f\theta_f$ due to the partial conservation of the axial current, which is explained below (see Eq. \eqref{pcac}). Therefore, as the quark operator can be rotated away and reabsorbed by the gluon operator $\mathcal{O}_g$, one usually forgets about the former and focuses in the gluon one. Experimental measurements \cite{thetaexp} lead to a a strong bound $\theta\lesssim 10^{-9}$. The particularly small value of this parameter represents a case of fine-tuning that is known as the \emph{strong $CP$ problem}. In the next section more details will be given concerning the $\theta$-term.

\medskip

Several solutions have been addressed to solve the strong $CP$ problem. The most accepted one consists in introducing a new global chiral $U(1)_{PQ}$ symmetry whose spontaneous breaking produces a Nambu-Goldstone (pseudoscalar) boson called axion \cite{PQ}. The axion interacts with the gluon fields through a $\theta$-like term that explicitly breaks the axial $U(1)_{PQ}$ symmetry. The minimum of the axion potential provides a relation where the axion condensate cancels the $\theta$-term. The resulting Lagrangian is invariant under $CP$ thus explaining the small bound on the effective $\theta$ parameter. Nowadays the axion is still considered as a viable solution of the strong $CP$ problem, and one of the plausible dark matter candidates. Several experimental groups are currently searching for axions (see \cite{axionexp} and references therein). While strong interactions preserve $P$ and $CP$ invariances at present time, this scenario suggests that in the early universe these symmetries could have been substantially violated leading to an asymmetry between matter and antimatter. This imbalance resulted in a substantial amount of residual matter that forms the present universe.

\medskip

We will now explore the possibility that strong interactions do not respect parity after all. Two different approaches have been carried out in order to investigate the possible violation of parity in QCD. On the one hand, analytical studies with effective theories suggest that parity may be broken in dense systems. A generalised sigma model containing the two lightest $SU(2)$ multiplets of scalar and pseudoscalar fields reveals that a meson lagrangian realizing all the QCD properties at low energies may lead to a phase where parity is spontaneously broken due to a neutral isotriplet pseudoscalar condensation \cite{anesp,anesp2}. It seems that the necessary conditions for this parity-breaking phase may be satisfied for a definite window of baryonic chemical potentials. On the other hand, it has been proposed that $P-$ and $CP-$ odd bubbles may appear in a finite volume due to large topological fluctuations \cite{kharzeev,zhitn} in a hot (rather than dense) medium. The presence of a large magnetic field or angular momentum may induce an electric field perpendicular to the reaction plane through the $\theta$-term and correspondingly, a net electric dipole moment. This is the Chiral Magnetic Effect that we already introduced in the previous chapter. This scenario can be regarded as a pseudo-equilibrium effect showing that a charge separation should be present in peripheral HIC. 

\section{Spontaneous parity breaking in dense systems}\label{SPB}

The simplest hadronic effective theory is the linear sigma model. Spontaneous chiral symmetry breaking (CSB) emerges due to a non-zero value for $\langle\sigma\rangle \sim \langle \bar q q\rangle /\Lambda^2$, $\Lambda\simeq 4\pi f_\pi$ with $f_\pi$ being the weak pion decay coupling constant. A generalised model may be built containing two $SU(2)$ multiplets of scalar/pseudoscalar fields that include the two lowest lying resonances in each channel. The presence of two multiplets constitutes the minimal model having the possibility of describing spontaneous parity breaking (SPB) \cite{anesp}. One should think of these two chiral multiplets as representing the two lowest-lying radial states for a given $J^{PC}$
\begin{equation}
H_j = \sigma_j {\bf I} + i \hat\pi_j, \quad j = 1,2,\text{ with }\hat\pi_j \equiv \pi^a_j \tau^a,
\end{equation}
where $\tau^a$ are Pauli matrices. This kind of parameterization preserves the parities of $\sigma_2$ and $\pi_2$ to be even and odd respectively (in the absence of SPB) and realizes manifestly the masslessness of Goldstone bosons. Current algebra techniques indicate that in order to relate this model to QCD one has to choose a real condensate for the scalar density, with its sign opposite to current quark masses, and avoid any parity breaking due to a v.e.v. of the pseudoscalar density. The introduction of a chemical potential does not change the phase of the condensate and therefore does not generate any parity breaking. This is just fine because in normal conditions parity breaking does not take place in QCD. The general effective potential of this model is given by
\begin{align}\label{effpot1}
\nonumber V_{\text{eff}}&=& \frac12 \text{Tr}\Bigg \{- \sum_{j,k=1}^2 H^\dagger_j \Delta_{jk} H_k +\lambda_1 (H^\dagger_1 H_1)^2 + \lambda_2 (H^\dagger_2 H_2)^2+\lambda_3 H^\dagger_1 H_1 H^\dagger_2 H_2\\
\nonumber && +\frac12 \lambda_4 (H^\dagger_1 H_2 H^\dagger_1 H_2 + H^\dagger_2 H_1 H^\dagger_2 H_1) +\frac12 \lambda_5 (H^\dagger_1 H_2 + H^\dagger_2 H_1) H^\dagger_1 H_1 \\
&&+\frac12 \lambda_6 (H^\dagger_1 H_2 + H^\dagger_2 H_1) H^\dagger_2 H_2 \Bigg \}+ {\cal O}\left (\frac{|H|^6}{\Lambda^2}\right )
\end{align}
and contains 9 real constants. The neglected terms are suppressed by powers of $\langle H\rangle^2/\Lambda^2$. Indeed, this is a good approximation if one assumes the v.e.v. of $H_j$ to be of the order of the constituent quark mass $\simeq 300$ MeV and the CSB scale $\Lambda\simeq1.2$ GeV.

\medskip

In order to guess the order of magnitude of the couplings, it may be useful to consider a particular model. In particular, a quasi-local quark model \cite{QQM} the general Lagrangian is given by
\begin{equation}
\mathcal L_{\text{QQM}}=\bar q i\partial\!\!\!/ q+\sum_{k,l=1}^2 a_{kl}\left [\bar q f_k(s) q\bar qf_l(s) q-\bar qf_k(s)\tau^a\gamma_5q\bar q f_l(s)\tau^a\gamma_5 q\right ],
\end{equation}
where $a_{kl}$ is a symmetric matrix of real coupling constants and $f_k(s)$ are a set of polinomial form factors depending on $s\equiv-\partial^2/\Lambda^2$, which specify the quasilocal interaction in momentum space. The polynomial form factors are usually taken to be orthogonal as the results do not depend on a concrete choice in the large-log approximation. For a convenient selection $f_1(s)=2-3s$ and $f_2(s)=-\sqrt 3s$, the values of the couplings $\lambda_i$ in \eqref{effpot1} are fixed for $i=2,...,6$: $\lambda_2=\frac{9N_c}{32\pi^2}$, $\lambda_3=\frac{3N_c}{8\pi^2}$, $\lambda_4=\frac{3N_c}{16\pi^2}$, $\lambda_5=-\frac{5\sqrt 3N_c}{8\pi^2}$ and $\lambda_6=\frac{\sqrt 3N_c}{8\pi^2}$.

\medskip

A further simplification may be performed by a (real) rotation the $H_j$ fields when the chemical potential and temperature are zero so that it is sufficient to select $\Delta_{ij}=\Delta\delta_{ij}$ and $\lambda_5=0$. The chiral multiplet $H_1$ is taken to be coupled to the quark fields and the global invariance of the model allows the parametrization
\begin{equation}\label{chipar}
H_1 (x) = \sigma_1 (x) \xi^2(x)= \sigma_1 (x) \exp\left(i\frac{\pi_1^a (x) \tau_a}{F_0}\right) ; \quad H_2 (x) = \xi(x)\Big(\sigma_2 (x) +i\hat\pi_2 (x)\Big)\xi (x).
\end{equation}
One could add some more terms to the potential \eqref{effpot1} that manifestly break parity but in this representation of the doublets, these operators identically vanish. There are also two more operators but they are redundant as they can be expressed as a linear combination of operators with constants $\lambda_3$ and $\lambda_4$.

\medskip

\begin{figure}[h!]
\centering
\includegraphics[scale=0.195]{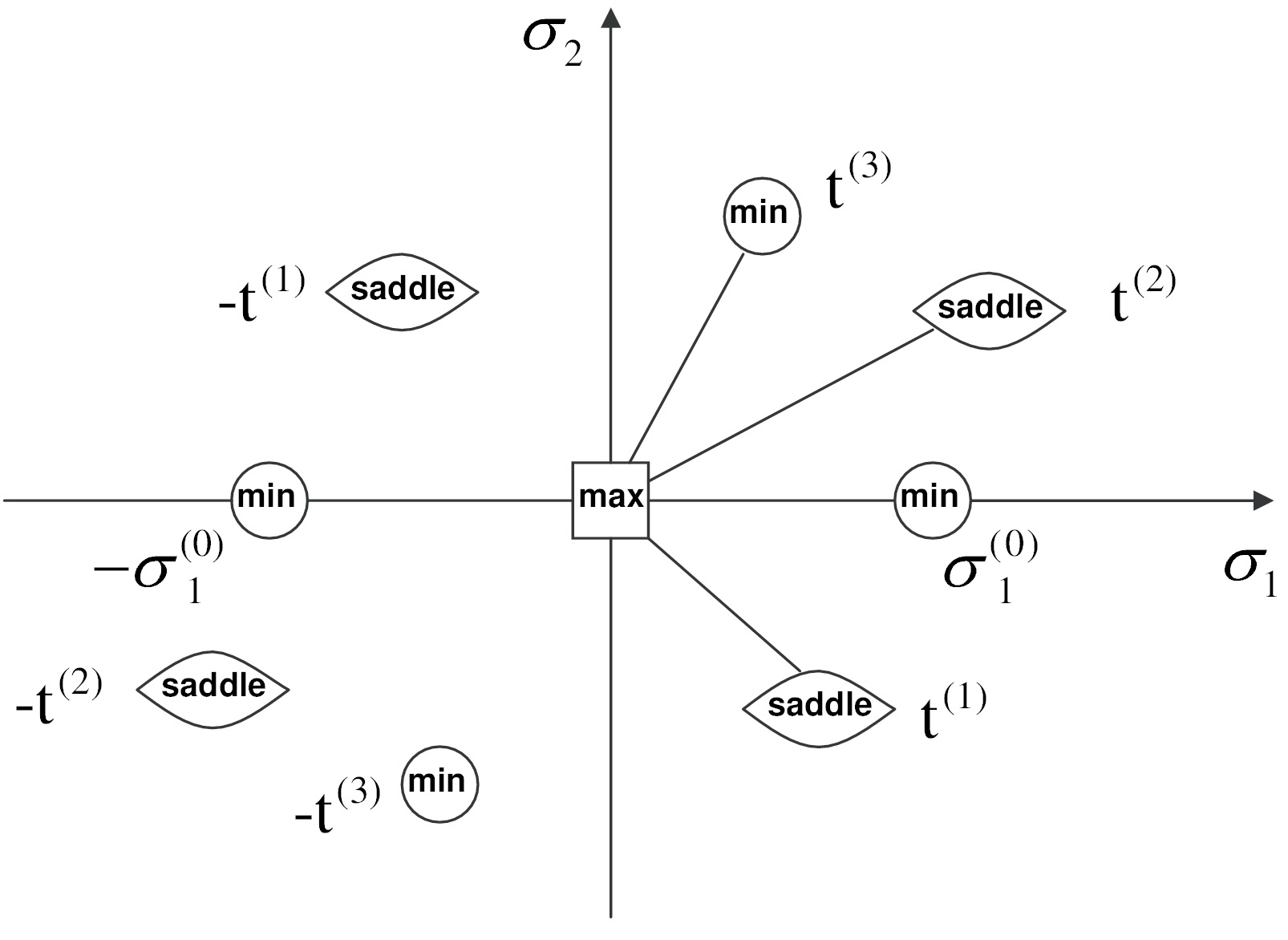}
\caption{Extrema of effective potential in the reduction basis \cite{anesp2}: the maximum is placed in the square, four minima are located in the circles and the corresponding four saddle points are depicted by the lentils. Which one corresponds to the true minimum depends on the actual value of the phenomenological constants.}\label{landsc_extrema}
\end{figure}

The potential \eqref{effpot1} may exhibit several solutions for $\sigma_1$ and $\sigma_2$. A necessary and sufficient condition for the absence of SPB is given by
\begin{equation}\label{ineq1}
(\lambda_3 - \lambda_4) \sigma_1^2 + \lambda_6 \sigma_1 \sigma_2 + 2\lambda_2 \sigma_2^2 > \Delta.
\end{equation}
In absence of SPB, the landscape of parameters shows a maximum of 9 extrema where only two of them at most are minima (see Fig. \ref{landsc_extrema}). The specific values of the constants determines which one is the true minimum of the theory.

It is assumed that the scalars under consideration are generated in the quark sector of QCD. The baryon chemical potential $\mu$ is transmitted to the meson sector via a local quark-meson coupling
\begin{equation}\label{yukawa}
\Delta {\cal L}= - (\bar q_R H_1 q_L +\bar q_L H_1^\dagger q_R)\longrightarrow - \bar q \sigma_1 q.
\end{equation}
After integrating out the constituent quarks, the temperature and chemical potential dependence are included in the gap equation of $\sigma_1$ through the one-loop contribution to the effective potential \eqref{effpot1}, $\Delta V_{\text{eff}}$. The saturation point to nuclear matter can be found at finite densities due to the presence of two multiplets, avoiding the chiral restoration that takes place in simpler models with only one multiplet like NJL. This effective potential is normalized to reproduce the baryon density for quark matter
\begin{equation}
\varrho_B=-\frac13\partial_\mu\Delta V_{\text{eff}}(\mu)=\frac{N_cN_f}{9\pi^2}p_F^3=\frac{N_cN_f}{9\pi^2}(\mu^2-\sigma_1^2)^{3/2},
\end{equation}
where the quark Fermi momentum is $p_F=\mu^2−\sigma_1^2$. Normal nuclear density is $\varrho_B\simeq 0.17$ fm$^{−3} \simeq$ (1.8fm)$^{−3}$ that corresponds to the average distance 1.8 fm between nucleons in nuclear matter. An accurate description of the condensation point, i.e. the critical density where nuclei become stable requires the introduction \cite{anesp,anesp2} of an isosinglet repulsive interaction driven by the $\omega$ meson, but this will not be considered here.

\medskip

\begin{figure}[h!]
\centering
\includegraphics[scale=0.4]{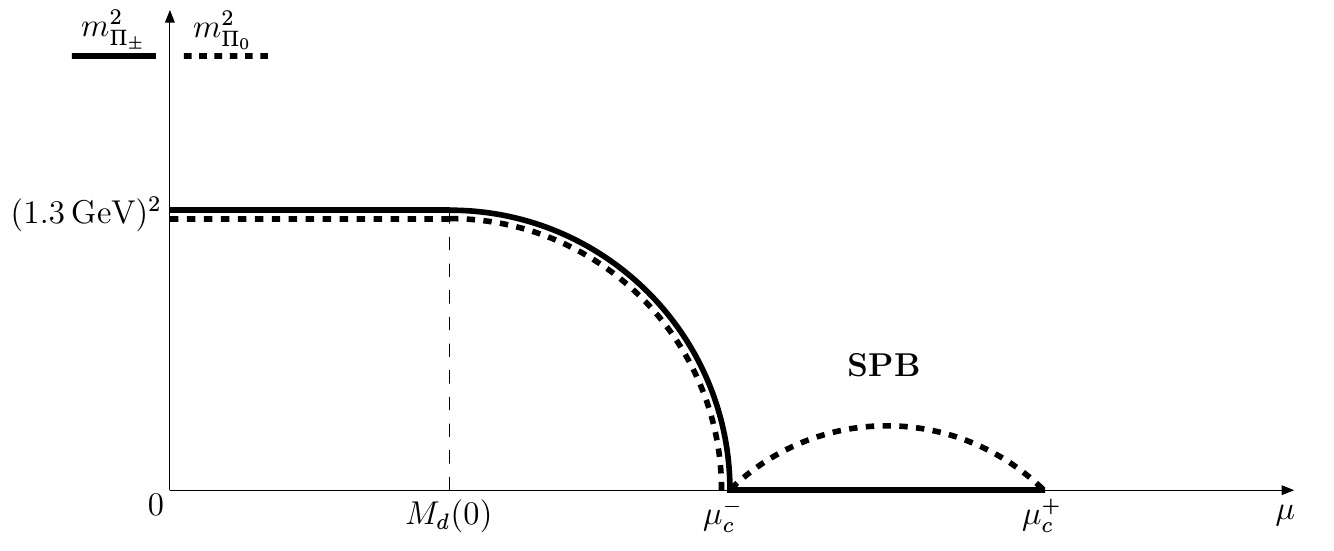}
\caption{Qualitative behaviour of excited pions depending on the quark chemical potential $\mu$ when the SPB phase is entered \cite{Afonin}. $M_d(0)$ is the constituent quark mass at zero $\mu$. Charged pions become massless in the SPB phase while the neutral component only aquires zero mass at the very transition point.}\label{afonin}
\end{figure}

A phase with SPB may take place due to condensation of $\pi^0_2$ together with non-trivial scalar condensates. A direct relation seems to connect the eventual presence of a SPB phase with the absence of chiral collapse in the model provided that $\Delta>0$. The transition from the CSB to the SPB phase is found to be possible at zero temperature, and at finite temperature with lower chemical potential. However, this relation strongly depends on the generic parameters of the model. In addition to the breaking of chiral symmetry that converts the triplet of light pions in massless Goldstone bosons, spontaneous condensation in the isotriplet channel breaks the isospin symmetry $SU(2)_V$ down to $U(1)$ and two charged excited $\pi'$ mesons are expected to be massless throughout this phase while the neutral state becomes massless only at the very transition phase. Notice that convexity around the minimum of the potential \eqref{effpot1} implies that all diagonal elements are non-negative. This gives positive masses for two scalar and one pseudoscalar mesons, whereas the triplet of pions and charged doublet of $\pi'$ mesons remain massless. In Fig. \ref{afonin} this behaviour is displayed.

\medskip

Once fixed the interaction to quark matter is fixed, there is no freedom in the choice of the kinetic term for scalar fields. Namely one cannot rotate two fields and rescale the doublet $H_1$ without changes in the chemical potential driver through $\Delta V_{\text{eff}}$. However the rescaling of $H_2$ is possible at the expense of an appropriate redefinition of other coupling constants and this freedom can be used to fix one of the constants which appear in the kinetic term. Thus the general kinetic term symmetric under $SU(2)_L\times SU(2)_R$ global reads
\begin{equation}
\mathcal L_{\text{kin}}=\frac14\sum_{j,k=1}^2A_{jk}\text{Tr}\left \{\partial_\mu H_j^\dagger \partial^\mu H_k\right \}.
\end{equation}
In general, $\pi_1$ and $\pi_2$ are mixed and their diagonalization leads to the physical mass states $\pi$ and $\Pi$. By means of the parametrization \eqref{chipar}, the mass of the heavy pion triplet in the parity-even phase can be written as
\begin{equation}
m_\Pi^2=\frac{-\Delta+(\lambda_3-\lambda_4)\sigma_1^2+\lambda_6\sigma_1\sigma_2+2\lambda_2\sigma_2^2}{A_{22}-\zeta^2},
\end{equation}
with
\begin{equation}
F_0^2=\sum_{j,k=1}^2A_{jk}\sigma_j\sigma_k, \quad \zeta\equiv\frac1{F_0}\sum_{j=1}^2A_{j2}\sigma_j.
\end{equation}
In the SPB phase, the expected mixing among scalar and pseudoscalar states is explicit in the kinetic term, where the notation $\langle\pi_2^0\rangle=\rho$ is used. This kinetic term has the following form
\begin{align}\label{kinetic}
\nonumber \mathcal L_{\text{kin}}&=\partial_\mu\pi^\pm\partial^\mu\pi^\mp+\frac12\left (1+\frac{A_{22}\rho^2}{F_0^2}\right )\partial_\mu\pi^0\partial^\mu\pi^0+(A_{22}-\zeta^2)\partial_\mu\Pi^\pm\partial^\mu\Pi^\mp\\
&+\frac12\left (A_{22}-\frac{F_0^2}{F_0^2+A_{22}\rho^2}\zeta^2\right )\partial_\mu\Pi^0\partial^\mu\Pi^0\\
\nonumber &+\frac12\sum_{j,k=1}^2\frac{A_{jk}F_0^2+\rho^2\text{det}(A)\delta_{1j}\delta_{1k}}{F_0^2+A_{22}\rho^2}\partial_\mu\Sigma_j\partial^\mu\Sigma_k-\frac{F_0\rho}{F_0^2+A_{22}\rho^2}\zeta\partial_\mu\Pi^0\sum_{j=1}^2A_{j2}\partial^\mu\Sigma_j.
\end{align}
It is interesting to note that even in the massless pion sector the isospin breaking occurs: neutral pions become less stable with a larger decay constant. Furthermore, in the charged meson sector the relationship between massless $\pi$ and $\Pi$ remain the same as in the symmetric phase. If a small quark mass is considered without an explicit isospin breaking, one finds two massless charged pions $\pi^\pm$ and three light pseudoscalars $\pi^0,\Pi^\pm$ with masses linear in the current quark mass.

\medskip

A relevant example to be examined corresponds to models with residual discrete chiral symmetry $Z_2\times Z_2$ after the breaking of $SU(2)_L\times SU(2)_R\to SU(2)_V$. This extra symmetry is related to the independent reflections $H_j\to-H_j$. This subset of particular models requires $\lambda_5=\lambda_6=A_{12}=0$ but $\lambda_4\neq0$. One can always fix $A_1=A_2$ redefining the other parameters. The gap equations and stability conditions are significantly simplified allowing to get a feeling of the possible scales involved in the system. A few plausible assumptions such as $F_0=100$ MeV and a constituent quark mass $M_d=\sigma_1=300$ MeV allow to get an insight into this model obtaining a reasonable mass spectrum in the parity-even phase provided that $\Delta\simeq 2.7F_0^2$, $\lambda_1\simeq 0.15$ and $\lambda_4\simeq 0.35$. With these values and taking $M_{d,\text{crit}}\simeq 1.8F_0$, one finds that SPB occurs at dense nuclear matter with a typical density $\varrho_{B,\text{crit}}\simeq 0.5$ fm$^{-3}\simeq 3\varrho_{B,\text{nuclear}}$. Thus the possibility of SPB emerges naturally for reasonable values of the meson physics parameters and low-energy constants.

\medskip

The parity-breaking effect triggered by an isotriplet pseudoscalar condensation has some characteristic footprints that could eventually lead to the detection of this phase. The isospin breaking could affect the heavy pion decay constant as shown in Eq. \eqref{kinetic} enhancing their electroweak decays. Higher-mass meson resonances (radial excitations) do not have a definite parity and therefore the same resonance can decay both in two and three pions (in general into even and odd number of pions). As already mentioned, at the very point of the phase transition leading to parity breaking one has six massless pion-like states. After crossing the phase transition, in the parity broken phase, the massless
charged $\Pi$ remain as Goldstone bosons enhancing charged pion production, whereas the additional neutral pseudoscalar state becomes massive. Finally, a reinforcement of long-range correlations in the pseudoscalar channel could be hunted in lattice simulations.

\medskip

Despite dealing with a simple model, the feature of parity breaking could be a rather generic feature of QCD at finite density. A rigorous proof does not exist in QCD yet due to the difficulties of dealing with non-vanishing chemical potentials in the lattice \cite{latt}. A vector chemical potential in gauge theories like QCD cannot be easily treated and therefore simpler models hopefully reproducing the main features of the theory may be useful. Needless to say, non-equilibrium effects are also notoriously difficult to study non-perturbatively. In spite of this, the conclusions of the analysis in \cite{anesp} seem rather robust: there is a range of densities where a parity violating vacuum is energetically favourable.

\medskip
\begin{figure}[h!]
\centering
\includegraphics[scale=0.39]{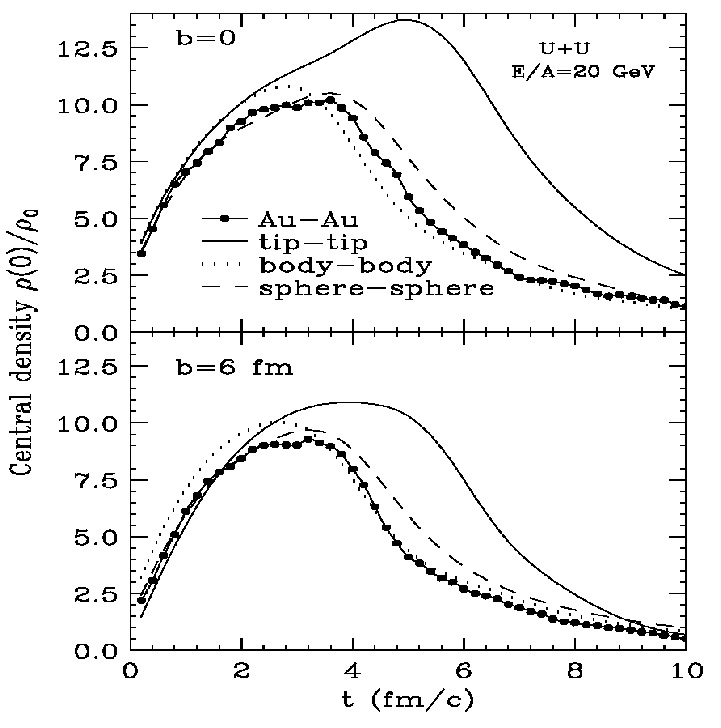}
\caption{Central density as a function of time for central Au+Au collisions (dotted line) and tip-on-tip U+U collisions (solid line) at 20A GeV as calculated with the ART transport code \cite{centrdensity} (upper plot). Same for semi-central collisions (b=6 fm) (lower plot).}\label{centraldensity}
\end{figure}

Such a parity-violating state could be produced in HIC in conditions which are expected at the FAIR facility or at the planned NICA accelerator. An illustrative simulation for the time evolution of the medium density in central and semicentral collisions is presented in Fig. \ref{centraldensity} based on a particular transport code \cite{centrdensity}. In the center-of-mass frame of the colliding ions the density fastly increases in the pre-equilibrium stages and then slowly decreases eventually entering in the hadron phase. Once entered this phase, the decreasing density suggests a time-dependent neutral pion condensate (if any) with a relatively smooth time dependence. The latter just produces a modification of quantum electrodynamics (QED) via parity-breaking term $\sim \langle \Pi(t) \rangle F_{\mu\nu}\widetilde F^{\mu\nu}$ that describes the neutral pion decay into two photons. If we assume that $\langle\Pi(t)\rangle\propto t$ the presence of pion condensation distorts the photon spectrum and its two transversal (circular) polarizations obtain different masses \cite{axion}. One of them becomes massive and growing with the photon 3-momentum and the other one appears as a tachyon. It is remarkable that the generation of photon mass is also predicted in the presence of a neutral pion condensate \cite{photonmass}. As a result, photons with momenta $|\vec k|>4m_\ell^2/\eta$ start to decay into lepton pairs $\gamma\to\ell^+\ell^-$, with $\eta\sim\alpha\langle\dot\Pi\rangle/f_\pi$. A hierarchy of thresholds appears for different leptons, mostly electrons and muons. The decay width $\sim\alpha\eta$ is suppressed by the fine structure constant $\alpha$ but may be considerably enhanced if the pion condensate derivative $\eta$ rapidly changes.

\medskip

However, this scenario is not completely accurate as it is known that vector mesons mix with photons and the calculation of dilepton rates only from photons is a crude approximation at low energies. We will investigate this effect in detail in Chapter \ref{ChVMD}. 

\section{Topological fluctuations as a source for parity breaking}

A remarkable feature of gauge theories like QCD is the existence of topologically non-trivial configurations of gauge fields \cite{instantons}, the so-called instantons. They correspond to a connection of different degenerate vacua which are related by topologically non-trivial gauge transformations. The amplitudes of such transitions vanish order by order in perturbation theory and therefore they are associated to non-perturbative phenomena. Typical calculations of quantum tunnelling between states with diferent topology suggested that these transitions were suppressed by factors of the order of $\exp(-2\pi/\alpha)$, where $\alpha$ is the interaction strength of the gauge theory.

\medskip

At high temperature the suppression factor disappears \cite{sphaleron} since there is enough energy to pass over the potential barrier that separates the topological inequivalent vacua. This phenomenon is called a sphaleron transition \cite{sphaleronqcd} and it could be responsible for the appearance of metastable domains where the QCD vacuum breaks $P$ and $CP$ \cite{kharzeev} in the vicinity of the deconfinement phase transition. $P$-even, but $CP$-odd metastable states have also been argued to exist in hot gauge theories \cite{bronoff}. Several dynamical scenarios for the decay of $P$-odd bubbles have been considered \cite{zhitn}, and numerical lattice calculations of the fluctuations of topological charge in classical Yang-Mills fields have been performed \cite{latttopcharg,lappi,polya,polikarpov}. Progress in understanding the vacuum structure of gauge theories, especially supersymmetric ones \cite{susyvac}, also points to the possible existence of $P$- and $CP$-odd vacuum states. A parity broken phase also exists in lattice QCD with Wilson fermions \cite{aoki}, but this phenomenon has been recognized as a lattice artifact.

\medskip

The experimental study of metastable $P$ and $CP$-odd domains at LHC or RHIC provides a unique link to the QCD phase transition in the early universe when probably $\theta\neq0$. Such a large $CP$ violation during the QCD phase transition might have lead to a separation of matter and anti-matter. Current experiments show that the very tiny amount of $CP$ violation from the electroweak part of SM is probably not sufficient to explain the baryon number density observed today.

\medskip

As it is well known, QCD with massless quarks shows an exact chiral $SU(2)_L\times SU(2)_R$ symmetry. However, the $U(1)_A$ symmetry is not satisfied at the quantum level leading to the axial anomaly. As a result, the colour- and iso-singlet axial current $J_{5,\mu} =\bar q \gamma_\mu\gamma_5 q$ is no longer conserved despite considering massless quarks. The local partial conservation of the axial current (PCAC) relation is afflicted with the gluon anomaly (for $N_f$ flavours)
\begin{equation}\label{pcac}
\partial^\mu J_{5,\mu} -2 i \bar q \hat m_q \gamma_5 q = \frac{N_fg^2}{8\pi^2} \text{Tr}\left (G^{\mu\nu}\widetilde G_{\mu\nu}\right ),
\end{equation}
where $\hat m_q$ is the quark mass matrix and $\widetilde G_{\mu\nu}\equiv \frac12\varepsilon_{\mu\nu\rho\sigma}G^{\rho\sigma}$ with $\varepsilon_{0123}=-\varepsilon^{0123}=+1$. Here $g$ denotes the QCD coupling constant with generators normalized as Tr$(t_at_b)=\delta_{ab}/2$. The term at the r.h.s. of Eq. \eqref{pcac} can be expressed as a total derivative of the Bardeen current $K^\mu$
\begin{equation}
K_\mu = \frac{g^2}2 \epsilon_{\mu\nu\rho\sigma}\text{Tr}\left(G^\nu\partial^\rho G^\sigma -i\frac 23 g G^\nu G^\rho G^\sigma\right), \qquad \partial_\mu K^\mu=\frac{g^2}4\text{Tr}\left (G^{\mu\nu}\widetilde G_{\mu\nu}\right ).
\end{equation}
As already mentioned in the previous section, including the vacuum structure is equivalent to adding a new $\theta$-term in the QCD Lagrangian
\begin{equation}\label{thetaterm}
\Delta \mathcal L_\theta=\theta \frac{g^2}{16\pi^2}\text{Tr}\left (G^{\mu\nu}\widetilde G_{\mu\nu}\right ),
\end{equation}
where unless $\theta$ vanishes, parity is violated. Discarding a total derivative from the QCD Lagrangian seems legitimate in case of an homogeneous in space and time $\theta\neq0$. However, local dependences may trigger local parity breaking (LPB) in the small domains quoted by \cite{kharzeev}. Gauge field configurations can be characterised by an integer topological (invariant) charge $T_5$ (equivalent to the winding number with this normalization) given by
\begin{equation}
T_5= \frac{g^2}{16\pi^2}\int^{t_f}_{t_i} dt\int_{\text{vol.}}d^3x \, \text{Tr} \left (G^{\mu\nu} \widetilde G_{\mu\nu}\right ) \in \mathbb Z,
\end{equation}
where the integration limits explicitly reflect that a finite volume is considered\footnote[1]{The topological current flux through the boundary is normally neglected. Then, the following definition of $T_5$ is also used
\begin{equation}
\nonumber T_5= \frac1{4\pi^2}\int^{t_f}_{t_i} dt\int_{\text{vol.}}d^3x \, \partial_0 K^0= \frac1{4\pi^2} \int_{\text{vol.}}d^3x \, \left [K^0(\vec x,t_f)-K^0(\vec x,t_i)\right ].
\end{equation}}. The integration over a finite space volume of the PCAC in Eq. \eqref{pcac} allows to associate a non-zero topological charge with a non-trivial quark axial charge $Q_5^q$
\begin{equation}\label{axcons}
\frac{d}{dt} (Q_5^q - 2 N_f T_5) \simeq 2 i \int_{\text{\small vol.}} d^3 x\, \bar q \hat m_q \gamma_5 q\ ,
\quad Q_5^q =\int_{\text{\small vol.}}d^3x\, \bar q\gamma_0\gamma_5q.
\end{equation}
Perturbative interactions between quarks and gluons cannot induce a difference between the number of right- and left-handed quarks. A mass term always will tend to wash out such difference \cite{ambjorn}. For massless quarks a topological charge induces a net chirality in a finite volume.

\medskip

It was pointed out that the presence of non-zero angular momentum in peripheral HIC may trigger a charge separation in $P$- and $CP$- odd domains \cite{kharzeev}. In such collisions, an induced electric field perpendicular to the reaction plane was found to appear of order $E\sim-(e\,\theta/2\pi)\,l$, where $l\sim b$ is the angular momentum in a collision at impact parameter $b$. Another way to understand this effect comes from considering the angular momentum of a peripheral collision as an equivalent magnetic field orthogonal to the reaction plane. The $\theta$-term can be written as $L_\theta\sim\theta\vec E\vec B$ so a coupling between the electric and magnetic fields arises. At non-trivial $\theta$ angle, a magnetic field induces a parallel electric field $\vec E\sim\theta \vec B$ through the axial anomaly and a corresponding electric dipole moment.

\medskip

A further explanation for this charge separation in peripheral HIC can be provided. In a domain with a non-trivial topological charge and massless quarks, it was proved from Eq. \eqref{axcons} that an imbalance of the left- and right-handed quarks arises, which is translated into a net helicity. If an electrodynamic magnetic field is applied, positively charged quarks will orient their spin along the direction of the magnetic field due to the magnetic moment interaction. Likewise, negatively charged quarks will orient in the opposite direction. Since a net helicity is conserved in the domain, the quark spins will preferentially orientate in the direction of their momenta. Therefore, positively charged quarks will move in the direction of the magnetic field, and negatively charged ones will move opposite to it. A resulting electromagnetic current will set up in the direction of the magnetic field. Due to this conception, the charge separation induced by an angular momentum or magnetic field in a small domain characterised with a non-trivial topological charge was called the Chiral Magnetic Effect (CME). For a review of this and other related 'environmental symmetry violations' in heavy ion collisions, see \cite{Liao:2014ava}.

\medskip

The CME could be used to determine whether a deconfined chirally symmetric phase of matter is created in HIC. Deconfinement is a necessary requirement for this effect to work since it requires that quarks can separate over distances much greater than the radius of the nucleon. Moreover, chiral symmetry restoration is essential, because a chiral condensate will tend to erase any asymmetry between the number of right- and left-handed fermions.

\medskip

An experimental observable measuring the CME charge separation with respect to the reaction plane was proposed shortly afterwards \cite{CMEexp}. More recently, it has been claimed that this signal is observed in the STAR experiment at RHIC \cite{star} (see Fig. \ref{star}), although the issue still remains controversial. The effect should be most visible for ultra-peripheral HIC where large angular momenta induce large magnetic fields contributing to the axial charge separation.

\section{Quasi-equilibrium treatment of a conserved quark axial charge}

We already mentioned that the two parity-violating operators presented in Eq.\eqref{pbreakoper} can be related by a $U(1)_A$ transformation. Indeed, due to the axial anomaly \eqref{pcac} the $\theta$-term in Eq. \eqref{thetaterm} can be transformed into
\begin{equation}\label{thetarotation}
\frac1{2N_f}\partial_\mu\theta \bar\psi\gamma^\mu\gamma^5\psi.
\end{equation}
For massless quarks, QCD with a topological charge $\langle T_5 \rangle\neq 0$ can be equally described at the Lagrangian level by a conserved topological chemical potential $\mu_\theta$ or by an axial chemical potential $\mu_5$ \cite{kharzeev,epjc,polikarpov}. Their relation arises from \eqref{axcons}
\begin{equation}
\langle T_5 \rangle = \frac{1}{2N_f} \langle Q_5^q \rangle \quad \Longleftrightarrow \quad \mu_5 = \frac{1}{2N_f} \mu_\theta.
\end{equation}
This quasi-equilibrium situation can be treated simply adding to the QCD Lagrangian $\Delta {\cal L}_{\text{\rm top}}= \mu_\theta T_5$ or, alternatively, $\Delta {\cal L}_q = \mu_5 Q_5^q$. The latter realization together with Eq. \eqref{thetarotation} allows to identify
\begin{equation}\label{mu5}
\mu_5=\frac1{2N_f}\partial_0\theta \quad \Longrightarrow \quad \mu_\theta=\partial_0\theta.
\end{equation}
Thus we assume that a time dependent but approximately spatially homogeneous pseudoscalar axion-like background is induced at the energy densities reached in HIC with $\partial_\mu\theta(x) \sim (\mu_5,0,0,0)$. We could refer the reader to \cite{axion} where it has been shown that a pseudoscalar field slowly evolving in time drastically changes the electromagnetic properties of the vacuum. In particular a photon propagating in this background with sufficiently high energy may decay on shell in medium into dileptons.

\medskip

For massive light quarks the axial charge is conserved provided that the quark mass term breaking chiral symmetry remains subdominant. The characteristic oscillation time is governed by inverse quark masses. Evidently for $u,d$ quarks the characteristic time $1/\hat m_q \sim 1/5$ MeV$^{-1} \sim 40$ fm is much bigger than the fireball lifetime ($\tau_{\text{fireball}}\simeq 5-10$ fm) and the left-right quark mixing can be neglected. But it is not the case for strange quarks as $1/ m_s \sim 1/150$ MeV$^{-1}$ $\sim 1$ fm $\ll \tau_{\text{fireball}}$ and even if a topological charge persists during the fireball lifetime, the mean value of the strange quark axial charge is around zero due to left-right oscillations.

\medskip

When nuclear matter is superdense, pseudoscalar condensation in the $I=1$ channel may occur too, as it has been outlined in the previous section \ref{SPB}. This situation corresponds to introducing a vector chemical potential $\mu$. Thus for light quarks in hot and dense nuclear matter, the matrix structure of the axial chemical potential $\mu_5$ in flavour space generically includes not only $SU(3)_f$ singlet but also neutral components of $SU(3)_f$ octet, that corresponds to including both a $\mu_5$ and a $\mu$ term, respectively. Note that the appearance of a condensate in the $\tau_3$ direction implies in addition a breaking of the isospin symmetry of the vacuum.

\medskip

This thesis is mostly concerned with the possibility of LPB associated to the appearance of an isosinglet axial chemical potential. This case is expected to be the dominant one in experiments where the temperature in the nuclear fireball is much larger than the chemical potential $T\gg\mu$, as it is the case for the LHC or RHIC. The issue of isotriplet condensation induced by finite density\footnote[2]{Finite density may contribute to isosinglet pseudoscalar condensation too.} will not be discussed in this approach. This scenario corresponds to the region where $\mu \gg T$ that will be explored in future experiments at FAIR and NICA. Although we will work with an isosinglet $\mu_5$ throughout, some indications will be provided along the text for the case that one could be interested in selecting an isotriplet pseudoscalar condensate.

\medskip

Unlike the case of having a dense system with a non-trivial $\mu$, an axial chemical potential is tractable on the lattice \cite{yama} and with other methods \cite{ChernLuo}. For central collisions and light quarks, local topological fluctuations may correspond to the existence of an ephemeral phase characterised by LPB with non-trivial axial chemical potential. In Fig. \ref{QCDphasediag}, the phase diagram of QCD is showed depending on the temperature and the baryon density. The hypothetical location of the LPB phase is displayed suggesting that this phenomenon may be important in heavy ion experiments where the produced system is very hot but not specially dense. We conjecture that the CME and topological fluctuations may be complementary effects revealing two facets associated to the formation of a thermodynamic phase where parity is locally broken. The red lines in the graphic show the compression and later expansion of the system.

\medskip

The treatment of $\mu_5$ is a non-trivial task due to the complex dynamics of HIC. During the first stages of the collision, nuclear matter is quickly compressed and heated with a very short thermalization time $\tau_{\text{heating}}<0.5$ fm (attempts to explain this process in holographic QCD can be found in \cite{holo}) and it may enter the LPB phase. After reaching the QGP, the system slowly expands and cools down reentering the LPB phase where large topological fluctuations may take place. Of course, the cooling time greatly exceeds the time taken for the LPB phase to form so most of the expected effects should come from this latter period. At this stage the nuclear matter cools down (for the fireball lifetime) until freeze-out. During this period the topological charge is supposedly conserved and the system may reach thermal equilibrium. The effect of having a non-trivial axial charge could trigger the main phenomenological effects that are discussed all along this thesis. It is not clear that the value of the axial chemical potential is uniform but we shall assume so here and consider an average $\mu_5$ which has to be understood as an effective value extracted from such dynamics.

\medskip
\begin{figure}[h!]
\centering
\includegraphics[scale=0.4]{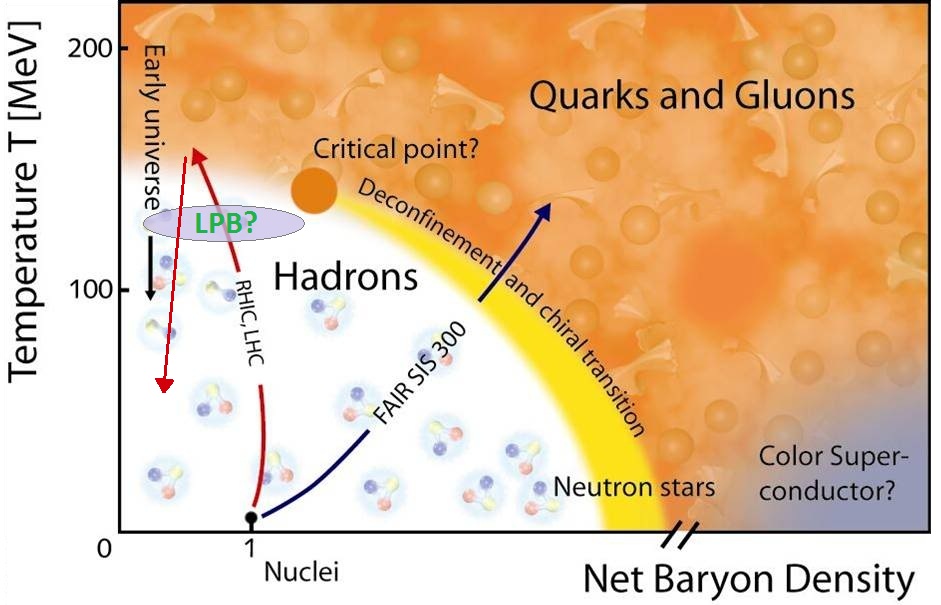}
\caption{QCD phase diagram depending on the temperature and the baryon density where the approximate situation of the hypothetical LPB phase is showed. For rather hot (but not dense) HIC, the system may enter this phase and remain there for a time similar to the nuclear fireball lifetime $\tau_\text{fireball}\simeq 5-10\ $ fm.}\label{QCDphasediag}
\end{figure}

If this new thermodynamic phase really exists, experiments should be able to encounter it. By means of both a vector and an axial chemical potential, we describe how the symmetry breaking mechanism works from the theoretical point of view and its implications. Finding observables that could be measured at the mentioned experiments constitutes one of the main goals of our research. This is a field in between high energy physics, nuclear physics and lattice numerical studies, and one where theory and experiment still work close together.					
\chapter{Parity breaking in the Nambu--Jona-Lasinio model}\label{ChNJL}
\graphicspath{{3-njl/figures/}{figures/}}

As already discussed, QCD with a vector chemical potential presents serious difficulties to be treated. Therefore, the study a simpler model reproducing the main symmetries of the QCD may provide some qualitative and quantitative ideas that could help in a further understanding of strong interactions.

\medskip

In the present chapter we shall consider the Nambu--Jona-Lasinio model (NJL) \cite{NJL,njl,NNJJLL,NNJJLL2,Ebert}, which shares interesting features with QCD. Although it may be implemented with an exact chiral symmetry with massless quarks, this symmetry is spontaneously broken due to the formation of a dynamical chiral (quark) condensate. Such phenomenon inspired many theories of the breaking of electroweak symmetry such as technicolor.

\medskip

In the NJL modelization, QCD gluon interactions among quarks are assumed to be replaced by some effective four-fermion couplings. The appearance of these dimension 6 operators makes the model a non-renormalizable effective theory valid up to some physical scale $\Lambda$. The absence of gluons permits the model to be bosonised, replacing the 4-fermion couplings by composed degrees of freedom that are normally associated with light mesons such as $\sigma$ and pions.

\medskip

Confinement is absent in the NJL model, but global symmetries can be arranged to be identical in both QCD and NJL. When the chiral condensate forms, the chiral symmetry is spontaneously broken into a diagonal subgroup $SU(2)_V$ since the condensate leads to a pairing of the left-handed and the right-handed flavours. The broken symmetries lead to massless pseudoscalar bosons which are the associated Goldstone bosons, in this case, pions.

\medskip

The NJL model has been used recently by some authors \cite{azcoiti} to investigate the nature of the Aoki phase in QCD \cite{aoki}. This is a phase in lattice QCD with Wilson fermions where parity and possibly isospin symmetry is broken. It does not survive the continuum (note that the NJL does not have a 'continuum limit' either). It is however conceivable that the introduction of the chemical potential may enlarge the scope of the Aoki phase and allow for a sensible continuum interpretation. This is what should happen if the effective theory analysis described in \cite{anesp} is correct. 

\medskip

Previously some authors have studied the effect of a vector chemical potential $\mu$ with three flavours \cite{ikkk} in the NJL model for a study the nonet meson characteristics. An analysis of the QCD phase diagram at finite temperature including an axial chemical potential $\mu_5$ has also been considered by several groups \cite{ChernLuo}. Nevertheless, the consequences of including both a vector and an axial chemical potentials have not been considered so far to our knowledge.

\medskip

In this work NJL constitutes a toy model for the implementation of parity-breaking effects. With the help of both a vector and an axial chemical potentials, we shall investigate the phase diagram of the model with special interest in the appearance of possible pseudoscalar condensates leading to spontaneous breaking of parity. It turns out that the inclusion of $\mu_5$ changes radically the phase structure of the model and shows that $\mu$ is not a key player in ushering a thermodynamically stable phase where parity is violated in the NJL model, but $\mu_5$ is. The introduction of $\mu_5$ constitutes an explicit source for parity violation. However, a spontaneous breaking of this symmetry points towards a region where pseudoscalar condensation may be thermodynamically stable. The characteristic values of the external drivers allowing for this possible new stable phase may provide some hints on the realistic values of such magnitudes reached in HIC.

\medskip

However, NJL is definitely not QCD and the present work does not attempt to draw definite conclusions on the latter theory; just to point out possible phases requiring further analysis.

\medskip

This chapter is organized as follows. In Sec. \ref{NJL2} the NJL Lagrangian with the incorporation of $\mu$ and $\mu_5$ will be introduced. We describe how an effective potential is extracted when one introduces some effective light meson states and integrates out the fermion degrees of freedom. In Sec. \ref{NJL3} we show the gap equations of the model and the conditions for their stability. After that, the different stable phases of this model are presented and discussed. We show in Sec. \ref{NJL4} that a phase with an isospin singlet pseudoscalar condensate in addition to a scalar condensate is possible. It turns out that the conditions for this phase to be stable and exhibit chiral symmetry breaking too are such that one gets an inverted mass spectrum with $m_\pi>m_{\eta_q}$ and $m_\sigma>m_{a_0}$, which is quite different from QCD. In Sec. \ref{NJL4} we also present the main results of this work with plots of the evolution of the scalar and pseudoscalar condensates together with the main features of the phase transition.

\section{Nambu--Jona-Lasinio Lagrangian with $\mu$ and $\mu_5$}\label{NJL2}

The starting point of this chapter is the NJL Lagrangian where we incorporate a vector and an axial chemical potentials $\mu$ and $\mu_5$, respectively. For two flavours and $N$ colours, we have \cite{njl}
\begin{equation}\label{lagNJL}
\mathcal L=\bar\psi(\partial\!\!\!\!\!\!\not\;\; + m 
- \mu\gamma_0 - \mu_5\gamma_0\gamma_5)\psi - 
\frac{G_1}N[(\bar\psi\psi)^2+(\bar\psi i\gamma_5\vec\tau\psi)^2]
-\frac{G_2}N[(\bar\psi\vec\tau\psi)^2+(\bar\psi i\gamma_5\psi)^2],
\end{equation}
with a full $U(2)_L\times U(2)_R$ chiral invariance in the case that $G_1=G_2$, while if these constants differ, the $U(1)_A$ symmetry breaks and only $SU(2)_L\times SU(2)_R\times U(1)_V$ remains. In the case of an isotriplet axial chemical potential would appear attached to the quark bilinear $\bar\psi\gamma_0\gamma_5\tau^3$. One may introduce two doublets of bosonic degrees of freedom $\{\sigma,\vec\pi\}$ and $\{\eta,\vec a\}$ by adding the following chiral invariant term
\begin{equation}
\Delta\mathcal L=\frac{Ng_1^2}{4G_1}(\sigma^2+\vec \pi^2)+\frac{Ng_2^2}{4G_2}(\eta^2+\vec a^2).
\end{equation}
These would be identified with their namesake QCD states (actually $\eta_q$ and $\vec a_0$ for the last two). Euclidean conventions will be used throughout this chapter. We bosonize the model following the same procedure as in \cite{NNJJLL}.

\medskip

After shifting each bosonic field with the quark bilinear operator that carries the corresponding quantum numbers
\begin{align}
\nonumber \sigma\to\sigma+\frac{2G_1}{Ng_1}\bar\psi \psi, \qquad \vec \pi\to\vec\pi+\frac{2G_1}{Ng_1}\bar\psi i\gamma_5\vec\tau\psi, \\
\eta\to\eta+\frac{2G_2}{Ng_2}\bar\psi i\gamma_5 \psi, \qquad \vec a_0\to\vec a_0+\frac{2G_2}{Ng_2}\bar\psi \vec\tau\psi,
\end{align}
the Lagrangian \eqref{lagNJL} may be rewritten as
\begin{align}
\nonumber \mathcal L=&\bar\psi[\partial\!\!\!\!\!\!\not\;\; + m -\mu\gamma_0 -\mu_5\gamma_0\gamma_5
+g_1(\sigma+i\gamma_5\vec\tau\vec\pi)+g_2(i\gamma_5\eta+\vec\tau\vec a)]\psi\\
&+\frac{Ng_1^2}{4G_1}(\sigma^2+\vec \pi^2)+\frac{Ng_2^2}{4G_2}(\eta^2+\vec a^2),
\end{align}
which shows a redundancy related to the coupling constants $g_{1,2}$ that appear attached to each doublet and it is eventually related to their wave function normalization. In other words, these coefficients will be absorbed in the fields once they are renormalised $g_1\sigma\to\sigma^r$. Without further ado we will take $g_1=g_2=1$.

\medskip

Integration of the fermions will produce a bosonic effective potential (or free energy) and will
allow to study the different phases of the model. We will work in the mean field approximation and accordingly neglect
fluctuations. The results will be exact in the large $N$ limit.
\begin{equation}\label{V}
V_{\text{eff}}=\frac{N}{4G_1}(\sigma^2+\vec \pi^2)+\frac{N}{4G_2}(\eta^2+\vec a^2)- \text{Tr}\log\mathcal M(\mu,\mu_5),
\end{equation}
where the trace is understood to be performed in the isospin and Dirac spaces in addition to a 4-momentum integration of the operator in the momentum space. We will assume that $\mu>0$, namely we consider a baryon (as opposed to antibaryon) finite density. The invariance under $CP$ of the action ensures that the free energy (\ref{V}) only depends on the modulus of $\mu$.

\medskip

We also define the fermion operator
\begin{equation}\label{fermoper}
\mathcal M(\mu,\mu_5)=\partial\!\!\!\!\!\!\not\;\; + (M+\vec\tau\vec a)- \mu\gamma_0 - \mu_5\gamma_0\gamma_5
+i\gamma_5(\vec\tau\vec\pi+\eta),
\end{equation}
with the introduction of a constituent quark mass $M\equiv m+\sigma$. 

\medskip

In appendix \ref{trick} we show that the dependence on both vector and axial chemical potentials does not change the reality of the fermion determinant. However, its sign remains undetermined, and in order to ensure a positive determinant, we shall consider an even number of 'colours'\footnote{The choice of an even number of colours, unlike QCD, is simply a technical restriction to ensure the fermion determinant to be positive definite.} $N$ so that one can safely assume
\begin{equation}\label{det+}
\det[\mathcal M(\mu,\mu_5)]=\sqrt{\det[\mathcal M(\mu,\mu_5)]^2}
\end{equation}
and hence, use the calculations in Appendix \ref{trick}. If we just retain the neutral components of the triplets, this determinant can be written in the following way
\begin{align}\label{Trlog}
\nonumber \log\det\mathcal M(\mu,\mu_5)=&\text{Tr}\log\mathcal M(\mu,\mu_5)\\
\nonumber =&\frac18\text{Tr}\sum_{\pm}\Bigg\{\log\left [-(ik_0+ \mu)^2+(|\vec k|\pm \mu_5)^2+M_+^2\right ]\\
&+\log\left [-(ik_0 + \mu)^2+(|\vec k|\pm \mu_5)^2+M_-^2\right ]\Bigg\},
\end{align}
where
\begin{equation}
M_\pm^2\equiv (M\pm a)^2+(\eta\pm\Pi)^2
\end{equation}
and
\begin{equation}
\text{Tr}(1)=8NT\sum_n\int\frac{d^3\vec k}{(2\pi)^3}\left [k_0\to \omega^F_n=\frac{(2n+1)\pi}\beta\right ].
\end{equation}
From now on, when we refer to the neutral pion condensate, we will write $\Pi$. Note that, as explained in appendix \ref{trick}, one is able to write the determinant as the trace of an operator that is the identity in flavour space in spite of the initial non-trivial flavour structure. This facilitates enormously the calculations.

\medskip

In the search for stable configurations in the potential \eqref{V} we will need the derivatives of the fermion determinant, which are basically given by the function $K_1$ that we define as
\begin{equation}\label{trace}
4NK_1=\text{Tr}\sum_\pm\frac 1{(ik_0+\mu)^2-[(|\vec k|\pm\mu_5)^2+M^2]},
\end{equation}
which is clearly divergent in the UV. We will deal with the NJL model using dimensional regularization (DR) and a 3-momentum cut-off ($\Lambda$) both at zero temperature \cite{ikkk2,ikkk3}. The function $K_1$ depending on the regulator can be written as follows
\begin{align}\label{K1DR}
\nonumber K_1^{\text{DR}}(M,\mu,\mu_5)=&\frac1{2\pi^2}\Bigg [\Theta(\mu-M)\left \{\mu\sqrt{\mu^2-M^2}+(2\mu_5^2-M^2)\log\left (\frac {\mu+\sqrt{\mu^ 2-M^2}}M\right )\right \}\\
&-\frac12 M^2+\frac1{2}(M^2-2\mu_5^2)\left (\frac1{\epsilon}-\gamma_E + 2 -\log\frac{M^2}{4\pi\mu_R^2}\right )\Bigg ],
\end{align}
\begin{align}\label{K1Lambda}
\nonumber K_1^\Lambda(M,\mu,\mu_5)=&\frac1{2\pi^2}\Bigg [\Theta(\mu-M)\left \{\mu\sqrt{\mu^2-M^2}+(2\mu_5^2-M^2)\log\left (\frac {\mu+\sqrt{\mu^ 2-M^2}}M\right )\right \}\\
&-\frac12 {M^2}+\frac12 (M^2-2\mu_5^2)\log\frac {4\Lambda^2}{M^2}-\Lambda^2\Bigg ].
\end{align}
The quadratically divergent term in the cut-off regularization can be reabsorbed in the couplings $G_{1,2}$. After the redefinition, the two results are then identical if we identify
\begin{equation}
\frac1\epsilon-\gamma_E + 2 ~\longleftrightarrow ~\log\frac{\Lambda^2}{\pi\mu_R^2}.
\end{equation}
However in both cases the logarithmic divergence cannot be absorbed \cite{ZJ} unless we include extra terms in the Lagrangian like $(\partial\sigma)^2$ and $\sigma^4$. This is of course a manifestation of the non-renormalizability of the model. For this reason, we shall assume the scale $\Lambda$ (or equivalently $\mu_R$) to represent a physical cut-off and write
\begin{align}\label{K1}
\nonumber K_1(M,\mu,\mu_5)=&\frac1{2\pi^2}\Bigg [\Theta(\mu-M)\left \{\mu\sqrt{\mu^2-M^2}+
(2\mu_5^2-M^2)\log\left (\frac {\mu+\sqrt{\mu^ 2-M^2}}M\right )\right \}\\
&-\frac{M^2}2+(M^2-2\mu_5^2)\log\frac{2\Lambda}{M}\Bigg ].
\end{align}
Note that $K_1$ increases with $\mu$ and decreases with $\mu_5$. The derivative of this function will also be used
\begin{align}\label{L1}
\nonumber L_1(M,\mu,\mu_5)\equiv& \frac 1M\frac{\partial K_1}{\partial M}=
-\frac 1{\pi^2}\Bigg[\Theta(\mu-M)\Bigg \{\frac{\mu\mu_5^2}{M^2\sqrt{\mu^2-M^2}}
+\log\left (\frac{\mu+\sqrt{\mu^2-M^2}}M\right)\Bigg \}\\
&+1-\frac{\mu_5^2}{M^2}-\log\frac{2\Lambda}{M}\Bigg].
\end{align}
It verifies the property $L_1(\mu_5=0)>0$.

\section{Search for stable vacuum configurations}\label{NJL3}
We will now explore the different phases that are allowed by the effective potential \eqref{V} by solving the gap equations and analysing the second derivatives to investigate the stable configurations of the different scalar and pseudoscalar condensates. The gap equations for the system read
\begin{align}
\nonumber \frac{\sigma}{2G_1}+\sum_{\pm}(M\pm a) K_1^\pm =0, \qquad \frac{\eta}{2G_2}+ \sum_{\pm}(\eta \pm \Pi) K_1^\pm =0\\
\frac{\Pi}{2G_1}+\sum_{\pm}\pm(\eta\pm\Pi) K_1^\pm=0, \qquad \frac{a}{2G_2}+ \sum_{\pm} \pm (M\pm a)K_1^\pm=0
\end{align}
where $K_1^\pm\equiv K_1(M_\pm,\mu,\mu_5)$ (the same convention applies to $L_1$). The second derivatives of the potential are
\begin{gather}
\nonumber V_{\sigma\sigma}=\frac{1}{2G_1}+\sum_{\pm}\left [(M\pm a)^2 L_1^\pm +K_1^\pm \right ], \qquad 
 V_{\eta\eta}=\frac{1}{2G_2}+\sum_{\pm}\left [(\eta \pm \Pi)^2 L_1^\pm +K_1^\pm\right ]\\
\nonumber V_{\pi\pi}=\frac{1}{2G_1}+\sum_{\pm}\left [(\eta \pm \Pi)^2 L_1^\pm +K_1^\pm\right ], \qquad 
V_{aa}=\frac{1}{2G_2}+\sum_{\pm}\left [(M\pm a)^2 L_1^\pm +K_1^\pm\right ]\\
\nonumber V_{\sigma\eta}=V_{\pi a}=\sum_{\pm}(M\pm a)(\eta \pm \Pi)L_1^\pm , \qquad 
V_{\sigma\pi}=V_{\eta a}=\sum_{\pm} \pm (M\pm a)(\eta\pm \Pi) L_1^\pm\\
V_{\sigma a}=\sum_{\pm}\pm \left [(M\pm a)^2 L_1^\pm + K_1^\pm \right ], \qquad V_{\eta\pi}=
\sum_{\pm}\pm \left [(\eta \pm \Pi)^2 L_1^\pm +K_1^\pm\right ]
\end{gather}
To keep the discussion simple we will assume in the subsequent that $a=0$. However, in Sec. 4 we will see that in a very tiny region of the parameter space there is evidence of the existence of a phase with $a\neq 0$.

\subsection{Chirally symmetric phase}
We will first consider the phase where none of the fields condenses (in the chiral limit with $m=0$ and $\mu=\mu_5=0$ for simplicity). The gap equations are automatically satisfied, while the second derivatives read in this case
\begin{gather}
\nonumber V_{\sigma\sigma}=V_{\pi\pi}=\frac{1}{2G_1}+2K_1, \qquad V_{\eta\eta}=V_{aa}=\frac{1}{2G_2}+2K_1,\\
V_{\sigma\eta}=V_{\sigma\pi}=V_{\sigma a}=V_{\eta\pi}=V_{\eta a}=V_{\pi a}=0.
\end{gather}
After absorbing the quadratic divergence from the cut-off regularization into the coupling constants as mentioned previously
\begin{gather}
\frac{1}{2G_i}-\frac{\Lambda^2}{\pi^2} = \frac{1}{2G_i^r}
\end{gather}
the stability conditions for this phase are $G_{1,2}^r>0$. For simplicity, we will drop the superindex $r$ throughout this study.

\subsection{Chirally broken phase}\label{CSB}
In this section we will explore the phase where the field $\sigma$, and only this field, condenses. The gap equations reduce just to one
\begin{gather}\label{chiSB}
K_1=-\frac1{4G_1}\left (1-\frac mM\right ).
\end{gather}
Let us first assume $\mu=\mu_5=0$. Then the condition for CSB after absorbing the quadratic divergence into the coupling constants (or right away in DR for that matter) reads
\begin{equation}\label{genericgap}
M^2\left (\frac{1}2-\log\frac {2\Lambda}M\right )=\frac{\pi^2}{2G_1}\left (1-\frac mM\right ).
\end{equation}

\begin{figure}[h!]
\centering
\includegraphics[scale=0.3]{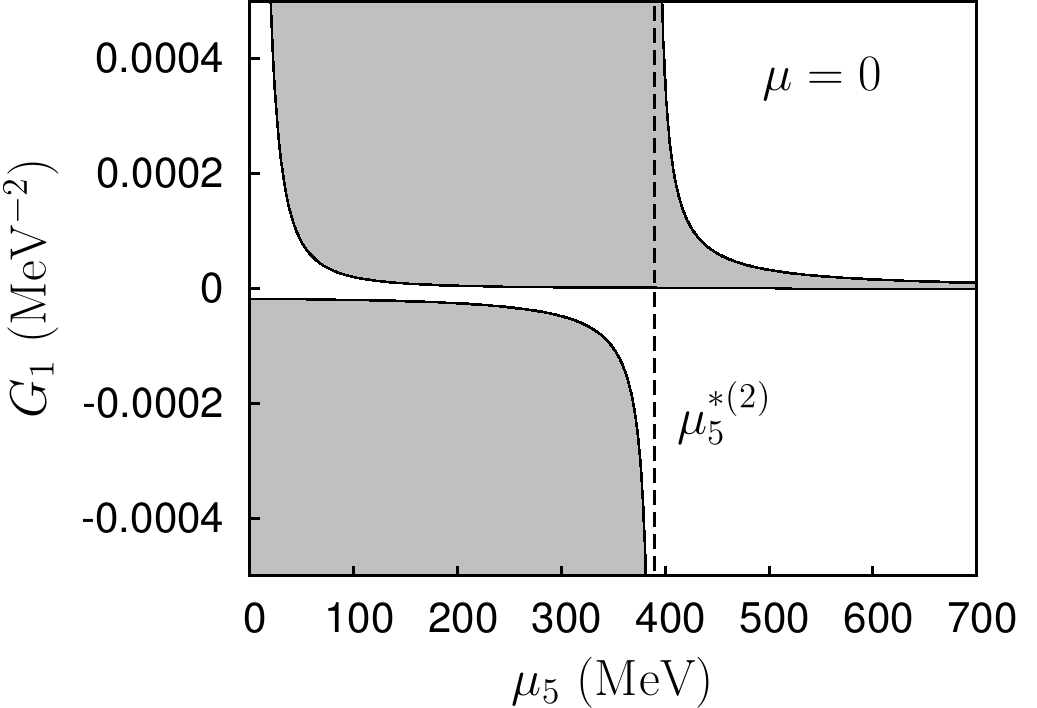}\;\;\includegraphics[scale=0.3]{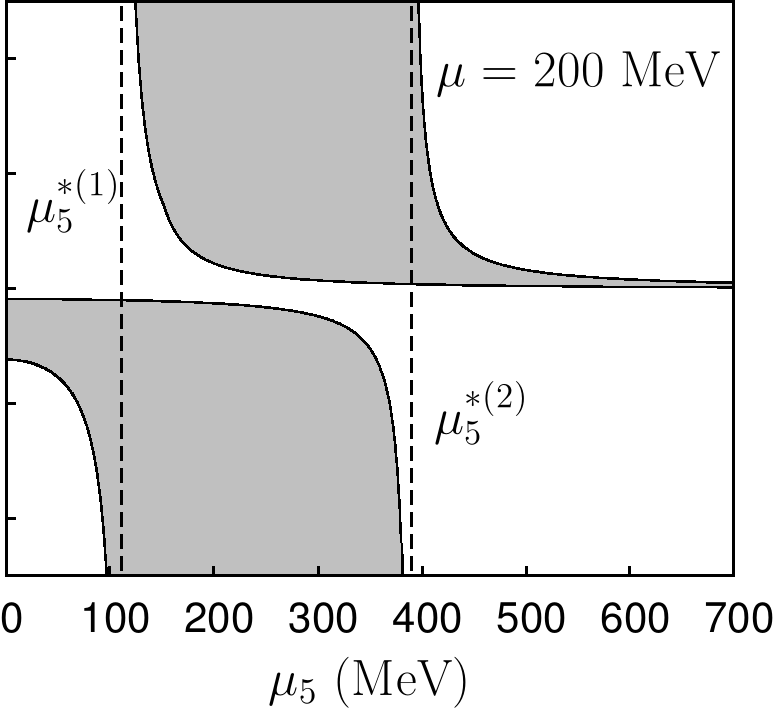}
\caption{Allowed region of $G_1$ as a function of $\mu_5$ with fixed $\mu$ for a stable CSB phase (dark region). The left panel shows $\mu=0$ while the right one corresponds to $\mu=200$ MeV. The figure corresponds to $m=0$ and $\Lambda=1$ GeV.}\label{CSB}
\end{figure}
In Fig. \ref{CSB} we show the region of $G_1$ that provides a stable CSB phase with $m=0$ for non-trivial values of the external drivers. All dimensional quantities scale with $\Lambda$, which we take to be $\Lambda=1$ GeV throughout.
Two discontinuities appear in the plot. The first one is found at
\[\left (\mu_5^{*(1)}\right )^2=\frac{\mu^2}2\left [\Theta(\mu-\mu^*)\left (1-\frac1{2\ln\frac{2\Lambda}\mu}\right )+\Theta(\mu^*-\mu)\frac1{\ln\frac\Lambda\mu}\right ]\]
with
\[\mu^*\equiv\exp\left [-\frac14\left (3-2\ln2+\sqrt{9+4\ln2+4\ln^22}\right )\right ]\Lambda\approx 0.265\Lambda,\]
while the second one can be written analytically only if $\mu<2\exp[-\frac14(1+\sqrt5)]\Lambda\approx 0.891\Lambda$. In this case, the second discontinuity is given by
\[\left (\mu_5^{*(2)}\right )^2=(3-\sqrt 5)\Lambda^2\exp\left [-\frac12\left (1+\sqrt 5\right )\right ]\approx (0.389\Lambda)^2.\]
For $\mu=0$ and $\mu=200$ MeV, the condition $\mu<0.891\Lambda$ is satisfied and the previous equation can be used to find the discontinuity, which is clearly independent of $\mu$. The limit $\mu\to 0$ reduces to $G_1<0$, a known result from a previous work on the NJL model in DR \cite{ikkk3}. Finally, note that the restriction for $G_2$ is simply $\frac{1}{G_2}>\frac{1}{G_1}$.

\medskip

A similar analysis could be done setting $m\neq 0$ but except for the complications appearing due to the inclusion of this extra parameter, no big differences would appear with respect to the curves presented in Fig. \ref{CSB}. Therefore, we omit the details of such calculation.

\medskip

Let us set $m\neq 0$ again. The meson spectrum for any value of the external chemical potentials is given by the second derivatives at the local minimum
\[\nonumber V_{\sigma\sigma}=\frac{m}{2G_1M}+2M^2L_1, \qquad V_{\eta\eta}=\frac{m}{2G_1M}+\frac12\left (\frac{1}{G_2}-\frac{1}{G_1}\right )\]
\[\nonumber V_{\pi\pi}=\frac{m}{2G_1M}, \qquad V_{aa}=\frac{m}{2G_1M}+\frac12\left (\frac{1}{G_2}-\frac{1}{G_1}\right )+2M^2L_1\]
\begin{equation}\label{spectr}
V_{\sigma\eta}=V_{\sigma\pi}=V_{\sigma a}=V_{\eta\pi}=V_{\eta a}=V_{\pi a}=0,
\end{equation}
where one has to use a bare quark mass $m$ of the same sign as the coupling $G_1$ so as to provide a positive pion mass. 

\medskip

The stability conditions read
\begin{equation}\label{condCSB}
\frac1{G_2}>\frac1{G_1}\left (1-\frac mM\right ),\qquad 2M^2L_1>\max\left[-V_{\pi\pi},-V_{\eta\eta}\right ].
\end{equation}
Let us set once again $\mu=\mu_5=0$. Then $L_1>0$ and the second covexity condition is always met if the first one is fulfilled. In this case the mass spectrum obeys the relation
\[m_\sigma^2-m_\pi^2=m_a^2-m_\eta^2>0\]
in analogy to the situation in QCD. In addition the following relation also holds
\[m_a^2-m_\sigma^2=m_\eta^2-m_\pi^2,\]
and the difference $m_\eta^2-m_\pi^2$ is positive (like the analogous one in QCD \cite{vw}) provided that $\frac{1}{G_2}-\frac{1}{G_1}>0$.

\medskip

\begin{figure}[h!]
\centering
\includegraphics[scale=0.28]{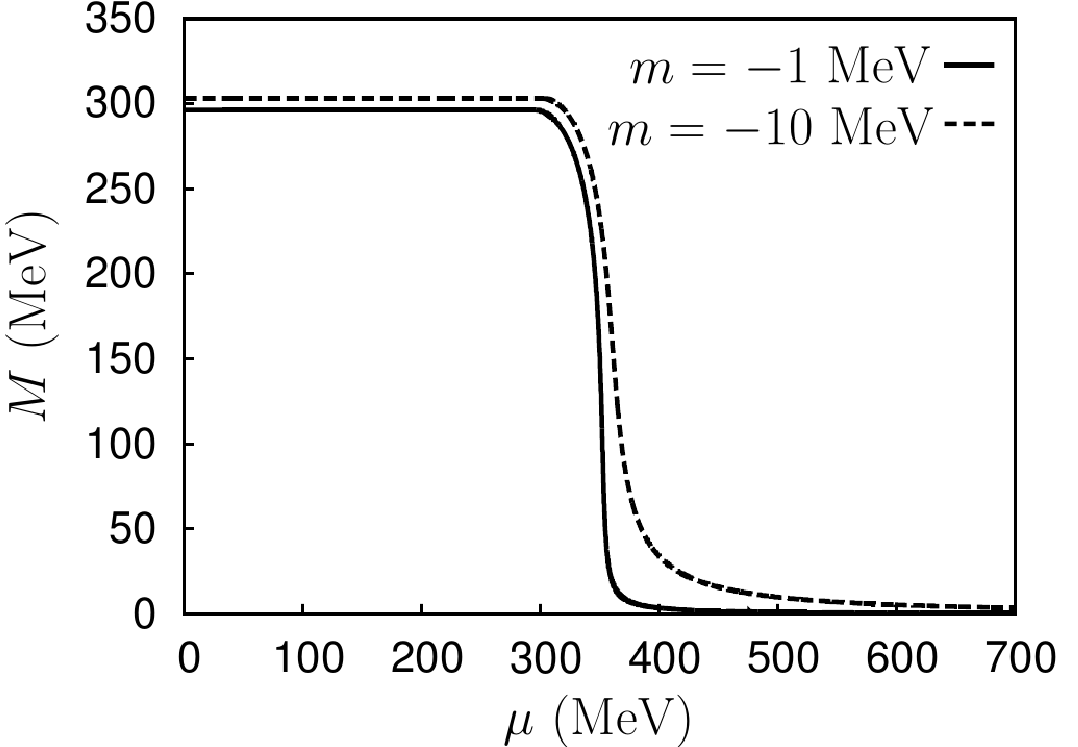}\;\;\includegraphics[scale=0.28]{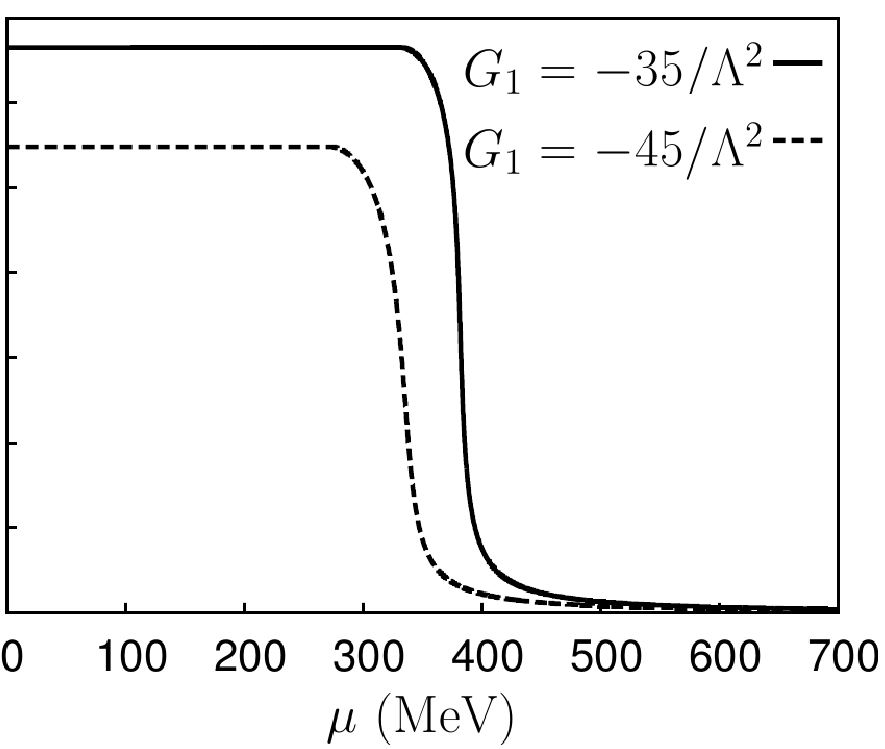}
\caption{Evolution of the constituent quark mass $M$ depending on $\mu$. For both plots we set $G_2=-45/\Lambda^2$ with $\Lambda=1$ GeV and $\mu_5=0$. In the left panel, we fixed $G_1=-40/\Lambda^2$ and plot for different values of $m$. In the right panel instead, we fixed $m=-5$ MeV in order to examine the variation of $G_1$. The transition becomes sharper as $m$ decreases.}\label{japos1}
\end{figure}

Let us now examine in detail the dependence of the chiral condensate on the external chemical potentials. In Figure \ref{japos1} we present the evolution of the constituent quark mass as a function of the vector chemical potential for different values of the current quark mass and coupling $G_1$ (left and right panels respectively) with $\mu_5=0$. Both the bare quark mass and the coupling $G_1$ are taken to be negative, as just explained above. There is chiral restoration around a certain value of the chemical potential that depends mostly on $G_1$; this phenomenon of chiral restoration is well known in the NJL model \cite{buba} and it is possibly the main reason that this simple model fails to reproduce correctly the transition to nuclear matter. The transition becomes sharper as the value $m=0$ is approached.

\medskip

\begin{figure}[h!]
\centering
\includegraphics[scale=0.45]{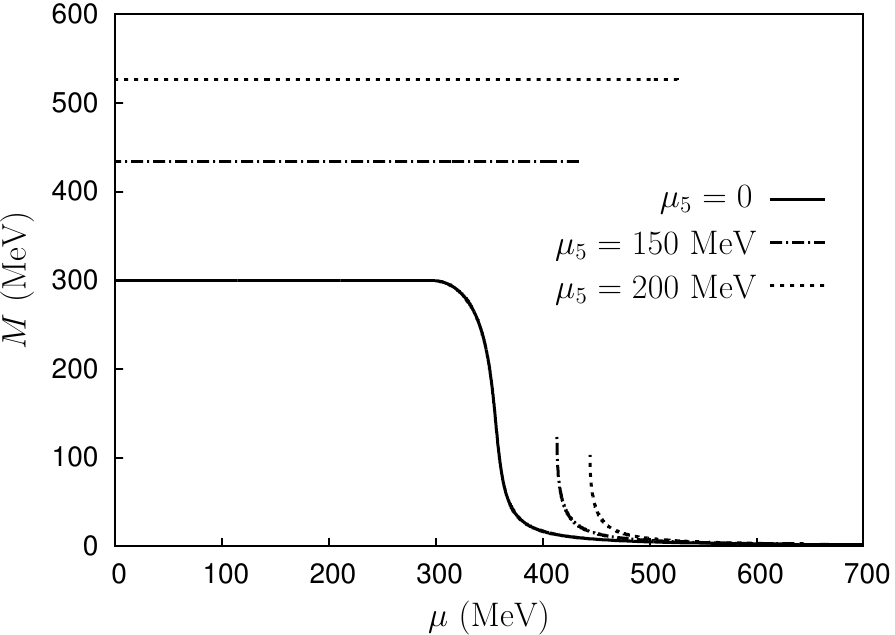}
\caption{Evolution of the constituent quark mass $M$ depending on $\mu$ for different values of the axial chemical potential $\mu_5$ setting $m=-5$ MeV, $G_1=-40/\Lambda^2$ and $G_2=-45/\Lambda^2$. The drawn lines correspond to locally stable phases and accordingly the absence of a continuous line in the cases where $\mu_5\neq0$ is due to the fact that the Hessian matrix is not positive definite. The transition to a chirally restored phase changes to a first order one as $\mu_5$ increases.}\label{japos2}
\end{figure}

In Fig. \ref{japos2} we observe the influence of the axial chemical potential $\mu_5$ on the restoration of chiral symmetry that always takes place in the NJL as $\mu$ increases. For high values of the axial chemical potential, the plateau appearing for $M>\mu$ acquires bigger values and spreads over a wider range of $\mu$. At some point, the solution of the gap equation shows a stable and a metastable solution that must necessarily flip thus implying a jump of the constituent quark mass at some value of the chemical potential where both solutions coexist. Between these solutions, another unstable solution exists, but is not shown in the plot since the Hessian matrix is not positive definite. The jump represents a first order phase transition from $\mu<M$ (=constant) to a non-constant $M$ smaller than the chemical potential.

\medskip

\begin{figure}[h!]
\centering
\includegraphics[scale=0.28]{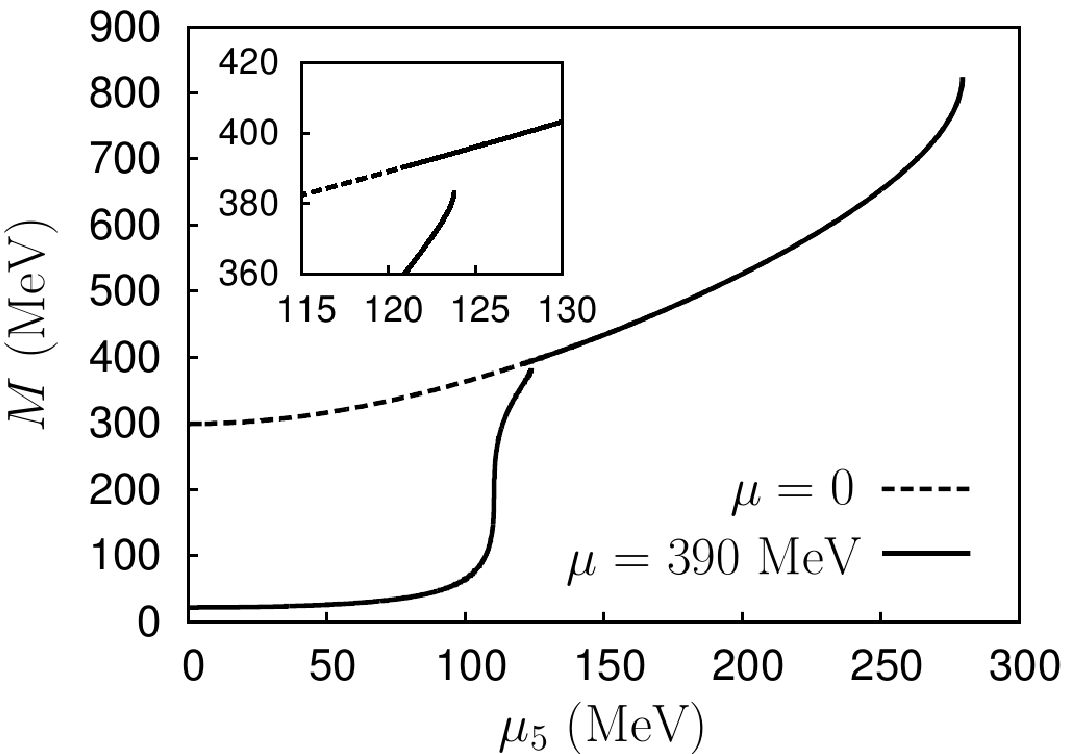}\;\;\includegraphics[scale=0.28]{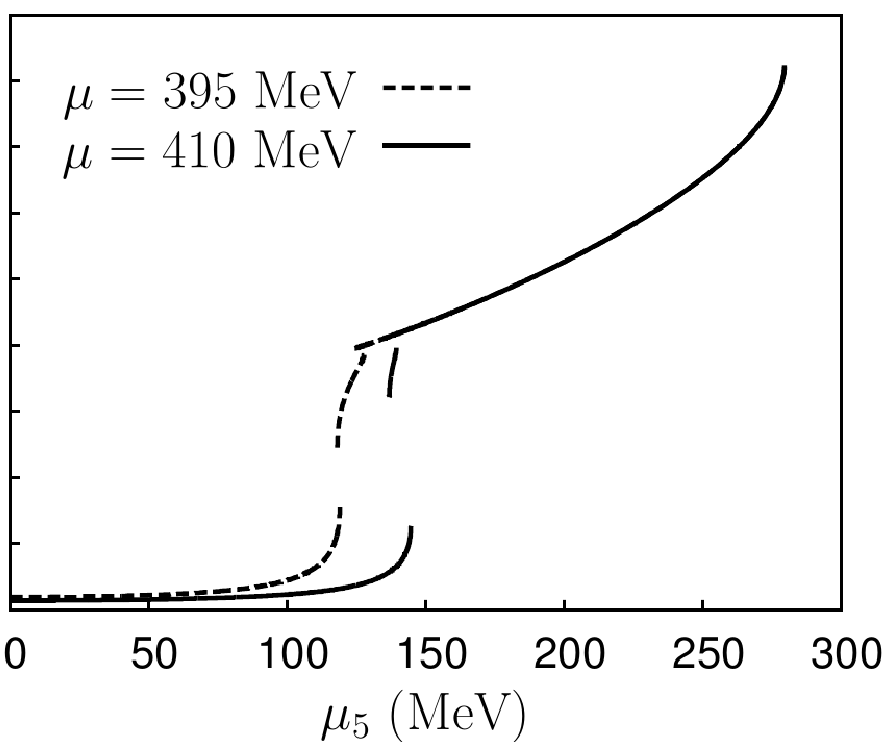}
\caption{Evolution of the constituent quark mass $M$ depending on $\mu_5$ for different values of the chemical potential $\mu$ setting $m=-5$ MeV, $G_1=-40/\Lambda^2$ and $G_2=-45/\Lambda^2$. Both graphics show the regions where all the second derivatives are positive. Certain values of $\mu_5$ exhibit coexisting solutions implying first order phase transitions. In the left panel, we show a plot for $\mu=0$ (or indeed for any $\mu<M$) and $\mu=390$ MeV. The second curve exhibits a small jump that is shown more detailed in the inset. The right panel corresponds to $\mu=395$ (two jumps) and 410 MeV (probably only one jump). This plot shows that the NJL with external drivers has a rather complex phase diagram.}\label{japos3}
\end{figure}

It may be helpful to show a plot of the same constituent quark mass depending on $\mu_5$ for different values of $\mu$. In the left panel of Fig. \ref{japos3} we display such evolution for $\mu=0$ and 390 MeV\footnote[2]{The solid and dashed lines were interchanged in \cite{njl}. This new version is corrected.}. The first curve is valid for any $\mu<M\approx 300$ MeV while the second one shows a small discontinuity that represents a first order phase transition within the CSB phase. A detail of the jump is presented in the inset. Note that both curves coincide after the jump and stop at $\mu_5\sim 280$ MeV since beyond this value, the phase becomes unstable, as presented previously in Fig. \ref{CSB}.

\medskip

In the right panel, we present the values of $\mu=395$ and 410 MeV, which correspond to qualitatively different cases. The curve for $\mu=395$ MeV shows two separate regions where the function is bivaluated. First, the lower and intermediate branches share some common values of $\mu_5$ even that it cannot be appreciated in the plot. Thus, a first order phase transition must take place within this region. The same behaviour happens for the intermediate and the upper branches, implying another first order phase transition. For bigger values of $\mu_5$ one recovers the tendency of $\mu=0$ as in the previous case. The curve $\mu=410$ MeV is somewhat similar to the previous one but now with a trivaluated region: for a certain small range of $\mu_5$ the three branches may be reached and therefore one or two jumps may take place. For bigger values of $\mu$, the intermediate branch disappears and only one jump may take place.

\medskip

All the jumps in Fig. \ref{japos3} are due to the presence of unstable regions that would connect the different branches of the same curve. Here, it can be shown that $V_{\sigma\sigma}<0$ is the responsible for these unstable zones. On the other hand, $V_{aa}$ is simply $V_{\sigma\sigma}$ with a positive shift and the restriction $V_{aa}>0$ does not add anything new.

\medskip

We want to stress that all the first order phase transitions just explained are a direct consequence of the addition of $\mu_5$ to the problem. No other assumptions are made beyond using the mean field approximation.

\section{Isosinglet pseudoscalar condensation and parity breaking}\label{NJL4}
Next we focus in the analysis of parity violating phases. It turns out that the only stable one corresponds to condensation in the isoscalar channel  \cite{njl}. Neutral pseudoscalar isotriplet condensation, either with or without CSB, does not lead to a stable termodynamical phase\footnote[3]{This is at variance with the QCD- inspired effective theory analysis of \cite{anesp} where the possibility of a condensation in the isotriplet channel was proven.}. Now, in addition to the scalar condensate $\sigma$ that was explored in the previous section we will allow for a non-vanishing isosinglet pseudoscalar condensate $\eta$. The gap equations now turn to be
\begin{equation}\label{PB}
M=\frac m{G_1}\frac1{\frac1{G_1}-\frac1{G_2}}, \qquad K_1=-\frac{1}{4G_2}.
\end{equation}
The first gap equation shows that the scalar condensate exhibits a remarkable independence on the external chemical potentials as it turns out to be constant once the parameters of the model are fixed. Unlikely the $\eta$ condensate does depend on the external drivers through the second equation. Moreover, from the first equation one finds that in the parity-breaking phase $m=0$ iff $G_1=G_2$; namely, the parity-breaking $\eta$ condensate is a stationary point of the effective potential \eqref{V} only when the chiral and $U(1)_A$ symmetries are explicitly preserved or broken at the same time in the NJL Lagrangian \eqref{lagNJL}. However this stationary point would not be a true minimum but a stationary point with two flat directions. The more general case where $m\neq 0$ and $G_1\neq G_2$ is thus the only possibility to have a genuine parity-breaking phase. We will see in a moment how as one takes the limit $m\to0$, the narrow window to have access to this phase disappears.

\medskip

The second derivatives read
\begin{gather}
\nonumber V_{\sigma\sigma}=\frac{1}2\left (\frac{1}{G_1}-\frac{1}{G_2}\right )+2M^2L_1, 
\qquad V_{\eta\eta}=2\eta^2L_1, \qquad V_{\sigma\eta}=2M\eta L_1\\
\nonumber V_{\pi\pi}=\frac{1}2\left (\frac{1}{G_1}-\frac{1}{G_2}\right )+2\eta^2L_1, 
\qquad V_{aa}=2M^2L_1, \qquad V_{\pi a}=2M\eta L_1\\
\nonumber V_{\sigma\pi}=V_{\sigma a}=V_{\eta\pi}=V_{\eta a}=0.
\end{gather}
We find that the Hessian matrix is not diagonal but has a block structure with two isolated sectors $\sigma-\eta$ and $\pi-a$ that reflect the mixing of states with different parity \cite{anesp,epjc}. This mixing will be carefully studied in Chapter \ref{ChSmodel}, where more degrees of freedom will be included. As already stressed, the focal point of this chapter is the analysis of the stable thermodynamic phases and therefore, the mixing effect will be omitted here. The determinants of these blocks are
\[\det(V^{\sigma,\eta})=\eta^2 L_1\left (\frac{1}{G_1}-\frac{1}{G_2}\right ), \qquad 
\det(V^{\pi, a})=M^2L_1\left (\frac{1}{G_1}-\frac{1}{G_2}\right ),\]
and thus, the resulting conditions for this phase to be stable reduce to
\begin{equation}\label{conditions}
L_1>0, \qquad \left (\frac{1}{G_1}-\frac{1}{G_2}\right )>0.
\end{equation}
The second of the previous conditions leads to a peculiar ordering of the physical meson spectrum. Recall that in the chiral symmetry breaking phase we had
\[m_a^2-m_\sigma^2=m_\eta^2-m_\pi^2=-\frac N2\left (\frac{1}{G_1}-\frac{1}{G_2}\right ),\]
and therefore, a stable parity-breaking phase is not compatible with a fit to the phenomenology. Thus parity breaking in the NJL model corresponds to a choice of parameters that makes this model quite different from QCD predictions \cite{vw}. In other words, the NJL model with a stable parity-breaking phase will have nothing to do with QCD. Note that the above differences are independent of the external drivers $\mu$ and $\mu_5$.

\medskip

The rest of the possible phases with a vanishing $a$ require $m=0$ to satisfy the gap equations; they are not true minima. 
In particular, there is no phase with parity breaking and $\sigma=0$.

\subsection{Transition to the parity-breaking phase}
In this section we will analyse the characteristics of the transition to the phase where parity is broken. First of all, let us define $M_0$ as the solution to $M_0=M(G_1,\mu=\mu_5=0)$ in the CSB phase given by Eq. \eqref{chiSB}. Recall the inequality $V_{\eta\eta}>0$ of the same phase given in Eq. \eqref{spectr} and the stability condition of the parity-breaking phase in Eq. \eqref{conditions}. Putting all of them together yields the following inequalities
\[0<\frac1{G_1}-\frac1{G_2}<\frac m{G_1M_0}.\]
The second inequality can be inserted in the first gap equation of the parity-breaking phase (see Eq. \eqref{PB}) to show that in this phase, $M>M_0$. The same set of inequalities can be rewritten as
\begin{equation}\label{window}
\frac1{G_1}\left (1-\frac{m}{M_0}\right )<\frac1{G_2}<\frac1{G_1},
\end{equation}
which means that $G_1$ and $G_2$ necessarily have the same sign, while in the CSB phase $G_2$ had no restriction and could have opposite sign. This set of inequalities represents the necessary condition to have a transition from the CSB to a parity-breaking phase, as they provide the stability conditions of both phases. Notice that the model allows a narrow window of $G_2$ (once $G_1$ is fixed) so that both phases may take place depending on the value of the external drivers. In the limit $m\to0$, this window closes and no parity breaking can be found. 

\medskip

Let us recall the gap equation in the CSB phase Eq. \eqref{chiSB} and assume $\mu=\mu_5=0$. Provided that Eq. \eqref{window} is satisfied, it follows that
\[K_1=-\frac1{4G_1}\left (1-\frac{m}{M_0}\right )>-\frac1{4G_2}.\]
In the parity-odd phase, the gap equation is $K_1=-\frac1{4G_2}$. In order to reach this phase from the parity-even one, $K_1$ has to decrease as a function of the external chemical potentials. From \eqref{chiSB} this statement is equivalent to an increasing chiral condensate $M(\mu,\mu_5)>M_0$ in the parity-even phase. All this discussion is in agreement with the fact that $M>M_0$ in the parity-odd phase, as we explained above.

\medskip

Let us point out the fact that the condition $L_1>0$ from the parity broken phase in Eq. \eqref{conditions} is stronger than the one from the CSB one in Eq. \eqref{condCSB} so the former will remain to provide stability to both phases. Let us describe how the phase transition takes place first for $\mu=0$ and finally for $\mu\neq 0$.

\subsection{Phase transitions with $\mu=0$}
Let us simplify the analysis by setting $\mu=0$ and let us study the dependence on $\mu_5$, which makes $M$ increase from its initial value $M_0$. At some critical value such that
\begin{equation}
M^c\equiv M(\mu_5^c)=\frac m{G_1}\frac1{\frac1{G_1}-\frac1{G_2}},
\end{equation}
where the critical value of the axial chemical potential is
\[(\mu_5^c)^2=\frac{M_c^2}2-\frac1{4\log \frac {2\Lambda}{M_c}}\left (M_c^2-\frac{\pi^2}{G_2}\right ),\]
$m_\eta$ vanishes, and from here on, we get into the parity-breaking phase via a second order phase transition, where $M$ remains frozen as discussed while the dependence on $\mu_5$ is absorbed into a non-vanishing $\eta$ condensate. The dependence of $K_1$ on $M_\pm^2$ will be now on $M_c^2+\eta^2$. Note that $(\mu_5^c)^2>0$ and therefore, a threshold in $M^c$ follows.

\begin{figure}[h!]
\centering
\includegraphics[scale=0.45]{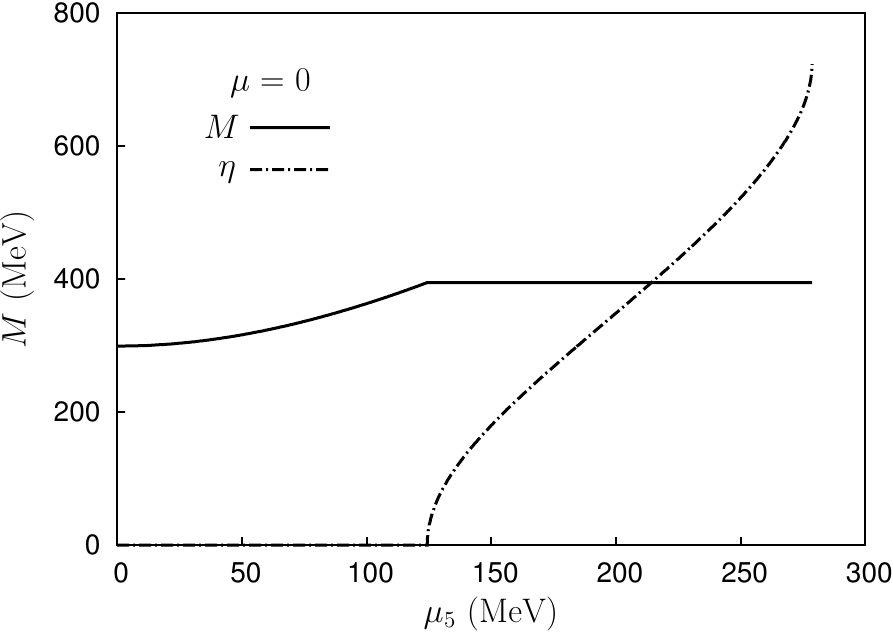}
\caption{$M$ and $\eta$ dependence on $\mu_5$ for $\mu<M_0$, $G_1=-40/\Lambda^2$, $G_2=-39.5/\Lambda^2$, $m=-5$ MeV and $\Lambda=1$ GeV.}\label{M,eta(mu5,mu=0)}
\end{figure}

In Fig. \ref{M,eta(mu5,mu=0)} we present a plot showing the evolution of $M$ and $\eta$ with respect to $\mu_5$ for $\mu=0$ (or any $\mu<M_0\approx 300$ MeV). In the CSB phase $M$ grows with $\mu_5$ up to the critical value $M^c$, the point where this magnitude freezes out and $\eta$ condensates acquiring non-trivial values, also growing with the axial chemical potential. At $\mu_5\simeq 0.28\Lambda$, this phase shows an endpoint and beyond, no stable solution exists. This point is the same one that we found in the CSB phase, meaning that the model becomes unstable at such value of $\mu_5$, no matter which phase one is exploring.

\subsection{Phase transitions with $\mu>0$} 
\begin{figure}[h!]
\centering
\includegraphics[scale=0.28]{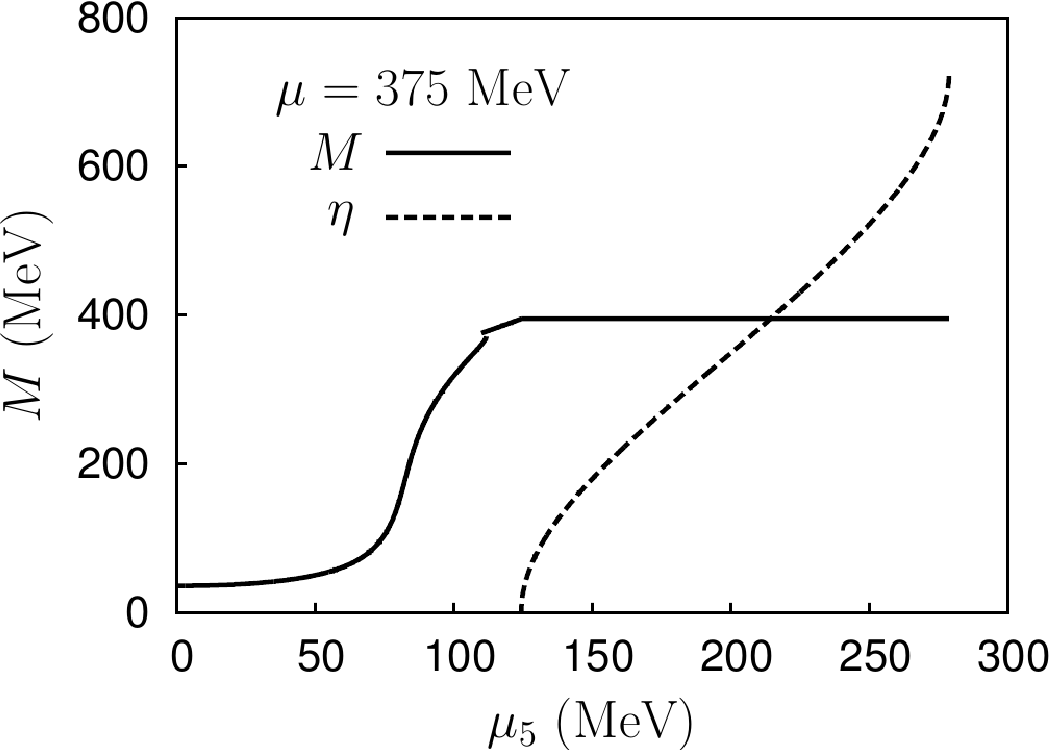}\;\;\includegraphics[scale=0.28]{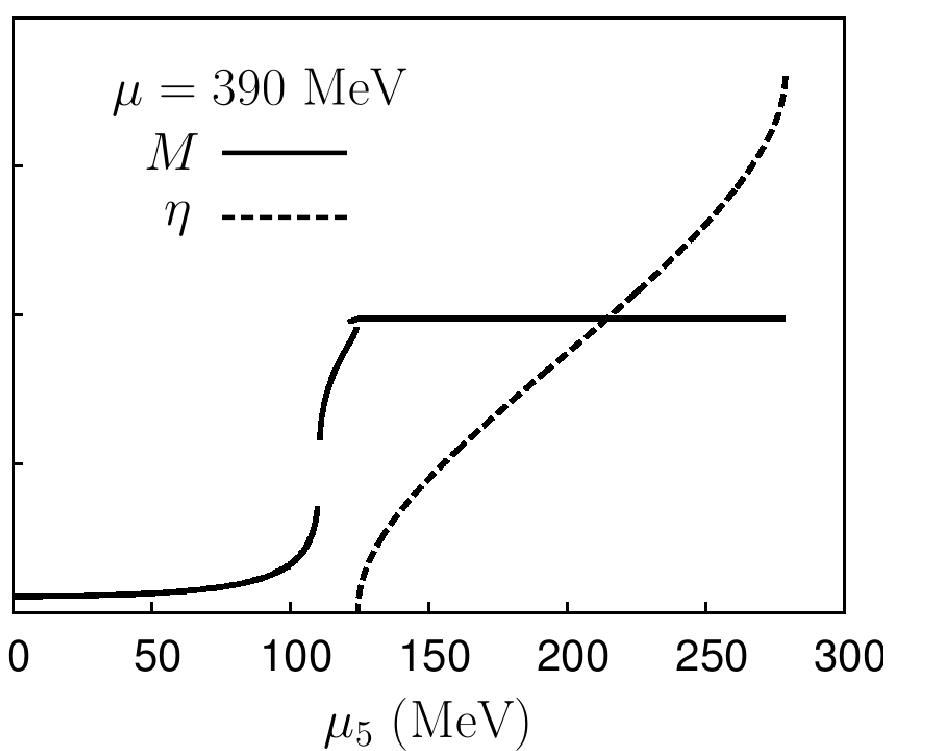}
\includegraphics[scale=0.28]{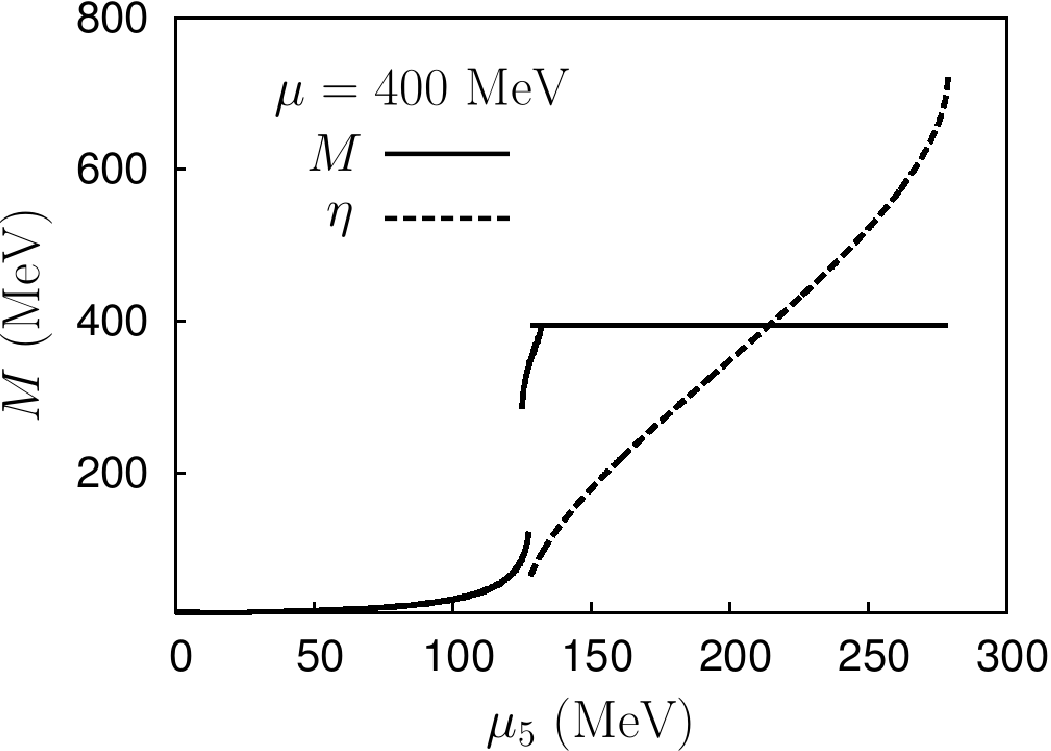}\;\;\includegraphics[scale=0.28]{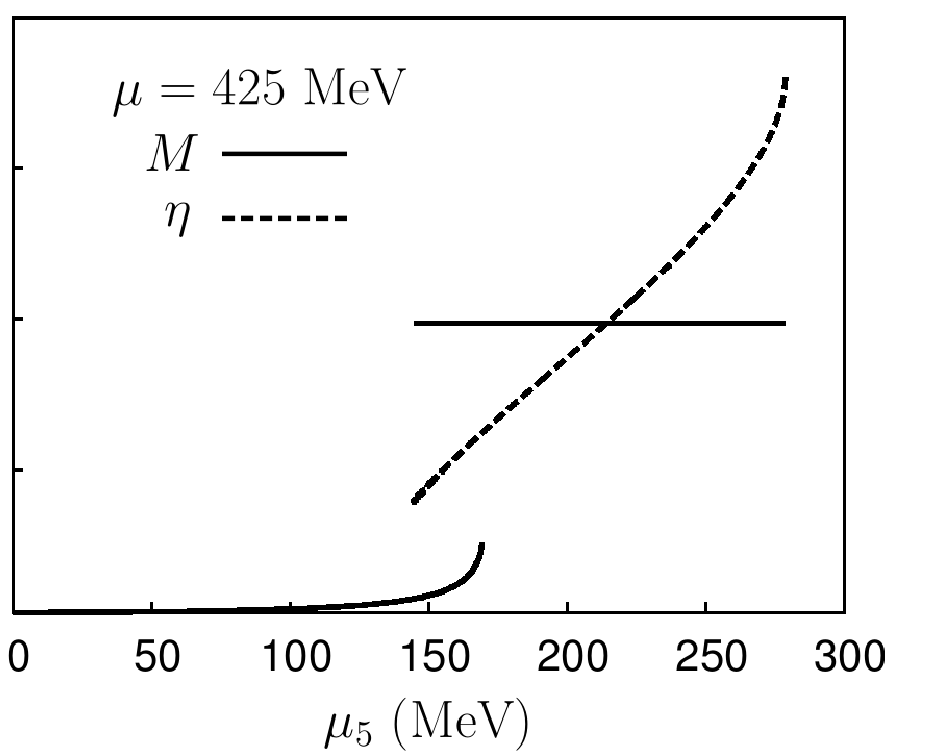}
\caption{$M$ and $\eta$ dependence on $\mu_5$ for $\mu=375$, 390, 400 and 425 MeV, $G_1=-40/\Lambda^2$, $G_2=-39.5/\Lambda^2$, $m=-5$ MeV and $\Lambda=1$ GeV. The graphics show the regions where all the second derivatives are positive. Certain values of $\mu_5$ exhibit coexisting solutions implying first order phase transitions. The first jump in the plot for $\mu=390$ MeV shows a very small region of $\mu_5$ where the function is not defined. This region is characterised by $V_{aa}<0$, thus suggesting a phase with a non-trivial scalar isotriplet condensate. This is the only region where we have found indications for a phase with $a\neq 0$. The landscape of first order phase transitions in the constituent quark mass is essentially the same as the one explained in Fig. \ref{japos3}.}\label{mu-evolut}
\end{figure}

The presence of both chemical potentials makes the function $K_1$ exhibit more complicated features. As $K_1$ decreases with $\mu_5$ and $\mu$ does the opposite job, $\mu_5$ needs larger values than $\mu$ to reach the parity-breaking phase. In Fig. \ref{mu-evolut}, we present a set of plots with the evolution of both $M$ and $\eta$ for non-vanishing values of the chemical potential. As before we take the value $\Lambda=1$ GeV to make the model in order to have some QCD-inspired intuition. Of course everything scales with $\Lambda$.

\medskip

In the upper panels, we set $\mu=375$ MeV (left) and $\mu=390$ MeV (right), both of them $M_0<\mu<M^c$, where jumps in $M$ are observed in the parity even phase together with tiny metastable regions. This behaviour is very similar to the one described in Fig. \ref{japos3} with the subtlety that we inverted the sign of $\frac1{G_1}-\frac1{G_2}$ and therefore, the parity-odd phase may be reached.

\medskip

In addition, this change of sign shifts the second derivative $V_{aa}$, which is the only responsible for the apparent big jump in the $\mu=390$ MeV window (the one with lower $\mu_5$). $V_{aa}$ becomes negative while all the other derivatives remain positive. If for a moment we forgot $V_{aa}$, the curve would be smoothly increasing and we would only have the other tiny jump close to the flat region of constant $M$, as in the CSB case. However, the fact that this second derivative becomes negative leads to a small range of $\mu_5$ where no stable solution exists. Hence, it seems natural to think that the system goes away from the phase with $a=0$ and acquires a non-trivial scalar isotriplet condensate. We emphasize this region is really tiny and depends crucially on the specific values for the parameters, even disappearing for $G_1>-30/\Lambda^2$. Both graphics show a smooth transition to the parity-odd phase, say, via a second order phase transition with the same characteristics of the previous section with $\mu=0$.

\medskip

In the lower panels of Fig. \ref{mu-evolut} we set $\mu=400$ MeV (left) and $\mu=425$ MeV (right) with $\mu>M^c$ and observe what we could more or less expect from Fig. \ref{japos3} with the same landscape of first order phase transitions in the parity-even phase. The main difference of these two latter values appears in the finite jump of $\eta$, implying now a first order phase transition towards the parity-breaking phase.

\medskip

\begin{figure}[h!]
\centering
\includegraphics[scale=0.45]{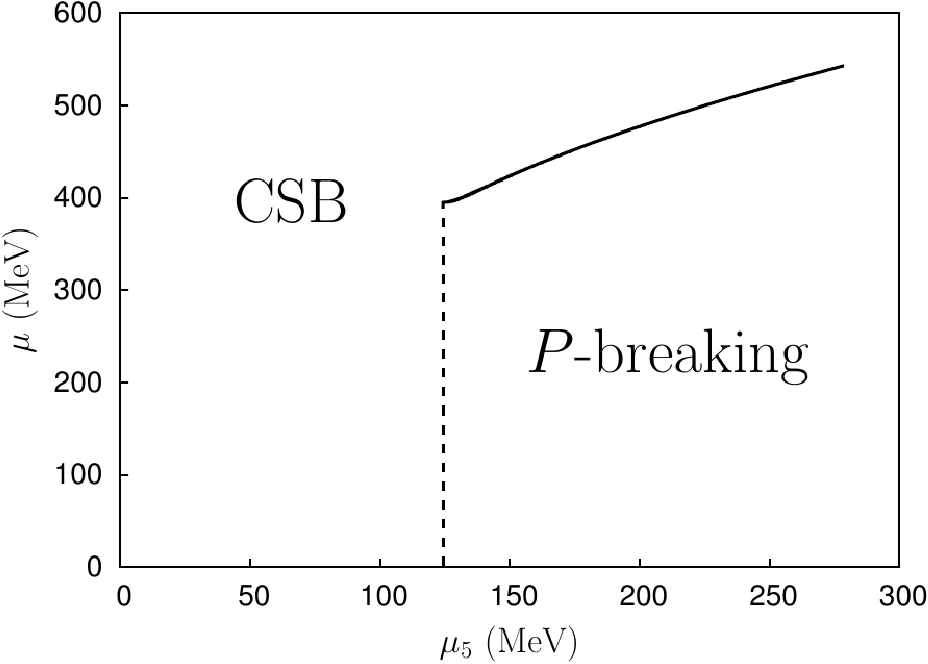}
\caption{Transition line from the CSB to the $P$-breaking phase with $G_1=-40/\Lambda^2$, $G_2=-39.5/\Lambda^2$, $m=-5$ MeV and $\Lambda=1$ GeV. The vertical dashed line is related to a second order phase transition while the solid one corresponds to a first order one.}\label{mu5-mu}
\end{figure}

Finally, we present the phase transition line in a $\mu^c(\mu_5^c)$ plot in Fig. \ref{mu5-mu}. For $\mu<M^c\approx 395$ MeV (or equivalently, for $\mu_5^c=\mu_5^c(\mu=0)$), the transition is smooth (second order) while beyond that there is a jump in the condensates (first order), as it was also observed in the previous figure.

\section{Summary}

In this chapter we worked in the continuum and explored in detail the different phases that arise in the Nambu--Jona-Lasinio model in the presence of both vector and axial chemical potentials at zero temperature. The axial chemical potential changes considerably the thermodynamical properties of the model. Interestingly, when the full $U(2)_L\times U(2)_R$ global symmetry is broken to $SU(2)_L\times SU(2)_R \times U(1)_V$ (i.e. $G_1 \neq G_2$) a phase where parity is spontaneously broken by the presence of an isosinglet condensate $\eta$ appears. However, we have not found any phase where parity and flavour symmetry are simultaneously broken thus indicating the presence of a non-zero value for $\langle \bar\psi \gamma_5 \tau^3\psi\rangle$. On the contrary we have found an extremely small region in the space of parameters where flavour symmetry is broken by a non-zero value of $\langle \bar\psi\tau^3\psi\rangle$ but parity is not broken yet.

\medskip

Demanding stability of such a phase however leads to a region of the parameter space where the spectrum has little resemblance to QCD. We have investigated all the properties of the transition from the parity-even to the parity-odd phase providing results on the evolution of both condensates, which exhibit finite jumps under certain conditions, and finally examining the phase transition line, where it was shown that for $\mu<M^c$ we have a 2nd order transition while for $\mu>M^c$, it corresponds to a 1st order one.

\medskip

The discussion presented here on the phase structure of the NJL model in the presence of external chemical potentials is rather general and, as discussed above, the model -in spite of its simplicity- exhibits an enormously rich phase structure. 	
\chapter{An effective theory for spin 0 mesons in a parity-breaking medium}\label{ChSmodel}

\graphicspath{{4-Sigmamodel/figures/}{figures/}}

After a careful study of a phase where parity breaks spontaneously in the NJL model due to the presence of $\mu_5\neq0$ (see also \cite{anesp}), it seems rather realistic to assume that such phase may really exist in QCD. If this is the case, the order of magnitude of the axial chemical potential reached in HIC should be similar to the one explored in the previous chapter in order the new phase to be thermodynamically stable. Another naive estimate would be assuming simply that $\mu_5\sim\tau_\text{heating}^{-1}$ so a formation time of the fireball $\tau_\text{heating}\sim 0.5-1$ fm leads to $\mu_5\sim 200-400$ MeV. On the other hand, we can also estimate $\mu_5$ by assuming that the change in the free energy has to be of order $f_\pi$ and the average change in the quark axial charge $\langle \Delta Q_5^q\rangle\sim 1$. Then, assuming a linear response, $\mu_5 \sim f_\pi\simeq 100$ MeV. Therefore, in the rest of this thesis we shall assume $\mu_5$ to take values in the range of hundreds of MeV. Note that this parameter may take different values from collision to collision so the magnitude we are dealing with has to be understood as an effective one. In any case, the dependence of the results on $\mu_5$ will be discussed everywhere it is needed. 

\medskip

It seems interesting to investigate how hadronic physics is modified by the presence of $\mu_5$. In this chapter we will address the treatment of the spin zero resonances while we postpone to Chapter \ref{ChVMD} the case of vector mesons.

\medskip

The appearance of an axial chemical potential as a consequence of local large topological fluctuations in a hot medium constitutes an explicit source of parity violation. It generates new in-medium effects on resonances that can be unambiguously predicted in terms of the axial chemical potential alone. As introduced in the previous chapter if parity breaks, the regular classification of particles into scalars and pseudoscalars is meaningless. The consequences are far reaching; parity ceases to be a guidance for allowing/forbidding strong interaction processes, and states that have opposite parities, but otherwise equal quantum numbers, mix. In this scenario two new processes are then likely to be most relevant inside the fireball thermodynamics: the decays $\eta,\eta^\prime \to \pi\pi$ that are strictly forbidden in QCD on parity grounds.

\medskip

The fireball created from a HIC is usually treated as a pion gas. The $\rho(770)$ meson plays a significant role in this context as its large width $\Gamma_{\rho\to\pi\pi}\simeq 150$ MeV allows it to decay and regenerate in-medium several times. As a consequence, this channel has been deeply analysed \cite{rapp} and suggested to reach thermal equilibrium within the pion gas in experiments like RHIC or LHC. In a similar way, the new decays $\eta,\eta^\prime \to \pi\pi$ in a parity-odd medium could eventually lead to large widths and a thermalization of the corresponding channels. In this chapter we shall address a detailed study of this effect.

\medskip

In order to introduce the axial chemical potential to an effective theory we have to recall that, just as the ordinary baryonic potential is introduced as the zero-th component of a spurious vector field, the axial chemical potential $\mu_5$ can be implemented as the time component of a fictitious axial-vector field. We follow the arguments for strange quark suppression already mentioned in Chapter \ref{Chfase} of parity-breaking effects in a finite volume. Let us recall that according to Eq. \eqref{pcac}, due to the unavoidable left-right oscillations the mean value of strange quark axial charge is washed out as the strange quark mass is comparable with the decay width of fireballs. Therefore, we will use axial chemical potential in the non-strange sector only.

\medskip

At the level of an effective meson Lagrangian, $\mu_5$ will appear using the spurion technique through the action of the covariant derivative
\begin{equation}\label{covderiv}
\partial_\mu \to
D_\mu=\partial_\mu -i\{\textbf{I}_q\mu_5\delta_{\mu 0},\cdot\}=\partial_\mu -2i\textbf{I}_q \mu_5\delta_{\mu 0}
\end{equation}
in the same way an external axial-vector field would be introduced through $D_\mu=\partial_\mu-i\{A_\mu,\cdot\}$. In the case one considered an isotriplet $\mu_5$, the non-strange $\textbf{I}_q$ should be replaced by the neutral Pauli matrix $\tau^3$ and the anticommutator structure should be preserved due to the non-trivial flavour structure of a general lagrangian.

\medskip

To get a rough estimate of the relevance of the previously forbidden processes $\eta,\eta^\prime \to \pi\pi$ we introduced $\mu_5$ to the Chiral Lagrangian. However, it can be checked that there is no contribution from dimension two operators, but dimension four terms (see e.g. \cite{chpt}) which lead to
\begin{equation}
\sim \frac{16\mu_5}{F_\Pi f_\pi^2} L \, \partial \eta \, \mbox{Tr}\left( \partial\hat\pi \partial\hat\pi\right);
\quad \Pi = \eta,\, \eta',
\end{equation}
where $L$ is a combination of the Gasser-Leutwyler constants $L_{1,2,3} \sim 10^{-3}$. In order to obtain a numerical estimate we take the average pion momenta to be $\sim m_\eta/2$. The effective coupling constant affecting this operator can be estimated to be $\sim 0.4$, which is large enough to induce substantial $\eta$ meson regeneration in the hot pion gas. A very rough estimate of the partial width for the exotic process under discussion gives $\Gamma_{\eta\to\pi\pi}\simeq 100\,\text{MeV} $, when we assume $\mu_5\simeq 150$ MeV, to be compared for instance to $\Gamma_{\rho\to\pi\pi}\simeq 150$ MeV. Clearly if $\rho$ mesons are in thermal equilibrium in the pion bath, so will the $\eta$. A similar analysis leads to an even larger width for the $\eta^\prime$ due to the above parity-breaking operator.

\medskip

However at this point, one should be aware that the previous estimate using the Chiral Lagrangian may not be reliable at all because the decay width grows quadratically with the axial chemical potential and a slight increase of $\mu_5$ would lead to abnormally large widths and a violation of unitarity. Thus, $\mu_5$ has to be understood as a perturbatively small parameter. In addition, one expects substantial mixing with the scalar partners of $\eta,\eta^\prime$ (i.e. $\sigma$, $f_0$, glueballs, etc.) with comparable masses. For this reason, we will build a more sophisticated effective theory in order to investigate this effect. In the next section, we introduce a generalised sigma model with the lightest meson states with different parities to account for the mixing due to the parity violation. 

\medskip

This chapter is organized as follows. In Sec. \ref{Smod2} a generalised sigma model is presented. Mass-gap equations for three v.e.v. of neutral scalar fields are derived and solved analytically. Then we determine the best fit of parameters of the model comparing with the experimental inputs for scalar and pseudoscalar meson spectral data \cite{PDG}. In Sec. \ref{Smod3} the axial chemical potential is introduced. We obtain the modification of the mass-gap equations and find the distortion of $a_0$- and $\pi$- meson spectra caused by the parity breaking. In Sec. \ref{Smod4} a more complicated mixing of three meson states, $\sigma$, $\eta$, $\eta'$, is investigated when the medium has an axial charge. At certain energies some particle states become tachyons (recall we are in-medium so this actually does not represent a fundamental problem). In Sec. \ref{Smod5} all decay widths are calculated for both the rest frame and for moving particles (we note that when an axial charge fills the medium, the Lorentz invariance is broken). In Sec. \ref{Smod6} the problem of isosinglet axial-vector meson condensation and its interference with the axial chemical potential is examined.

\section{Generalised sigma model}\label{Smod2}
Our starting point will be the following Lagrangian, invariant under $SU(3)_F$ \cite{epjc}
\begin{align}\label{lageff}
\nonumber \mathcal L=&\frac 14 \text{Tr}\left (D_\mu HD^\mu H^\dagger\right )+
\frac b2 \text{Tr}\left [M(H+H^\dagger)\right ]+
\frac{M^2}2 \text{Tr}\left (H H^\dagger\right ) \\
\nonumber &-\frac{\lambda_1}2 \text{Tr}\left [(HH^\dagger)^2\right ]-\frac{\lambda_2}4 \left [\text{Tr}\left (H H^\dagger\right )\right ]^2 +\frac c2 (\text{det}H + \text{det}H^\dagger)\\
& +
\frac{d_1}2 \text{Tr}\left [M(H H^\dagger H+H^\dagger H H^\dagger)\right ]+
\frac{d_2}2 \text{Tr}\left [M(H+H^\dagger)\right ] \text{Tr}\left (H H^\dagger\right )
\end{align}
where $H=\xi \Sigma\xi$, $\xi=\exp\left(i\frac{\Phi}{2f}\right )$, $\Phi=\lambda^a\phi^a$ and $\Sigma=\lambda^b\sigma^b$. This model can be confronted to similar models in \cite{boguta,shechter,rischke} with the important difference (see below) in the trilinear vertices with couplings $d_1, d_2$. These interaction terms are needed if a fit to phenomenology wants to be performed accurately. The piece with the determinant is responsible for providing the breaking of $U(1)_A$ and a non-degenerate mass to $\eta$ and $\eta'$ in the chiral limit. The neutral v.e.v. of the scalars are defined as $v_i=\langle\Sigma_{ii}\rangle$ where $i=u,d,s$, and satisfy the following gap equations:
\begin{equation}
M^2v_i-2\lambda_1 v_i^3-\lambda_2\vec v^2v_i+c\frac{v_uv_dv_s}{v_i}=0.
\end{equation}
Let us stress that the parameter $f$, which is commonly identified with the pion decay constant, appears attached to the fields throughout and therefore, it will be absorbed in the field renormalization. Here the role of the pion decay constant is assigned to the chiral quark condensates $v_i$.

\medskip

For further purposes we need the non-strange meson sector and $\eta_s$. In terms of v.e.v. and physical scalar and pseudoscalar states, the parametrization used here is
\begin{equation}\label{Phi}
\Phi=\begin{pmatrix}
\eta_q+\pi^0 & \sqrt{2}\pi^+ & 0\\
\sqrt 2 \pi^- & \eta_q-\pi^0 & 0\\
0 & 0 & \sqrt 2\eta_s
\end{pmatrix}, \qquad \Sigma=\begin{pmatrix}
v_u+\sigma+a_0^0 & \sqrt 2 a_0^+ & 0\\
\sqrt 2 a_0^- & v_d+\sigma-a_0^0 & 0\\
0 & 0 & v_s
\end{pmatrix}.
\end{equation}
In the scalar sector, besides considering different flavour vacuum expectation values, we included in our model the isosinglet $\sigma$ meson and the isotriplet $a_0(980)$. In the pseudoscalar sector, we introduced pions and the strange and non-strange components of the $\eta$ and $\eta'(958)$ mesons. $\eta_q$ corresponds to the same degree of freedom that we introduced in the NJL model in Chapter \ref{ChNJL}. That description was rather poor for an accurate fit to phenomenology since the strange sector was not accounted for. Therefore a more precise phenomenological treatment was postponed to this section. Indeed $\eta_q$ and $\eta_s$ are not mass eigenstates of the strong interaction as they appear to be mixed mainly due to the determinant term in \eqref{lageff}. However, dealing with these states will simplify considerable our analysis. The mixing among them leading to the genuine mass eigenstates $\eta$ and $\eta'$ is defined by a rotation of angle $\psi$
\begin{equation}
\begin{pmatrix}
\eta_q\\
\eta_s
\end{pmatrix}=\begin{pmatrix}
\cos\psi & \sin\psi\\
-\sin\psi & \cos\psi
\end{pmatrix}\begin{pmatrix}
\eta\\
\eta'
\end{pmatrix}.
\end{equation}

\medskip

As stated in Chapter \ref{Chfase}, an axial charge can only be effectively generated for nearly massless quarks. Due to the non-negligible strange quark mass, even that an axial charge is generated in the strange sector from a HIC, it will not be transmitted to hadron physics. Therefore we exclude kaons and their scalar partners $\kappa$ from the analysis as they will not be affected by a non-strange $\mu_5$. Let us take for the time being $\mu_5=0$ and assume $v_u=v_d=v_s=v_0\equiv f_\pi\approx 92$ MeV because we only consider the effect of masses perturbatively. As a function of the Lagrangian parameters, the main physical magnitudes derived from this model are
\begin{align}
\nonumber v_0&=\frac{c\pm\sqrt{c^2+4M^2(2\lambda_1+3\lambda_2)}}{2(2\lambda_1+3\lambda_2)},
\qquad m^2_\pi= \frac 2{v_0}(b+(d_1+3d_2)v_0^2)m,\\
\nonumber m^2_a&=2(-M^2+ 3(2\lambda_1+\lambda_2)v_0^2+cv_0-2(3d_1+2d_2)mv_0-2d_2m_sv_0),\\
\nonumber m^2_\sigma&= 2(-M^2+(6\lambda_1+7\lambda_2)v_0^2-cv_0-6(d_1+2d_2)mv_0-2d_2 m_sv_0),\\
\nonumber m^2_{\eta,\eta'}&=\frac{m^2_\pi}{2m}(m+m_s)+3cv_0\mp\sqrt{8c^2 v_0^2+\left [\frac{m^2_\pi}{2m}(m-m_s)+cv_0\right ]^2},\\
\nonumber \Gamma^2_a&=\frac{(-4m^2_\eta m^2_\pi+(-m^2_a+m^2_\eta+m^2_\pi)^2)(m^2_a-m^2_\eta +4d_1mv_0)^4}{(48 m^3_a \pi v_0^2)^2},
\end{align}
\begin{align}
\nonumber \Gamma^2_\sigma&=\frac{9(m^2_\sigma-4 m^2_\pi)(m^2_\sigma-m^2_\pi+4(d_1+2d_2)mv_0)^4}{(32m^2_\sigma \pi v_0^2)^2},\\
\sin(2\psi)&=\frac{4\sqrt{2}cv_0}{m_{\eta'}^2-m_\eta^2},
\end{align}
with $m/m_s=(m_u+m_d)/(2m_s)\simeq 1/25$. The angle $\psi$ is not really necessary for our subsequent discussion but we can use it as a test of the model. $v_0$ is found via gap equations. These equations are inserted in MINUIT in order to find the minimum of the $\chi^2$ estimator using the following experimental numbers (in MeV):
\begin{gather}
\nonumber v_0^{\text{exp}}=92\pm 5, \quad m_\pi^{\text{exp}}=137\pm 5, \quad m_a^{\text{exp}}=980\pm 50,\\
\nonumber m_\sigma^{\text{exp}}=600\pm 120, \quad m_\eta^{\text{exp}}=548\pm 50, \quad m_{\eta'}^{\text{exp}}=958\pm 100,\\
\Gamma_a^{\text{exp}}=60\pm 30, \quad \Gamma_\sigma^{\text{exp}}=600\pm 120.
\end{gather}
The $\sigma$ mass is assumed to be relatively large as we are not considering any glueballs, so this $\sigma$ is a sort of average of the real $\sigma$ and other $0^+$ light states. We have assigned generous error bars to include the uncertainties in the model itself. In the minimization process there are several control variables:
\begin{itemize}
\item The value of the potential at the minimum has to be $V(v_0)<0$ since $v=0$ is an extremal point with $V(v=0)=0$ but in the case the latter is a minimum, the true vacuum has to have a lower energy. Also, there is a control of the third extremal, which has to be higher than the true minimum, given by
\begin{equation}
V(v_0)=\frac 14v_0^2(-6M^2+3(2\lambda_1+3\lambda_2)v_0^2-4cv_0).\nonumber
\end{equation}

\item The Hessian matrix has degenerate eigenvalues and there are only two different eigenvalues whose positivity should be provided
\begin{align}
(V'')_1(v_0)&=-M^2+3(2\lambda_1+\lambda_2)v_0^2+cv_0,\nonumber\\
(V'')_2(v_0)&=-M^2+3(2\lambda_1+3\lambda_2)v_0^2 -2cv_0\nonumber.
\end{align}
\end{itemize}
The final result of the minimization process is given in Table \ref{table}.
\medskip
\begin{table}\centering
\begin{tabular}{cccc}
\hline
Magnitude & \texttt{MINUIT} value (MeV) & Experimental value (MeV) & Error\\\hline
$v_0$ & 92.00 & 92 & $-3.52\times 10^{-7}$\\
$m_\pi$ & 137.84 & 137 & $+6.10\times 10^{-3}$\\
$m_a$ & 980.00 & 980 & $-1.26\times 10^{-6}$\\
$m_\sigma$ & 599.99 & 600 & $-1.66\times 10^{-5}$\\
$m_\eta$ & 497.78 & 548 & $-9.16\times 10^{-2}$\\
$m_{\eta'}$ & 968.20 & 958 & $+1.06\times 10^{-2}$\\
$\Gamma_a$ & 60.00 & 60 & $+2.04\times 10^{-5}$\\
$\Gamma_\sigma$ & 600.00 & 600 & $+6.81\times 10^{-6}$ \\\hline
\end{tabular}
\caption{Comparison of the experimental data with the values obtained in the model using the \texttt{MINUIT} routine to find the minimum of the $\chi^2$ estimator.}\label{table}
\end{table}

All the requirements concerning the control parameters are satisfied at this global solution. The fit makes the cubic (in $H$) terms in (\ref{lageff}) actually more relevant than the linear one. As a last point, the $\psi$ angle is treated as a prediction. Experimentally \cite{PDG}, $\psi\simeq -18^{\circ}+\arctan\sqrt 2\simeq -18^{\circ}+54.7^{\circ}\approx 36.7^{\circ}$, while our result is $\psi_{\texttt{MINUIT}}\approx 35.46^{\circ}$, in excellent agreement.

\section{Introducing the axial chemical potential}\label{Smod3}
\begin{figure}[h!]
\centering
\includegraphics[scale=0.3]{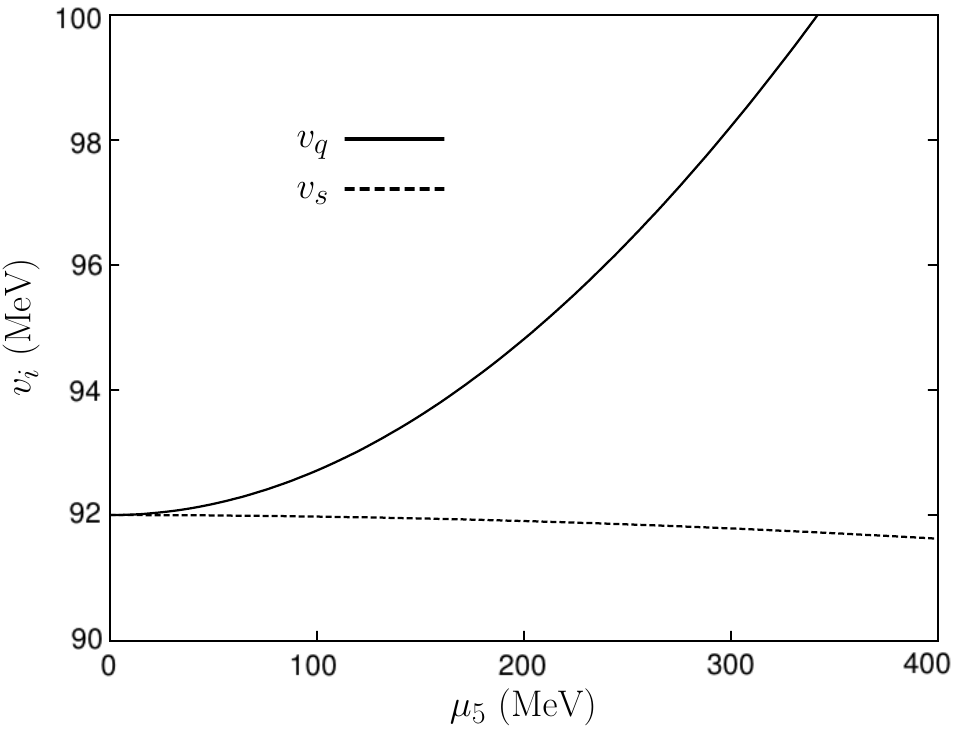}
\caption{$v_q$ and $v_s$ dependence on $\mu_5$.}\label{vqvs}
\end{figure}

The introduction of the axial chemical potential $\mu_5$ is carried out following the prescription described by the spurion technique in Eq. \eqref{covderiv}. An extra piece which is proportional to $\mu_5$ appears in the $P$-even Lagrangian
\begin{equation}\label{chempot}
\Delta \mathcal L=\frac i2\mu_5 \text{Tr}\left [\textbf{I}_q\left (\partial_0 HH^\dagger-H\partial_0 H^\dagger\right )\right ]+\mu_5^2\text{Tr}\left (\textbf{I}_q HH^\dagger\right ). 
\end{equation}
For non-vanishing $\mu_5$, we will assume isospin symmetry and thus, we impose to our solutions to have $v_u=v_d=v_q\neq v_s$. The corresponding gap equations are
\begin{align}
M^2-2(\lambda_1+\lambda_2)v_q^2-\lambda_2 v_s^2+cv_s +2\mu_5^2 &=0,\\
v_s(M^2 -2\lambda_2v_q^2-(2\lambda_1+\lambda_2)v_s^2)+cv_q^2&=0.
\end{align}
The correct solution of this system of coupled equations is taken imposing the correct limit to the chiral condensates $v_q,v_s\to v_0$ when $\mu_5\to 0$ (see Figure \ref{vqvs} for the $\mu_5$ evolution of the solution). This solution can be checked to be the global minimum independently on the value of the axial chemical potential.

\medskip

As an example of such a mixing let us consider the two isotriplets of opposite parity $\pi$ and $a_0$. After normalization of the states, let us consider the piece of the effective Lagrangian that is bilinear in the $\pi$ and $a_0$ fields
\begin{equation}
\mathcal L=\frac 12 (\partial a_0)^2+\frac 12 (\partial \pi)^2-\frac12 m_1^2 a_0^2-\frac12 m_2^2 \pi^2-4\mu_5 a_0 \dot{\pi},
\end{equation}
where
\begin{align}
\nonumber m_1^2&=-2(M^2-2(3\lambda_1+\lambda_2)v_q^2-\lambda_2 v_s^2-cv_s+2(3d_1+2d_2)mv_q+2d_2m_sv_s+2\mu_5^2),\\
m_2^2&=\frac{2m}{v_q}\left [b+(d_1+2d_2)v_q^2+d_2v_s^2\right ].
\end{align}
Notice that the resulting Lagrangian is not Lorentz invariant, which is obvious from \eqref{covderiv}. We will perform a diagonalization in momentum space, so the Lagrangian operator is written as
\begin{equation}
\mathcal L=-\frac 12 \begin{pmatrix}
a_0^*(k) & \pi^*(k)
\end{pmatrix}\begin{pmatrix}
-k^2+m_1^2 & 4i\mu_5 k_0\\
-4i\mu_5 k_0 & -k^2+m_2^2
\end{pmatrix}\begin{pmatrix}
a_0(k)\\
\pi(k)
\end{pmatrix}.
\end{equation}
Recall that fields in the momentum representation satisfy $A^*(k)=A(-k)$. Note also the fact that the mixing term has been rewritten as $-4\mu_5 a_0 \dot{\pi}=-2\mu_5(a_0 \dot{\pi}-\dot{a}_0 \pi)$ in order the matrix to be hermitian. The eigenvalues are $\frac 12(k^2-m_{\text{eff}}^2)$, where the (energy dependent) effective masses are
\begin{equation}\label{m2effpi}
m^2_{\text{eff}\ \pm}(k_0)=\frac12 \left [m_1^2+m_2^2 \pm \sqrt{(m_1^2-m_2^2)^2+(8\mu_5 k_0)^2}\right ].
\end{equation}
The new eigenstates $X_{1,2}$ are defined through
\begin{gather}
a_0=\sum_{j=1}^2 C_{aj}X_j,\quad \pi=\sum_{j=1}^2 C_{\pi j}X_j, \\
\nonumber C_{a1}=iC_{\pi 2}=C_+, \quad C_{a2}=-iC_{\pi 1}=-C_-,
\end{gather}
with
\begin{equation}
C_\pm=\frac 1{\sqrt 2}\sqrt{1\pm\frac{m_1^2-m_2^2}{\sqrt{(m_1^2-m_2^2)^2+(8\mu_5 k_0)^2}}}.
\end{equation}
We can also use the notation $X_1,X_2\equiv \tilde a,\tilde\pi$, indicating that $X_1$ (resp. $X_2$) is the state that when $\mu_5=0$ goes over to $a_0$ (resp. $\pi$).

\medskip
\begin{figure}[h!]
\centering
\includegraphics[scale=0.28]{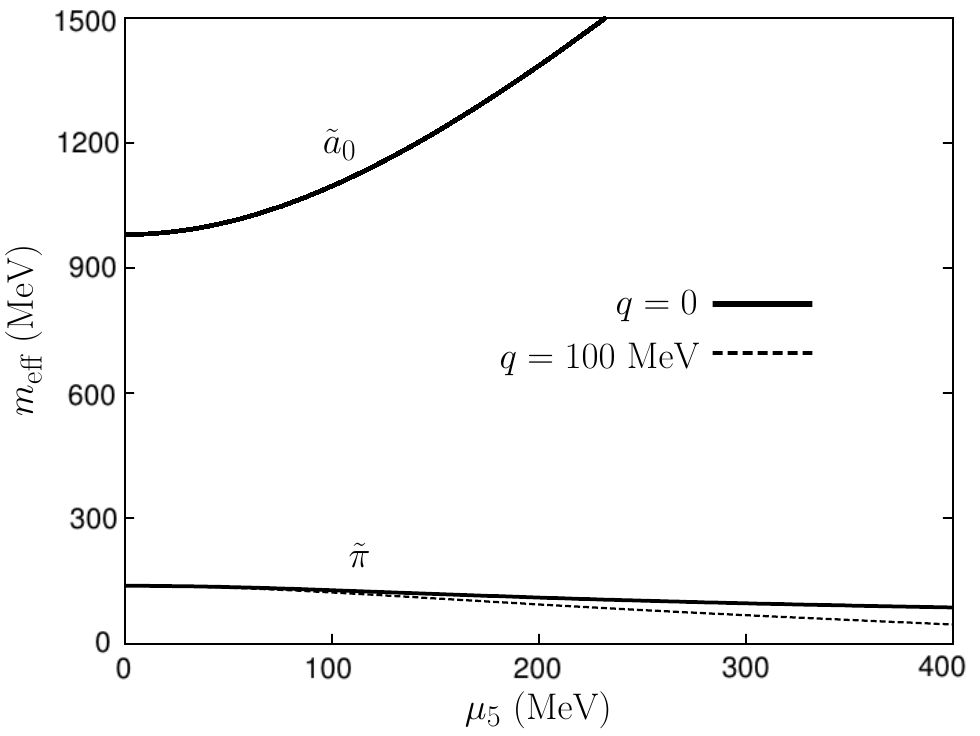}\includegraphics[scale=0.28]{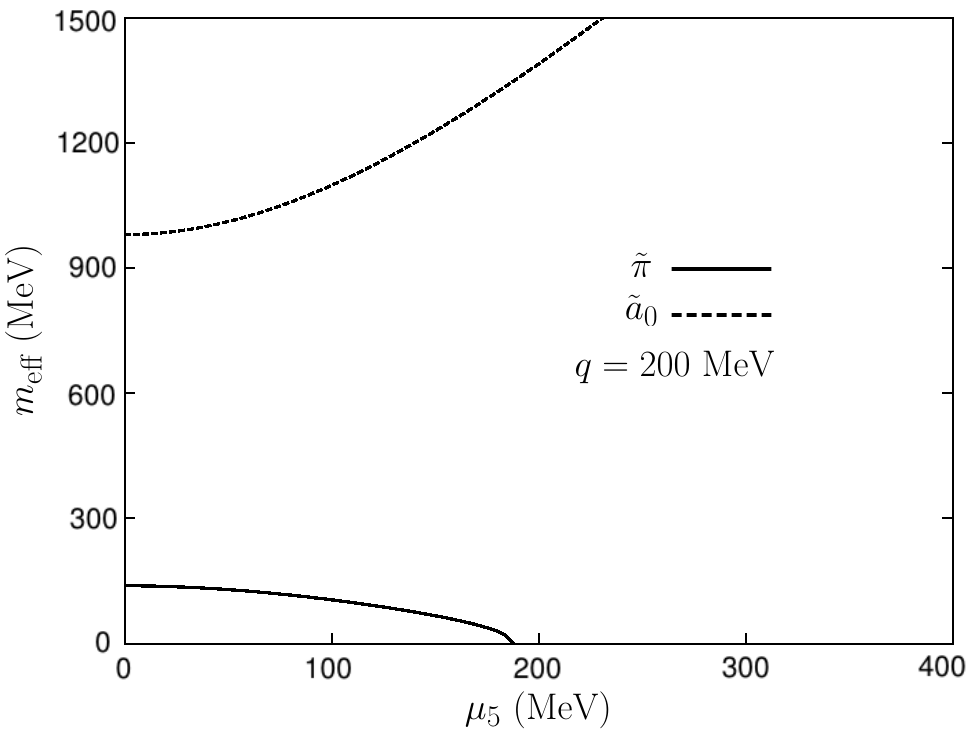}
\vspace{-2em}
\caption{Effective mass dependence on $\mu_5$ for $\tilde\pi$ and $\tilde{a}_0$. Left panel: comparison of masses at rest and at low momentum $q=100$ MeV. Right panel: masses at $q=200$ MeV, where the $\tilde\pi$ mass goes tachyonic, as discussed in the text. For such momenta, the variations in $\tilde a_0$ are almost invisible and only slightly visible for large values of $\mu_5$ for $\tilde \pi$.}\label{meffpi}
\vspace{-1em}
\end{figure}

In Fig. \ref{meffpi}, we present the results for the evolution of $\tilde \pi$ and $\tilde{a}_0$ effective masses with respect to the axial chemical potential $\mu_5$ for different values of the 3-momentum. As stressed, both states tend to the known physical ones in the limit $\mu_5\to0$. In general terms, we find a growing mass depending on $\mu_5$ for $\tilde{a}_0$ while the opposite holds for $\tilde\pi$. A remarkable feature of this model is the appearance of in-medium tachyonic states at high energies (or momenta). It is evident from Eq. \eqref{m2effpi} that for energies higher than a critical value $k_0,|\vec k|>m_1m_2/(4\mu_5)\equiv k^c_{\tilde\pi}$, the square root term dominates, thus leading to a negative squared mass for pions, as shown in the right panel of Fig. \ref{meffpi}. Such a behaviour does not represent a serious physical obstacle since it can be checked that the energies are always positive and no vacuum instabilities appear. The group velocity can also be checked to be smaller than the speed of light independently of $\mu_5$. On the other hand, the $\tilde{a}_0$ mass shows an important enhancement, but in this model $\mu_5$ has to be understood as a perturbatively small parameter, and very high values are beyond the domain of applicability of the effective Lagrangian. A better treatment of $\tilde a_0$ would require the inclusion of heavier degrees of freedom such as $\pi(1300)$ for instance.

\section{Mixing $\eta-\sigma-\eta'$}\label{Smod4}
A similar analysis applies to the isosinglet case. We shall consider three states here: $\eta$, $\eta^\prime$ and $\sigma$. As before, the starting point will be the piece of the effective Lagrangian (\ref{lageff}) that after the inclusion of $\mu_5$ is bilinear in the fields, i.e. the properly normalized kinetic part
\begin{equation}
\mathcal L=\frac 12[(\partial\sigma)^2+(\partial \eta_q)^2+(\partial \eta_s)^2]-\frac 12m_3^2 \sigma^2-\frac 12m_4^2\eta_q^2-\frac 12m_5^2\eta_s^2-4\mu_5\sigma\dot{\eta}_q-2 \sqrt{2}cv_q\eta_q\eta_s.
\end{equation}
The constants appearing in the previous equation are given by\footnote{There were two miswritten numerical factors in \cite{epjc}. This version is corrected.}
\begin{align}
\nonumber m_3^2&=-2(M^2-6(\lambda_1+\lambda_2)v_q^2-\lambda_2 v_s^2+cv_s+6(d_1+2d_2)mv_q+2d_2m_sv_s +2\mu_5^2),\\
\nonumber m_4^2&=\frac {2m}{v_q}\left [b+(d_1+2d_2)v_q^2+d_2v_s^2\right ]+4cv_s,\\
m_5^2&=\frac{2m_s}{v_s}[b+2d_2v_q^2+(d_1+d_2)v_s^2]+\frac{2cv_q^2}{v_s}.
\end{align}
In this case two different mixing pieces appear: the first one is indeed proportional to the axial chemical potential $\mu_5$ and the other is proportional to the coupling $c$ that breaks the axial anomaly due to the determinant term in Eq. \eqref{lageff} as already mentioned before. In matrix form, the previous lagrangian can be written as
\begin{equation}
\mathcal L=-\frac 12\begin{pmatrix}
\sigma^*(k) & \eta_q^*(k) & \eta_s^*(k)
\end{pmatrix}\begin{pmatrix}
-k^2+m_3^2 & 4i\mu_5 k_0 & 0\\
-4i\mu_5 k_0 & -k^2+m_4^2 & 2 \sqrt{2}cv_q\\
0 & 2 \sqrt{2}cv_q & -k^2+m_5^2
\end{pmatrix}\begin{pmatrix}
\sigma(k) \\
\eta_q(k) \\
\eta_s(k)
\end{pmatrix}.
\end{equation}
The equation for the eigenvalues (effective masses) is now a cubic one and the solution is determined numerically
\begin{align}
-8c^2v_q^2(m_{\text{eff}}^2-m_3^2)&+(m_{\text{eff}}^2-m_5^2)\\
\nonumber &\times \left [\left (m_{\text{eff}}^2-\frac 12(m_3^2+m_4^2)\right )^2-\frac 14 (m_3^2-m_4^2)^2-(4\mu_5 k_0)^2\right ]=0.
\end{align}
As before the eigenstates are defined via
\begin{align}\label{eigens}
\sigma=\sum_j C_{\sigma j}X_j, \qquad \eta_q=\sum_j C_{\eta_q j}X_j, \qquad \eta_s=\sum_j C_{\eta_s j}X_j ,
\end{align}
where
\begin{gather}
\nonumber C_{\sigma j}=\frac{4i\mu_5 k_0 (m_5^2-m_j^2)}{N_j\prod_{k\neq j}(m_j^2-m_k^2)}, \quad C_{\eta_q j}=\frac{(m_5^2-m_j^2)(m_3^2-m_j^2)}{N_j\prod_{k\neq j}(m_j^2-m_k^2)},\\
C_{\eta_s j}=\frac{-2\sqrt{2}cv_q(m_3^2-m_j^2)}{N_j\prod_{k\neq j}(m_j^2-m_k^2)}, \quad \frac 1{N_j}=\sqrt{1+\frac{(4\mu_5 k_0)^2}{(m_j^2-m_3^2)^2}+\frac{8c^2v_q^2}{(m_j^2-m_5^2)^2}}.
\end{gather}
$N_j$ is the proper (eigen)field normalization factor.

\medskip
\begin{figure}[h!]
\centering
\includegraphics[scale=0.28]{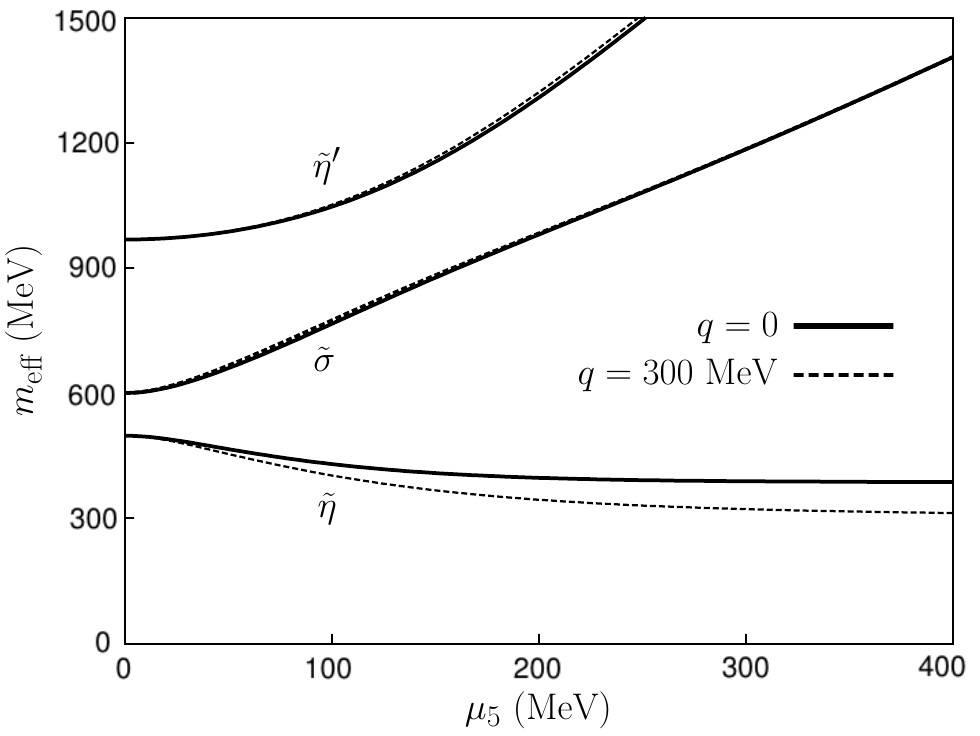}\includegraphics[scale=0.28]{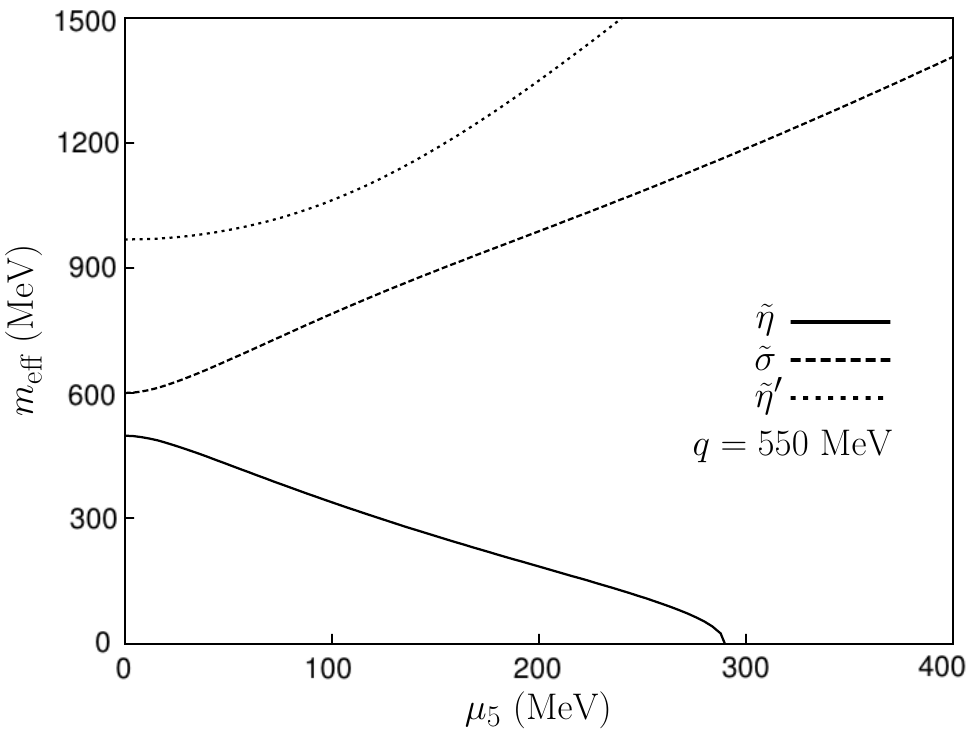}
\vspace{-2em}
\caption{Effective mass dependence on $\mu_5$ for $\tilde\eta$, $\tilde\sigma$ and $\tilde\eta'$. Left panel: comparison of masses at rest and at low momentum $q=300$ MeV. Right panel: masses at $q=550$ MeV, where the $\tilde\eta$ mass goes tachyonic, as discussed in the text. As in the previous example, for this range of momenta, the variations in the heavier degrees of freedom $\tilde \sigma$ and $\tilde\eta'$ are almost invisible and only slightly visible for large values of $\mu_5$ for $\tilde \eta$.}\label{meffeta}
\end{figure}

\begin{figure}[h!]
\centering
\includegraphics[scale=0.3]{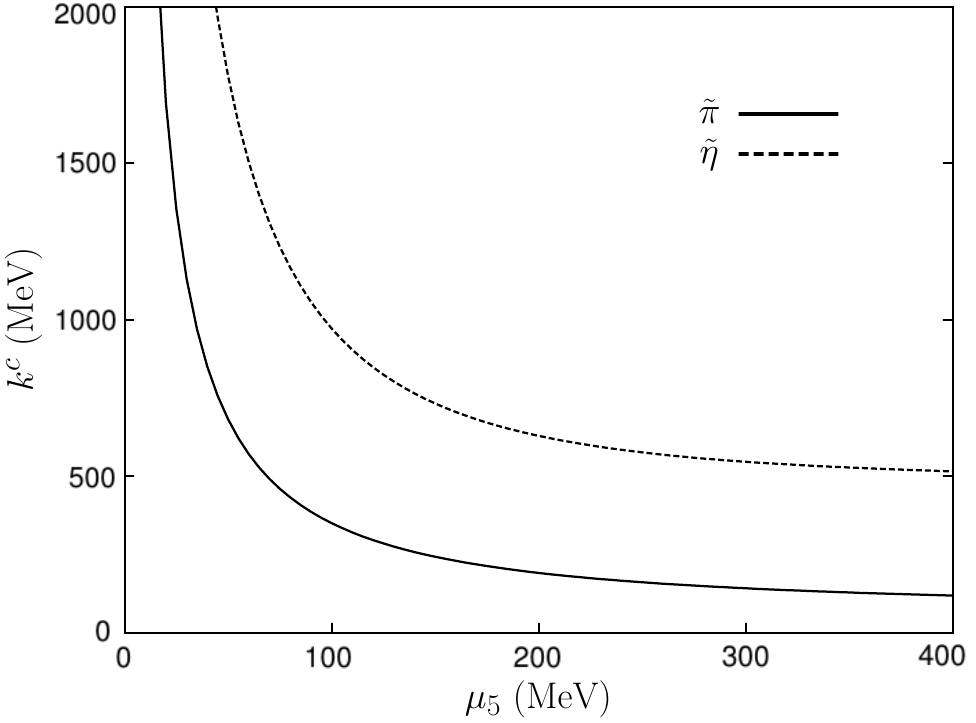}
\vspace{-0.5em}
\caption{$\mu_5$-dependence of the tachyon critical energy for isotriplet $k^c_{\tilde\pi}$ and isosinglet case $k^c_{\tilde\eta}$.}\label{kcrit}
\vspace{-1em}
\end{figure}

The $\mu_5$-dependence of the effective masses is plotted in Fig. \ref{meffeta} for different values of the 3-momentum. As in the previous example, the heavy degrees of freedom $\tilde\sigma$ and $\tilde\eta'$ become heavier as one increases $\mu_5$ while the opposite applies for the light $\tilde\eta$. The latter state eventually becomes tachyonic for high energies or momenta $k_0,|\vec k|>k^c_{\tilde\eta}$ with $k^c_{\tilde\eta}\equiv m_3/(4\mu_5m_5)\sqrt{m_4^2m_5^2-8c^2v_q^2}$, as shown in the right panel of Fig. \ref{meffeta}. As in the previous example, vacuum is stable and group velocities are checked to be smaller than the speed of light. The tachyon critical energy presents a similar behaviour as the one in the triplet case. In Fig. \ref{kcrit}, both isotriplet $k^c_{\tilde\pi}$ and isosinglet critical energies $k^c_{\tilde\eta}$ are plotted together.

\section{New interactions and decay widths}\label{Smod5}
After the inclusion of $\mu_5$ the cubic couplings present in the effective Lagrangian \eqref{lageff} containing one isosinglet and two isotriplets are
\begin{align}
\nonumber \mathcal L_{\sigma aa}&=2[(3d_1+2d_2)m-2(3\lambda_1+\lambda_2)v_q]\sigma \vec a^2,\\
\nonumber \mathcal L_{\sigma\pi\pi}&=\frac 1{v_q^2}\left [(\partial\vec\pi)^2v_q-(b+3(d_1+2d_2)v_q^2+d_2v_s^2)m\vec\pi^2\right ]\sigma,\\
\nonumber \mathcal L_{\eta a\pi}&=\frac 2{v_q^2}[\partial \eta_q \vec a\partial \vec\pi v_q-(b+
(3d_1+2d_2)v_q^2+d_2v_s^2)m\eta_q\vec a\vec\pi],\\
\mathcal L_{\sigma a\pi}&=-\frac{4\mu_5}{v_q}\sigma\vec a\dot{\vec{\pi}}, \quad \mathcal L_{\eta aa}=-\frac{2 \mu_5}{v_q}\dot \eta_q \vec a^2,\quad \mathcal L_{\eta\pi\pi} = 0.
\end{align}

As seen in the previous expressions, interactions (and therefore decays) that are normally forbidden on parity conservation grounds are now possible with a strength proportional to the parity-breaking parameter $\mu_5$. However, the previous interaction terms are not physical because the genuine eigenstates are now $X_i$ rather than the original fields $\pi$, $a_0$, etc. Going to the physical basis requires using the diagonalization process defined in the previous section.

\medskip

Our ultimate purpose is to check the relevance of dynamically generated parity breaking through topological charge fluctuations in heavy ion collisions. It is natural then to ask how the previously derived masses and vertices may influence the physics in the hadronic fireball.

\medskip

It should be clear that the influence may be very important if $\mu_5$ is such that the induced parity-breaking effects are significant. After the initial collision of two heavy ions in a central or quasi-central process a fireball is formed. This fireball could be described in rather simplistic terms as a hot and dense pion gas. Pion-pion interaction is dominated by $\sigma$ and $\rho$ -particle exchanges and processes such as $\eta\to \pi\pi$ or $\eta^\prime \to \pi\pi$ are forbidden. If parity is no longer a restriction, these two processes, or rather processes such as $X_i \to \tilde\pi \tilde\pi$ ($i=3,4,5$) are for sure relevant and the new eigenstates produced due to parity breaking could thermalize inside the fireball.

\medskip

Let us now try to be more quantitative. It should be clear from the mass evolution as a function of $\mu_5$ (Figure \ref{meffpi}) that $\tilde\pi$ is the lightest state and it dominates the partition function in the fireball. To get an estimate of the relevance of the new states $X_i$ we will compute their width to distorted pions $\tilde \pi$. A non-negligible decay width comparable (or higher) to the inverse fireball lifetime would correspond to a state with a mean free path smaller than the size of the fireball. As a consequence, such state would decay in-medium into (distorted) pions meaning that it could be regenerated as in the case of the $\rho$ meson, even opening the possibility of reaching thermal equilibrium. To do so we need the following $S$-matrix element corresponding to $X_i(q) \to \tilde\pi^+(p) \tilde\pi^-(p')$
\begin{align}\label{iM}
\nonumber i\mathcal M=&4i[(3d_1+2d_2)m-2(3\lambda_1+\lambda_2)v_q]C_{\sigma i} C_{a2}^+C_{a2}^- +\frac{4\mu_5}{v_q}C_{\sigma i}(E_{p'} C_{a2}^+ C_{\pi 2}^- +E_p C_{a2}^- C_{\pi 2}^+) \\
\nonumber &-\frac i{v_q^2}\left [(m_{X_i}^2-m_{\tilde\pi^+}^2-m_{\tilde\pi^-}^2) v_q+2(b+3(d_1+2d_2)v_q^2+d_2v_s^2)m\right ]C_{\sigma i}C_{\pi 2}^+ C_{\pi 2}^-\\
\nonumber &-\frac{4 \mu_5 E_q}{v_q}C_{\eta_q i} C_{a2}^+C_{a2}^- +\frac {i}{v_q}(m_{\tilde\pi^+}^2-m_{\tilde\pi^-}^2)C_{\eta_q i}(C_{a2}^-C_{\pi 2}^+ -C_{a2}^+C_{\pi 2}^-) \\
&-\frac {i}{v_q^2}[2(b+(3d_1+2d_2)v_q^2+d_2v_s^2)m-m_{X_i}^2v_q]C_{\eta_q i}(C_{a2}^+C_{\pi 2}^- +C_{a2}^-C_{\pi 2}^+).
\end{align}
We are dealing with a relativistic non-invariant theory and therefore the widths do not a priori transform as one would naively think. We shall compute them first at rest, then for different values of the 3-momentum of the decaying particle.

\subsection{Widths at rest}

\begin{figure}[h!]
\centering
\includegraphics[scale=0.37]{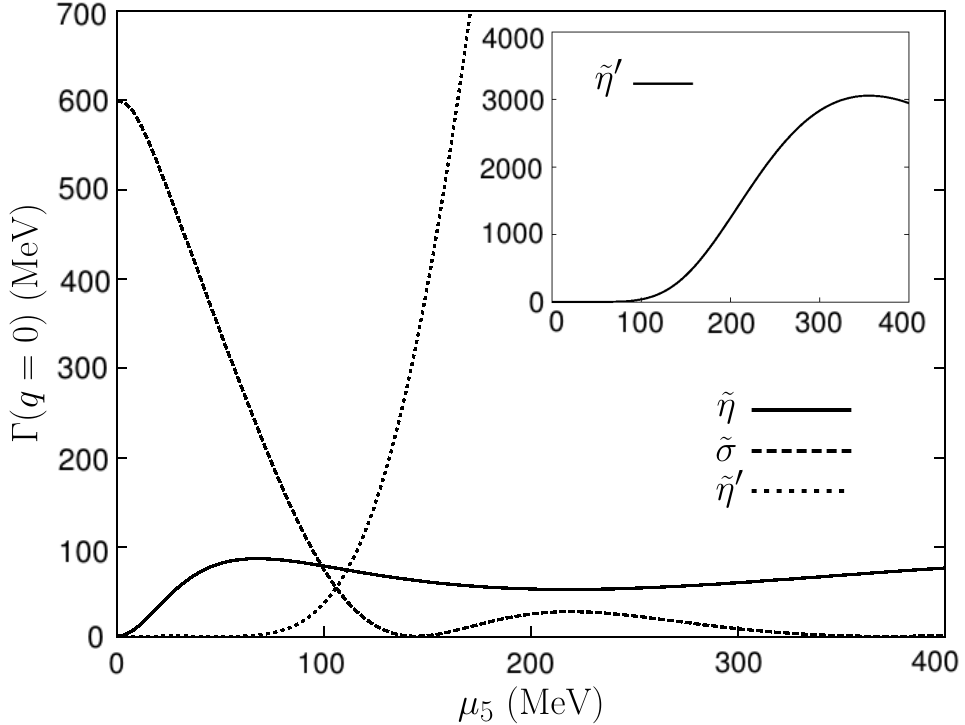}
\vspace{-0.5em}
\caption{$\tilde\eta$, $\tilde\sigma$ and $\tilde\eta'$ widths at rest depending on $\mu_5$. Down to $\mu_5=50$ MeV, $\tilde\eta$ acquires a width of order 60 MeV, with a characteristic mean free path smaller than the typical fireball size of $5 - 10$ fm and hence implying that thermalization may occur in this channel. Nevertheless, $\tilde \sigma$ shows a pronounced fall and beyond $\mu_5=100$ MeV, it becomes a stable channel. Inset: Detail of $\tilde\eta'$ width reaching the GeV scale, a clear violation of unitarity since we do not include heavier degrees of freedom in our model.}\label{width0}
\vspace{-1em}
\end{figure}

The width of $X_i$ is calculated from the amplitude shown in Eq. \eqref{iM} since we do not include further decaying processes. Recall that all masses are energy dependent. At the $X_i$ rest frame, $\vec q=\vec 0$ and $E_p=E_{p'}=m_{X_i}(\vec q=\vec 0)/2\equiv m^{X_i}_0/2$. Here, a momentum-dependent effective mass is taken instead of an energy-dependent one since we assume the decaying particle to be on-shell so both $m(\vec k)$ and $m(k_0)$ coincide. Thus, the rest width is given by
\begin{equation}
\Gamma_{X_i\to \tilde\pi\tilde\pi}=\frac 32\frac 1{2m^{X_i}_0}|\mathcal M|^2\frac{1}{4\pi}\frac{p_{\tilde\pi}}{m^{X_i}_0\frac{dp_{\tilde\pi}^2}{dE_p^2}}, \quad \frac{dp_{\tilde\pi}^2}{dE_p^2}=1+\frac{(4\mu_5)^2}{\sqrt{(m_1^2-m_2^2)^2+(8\mu_5 E_p)^2}}
\end{equation}
where $p_{\tilde\pi}(E_p)=\sqrt{E_p^2-m_{\tilde\pi}^2(E_p)}$ and the factor 3/2 accounts for the decay both to neutral and charged $\tilde\pi$. The results of the widths are shown in Figure \ref{width0}.

\medskip

The new state $\tilde\eta$ exhibits a smooth behaviour with an average value $\sim 60$ MeV, corresponding to a mean free path $\sim 3$ fm, which is smaller than the typical fireball size $L_\text{fireball}\sim 5 - 10$ fm. Hence, the thermalization of this channel via regeneration of $\tilde\pi$ within the gas seems to be possible.

\medskip

Another striking point concerning $\tilde\sigma$ takes place down to $\mu_5\sim 100$ MeV, when
the decay width decreases dramatically leading to scenarios where this state becomes stable.
The visible bumps in these two latter channels seem to reflect the tachyonic nature of the decaying $\tilde\pi$.

Finally, we present in the inset of Fig. \ref{width0} the detail of the $\tilde\eta'$ width, that grows up to the GeV scale, showing clear violations of unitarity. As in the case of $\tilde a_0$, more hadronic degrees of freedom are needed to obtain a reliable result, such as $f_0(980)$, etc.

\subsection{Decay widths of moving particles}

Next, let us compute the width when the decaying particle is not at rest. As explained before in a non-invariant relativistic theory this cannot be obtained from the one at rest by simply taking into account the time dilatation effect. The explicit calculation of such width is given by
\begin{equation}
\Gamma_{X\to \tilde\pi\tilde\pi}=\frac 32\frac 1{2E_q}\frac{1}{8\pi q}\int \frac{|\mathcal M|^2pdp }{E_p\frac{dp_{\tilde\pi}^2}{dE_p^2}}.
\end{equation}
In our numerical calculation we controlled that all the magnitudes are implemented coherently, like for example $|\cos\theta|<1$ and $E_{p'}>0$. Of course, the limit $q\to 0$ coincides with the calculation at rest performed before.

\medskip

\begin{figure}[h!]
\centering
\includegraphics[scale=0.28]{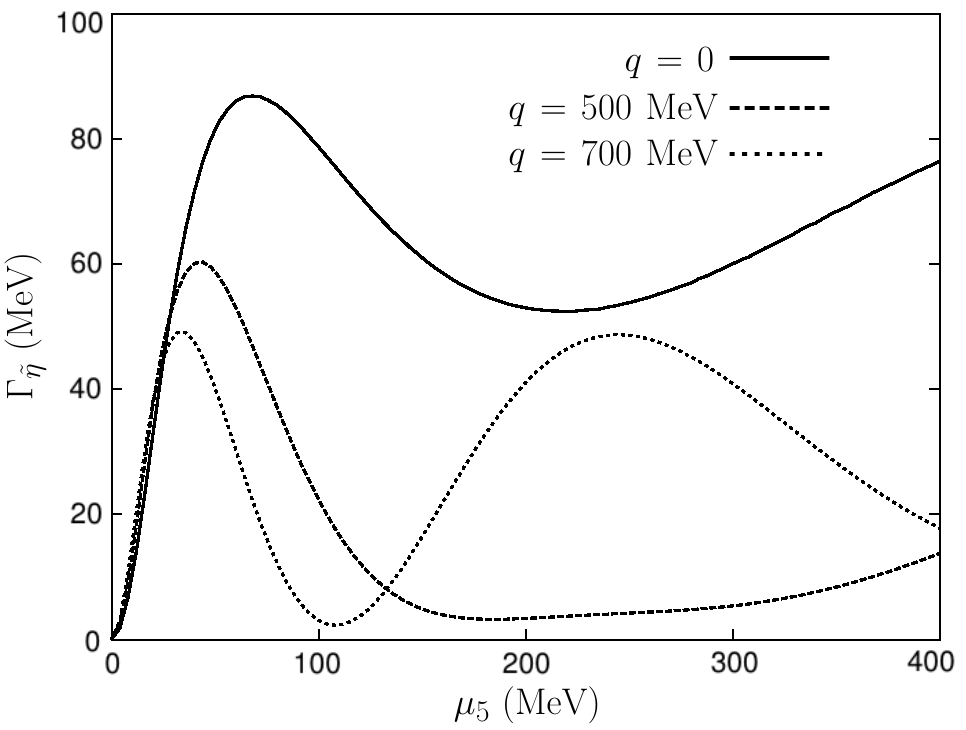}\includegraphics[scale=0.28]{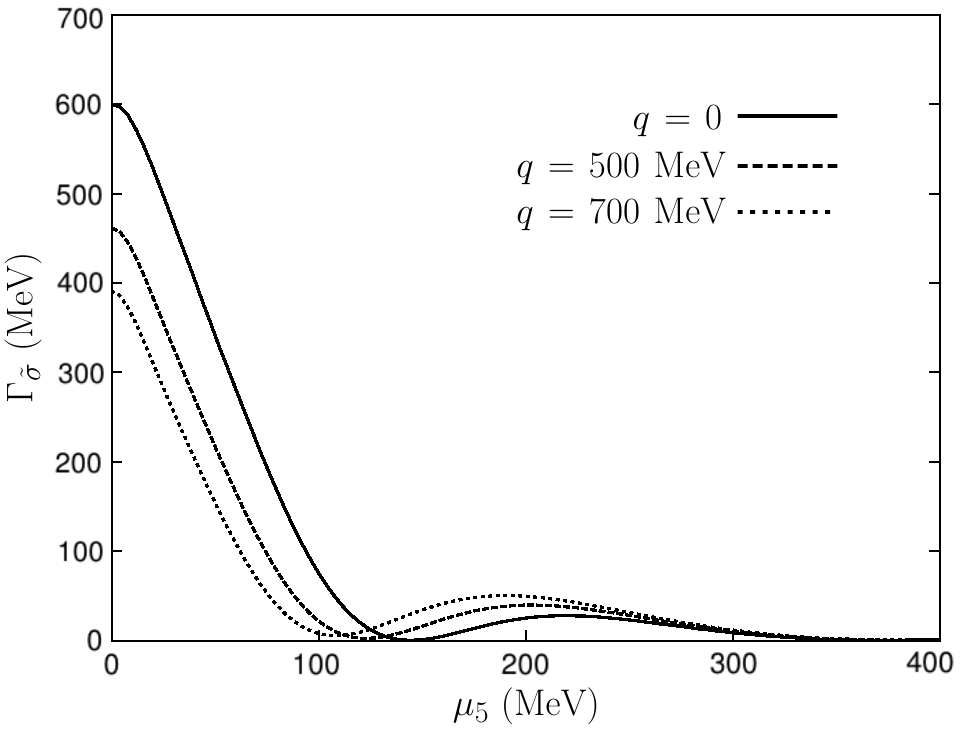}
\caption{$\tilde\eta$ (left) and $\tilde\sigma$ (right) widths depending on $\mu_5$ for different values of the incoming 3-momentum: $q=0,500,700$ MeV. The first plot shows a non-trivial dependence on $q$ (see text) while the second one shows a fall that is mainly due to the Lorentz factor.}\label{widthq}
\end{figure}

In the $\tilde\eta$ channel (see the left section of Fig. \ref{widthq}), one may observe the time dilatation effect by which the width should decrease as the 3-momentum grows. However, this cannot explain the small variations at low 3-momenta with respect to the width at rest, namely, two initial bumps at $\mu_5\sim 80$ MeV and 550 MeV (the latter being beyond the plot range) that slowly separate as one increases $q$. The two-dimensional representation $\Gamma_{\tilde\eta}(\mu_5,q)$ exhibits a saddle point around $\mu_5^*\sim 240$ MeV and $q^*\sim 500$ MeV, and in consequence, for large 3-momenta, a third intermediate bump appears opening the possibility of creating two different $\tilde\pi$ tachyons at the same time. The latter maximum grows fast as one increases $q$ and becomes the global one when the 3-momentum goes beyond $q\gtrsim 700$ MeV. 

\medskip

On the other hand, in the $\tilde\sigma$ and $\tilde\eta'$ channels no huge differences arise when boosting the decaying particle. In the right panel of Fig. \ref{widthq}, we show the $\tilde\sigma$ width for different values of $q$ and the most salient behaviour is, as in the previous case, the separation of the two minima as one increases $q$.

\section{Axial-vector meson condensation}\label{Smod6}
The introduction of an axial chemical potential into the quark-meson model interferes with the flavour-singlet axial-vector channel as this potential is just the time component of some axial-vector field, as discussed at the beginning of this chapter. Therefore if one includes the coupling of the singlet axial-vector quark current with the corresponding meson field $h_\mu$ one expects mixing and renormalization of the bare axial chemical potential due to condensation of the time component of the axial-vector field $h^\mu \simeq \langle h^0\rangle \delta^{0\mu}$. This phenomenon is in full analogy to the condensation of the time component of the $\omega$ meson field when a baryon chemical potential enters the Lagrangian \cite{walec} which is quite important to understand the repulsive nuclear forces in this channel.

\medskip

Let us elucidate this phenomenon in more details. The relevant Lagrangian for axial-vector mesons reads
\begin{equation}
\Delta\mathcal L=-\frac 14 h_{\mu\nu}h^{\mu\nu}+\frac 12m_h^2h_\mu h^\mu+ \bar q\gamma_\mu\gamma_5 (g_hh^\mu+\delta^{\mu0}\mu_5)\textbf{I}_q q,
\end{equation}
where $h$ stands for the $SU(2)$ singlet axial-vector meson $h_1(1170)$ \cite{PDG} and $g_h$ denotes its coupling to the quark current. We assume the condensation of $h^\mu \simeq \langle h^0\rangle \delta^{0\mu}$ and thus, the effective chemical potential $\bar\mu_5\equiv\mu_5+g_h\langle h_0\rangle$ can be defined. The equation of motion for $h$
\begin{equation}
 \frac{\delta\Delta\mathcal L}{\delta h_0} = m^2_h \langle h_0\rangle + g_h \langle \bar q\gamma_0\gamma_5\textbf{I}_q q\rangle = 0
\end{equation}
determines the relation among the effective non-strange axial charge density $\rho_5 = \langle \bar q\gamma_0\gamma_5 \textbf{I}_q q\rangle$ and the isosinglet axial-vector condensate.
\begin{equation}\label{deltav}
\rho_5(\bar\mu_5)=\frac{\mu_5-\bar\mu_5}{G_h} = \frac{\delta\Delta\mathcal L}{\delta\bar\mu_5},\quad \Delta V=-\frac 12m_h^2\langle h_0\rangle^2=-\frac 12\frac{(\bar\mu_5-\mu_5)^2}{G_h},\qquad
\end{equation}
where $G_h=g_h^2/m_h^2$. After including $\Delta V$ (Eq. \eqref{deltav}) in Eq. \eqref{chempot} the stationary point equation (or equivalently, the equation of motion for the axial-vector condensate) can be derived
\begin{equation}
\frac{\delta \mathcal L}{\delta \bar\mu_5}=\frac{\delta}{\delta \bar\mu_5}\left [2\bar\mu_5^2 v_q^2+\frac 12\frac{(\bar\mu_5-\mu_5)^2}{G_h}\right ]=0
\end{equation}
that allows to relate the bare and effective axial chemical potentials
\begin{equation}\label{renchem}
\bar\mu_5\left [1+4G_h v_q^2(\bar\mu_5)\right ]=\mu_5. 
\end{equation}
We stress that in the mass-gap equations for $v_q, v_s$ the effective axial chemical potential $\bar\mu_5$ must be used. The relation \eqref{renchem} is smooth against the decoupling of axial-vector mesons $g_h \to 0$. It determines unambiguously the axial charge density, $\rho_5(\bar\mu_5)=4\bar\mu_5 v_q^2(\bar\mu_5)$, which exhibits, in general, lower values when affected by axial meson forces, $|\bar\mu_5| < |\mu_5|$.

\section{Summary}

In this chapter we have analysed the physics of the lightest scalar and pseudoscalar mesons in the presence of a parity-odd medium driven by an axial chemical potential. By means of a generalised sigma model, we introduced some of the main resonances with spin zero in the $SU(3)$ representation. The presence of an axial chemical potential as an explicit source for parity violation induces a mixing among the states of different parity and same flavour content. The properties of in-medium mesons are energy dependent due to the breaking of Lorentz invariance when $\mu_5$ is implemented as a time-dependent axial-vector field. Then, meson physics is frame-dependent with the natural consequence that masses and decay widths do depend non-trivially on the momentum of the corresponding particle.

\medskip

We have given convincing arguments that, if $\mu_5\neq 0$ the pion gas in the fireball forming after a central heavy ion collision may actually not be made of the usual pions, but instead of some states of non-defined parity and energy-dependent effective mass. In addition the lightest spin zero states in each isospin channel become tachyons for high enough energies. As a consequence of this feature, the decay widths of the singlet states into distorted pions show non-trivial dependences on the axial chemical potential. Perhaps more importantly, the distorted $\eta$ meson seems to be in thermal equilibrium with the 'pion' gas, as indicated by the characteristic large width, completely different from the ones in vacuum.
\chapter{Vector Meson Dominance approach to local parity breaking}\label{ChVMD}

\graphicspath{{5-VMD/figures-VMD/}{5-VMD/figures-angles/}{figures-VMD/}{figures-angles/}}

It is natural to ask whether the mixing of states of different parities as studied in the previous chapter occurs in the vector/axial-vector sector. There are many models dealing with vector particles phenomenologically. If we assume that the vector mesons appear as part of a covariant derivative (as postulated e.g. in hidden symmetry models \cite{bando}), no mixing term can be generated by operators of dimension 4 if $\mu_5$ is an isosinglet. However, such a mixing is not forbidden on (global) symmetry grounds if $\mu_5$ is present, appearing as the time component of an axial-vector field (see e.g. \cite{rischke}). This means that this coupling is very much model dependent and, unfortunately, not much phenomenological information is present. Therefore, in the present chapter we will omit the mixing of vector and axial-vector mesons due to parity violation.

\medskip

The parity-breaking effect will be introduced through the Chern-Simons term assuming a time-dependent pseudoscalar (axion-like) condensate as pointed out in Chapter \ref{Chfase}. As we will see, this form of LPB can be incorporated by adding a parity-odd term in the Lagrangian; however, this breaking, while being technically 'soft', can be numerically large, leading to important consequences. We shall review in detail how in-medium vector mesons are influenced by this effect and discuss possible ways to verify its existence.

\medskip

We are interested in providing experimental signatures that allow to detect the LPB effect in the vector meson sector. Their hadronic decays are not useful channels as the decay products will be severely affected by the strongly interacting medium created within the HIC fireball. The cleanest way to extract reliable properties of in-medium vector mesons is through their electromagnetic decays. If we want to investigate LPB in HIC with the help of electromagnetic probes, we have to account for the photon contribution to the singlet axial anomaly. Following the lines of Eq. \eqref{axcons} an electromagnetic topological piece may to be added to the quark axial charge via
\begin{equation}\label{Q5tilde}
Q_{5}^q \, \rightarrow \, \tilde Q_5=Q_{5}^q- T^{\text{em}}_5,\quad T^{\text{em}}_5 =\frac{N_c e^2}{8\pi^2}\int_{\text{vol.}}d^3x \, \varepsilon_{jkl} \, \text{Tr}\left (\hat A^j\partial^k \hat A^l\right).
\end{equation}
Now $\mu_5$ is conjugated to the conserved $\tilde Q_5$ in the limit of vanishing quark masses but not to $Q_5$ itself \cite{rubakov}. We will bosonize QCD in the light vector meson sector using the Vector Meson Dominance model prescription \cite{vmd,vmd2} with vector meson and photon fields appearing in the quark covariant derivative. 

\medskip

The VMD mechanism induces a mixing among photons and vector mesons independently on the preservation or violation of parity. This effect allows to reproduce the electromagnetic decays of vector mesons. In this chapter we will focus on the phenomenology related to the decay of vector mesons into lepton pairs (dileptons). Experimental results from different collaborations \cite{star,NA60,phenix} are available so that a comparison with our theoretical predictions can be performed.

\medskip

As described in Chapter \ref{ChIntro}, in central HIC an abnormal dilepton excess has been observed \cite{NA60,phenix}. In the PHENIX experiment, for instance, an excess is seen for dileptons in the range $M < 1.2$ GeV, for centrality $0 - 20\%$ and $p_T < 1$ GeV \cite{phenix}. Theorists have so far been unable to account for the total dilepton excess convincingly. Thermal effects inducing vector resonance broadening and/or mass dropping fall short of providing a full explanation \cite{rapp,renkr,brown,BratkCass,zahed}. On the other hand, these effects seem to be able to explain most of the abnormal yield for NA60\cite{NA60,NA602}. The study of vector meson decays into dileptons within a parity-odd medium may be a helpful tool in order to investigate such dilepton excess around the $\rho-\omega$ resonance peak. The dilepton excess is conspicuously absent for peripheral HIC (where the CME should be more visible). We conjecture that the CME and the 'anomalous' dilepton excess may be complementary effects revealing two facets associated to the formation of a thermodynamic phase where parity is locally broken. We investigate whether LPB modifications of vector mesons may induce a large dilepton excess in certain HIC. We will briefly discuss how other hadronic decays may be relevant for the dilepton enhancement due to their substantial modification induced by LPB.

\medskip

The modifications of $\rho$ and $\omega$ induced by LPB will reveal distorted spectral functions. Distortion of the $\rho-\omega$ spectrum is well known to happen in HIC. Conventional explanations exist \cite{rapp,renkr,BratkCass,zahed} and if LPB occurs, all effects need to be treated jointly as long as they do not represent double counting. Due to the precision of current experiments, the distorted spectral function itself may be difficult to be accurately measured. In our analysis, we will see that this distortion is characterised by a polarization asymmetry around the $\rho-\omega$ resonance position. This feature could, in principle, be detected. Therefore, we will address a careful treatment of the polarization distribution associated to the lepton pairs produced in the decay of these mesons and to give possible signatures for experimental detection of LPB that could potentially be seen by the PHENIX and STAR collaborations are discussed.

\medskip

Current HIC experiments have investigated possible polarization dependencies in dilepton angular distributions with no significant results \cite{na60pol}. We claim that the conventional angular variables used in such studies are not efficient enough to convey all the information related to the parity-breaking effect. The dilepton invariant mass plays a key role in this game, which is normally missed. If not considered, dilepton polarization dependencies wipe out and no net effect can be extracted from experimental data. We will see that a combined analysis of some characteristic angles together with the dilepton invariant mass constitutes the appropriate framework to investigate the possible polarization asymmetry predicted by LPB.

\medskip

This chapter is organised as follows: in Section \ref{VMD1}, we introduce the Vector Meson Dominance model approach to LPB and derive in detail the distorted dispersion relation that affects the lightest meson states $\rho$ and $\omega$. In Section \ref{dilepLPB} their spectral functions are presented together with some considerations of a possible partial explanation of the abnormal dilepton production in the vicinity of their resonance peak position. An experimentally oriented study of the polarization asymmetry found in the spectral functions is discussed in Section \ref{angles}. Two different analysis of the dilepton angular distribution are presented in order to reflect how the LPB effect may be experimentally detected.

\section{Summary of the Vector Meson Dominance model}

Before the formulation of QCD, hadron physics was deeply investigated through a variety of models incorporating approximate symmetries. The interaction between photons and hadronic matter is a subject of a particular interest in this work that the Vector Meson Dominance model (VMD) can address satisfactorily (for reviews, see \cite{VMDreview}). The main idea behind this model is assuming that the quark loops appearing in the vacuum polarization of the photon can be described exclusively by vector mesons. This approximation results to be very accurate in the regions around the vector meson masses. In QED the photon self-energy can be reasonably well approximated using bare propagators and vertices for electron-positron loops due to higher order suppressions. However, the strong coupling of hadronic constituents requires considering the dressing of quark loops in this context.
  
\medskip

The success of the principles of quantum field theories in QED led to a formulation of strong interactions based in the same principles. In particular, the notion of conserved currents, the gauge principle and the universality of couplings should be applied to strong-interaction physics. The VMD model was introduced as the theory of the strong interaction mediated by vector mesons and based on a non-abelian Yang-Mills theory \cite{vmd}. Its simplest realization conceives the hadronic contribution to the photon polarization as a propagating vector meson that replaces the quark loop. This assumption is equivalent to considering the hadronic electromagnetic current operator (in the neutral isotriplet channel) given by the current-field identity
\begin{equation}\label{rhocurrent}
j_\mu^{\text{em}}(x)\Big|_{I=1}=\frac{m_\rho^2}{g_\rho}\rho_\mu^0(x)
\end{equation}
is proportional to the vector meson field operator. As the electromagnetic current has null divergence, so will the field $\rho_\mu$, both its neutral and its charged components so $\partial_\mu\vec\rho^\mu=0$. The breaking of gauge invariance due to vector meson masses was ignored since local gauge transformations were considered as the way to create interactions with universal couplings among nucleons and vector mesons, say $g_\rho \equiv g_{\rho NN} = g_{\rho\pi\pi} = g_{\rho\rho\rho}$.

\medskip

VMD was particularly successful in the description of the electromagnetic form factors of the nucleons. The nucleon was regarded as a core surrounded by a pion cloud and the form factor measurements were interpreted as an evidence for an isoscalar vector meson $\omega$ and for an isovector meson $\rho^0$ decaying into 3 and 2 pions, respectively. This picture of the nucleon form factors was subsequently generalised to all photon-hadron interactions. Based on the previous considerations, the electromagnetic current was associated to a linear combination of vector meson fields. A complete realization of the VMD model involves other light resonances such as $\omega$ and $\phi$ to be incorporated in the game equivalently to the $\rho$ meson in Eq. \eqref{rhocurrent} through the extension of the current-field identity
\begin{equation}\label{CFI}
j_\mu^{\text{em}}=\frac{m_\rho^2}{g_\rho}\rho_\mu^0+\frac{m_\omega^2}{g_\omega}\omega_\mu+\frac{m_\phi^2}{g_\phi}\phi_\mu,
\end{equation}
which generalises Eq. \eqref{rhocurrent}. With this notation the $SU(3)$ ratios $g_\rho:g_\omega:g_\phi=1:3:-3/\sqrt 2$ hold. This treatment implies that the amplitudes of photon interactions are proportional to the corresponding vector meson processes. In this sense, a virtual photon is understood to fluctuate to a vector meson which is the responsible for interactions with hadronic matter. A parallel derivation of the $\rho$ meson coupling can be obtained without any recurrence to gauge invariance. 

\medskip

Another way to derive Eq. \eqref{CFI} is from the hadronic electromagnetic current of the SM in terms of the quark fields in the $SU(3)$ representation \cite{weise} given by
\begin{equation}
j_\mu^{\text{em}}=\frac23\bar u\gamma_\mu u-\frac13\bar d\gamma_\mu d-\frac13\bar s\gamma_\mu s,
\end{equation}
which can be rewritten identically as
\begin{equation}
j_\mu^{\text{em}}=\frac1{\sqrt2}j_\mu^\rho+\frac1{3\sqrt 2}j_\mu^\omega-\frac13j_\mu^\phi,
\end{equation}
where the vector meson currents can be expressed in terms of their quark content via
\begin{align}
\nonumber j_\mu^\rho&=\frac1{\sqrt 2}\left (\bar u\gamma_\mu u-\bar d\gamma_\mu d\right ),\\
\nonumber j_\mu^\omega&=\frac1{\sqrt 2}\left (\bar u\gamma_\mu u+\bar d\gamma_\mu d\right ),\\
j_\mu^\phi&=\bar s\gamma_\mu s.
\end{align}
With this definition the matrix element can be expressed as
\begin{equation}
\langle 0|j_\mu^{\text{em}}|V\rangle=\frac{M_V^2}{g_V}\epsilon_\mu^{(V)},
\end{equation}
where $\epsilon_\mu^{(V)}$ is a polarization vector and $g_\omega \simeq g_\rho \equiv g \simeq 6 < g_\phi \simeq 7.8$ and $M^2_V = 2 g^2_V f_\pi^2$. We will use this notation for the couplings throughout this analysis.

\medskip

The fact that the electromagnetic current is made of vector meson fields suggests an underlying lagrangian with a mixing term proportional to $\rho_\mu A^\mu$ but this piece breaks gauge invariance. One of the consistent ways to introduce both gauge invariance and the photon-vector meson mixing is given by the current-mixing form \cite{KLZ} with an explicit mixing of vector mesons and photons both in the mass and kinetic terms. In the limit of universality $g_\rho=g_{\rho\pi\pi}=g_{\rho NN}=\dots$, an alternative description of VMD can be equivalently used with a mixing of vector mesons with photons in the mass term and a non-trivial photon mass (mass-mixing form) \cite{vmd}. In any of these formulations the photon remains massless when one computes its propagator to all orders. Despite the second formulation is not as elegant as the first due to a photon mass term, it has been established as the most popular representation of VMD \cite{vmd,vmd2}. From now on we will use this representation. Quark-meson interactions are described by \cite{avep}
\begin{align}\label{veclagr}
{\cal L}_{\text{int}} = \bar q \gamma_\mu V^\mu q;\quad V_\mu \equiv - e A_\mu Q +
\frac12 g_\omega \omega_\mu \mathbf{I}_{q} + \frac12 g_\rho \rho_\mu \lambda_3 + \frac1{\sqrt 2}g_\phi \phi_\mu \mathbf{I}_{s}, 
\end{align}
where
$Q= \frac{\lambda_3}{2} + \frac16 \mathbf{I}_{q} - \frac13 \mathbf{I}_{s}$, $\mathbf{I}_{q}$ and $\mathbf{I}_{s}$ are the identities in the non-strange and strange sector, respectively; and $\lambda_3$ is the corresponding Gell-Mann matrix. Then the Maxwell and mass terms of the VMD lagrangian are
\begin{eqnarray}\label{eqvdm}
&\!\!\!{\cal L}_{\text{kin}} = - \frac14 \left(F_{\mu\nu}F^{\mu\nu}+ \omega_{\mu\nu}\omega^{\mu\nu}+
 \rho_{\mu\nu}\rho^{\mu\nu} +
 \phi_{\mu\nu}\phi^{\mu\nu}\right)+ \frac12 V_{\mu,a} \hat m^2_{ab} V^\mu_b, \nonumber\\
&\!\!\! \hat m^2 =
m_V^2\left(\begin{array}{ccccccc}
\frac{4 e^2}{3g^2} & &-\frac{e}{3g} && -\frac{e}{g}&& \frac{\sqrt 2 eg_\phi}{3g^2} \\
 -\frac{e}{3g}&& 1 && 0 &&0 \\
 -\frac{e}{g} && 0 && 1 &&0 \\
 \frac{\sqrt 2 eg_\phi}{3g^2} && 0 && 0 &&\frac{g_\phi^2}{g^2} \\
\end{array}\right),\ \mbox{\rm det}\left( \hat m^2\right) = 0,
\end{eqnarray}
where $(V_{\mu,a})\equiv \left(A_\mu,\, \omega_\mu, \, \rho_\mu^0 \equiv \rho_{\mu}, \, \phi_\mu\right)$ and $m_V^2 \equiv m^2_\rho = 2 g^2_\rho f_\pi^2\simeq m^2_\omega$. This matrix reflects the VMD relations at the quark level \cite{vmd,vmd2}.

\medskip

The pion form factor $F_\pi(q^2)$ represents the contribution from the intermediate vector mesons connecting the photon to pions in a process like $\gamma\to\pi^+\pi^-$. The form factor is the multiplicative deviation from a point-like behaviour of photon-pion coupling. In the limit of zero momentum transfer ($q^2\to 0$), the photon is only affected by the charge of the pions and therefore one expects $F_\pi(0)=1$. This relation is automatically satisfied by the pion form factor derived from the current-mixing Lagrangian \cite{KLZ} while for the mass-mixing scenario \cite{vmd}, it is necessary to impose universality via $g_\rho=g_{\rho\pi\pi}$. This is the basis of the argument for universality given in \cite{vmd}, where the photon couples to the $\rho$ due to the imposition of local gauge symmetry. Therefore $g_{\rho\pi\pi}$ must equal $g_\rho$. This is a direct consequence of assuming complete $\rho$ dominance of the form factor.

\section{Vector Meson Dominance Lagrangian in the presence of $P$-breaking background}\label{VMD1}

Once we introduced the VMD model in the $SU(3)$ representation, let us recall the arguments given in Chapter \ref{Chfase} about the approximate decoupling of the strange quark flavour from the rest due to intensive left-right oscillations as $m_s \gg \tau_{\text{fireball}}^{-1}$. Hence, even if an axial charge appears in the strange sector, its average will be essentially zero. In addition the role of the $\phi$ meson is negligible as its typical mean free path $\sim 40$ fm makes it less sensitive to medium effects. In this sense, we will deal with a non-strange axial chemical potential like in the previous chapters and decouple the strange sector reducing the symmetry of our initial VMD Lagrangian \eqref{eqvdm} from $SU(3)$ to $SU(2)$. This simplification does not introduce significant changes to the Lagrangian except for the reduction of the group matrix $\lambda_3$ to $\tau^3$ and the mass matrix
\begin{align}\label{VMDmassSU2}
\hat m^2&=m_V^2\begin{pmatrix}
\frac{10e^2}{9g^2} & -\frac e{3g} & -\frac eg\\
-\frac e{3g} & 1 & 0\\
-\frac eg & 0 & 1
\end{pmatrix}, \qquad \det(\hat m^2)=0,
\end{align}

\medskip

As previously discussed, parity-breaking effects cannot be transmitted to the vector sector by means of an axial chemical potential playing the role of the time component of a spurious axial-vector field. Therefore, we will neglect the mixing among states of different parity and only use the parity-odd Chern-Simons term to include the effect an axial quark charge
\begin{align}\label{CS}
{\cal L}_{\text{CS}}\,= -\frac14 \varepsilon^{\,\mu\nu\rho\sigma}\, \mbox{\rm Tr}\left [ \,\hat\zeta_\mu \, V_\nu\, V_{\,\rho\sigma}\right ]= \frac12 \mbox{\rm Tr}\left [\,\hat\zeta \,\epsilon_{jkl}\, V_{j} \,\partial_k V_{l} \right ]
= \frac12 \,\zeta\,\epsilon_{jkl}\, V_{j,a} \,\hat N_{ab}\,\partial_k V_{l,b},
\end{align}
where a time-dependent pseudoscalar axion-like background is considered in order to accommodate an axial chemical potential associated to large topological fluctuations (recall Eq. \eqref{mu5}). By means of the PCAC \eqref{pcac} and the introduction of electromagnetism in the investigation of LPB \eqref{Q5tilde} it is possible to relate the gluon and the photon contributions and therefore relate the VMD bosonization with the $\theta$ angle appearing in the QCD Lagrangian \eqref{thetaterm}. With this definition, the relation $\zeta=N_c\;g^2\mu_5/8\pi^2 \simeq 1.37\mu_5$ holds. The presence of this term in the VMD Lagrangian induces an additional mixing among photons and vector mesons due to LPB. Note that in this chapter, $\zeta$ will appear naturally in the vector meson dispersion relation and thus, we will refer to either this magnitude or the axial chemical potential indifferently as they are proportional.

\medskip

For an isosinglet pseudoscalar background $\hat\zeta=2\zeta/g^2 \mathbf{I}_{q}$, the mixing matrix reads
\begin{align}\label{tab1}
(\hat N^\theta) \, \simeq\, \left(\begin{array}{ccccc}
\frac{10e^2}{9g^2} & &-\frac{e}{3g}& &-\frac{e}{g}\\
-\frac{e}{3g}& & 1 & & 0\\
-\frac{e}{g}& & 0 & & 1\\
\end{array}
\right)=\frac{(\hat m^2)}{m_V^2}\Bigg|_{SU(2)_f},\ \ \mbox{\rm det} (\hat N^\theta) = 0. 
\end{align}
In the case of an isotriplet pion-like condensate with $\hat\zeta=10\zeta/(3g^2)\tau^3$, the corresponding mixing matrix is
\begin{align}\label{tab2}
(\hat N^\pi)\, \simeq\, \left(\begin{array}{ccccc}
\frac{2e^2}{3g^2} & &-\frac{e}{g}& &-\frac{e}{3g}\\
-\frac{e}{g}& & 0 & & 1 \\
-\frac{e}{3g}& & 1 & & 0 \\
\end{array}
\right),\ \ \mbox{\rm det}\left(\hat N^\pi\right) = 0.
\end{align}

The VMD coefficients in \eqref{eqvdm}, \eqref{VMDmassSU2}, \eqref{tab1} and \eqref{tab2} are obtained from the anomalous Wess-Zumino action \cite{truhlik} and related to the phenomenology of radiative decays of vector mesons \cite{radec}. The ratios of matrix elements for an isotriplet condensate in \eqref{tab2} are in direct agreement with the experimental decay constants for the processes $\pi^0\to\gamma\gamma$, $\omega\to\pi^0\gamma$, $\rho^0\to\pi^0\gamma$ \cite{truhlik} and for the decay $\omega\to\pi\pi\pi$ \cite{weise} taken from \cite{PDG}. Likewise the VMD coefficients in \eqref{tab1} can be estimated from the experimental decay constants for the processes $\eta \to \gamma\gamma$, $\eta' \to \gamma\gamma$, $\omega \to \eta\gamma$ and $\rho^0 \to \eta\gamma $ after taking into account a strong $\eta_8 - \eta_0$ mixing \cite{mixing}. Only the ratio of the decay widths $\omega \to \eta\gamma$, $\rho^0 \to \eta\gamma $ is a little sensitive to the mixing and confirms the off-diagonal elements of \eqref{tab1}. Armed with the above information one can easily determine the modifications on the vector meson spectrum and their eventual relevance.

\medskip

With an isosinglet $\mu_5$ the mass-shell equations for vector mesons read
\begin{align}\label{KV}
\!\!\!\! K^{\mu\nu}_{ab} V_{\nu, b} = 0;\quad k^\nu\,V_{\nu, b} = 0, \quad
K^{\mu\nu} \equiv g^{\mu\nu} (k^2 \mathbf{I} - \hat m^2) - k^\mu k^\nu \mathbf{I} +
i \varepsilon^{\,\mu\nu\rho\sigma}\,\zeta_\rho k_\sigma \hat N^\theta
\end{align}
and select three physical polarizations vectors $(\varepsilon_+,\varepsilon_-, \varepsilon_L)$ for massive vector fields that couple to conserved fermion currents. Following the lines of \cite{axion} let us now define 
\begin{equation} S^{\,\nu}_{\;\lambda}\ \equiv\
\varepsilon^{\,\mu\nu\alpha\beta}\,\zeta_{\alpha}\,k_{\beta}\,\varepsilon_{\,\mu\lambda\rho\sigma}\,\zeta^{\rho}\,k^{\sigma}, \end{equation}
which satisfies the following properties
\begin{equation}
S^{\,\nu}_{\;\lambda}\,\zeta^{\lambda}=S^{\,\nu}_{\;\lambda}\,k^{\lambda}=0,\quad S^{\,\mu\nu}\,S_{\,\nu\lambda}=\frac{S}{2}\,S^{\,\mu}_{\;\;\lambda},\quad S=S^{\,\nu}_{\;\;\nu}= 2 [(\zeta\cdot k)^2-\zeta^2 k^2].
\end{equation}
For a time-like and spatially isotropic pseudoscalar condensate we find $S=2\zeta^2\,\vec k^2>0$. It is convenient to introduce the two orthonormal hermitian projectors
\begin{equation}
P^{\,\mu\nu}_{\,\pm}\equiv \frac{S^{\,\mu\nu}}{S}\; \pm\;\frac{\rm{i}}{\sqrt{2S}}\,
\varepsilon^{\,\mu\nu\alpha\beta}\,\zeta_{\alpha}\,k_{\beta} ,
\end{equation}
which satisfy the following properties:
\begin{eqnarray}\label{projectors}
\nonumber P^{\,\mu\nu}_{\,\pm}\;\zeta_{\nu}&=&P^{\,\mu\nu}_{\,\pm}\;k_{\nu}=0 , \qquad
g_{\,\mu\nu}\,P^{\,\mu\nu}_{\,\pm}=1 ,\\
P^{\,\mu\lambda}_{\,\pm}\,P_{\,\pm\,\lambda\nu}&=&
P^{\,\mu}_{\,\pm\,\nu} ,\qquad
P^{\,\mu\lambda}_{\,\pm}\,P_{\,\mp\,\lambda\nu} = 0 ,\qquad
P^{\,\mu\nu}_{+} + P^{\,\mu\nu}_{-} \ =\ \frac{2}{S}\, S^{\,\mu\nu}.
\end{eqnarray}

A pair of complex and space-like chiral polarization vectors can be constructed by means of a constant and space-like 4-vector $\epsilon^\mu$ in such a manner that we can set
\begin{equation}
\varepsilon^{\,\mu}_{\pm}(k)\ \equiv\ \left[\,\frac{\vec k^2-(\epsilon\cdot k)^2}{2\vec k^2}\,\right]^{-1/2}\, P^{\,\mu\nu}_{\,\pm}\;\epsilon_{\,\nu} ,
\end{equation}
which satisfy the orthogonality relations
\begin{equation}
-\,g_{\,\mu\nu}\;\varepsilon^{\,\mu\,\ast}_{\pm}(k)\,\varepsilon^{\,\nu}_{\pm}(k)=1 ,\qquad\quad g_{\,\mu\nu}\,\varepsilon^{\,\mu\,\ast}_{\pm}(k)\,\varepsilon^{\,\nu}_{\mp}(k)=0 ,
\end{equation}
as well as the closure relation
\begin{eqnarray}
\varepsilon^{\,\mu\,\ast}_{+}(k)\,\varepsilon^{\,\nu}_{+}(k) +\varepsilon^{\,\mu\,\ast}_{-}(k)\,\varepsilon^{\,\nu}_{-}(k)\ +\ {\rm c.c.} = -\frac{4}{S}\,{S^{\,\mu\nu}}.
\end{eqnarray}

The longitudinal polarization $\varepsilon^\mu_L$ is orthogonal to $k_\mu$ and to $\zeta_\mu$
\begin{equation}
\varepsilon^\mu_L = \dfrac{\zeta^\mu k^2 - k^\mu (\zeta \cdot k)}{\sqrt{k^2 \big((\zeta \cdot k)^2 - \zeta^2 k^2\big)}},\quad
\varepsilon_{L}\cdot \varepsilon_L = - 1,
\end{equation}
for $k^2 > 0$. Its projector can be easily found through the relation
\begin{equation}
\sum_\epsilon P^{\mu\nu}_\epsilon=g^{\mu\nu}-\frac{k^\mu k^\nu}{k^2} \quad \Longrightarrow \quad P_L^{\mu\nu}=g^{\mu\nu}-\frac{k^\mu k^\nu}{k^2}-2\frac{S^{\mu\nu}}S
\end{equation}
where the sum in $\epsilon$ is understood to be performed over all the physical polarizations. The expression of a projector as a function of the polarization vectors is given by
\begin{equation}\label{p=-ee}
\varepsilon^\mu_\epsilon\varepsilon^{*\nu}_{\epsilon'}=-P_\epsilon^{\mu\nu}\delta_{\epsilon \epsilon'}.
\end{equation}
Taking into account the relations \eqref{projectors} we may rewrite \eqref{KV} as
\begin{equation} 
 K^{\mu\nu} \varepsilon_{\nu}(k) = \left (g^{\mu\nu} (k^2 \mathbf{I} - \hat m^2) - k^\mu k^\nu \mathbf{I} +\sqrt{(\zeta \cdot k)^2 - \zeta^2 k^2} \left (P^{\mu\nu}_{+}-P^{\mu\nu}_{-}\right ) \hat N^\theta\right ) \varepsilon_{\nu}(k)= 0.
\end{equation}
The mass spectrum can be found after simultaneous diagonalization of $ \hat m^2|_{SU(2)}\propto \hat N^\theta $ with $\zeta_\mu \simeq (\zeta, 0,0,0)=$ constant
\begin{align}
\nonumber \hat N &= \mbox{\rm diag}\left[0,\, 1,\, 1+ \frac{10 e^2}{9g^2}\right] 
\simeq \mbox{\rm diag} \left[0,\, 1 ,\, 1\right],\\
\nonumber\hat m^2 &= m_V^2 \, \mbox{\rm diag} \left[0,\, 1,\, 1+ \frac{10 e^2}{9g^2}\right] \simeq m_V^2 \,\mbox{\rm diag} \left[0,\, 1,\, 1\right].
\end{align}
We notice that in the case of pure isosinglet pseudoscalar background, a null eigenvalue is found corresponding to massless photons that are not distorted by the mixing with massive vector mesons or the parity-breaking effect.

\medskip

In turn massive vector mesons do exhibit a distortion due to the violation of parity. Due to the orthogonality of the longitudinal polarization vector $\varepsilon_L$ to the $\pm$ projectors \eqref{projectors}, its mass is the same as in vacuum. However, the transversal polarizations satisfy
\begin{align}\label{mvec}
K^{\mu}_{\nu}\varepsilon^\nu_\pm = \Big(k^2 \mathbf{I} - \hat m^2 \pm \sqrt{(\zeta \cdot k)^2 - \zeta^2 k^2}\ \hat N \Big) \varepsilon^\mu_\pm;\quad
 m^2_{V,\pm}\equiv k_0^2 - \vec k^2 = m_V^2 \mp \zeta |\vec k| . 
\end{align}
We may combine the longitudinal and transverse polarizations into a unique expression \cite{avep}
\begin{equation}\label{disp.rel}
m_{V,\epsilon}^2=m_V^2-\epsilon\zeta|\vec k|,
\end{equation}
where $\epsilon$ accounts for the polarization of the vector meson ($\epsilon=\pm1$ for the transverse polarizations and $\epsilon=0$ for the longitudinal one). The breaking of Lorentz invariance through the 3-momentum dependence in the dispersion relation is due to the time-dependent background. The parity-breaking effect is unambiguously reflected in the polarization dependence of masses $m^2_{V,+} < m^2_{V,L}< m^2_{V,-}$. For large enough $|\vec k|\geq m_V^2/\zeta$ ($\simeq 1.5$ GeV for $\zeta=400$ MeV corresponding to $\mu_5\approx 290$ MeV) vector meson states with positive polarization become tachyons in the same way we observed in the scalar sector in Chapter \ref{ChSmodel}. However their group velocity remains less than the speed of light provided that $\zeta<2m_V\approx 1.6$ GeV. For higher values of $\zeta$ the vacuum state becomes unstable, namely, polarization effects give an imaginary part for the vacuum energy. However, this threshold is much above the scales we are considering to obtain a stable parity-breaking phase as we showed in Chapter \ref{ChNJL}. Note that the position of the resonance poles for $\pm$ polarized mesons is moving with wave vector $|\vec k|$. Therefore they appear as broadened resonances, leading to an enhancement of their spectral function away from their nominal vacuum resonance position and a reduction at the very peak. In this work we will only consider $|\zeta|$ since a change of sign simply interchanges the $\pm$ polarizations but does not affect the separation of the polarization-dependent vector meson masses.

\section{Dilepton production rate in a $P$-breaking medium}\label{dilepLPB}
In this section we want to link the in-medium modifications of vector mesons with the phenomenology associated to them. Our aim consists in finding possible experimental signatures of parity violation effects through the decays of such distorted vector mesons. Hadronic decays may be severely distorted due to the new dispersion relations of scalar and pseudoscalar mesons as we discussed in Chapter \ref{ChSmodel}. The cleanest way to extract reliable properties of in-medium vector mesons is through their electromagnetic (rather than strong) interactions. In what follows we will consider the modifications in the dilepton spectrum due to the $\rho$ and $\omega$ decays.

\medskip

The dilepton production from the $V(k)\to \ell^-(p)\ell^+(p')$ decays is governed by
\begin{align}\label{genericproduct}
\nonumber \frac{dN_V}{dM}=&\int \frac{d\tilde M}{\sqrt{2\pi}\Delta}\exp\left [-\frac{(M-\tilde M)^2}{2\Delta^2}\right ]c_V\frac{\alpha^2}{24\pi^2\tilde M}\Theta(\tilde M-n_Vm_\pi)\left (1-\frac{n_V^2 m_\pi^2}{\tilde M^2}\right )^{3/2}\\
\nonumber &\times \int\frac{d^3\vec k}{E_k}\frac{d^3\vec p}{E_p}\frac{d^3\vec p'}{E_{p'}}\delta^4(p+p'-k)\sum_{\epsilon}\frac{m_{V,\epsilon}^4\left (1+\frac{\Gamma_V^2}{m_V^2}\right )}{\left (\tilde M^2-m_{V,\epsilon}^2\right )^2+m_{V,\epsilon}^4\frac{\Gamma_V^2}{m_V^2}}\\
&\times P^{\mu\nu}_{\epsilon}(\tilde M^2g_{\mu\nu}+4p_\mu p_\nu)\frac1{e^{\tilde M_T/T}-1},
\end{align}
where $V=\rho,\omega$ and $n_V=2,0$ respectively\footnote{$n_\omega=0$ is taken since we do not include the threshold to 3 pions.}, and $\tilde M>n_Vm_\pi$. Note that the coupling to pions is only relevant for the position of the threshold and therefore we do not complicate the calculations by considering distorted pions as in the previous chapter. A dilepton invariant mass smearing of width $\Delta$ is taken into account to relate the measured $M$ with the physical $\tilde M$. From now on this smearing will be omitted (together with dropping the tilde of $\tilde M$) for the sake of simplicity but considered in our numerical calculations. $M_T$ is the transverse mass $M_T^2=M^2+\vec k_T^2$ where $\vec k_T$ is the vector meson transverse momentum and $M$ is the dilepton invariant mass. The projectors $P_\epsilon^{\mu\nu}$ have been detailed in \eqref{projectors}. $p$ and $p'$ correspond to the lepton and anti-lepton momenta. A Boltzmann distribution is included with an effective temperature $T$ \cite{phenix,lmlt} that may in fact depend on the range of $M$, $p_T$ and centrality. A simple thermal average is appropriate only for central collisions and moderate values of $p_T$. No serious attempt will be made here to extrapolate to peripheral processes, but an appropriate mixture of distorted and non-distorded vector mesons should be adequate. The constants $c_V$ normalize the contribution of the respective resonances parametrizing in an effective way their total cross sections. Since they are not known with precision, particularly their off-shell values, the relative weights are used as free parameters in the 'cocktail' of hadronic sources used to fit the experimental data \cite{CERES,HELIOS,scomparin-arnaldi2,NA60,phenix}. The usual 'cocktail' contains weights normalized to the peripheral collisions result (roughly agreeing with existing pp and p-nucleus data). Finally, modified form factors due to LPB are included according to our previous discussion.

\medskip

In order to make our plots meaningful we should implement some values relevant for the main HIC experiments. A prediction of the parity-breaking effect for the NA60 experiment is certainly appealing. However, their earlier results \cite{scomparin-arnaldi2} are not corrected for the detector acceptance and therefore a valid comparison with our theoretical predictions cannot be done. Their newest data \cite{NA602,Specht:2010xu} are already acceptance corrected but they do not exactly correspond to the same variables we are describing. For these reasons, we will compare our results with NA60 data but we understand that this is a crude connection.

\medskip

The temperature appearing in the previous formula is not the 'true' temperature of the hadron gas; it rather corresponds to the effective temperature $T_\text{eff}$ of the hadrons that best describes the slope of the multipliplicity distribution $d^3N/dp^3(M_T)$ for a given range of invariant masses. In fact, one should not confuse the theoretical and the phenomenological approaches for the calculation of multiplicities in ultra-relativistic collisions. The first one is the direct contribution to the invariant one-particle multiplicity distribution given by
\begin{equation}
\frac 1{M_T}\frac{dN}{dy\ dM_T}=AE\int_V d^3x\exp\left [-\frac{p\cdot u(x)}T\right ]=AE\int_V d^3x\exp\left [-\frac{\gamma(x)[E-\vec p\, \vec u(x)}T\right ],
\end{equation}
where $T$ is the real temperature, $A$ is an arbitrary normalization constant and $u(x)$ corresponds to the collective 4-velocity. The (phenomenological) transverse one-particle multiplicity distribution depends on the specific type of particle and reaction and is governed by a simpler expression
\begin{equation}
\frac 1{M_T}\frac{dN}{dy\ dM_T}=A'\exp\left (-\frac{M_T}{T_{\text{eff}}}\right ),
\end{equation}
$T_{\text{eff}}$ being an effective temperature or slope parameter and $A'$ a different arbitrary normalization constant. In the limit where the particle velocity is large compared an average collective velocity $v$ and specializing to the transverse direction, one may neglect the pre-exponential energy dependence and obtain the relation among the two temperatures \cite{Nix}
\begin{equation}
T_{\text{eff}}=\frac{M_T}{M_T-p_Tv}\sqrt{1-v^2}\,T.
\end{equation}
For massless particles, this expression is reduced to
\begin{equation}
T_{\text{eff}}=\sqrt{\frac{1+v}{1-v}}\,T.
\end{equation}
For a typical value of an average transverse collective velocity $v=0.4c$, a temperature $T=150$ MeV leads to an effective temperature (or slope) of the order of $T_{\text{eff}}\simeq 230$ MeV. Obviously $T_\text{eff} > T$ and therefore it is quite possible that $T_\text{eff}$ exceeds the deconfinement temperature while one is still dealing with hadrons.

\medskip

\begin{figure}[h!]
\centering
\includegraphics[scale=0.35]{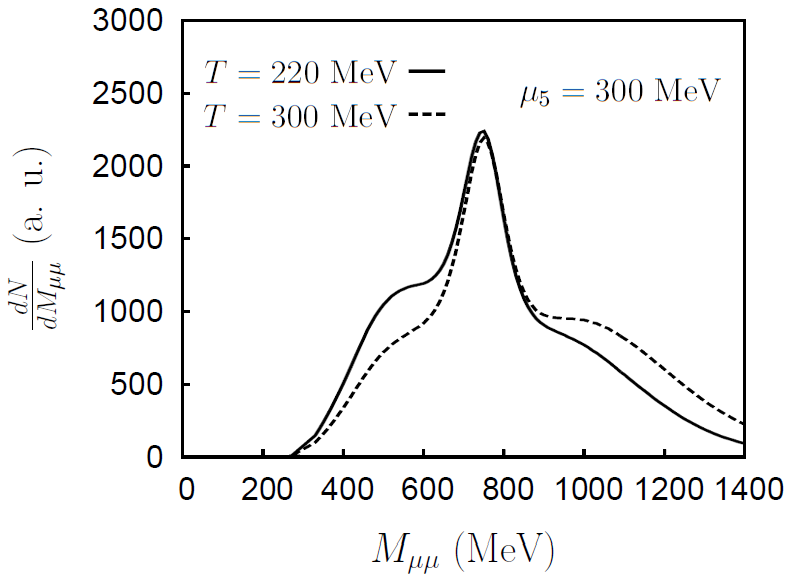}\qquad \includegraphics[scale=0.35]{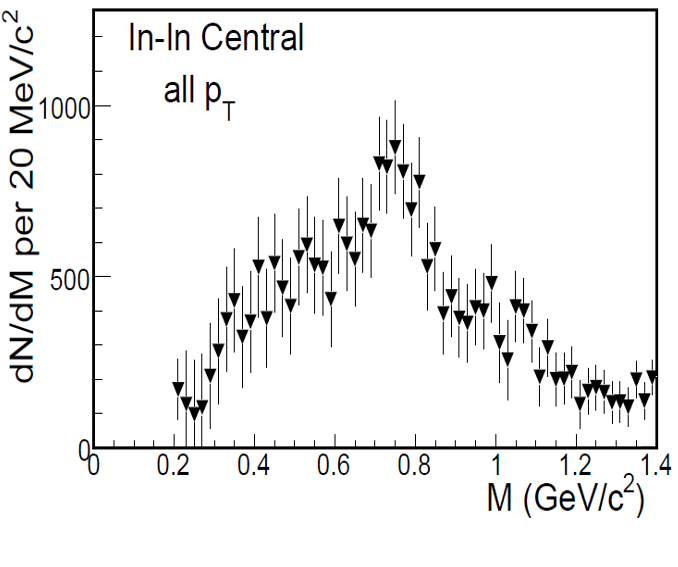}
\caption{The $\rho$ spectral function is presented depending on the invariant mass $M$ with $\mu_5=300$ MeV and $T=220,300$ MeV (left). For comparison NA60 experimental data \cite{scomparin-arnaldi2,NA60} is shown in the right panel. Experimental results are not corrected for NA60 experimental acceptance. We use arbitrary units in our fit.}\label{na60}
\end{figure}

NA60 has obtained accurate results for the $\rho$ spectral function \cite{scomparin-arnaldi2,NA60} by measuring the $\mu^+\mu^−$ spectrum with unprecedented precision and by carefully subtracting the contributions from the 'cocktail' except the $\rho$ itself. If one implements the NA60 acceptance with an experimental cut in rapidity ($3.3<y<4.3$ in the laboratory frame which is translated to $0.42<y<1.42$ in the center of mass frame where we work) together with a temperature $220\lesssim T\lesssim 300$ MeV \cite{scomparin-arnaldi2,NA60} (for all $p_T$), we obtain a $\rho$ spectral function similar to the experimental one using $m_\rho$ = 750 MeV. In Fig. \ref{na60} (left) we show two plots with $\mu_5=300$ MeV and $T=220,300$ MeV in order to compare them with experimental data (right). Of course, other explanations for such distortion \cite{rapp,renkr,BratkCass,zahed} would need to be also considered if one attempted to reproduce the data.

\medskip

The noticeable discrepancy below $M\lesssim600$ MeV may be due to a broad variety of facts. First, we recall that NA60 data are not corrected for the detector acceptance, subject that we did not pursue and therefore, strict comparison is misleading. Next, thermal effects inducing $\rho$ broadening and other effects have not been accounted for. And finally, it may be revealing that the upper kinematical limit for the Dalitz process $\omega\to\mu^+\mu^-\pi^0$ is 643 MeV, suggesting that this process could play a significant role as its contribution is subtracted from the total dilepton yield assuming a vacuum contribution. In fact, Dalitz decays are thought to take place after hadron freezout and therefore they should not be affected by in-medium effects, but the possibility that the parity-breaking phase still survives at lower temperatures is an interesting subject to be explored. We address this possibility in Appendix \ref{AppDalitz}.

\medskip

Another remarkable consideration that can also be extracted from Fig. \ref{na60} is the dilepton production near the threshold. While in our computation we implemented the two-pion threshold that corresponds to $2m_\pi\approx 280$ MeV, the experimental data seem to reflect a different behaviour showing a threshold due to the muon mass corresponding to $2m_\mu\approx 210$ MeV. The presence of dimuon events within these two values could entail the disappearance of the in-medium two-pion threshold. In any case, the main results of our work are not affected by this subtlety.

\medskip

We will implement now the experimental cuts of the PHENIX experiment ($p_T>200$ MeV and $|y_{ee}|<0.35$ in the center of mass frame) and $T\simeq 220$ MeV together with a Gaussian invariant mass smearing of width 10 MeV \cite{phenix,masssmear}. We have not implemented the single electron cut $|y_{e}|<0.35$ because in practice it makes no visible difference with simply imposing a cut on the dilepton pair momentum as a whole but we will check this issue explicitly more below. The value of the temperature is extracted from the measurement of direct photons, but it has to be understood as a representative value. In fact, NA60 reports even higher temperatures when the $\rho$ spectral function is decomposed into the peak and the continuum contributions \cite{scomparin-arnaldi2,NA60}. Note that with such value for the effective temperature, vector mesons with high energy will be produced with very low probability like the tachyonic states with $+$ transverse polarization which require at least $|\vec k|>1.5$ GeV.

\medskip

\begin{figure}[h!]
\centering
\includegraphics[scale=0.35]{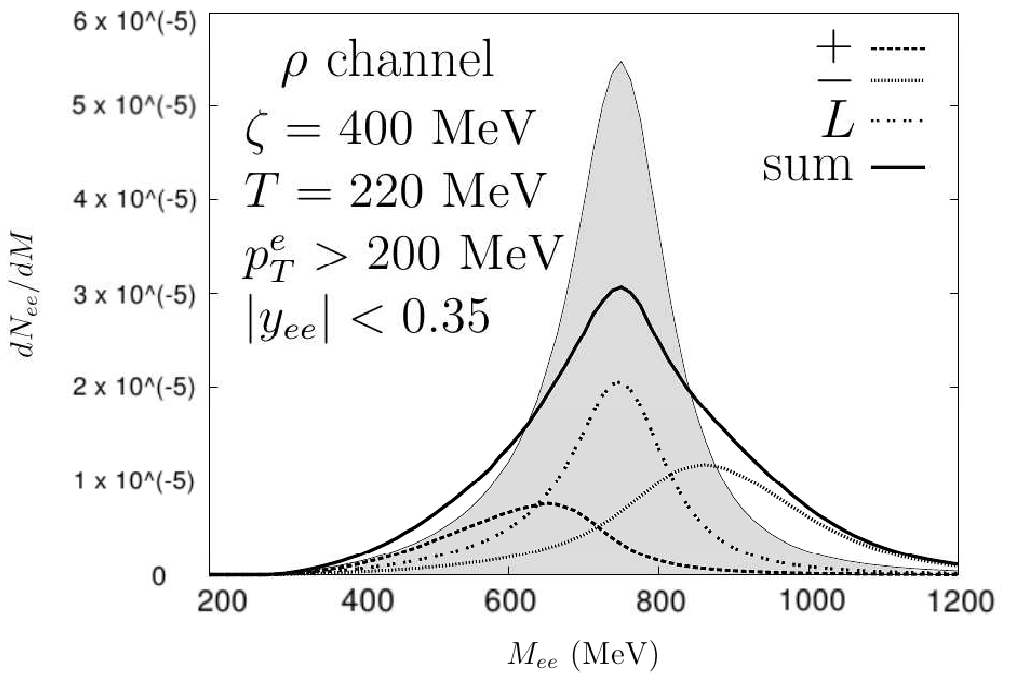}
\caption{The polarization splitting of the $\rho$ contribution to dilepton production is shown for LPB with $\zeta = 400$ MeV ($\mu_5\simeq290$ MeV). The comparison with a parity-even medium with $\zeta = 0$ (shaded region) is presented. The vertical units are taken to coincide with PHENIX experimental data \cite{phenix}, as well as experimental detector cuts and temperature.}\label{polasym}
\end{figure}

Then one is lead to the dilepton production result that is shown in Figure \ref{polasym}, where the $\rho$ spectral function is presented together with the separate contributions of each polarization for $\zeta=400$ MeV (corresponding to $\mu_5=290$ MeV) and a comparison with $\zeta=0$ (no LPB). The $\pm$ transversely polarized mesons exhibit two different peaks aside the vacuum resonance. Their combination with the longitudinal polarization results in a distorted spectral function that could be experimentally measured.

\medskip

A remarkable feature of the $\rho$ spectral function in Fig. \ref{polasym} is the polarization asymmetry observed aside the vacuum peak. Longitudinally polarized mesons seem to dominate the vacuum peak while the transverse ones are more important in the vicinity of the resonance. Therefore a measure of dilepton polarization event-by-event could reveal in an unambiguous way the existence of LPB, confirming the hypothesis of pseudoscalar condensate formation in HIC. We will come back to this important characteristic in Section \ref{angles} where we will deeply investigate how this effect may be experimentally detected. Let us stress that similar results may be obtained modifying $\zeta$ so when it increases, the circularly polarized peaks appear more separated from the vacuum one. If the effective temperature $T$ increased, we would see an enhancement of the upper mass tail. However, the main qualitative results from this plot are not affected by (reasonable) temperature modifications.
 
\medskip

\begin{figure}[h!]
\centering
\includegraphics[scale=0.35]{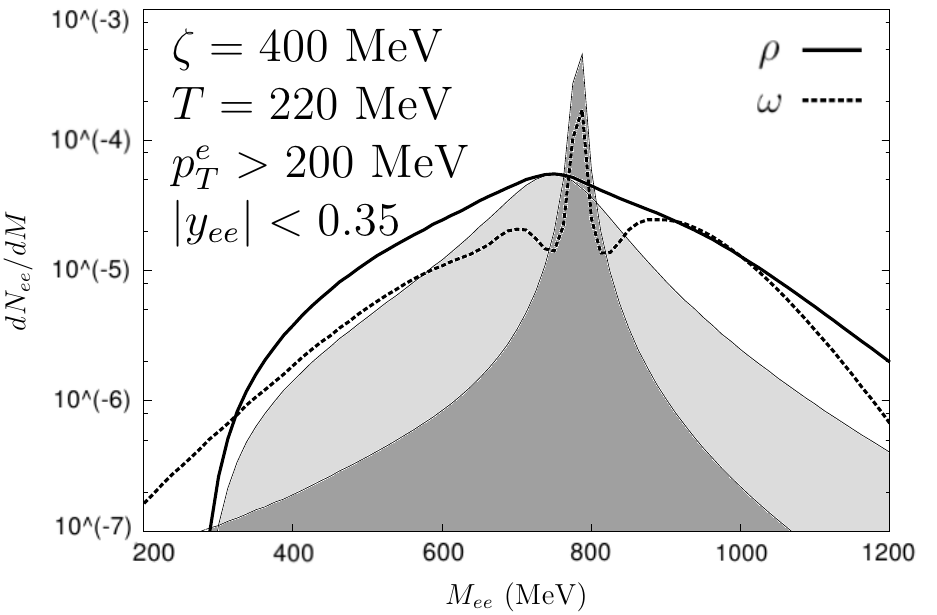}
\caption{The in-medium contribution in the $\rho$ and $\omega$ channels (solid and dashed line, respectively) is presented for $\zeta=400$ MeV together with their vacuum contributions (light and dark shaded regions, respectively). The in-medium $\rho$ yield is enhanced by a factor 1.8 (see text). The vertical units are taken to coincide with PHENIX experimental data \cite{phenix}, as well as experimental detector cuts and temperature.}\label{dilepenh}
\end{figure}

The simulation of $\omega$ meson production \cite{phenixomega} shows that a significant fraction of them decay inside of the nuclear fireball and therefore LPB distorted $\omega$ mesons may also play a significant role in this game. In Fig. \ref{dilepenh} the separate contributions of $\rho$ and $\omega$ are presented for $\zeta=400$ MeV to be compared with their vacuum results. An enhancement factor 1.8 is included for the ratio $\rho/\omega$ in central HIC, particularly at low $p_T$. This enhancement is due to the $\rho$ meson regeneration into pions within the fireball (a plausible consequence of the '$\rho$ clock' effect \cite{Specht:2010xu,Heinz:1991fn}) and the in-medium $\omega$ suppression \cite{omegasupp}. There are no published reports of a direct determination of the $\rho/\omega$ ratio at PHENIX and we have decided to use the above value that is very close to the average value for the ratio of the respective cross sections reported by NA60 \cite{scomparin-arnaldi2}. We note that the $\eta/\omega$ ratios measured both by NA60 and PHENIX collaborations roughly agree. Note too that due to the assumed effective thermal distribution and the fact that the $\rho$ and $\omega$ are nearly degenerate in mass the ratio $c_\rho/c_\omega$ is identical to the ratio of the respective cross-sections. However it should be stated right away that our conclusions do not depend substantially on the precise value of the ratio $c_\rho/c_\omega$.

\medskip

\begin{figure}[h!]
\centering
\includegraphics[scale=0.25]{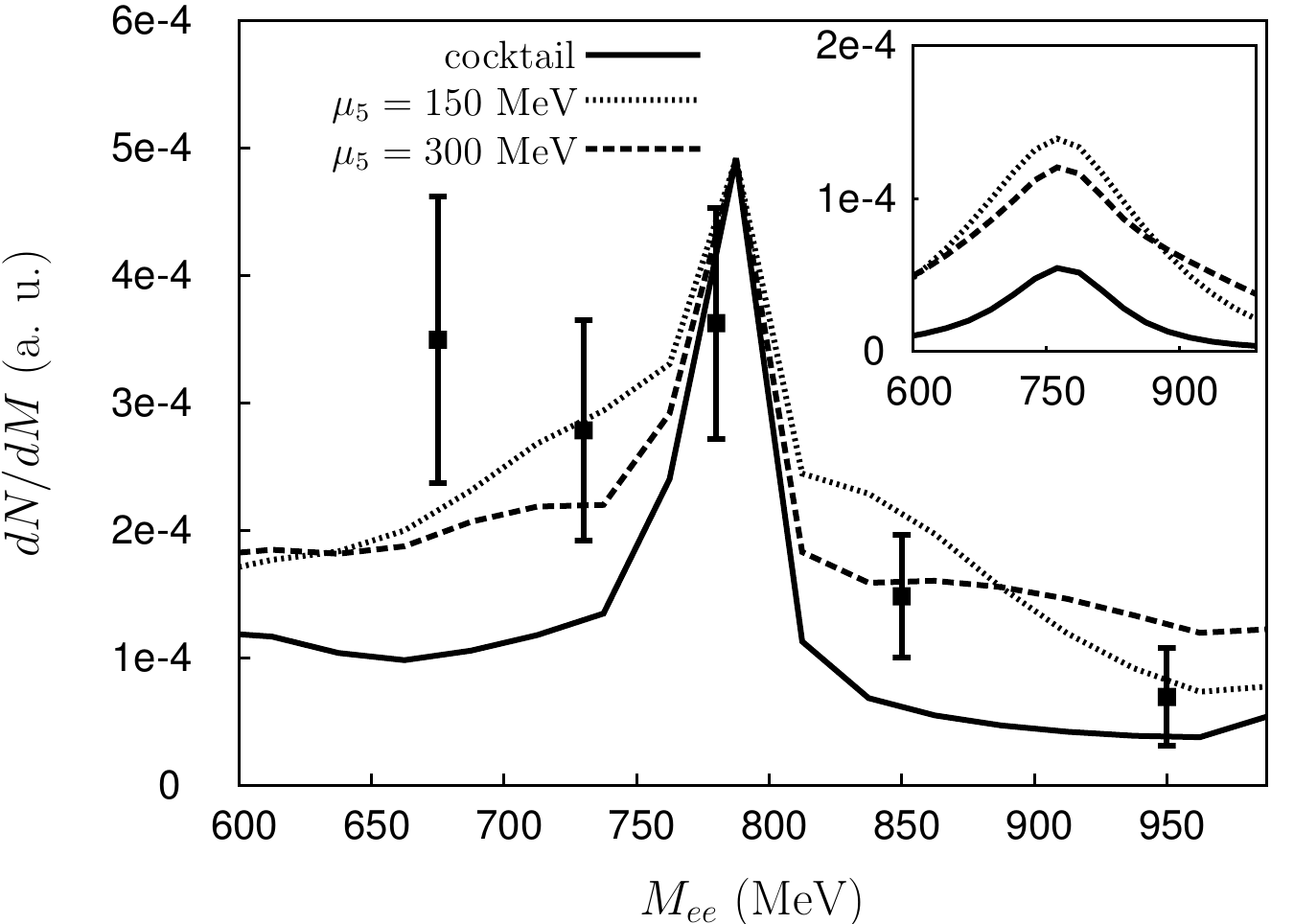}
\caption{The meson contribution to dilepton production is shown for parity symmetric nuclear matter $\mu_5=0$ (solid line) and for local parity breaking with $\mu_5=150$ and 300 MeV (dotted line and discontinuous line, respectively) compared to the PHENIX
measurements for $p_T^e>200$ MeV, $|y_{ee}|<0.35$ and $T=220$ MeV. The $\mu_5=0$ line is just the 'cocktail' contribution as quoted by PHENIX \cite{phenixdata}. The $\mu_5\neq 0$ results show an enhancement of the ρ/ω ratio by a factor 1.8 (see text) normalizing the $\omega$ contribution to fit the 'cocktail' peak. Data for $M < 650$ MeV is not shown as the contribution from Dalitz processes has not been included and they are expected to be relevant. Inset: the $\rho$ spectral function for the same values of the axial chemical potential is compared to the cocktail one with the same assumptions.}\label{phcockt}
\end{figure}

The parity-breaking effect shows an important increase in the dilepton production away from the $\rho-\omega$ peak due to the mass shifting of the circularly polarized resonances. Such enhancement suggests the appealing possibility of perhaps partially explaining the abnormal dilepton yield found in central HIC at low $p_T$ in the low invariant mass region quoted by PHENIX and STAR \cite{NA60,phenix}. However, direct comparison with PHENIX or STAR experimental data is not possible due to the lack of results for the $\rho$ and $\omega$ spectral functions. The measurement of the PHENIX dielectron continuum has a much poorer precision compared to NA60 and a numerical subtraction of the 'cocktail' sources in order to determine different individual contributions is meaningless. In this sense, the only possible comparison of our results with PHENIX data consists in removing the vector meson contributions included in the 'cocktail' and replace them by the distorted spectral functions previously presented. We will only focus in the region near the $\rho-\omega$ resonance peak. Fig. \ref{phcockt} shows the predicted $e^+e^−$ yield by the 'cocktail' of hadronic sources without any parity-breaking effect and for local parity breaking with $\mu_5=150$ and 300 MeV as well as PHENIX experimental data. As already stressed, we apply an enhancement factor 1.8 to the ratio $\rho/\omega$ and the $\omega$ is normalized to fit the resonance peak. The precision of the data does not allow for clear conclusions but it is clear that LPB noticeably improves the agreement to the data as compared to the 'cocktail'. In addition, if the $\omega$ peak was accurately measured, we could provide a more definite prediction of our results, i.e., we normalized the peak to coincide with 'cocktail', which might not coincide with the experiment. Finally, in the inset of Fig. \ref{phcockt} we present the $\rho$ spectral function for the same values of $\mu_5$, which has notably increased as compared to the vacuum case.

\medskip

Dileptons are produced in a variety of decays. The dominant processes in the range of invariant masses $200$ MeV $< M < 1200$ MeV are assumed to be the $\rho$ and $\omega$ decays into lepton pairs, the $\omega\to \pi^0 \ell^+\ell^-$ Dalitz decay and the $\eta, \eta^\prime \to\gamma\ell^+\ell^-$ Dalitz decays. These are the basic ingredients of the hadronic 'cocktail' \cite{rapp,NA60,phenix}, conventionally used to predict the dilepton yield\footnote[2]{In addition there is a substantial $c\bar c$ background that is not affected by the present considerations.}. Naturally, the dispersion relations of $\rho$, $\omega$, etc. are modified due to conventional in-medium thermal effects such as a dropping mass and/or broadening resonance. These investigations claim to explain the abnormal dilepton yield but the issue is still unclear and it may not be excluded that the distortion due to parity breaking may play a role.

\medskip

We have to retain the very important point that a number of light hadronic states will be in thermal equilibrium (so their respective abundances will be governed by the Boltzmann distribution) and regeneration of these resonances will take place in the fireball similarly to the one taking place for $\rho$ mesons. They will be much more abundant in the fireball if there is LPB and the decays of these resonances could be an important source of dileptons so far unaccounted for.

\medskip

The Dalitz processes are described by the Kroll-Wada formula \cite{landsberg} that includes the contribution of vector mesons and it remains valid in the case of LPB provided that we replace the vector meson masses by the values in \eqref{mvec} according to the intermediate meson polarization ($L,\pm$). The fact that the decaying meson is distorted by the medium complicates analytically and numerically the calculation due to the lack of Lorentz invariance generated by a polarization and momentum dependent meson mass. We address the calculations of Dalitz decays in Appendix \ref{AppDalitz}.

\section{Dilepton polarization analysis in $V\to\ell^+\ell^-$ decays}\label{angles}

In this section we want to address a careful analysis of the polarization distribution associated to the lepton pairs produced in the decay of vector mesons. Motivated by the results from Fig. \ref{polasym} where a clear polarization asymmetry was found aside the $\rho$ vacuum resonance peak, we will study the angular distribution of the outgoing leptons searching for polarization dependences.

\medskip

Current angular distribution analysis \cite{Bratk,CS} are not the most suitable way to extract all the information related to the parity-breaking effects we present \cite{polar}. First, they omit the non-trivial dependence on $M$ where one could be able to see the resonances associated to the transverse polarized mesons. And second, the usual angular variables considered are not able to isolate the different polarizations. We will perform a two-dimensional study of the decay product with the dilepton invariant mass $M$ and a certain angle that we will introduce below. Our aim is to elucidate which other angular variables of the decay process should be taken into account when experimental data is analysed. 

\medskip

From now on the overall constants $c_V$ in Eq. \eqref{genericproduct} are chosen in such a way that after integrating the entire phase space the total production at the vacuum resonance peak is normalized to 1 unless otherwise is stated. This choice will help us to quantify the number of events found when the phase space is restricted in the following discussion. We define the normalized number of events as
\begin{equation}
N_{\theta}(M)=\frac{\int_{\Delta\theta}\frac{dN}{dMd\cos\theta}\left (M,\cos\theta\right )d\cos\theta}{\int_{-1}^{+1}\frac{dN}{dMd\cos\theta}\left (M=m_V,\cos\theta\right )d\cos\theta},
\end{equation}
being $\theta$ one of the two different angular variables that we will consider in the next sections.

\subsection{Case A}
The first variable we will investigate is the angle $\theta_A$ between the two outgoing leptons in the laboratory frame. Some basic algebraic manipulations lead us from Eq. \eqref{genericproduct} to
\begin{align}
\nonumber \frac{dN_V}{dMd\cos\theta_A}=&c_V\frac{\alpha^2}{6\pi M}\left (1-\frac{n_V^2m_\pi^2}{M^2}\right )^{3/2}\int\frac{p^2p'^2dpd\cos\theta d\phi}{E_p\sqrt{(M^2-2m_\ell^2)^2-4m_\ell^2(E_p^2-p^2\cos^2\theta_A)}}\\
&\sum_{\epsilon}\frac{m_{V,\epsilon}^4\left (1+\frac{\Gamma_V^2}{m_V^2}\right )}{\left (M^2-m_{V,\epsilon}^2\right )^2+m_{V,\epsilon}^4\frac{\Gamma_V^2}{m_V^2}} P^{\mu\nu}_{\epsilon}(M^2g_{\mu\nu}+4p_\mu p_\nu)\frac1{e^{M_T/T}-1},
\end{align}
where $\theta$ is the angle between $p$ and the beam axis (not to be confused with $\theta_A$ previously defined) and $\phi$ is the azimuthal angle of $p'$ with respect to $p$. The lepton energies are given by $E_{p}^2-p^2=E_{p'}^2-p'^2=m_\ell^2$ and $p'=|\vec p'|(\theta_A,p,M,m_\ell)$ can be found from the decay kinematics that lead to the following equation:
\begin{equation}
E_pE_{p'}-pp'\cos\theta_A=\frac{M^2}2-m_\ell^2.
\end{equation}

\medskip

\begin{figure}[h!]
\centering
\includegraphics[scale=0.35]{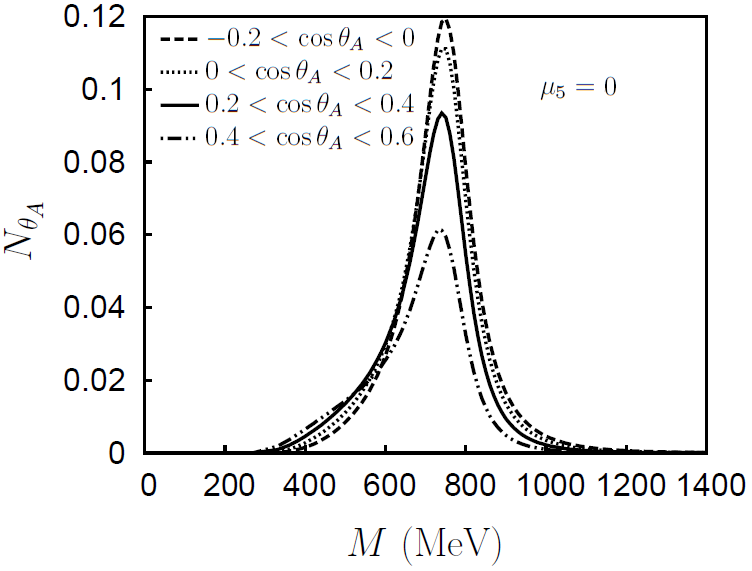}\;\;\includegraphics[scale=0.35]{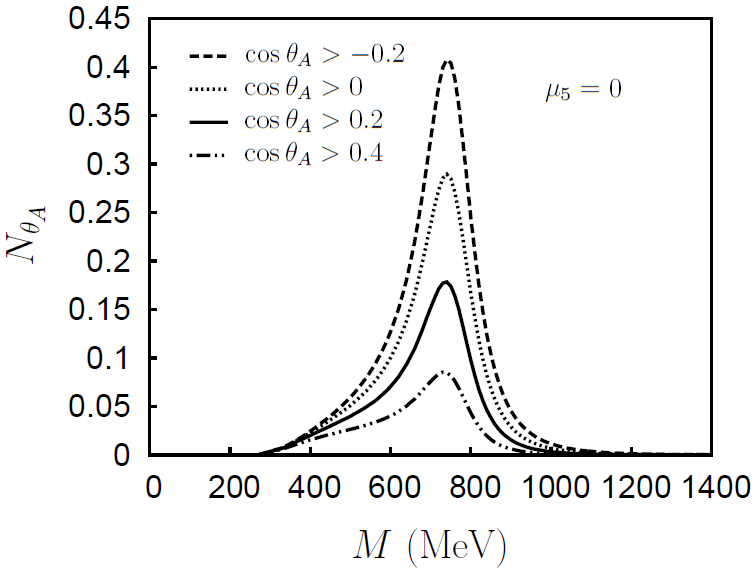}
\includegraphics[scale=0.35]{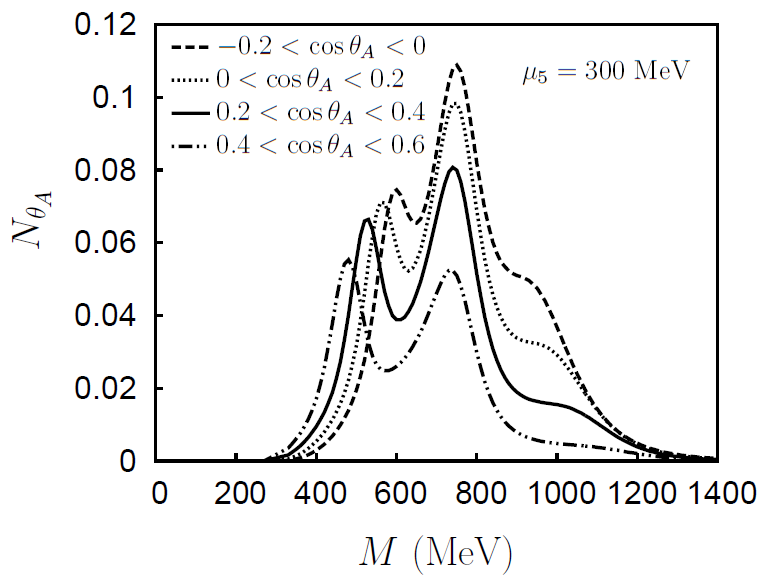}\;\;\includegraphics[scale=0.35]{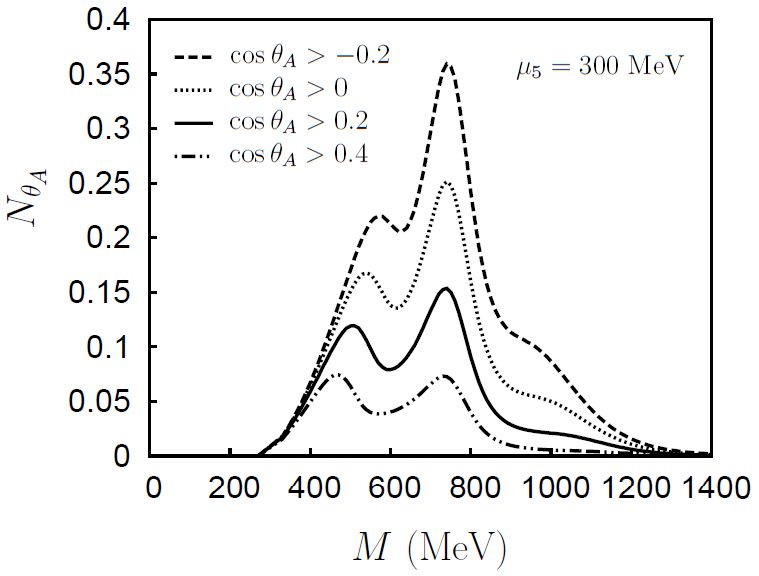}
\caption{The $\rho$ spectral function is presented depending on the invariant mass $M$ in vacuum ($\mu_5=0$) and in a parity-breaking medium with $\mu_5=300$ MeV (upper and lower panels, respectively) for different ranges of the angle between the two outgoing leptons in the laboratory frame $\theta_A$. We display the curves corresponding to $\cos\theta_A\in [-0.2,0],[0,0.2],[0.2,0.4],[0.4,0.6]$ and $[0.6,0.8]$ in the left panels, and $\cos\theta_A\geq -0.2,0,0.2,0.4$ in the right ones. The total production at the vacuum peak is normalized to 1 when the entire phase space is considered. Results are presented for the experimental cuts quoted by PHENIX \cite{phenix}.}\label{a}
\end{figure}

In the left panels of Fig. \ref{a} we present the results of the dilepton production in the $\rho$ channel as a function of its invariant mass $M$ integrating small bins of $\Delta\cos\theta_A=0.2$. The different curves displayed in the plot correspond to $\cos\theta_A\in [-0.2,0],[0,0.2],[0.2,0.4],[0.4,0.6]$ and $[0.6,0.8]$. The upper panel corresponds to the vacuum case with $\mu_5=0$ while the lower one represents a parity-breaking medium with $\mu_5=300$ MeV. The LPB effect produces a secondary peak corresponding to a transverse polarization in the low invariant mass region in addition to the vacuum resonance. This graphic does not imply a larger abundance of transversely polarized mesons with respect to the longitudinal ones but the fact that transverse mesons can be selected (and almost isolated) with a suitable angular coverage. If the effective temperature $T$ increased, we would see an enhancement of the upper mass tail making the secondary peak smaller in comparison to the vacuum one. As $\cos\theta_A$ increases, the outgoing leptons are more collimated, meaning a higher meson 3-momentum and from the dispersion relation in Eq. \eqref{disp.rel}, a larger separation between the two peaks together with a Boltzmann suppression. Note that the highest peaks in the figures with $\mu_5\neq0$ have a value $\simeq 0.1$, this is a $10\%$ of the total production at the vacuum peak when the total phase space is considered (except for the detector cuts). We took a representative value $\mu_5=300$ MeV so as to show two different visible peaks. If the axial chemical potential acquires smaller values, the secondary peak approaches the vacuum one. The smaller $\mu_5$ is taken, the more collinear the leptons have to be in order to obtain a large meson momentum and thus, a visible and significant secondary peak (recall Eq. \eqref{disp.rel}).

\medskip

Dealing with quantities smaller than $10\%$ may be tricky. As the main information related to LPB is focused in $\cos\theta_A\approx 1$, in the right panels of Fig. \ref{a} we show a somewhat more experimentally oriented set of plots where we integrated $\cos\theta_A$ down to -0.2, 0, 0.2 and 0.4. The upper panel corresponds to vacuum and the lower one represents a parity-breaking medium with $\mu_5=300$ MeV. The number of events grows when more and more separated leptons are considered but the secondary peak smears and becomes much less significant. Therefore, an optimization process should be performed in every different experiment so as to find the most suitable angular coverage.

\medskip

\begin{figure}[h!]
\centering
\includegraphics[scale=0.18]{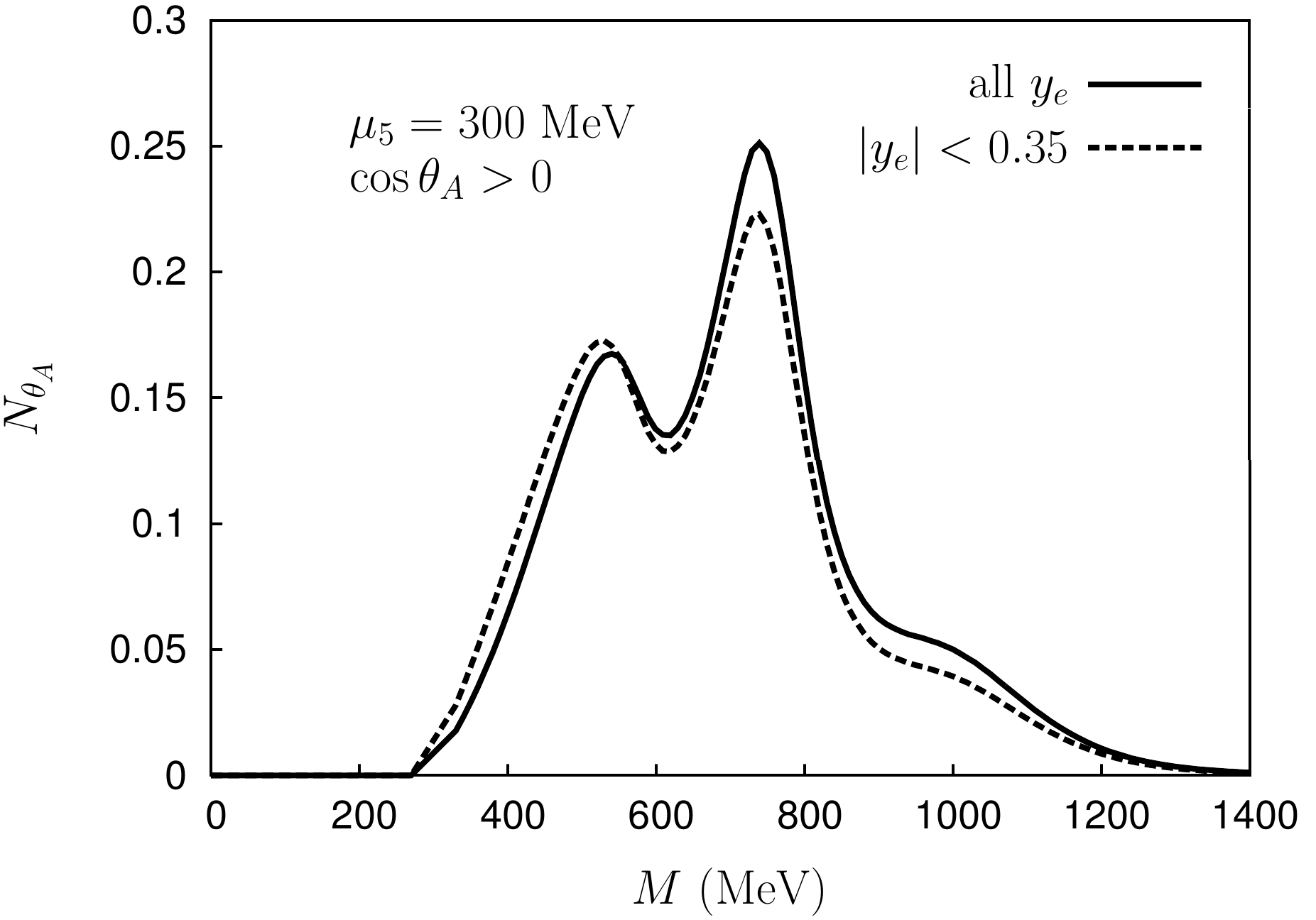}
\caption{One of the curves presented in Fig. \ref{a} is displayed (solid line) together with its modification when an additional single-electron cut $|y_e|<0.35$ is included (dashed line). No significant changes are observed concerning the significance of the secondary peak that appears due to a non-trivial $\mu_5$.}\label{ye_cut}
\end{figure}

At this point, we recover the discussion about introducing a realistic cut in single-electron rapidity. In principle, all accelerator experiments have to account for this cut (independently on the particle type) in order not to accumulate a large number of tracks and energy amounts in the forward direction that would be rather tricky to isolate. However, for the PHENIX experiment there is no clear information about this cut \cite{phenix} and therefore we decided to omit it. In Fig. \ref{ye_cut} we show one of the curves already presented in Fig. \ref{a} including a single-electron cut $|y_e|<0.35$ and without it. One may observe that the qualitative behaviour already discussed holds if this cut is added. In fact, the vacuum peak slightly decreases whereas the secondary peak due to LPB does not substantially change. Indeed, this small distortion does not affect the results already explained regarding the $\rho$ spectral function measured by NA60 or the improvement of the PHENIX cocktail.

\medskip

\begin{figure}[h!]
\centering
\includegraphics[scale=0.35]{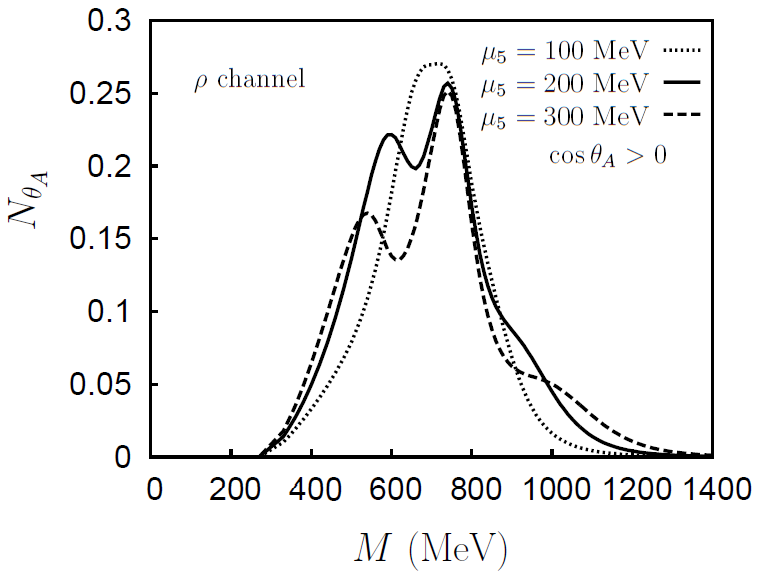}\includegraphics[scale=0.35]{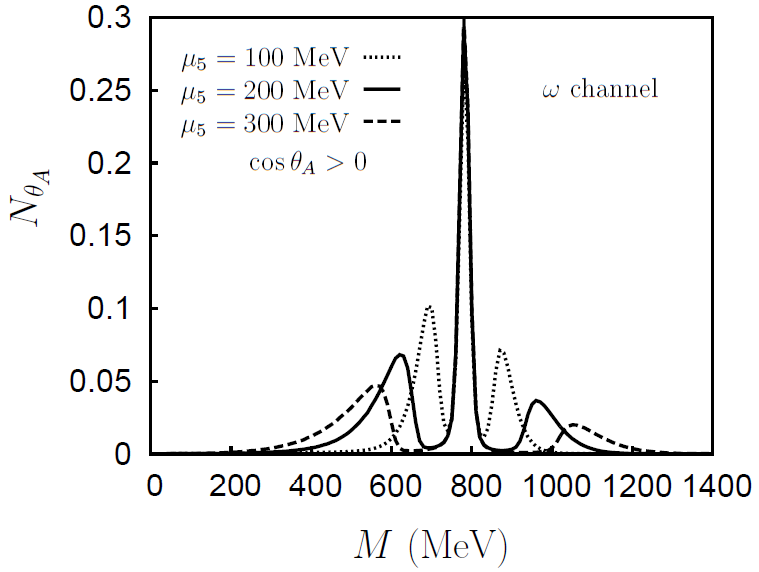}\\
\includegraphics[scale=0.45]{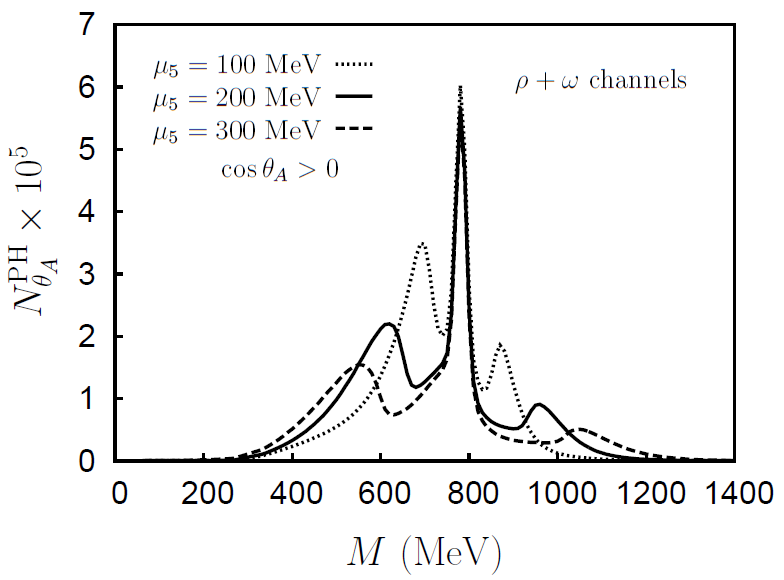}
\vspace{-0.5em}
\caption{The $\rho$ (upper-left panel) and $\omega$ (upper-right panel) spectral functions and their combination (lower panel) are showed depending on the invariant mass $M$ and integrating the forward direction $\cos\theta_A\geq 0$ for $\mu_5=100,200$ and 300 MeV. In the upper panels the total production at the vacuum peak is normalized to 1 when the entire phase space is considered whereas the lower panel is normalized to PHENIX data. Results are presented for the experimental cuts quoted by PHENIX \cite{phenix}.}\label{arw}
\end{figure}

Another issue that could be experimentally addressed is the analysis of $\mu_5$ for a particular (and fixed) coverage of $\theta_A$. If the secondary peaks due to the transversally polarized mesons were found, their positions would be an unambiguous measurement of $\mu_5$ (more precisely, $|\mu_5|$). In Fig. \ref{arw} we integrate the forward direction of the two outgoing leptons, i.e. $\cos\theta_A\geq 0$, and examine how the transverse polarized peaks move with respect to the vacuum one when we change the values of the axial chemical potential. The $\rho$ and $\omega$ spectral functions are displayed in the upper panels for $\mu_5=100$, $200$ and 300 MeV. The same tendency is found in both graphics except for the fact that both transverse peaks are observed in the $\omega$ channel (one below and one above the vacuum resonance). For small values like $\mu_5\simeq 100$ MeV the vacuum peak hides the transverse one due to the $\rho$ width, being impossible to discern if one or two resonances are present. In the $\omega$ channel all the peaks are visible even for such small $\mu_5$. The latter channel could be the most appropriate one in order to search for polarization asymmetry. For completeness, we also present their combination in the lower panel normalized to PHENIX data. The normalized number of events are defined in this case as
\begin{equation}
N_{\theta}^{\text{PH}}(M)=\frac{dN^{\text{PH}}}{dM}\left (M=m_V\right )\frac{\int_{\Delta\theta_A}\frac{dN}{dMd\cos\theta}\left (M,\cos\theta\right )d\cos\theta}{\int_{-1}^{+1}\frac{dN}{dMd\cos\theta}\left (M=m_V,\cos\theta\right )d\cos\theta},
\end{equation}
where $\frac{dN^{\text{PH}}}{dM}(M)$ is the theoretical spectral function used in the PHENIX hadronic cocktail. As it was discussed in Section \ref{dilepLPB} an enhancement factor 1.8 is included for the ratio $\rho/\omega$ due to the $\rho$ meson regeneration into pions within the HIC fireball \cite{Specht:2010xu,Heinz:1991fn} and the in-medium $\omega$ suppression \cite{omegasupp}. As already mentioned, our conclusions do not depend substantially on the precise value of this ratio as the experimental signal that we propose is amply dominated by the $\omega$ when the two resonances are considered together (see. Fig. \ref{arw}).

\subsection{Case B}
The second variable we present is the angle $\theta_B$ between one of the two outgoing leptons in the laboratory frame and the same lepton in the dilepton rest frame. The angle is exactly the same if one chooses the lepton with positive or negative charge. The dilepton production is given by
\begin{align}
\nonumber \frac{dN_V}{dM}=&c_V\frac{\alpha^2}{6\pi M}\left (1-\frac{4m_\pi^2}{M^2}\right )^{3/2}\int\frac{k^2dkd\cos\theta}{E_k}\frac{p^2d\cos\psi d\phi}{\sqrt{M^4-4m_\ell^2(E_k^2-k^2\cos\psi^2)}}\\
&\sum_{\epsilon}\frac{m_{V,\epsilon}^4\left (1+\frac{\Gamma_V^2}{m_V^2}\right )}{\left (M^2-m_{V,\epsilon}^2\right )^2+m_{V,\epsilon}^4\frac{\Gamma_V^2}{m_V^2}} P^{\mu\nu}_{\epsilon} (M^2g_{\mu\nu}+4p_\mu p_\nu)\frac1{e^{M_T/T}-1},
\end{align}
where $k(p)$ and $E_k(E_p)$ are the meson (selected lepton) momentum and energy. $\psi$ and $\phi$ are the polar and azimuthal angles respectively of the lepton with respect to the meson momentum. The beam direction forms an angle $\theta$ with the vector meson. One may extract $p$ from the decay kinematics. All these variables are defined in the laboratory frame. The new angle $\theta_B$ is defined via
\begin{equation}
\frac M2\cos\theta_B\sqrt{M^2-4m_\ell^2}=p(M\sin^2\psi+E_k\cos^2\psi)-kE_p\cos\psi.
\end{equation}
Regarding the fact that we would like to perform experimental cuts in $\theta_B$, our numerical computations integrate the whole phase space of the decay and reject the regions with unwanted values of $\theta_B$ instead of treating this angle as an integration variable, which makes the calculations more complicated.

\medskip

Events with $\cos\theta_B\simeq 1$ correspond to one lepton being produced in approximately the same direction in the laboratory frame and in the meson rest frame, implying that in the laboratory frame the vector meson is almost at rest. Vector mesons with low momentum are not suppressed by the Boltzmann weight but do not carry relevant information about $\mu_5$. Therefore the opposite limit (high momenta) will be the important one for our purposes.

\medskip

\begin{figure}[h!]
\centering
\includegraphics[scale=0.35]{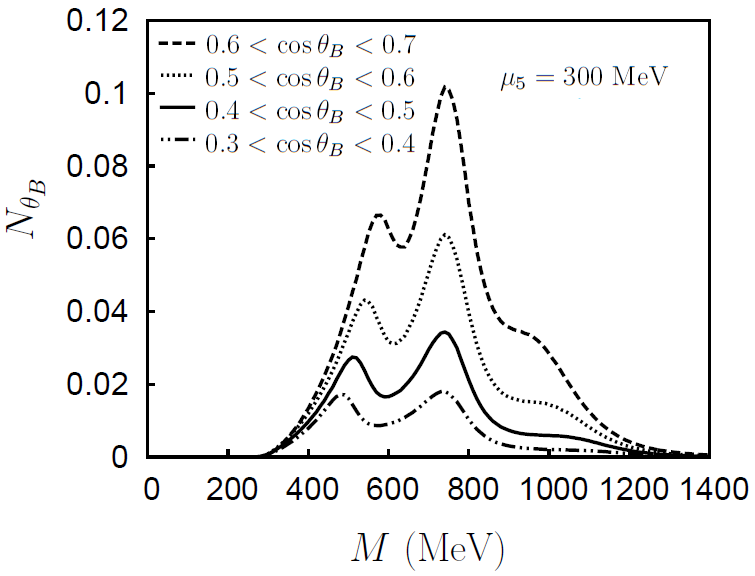}\includegraphics[scale=0.35]{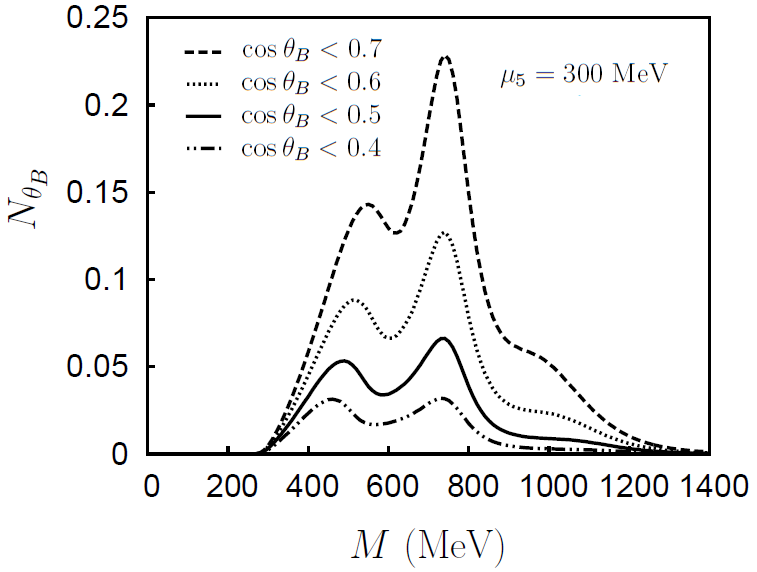}
\caption{The $\rho$ spectral function is presented depending on the invariant mass $M$ for different ranges of the angle $\theta_B$ between one of the outgoing leptons in the laboratory frame and the same lepton in the dilepton rest frame for fixed $\mu_5=300$ MeV. We display the curves corresponding to $\cos\theta_B\in [0.3,0.4],[0.4,0.5],[0.5,0.6]$ and $[0.6,0.7]$ in the left panel, and $\cos\theta_B\leq 0.4,0.5,0.6,0.7$ in the right one. The total production at the vacuum peak is normalized to 1 when the entire phase space is considered. Results are presented for the experimental cuts quoted by PHENIX \cite{phenix}.}\label{beta}
\end{figure}

In the left panel of Fig. \ref{beta} the $\rho$ spectral function is presented in small bins of $\Delta\cos\theta_B=0.1$. The curves correspond to $\cos\theta_B\in[0.3,0.4], [0.4,0.5], [0.5,0.6]$ and $[0.6,0.7]$ with a fixed $\mu_5=300$ MeV. At first glance, the plot looks as the one showed in the previous section for $\theta_A$ with a similar secondary peak below the vacuum resonance. The main difference with the previous section is the number of events. The analysis of $\theta_B$ is more sensitive to the Boltzmann suppression than the previous case with $\theta_A$. In the right panel of Fig. \ref{beta} it may be readily checked that the number of events at the highest secondary peak is $0.14$, whereas in Fig. \ref{a} it corresponds to $0.22$, a considerable enhancement of around a $60\%$.

\medskip

A phenomenological analysis of $\mu_5$ depending on the position of the secondary peak as well as a comparison with the vacuum contributions may be equally described in this section but no new features are found so we omit the corresponding details.

\section{Summary}

In this chapter we have seen that in the presence of an isosinglet time-dependent pseudoscalar background vector mesons are severely distorted. They acquire an effective mass depending on their polarization, 3-momentum and the axial chemical potential. Such dispersion relation predicts that massive vector mesons split into three polarizations with different masses. The resonance poles associated to the transverse mesons appear separated of vacuum peak implying a polarization asymmetry. We have computed this effect and found that it naturally tends to produce an overabundance of lepton pairs in the vicinity of the $\rho-\omega$ resonance peak. Thus LPB seems relevant to explaining the PHENIX/STAR 'anomaly'.

\medskip

As possible signatures for experimental detection of LPB we presented a description in two angular variables that are not considered in the literature when angular distribution analysis are investigated. The first one is the angle between the two outgoing leptons produced from the meson decay in the laboratory frame. The second one is defined as the angle between one of the leptons at the laboratory frame and the same lepton in the dilepton rest frame. These angles are the most suitable ones so as to extract information about LPB. Without a careful choice of angular observables the effects of LPB on the different polarizations can be easily missed. We claimed that a two-dimensional study of the decay product with the variables angle-dilepton mass allows to distinguish at least two of the resonance poles and confirm the parity-breaking hypothesis. We displayed the $\rho$ and $\omega$ spectral functions in order to illustrate this effect and discuss the most efficient ways to search for polarization asymmetry.

\medskip

The study presented here is still not fully realistic and several improvements could be made such as e.g. including a complete hydrodynamical treatment, a non-trivial baryon chemical potential, finite volume effects, thermal broadening of the resonances and a more detailed study of the time and (effective) temperature dependence. Yet none of these are expected to erase the traces of LPB if the latter is present and we would like to encourage the experimental collaborations to actively search for this interesting possibility.
\chapter{Conclusions and outlook}\label{Chconcl}

The high temperatures and densities reached at current HIC experiments allow to access new regions of the QCD phase diagram that had not been investigated before. In this work we have studied the possibility that QCD breaks parity in the nuclear fireball by means of an isosinglet axial chemical potential $\mu_5$. We saw that a phase where this symmetry is spontaneously broken may appear in the Nambu--Jona-Lasinio model, a feature that NJL may possibly share with QCD. The order of magnitude of $\mu_5$ that allows to enter this phase is the one expected to occur in HIC experiments. Armed with this tool as the only input for parity violation, we studied how in-medium mesons are distorted. When parity is broken, the distinction between scalars and pseudoscalars is meaningless. We found the new eigenstates of QCD in the scalar sector and investigated their energy-dependent masses and decay widths. Interestingly, the distorted $\tilde\eta$ meson may reach thermal equilibrium in the \emph{pion} gas forming the fireball. In the vector sector, the parity-breaking effect induces a mixing between the different vector mesons and photons. The light vector meson states $\rho$ and $\omega$ are distorted in a similar way as spin zero resonances. Their masses are energy- and polarization-dependent, which implies an overproduction of lepton pairs from their decay. A crude comparison of our in-medium $\rho$ spectral function with NA60 data reflects a similar behaviour. In the case of the PHENIX experiment, we improved the 'cocktail' of hadronic sources with the distorted $\rho$ and $\omega$ contributions and observed a better agreement with the data around the resonance peak. We also described how this abnormal yield may be detected experimentally by restricting the angular coverage of the final lepton states. In this way, the possibility that QCD breaks parity in HIC has been carefully discussed together with the main implications derived of this effect. More work is needed before definite conclusions can be drawn but we believe that sufficient evidence is accumulated at present to bring these tentative conclusions to the attention of the interested readers. The possible detection of this phenomenon constitutes an encouraging source of motivation to continue working in these lines with the conviction that QCD still hold many surprises for us.

\medskip

In Chapter \ref{Chfase} we presented the possibility that the discrete symmetries of gauge theories like parity or charge conjugation may be broken in a hot and/or dense medium. Two different approaches were described in detail. The first one corresponds to an analysis of dense systems with effective Lagrangian techniques. It was shown that a general sigma model containing two $SU(2)$ multiplets of scalar/pseudoscalar fields that include the two lowest lying resonances in each channel may induce a stable pseudoscalar condensate at non-trivial baryon chemical potential. Of course, the necessary conditions of this phase along with its stability depend on the exact values of the generic parameters.

\medskip

The second hypothesis of parity violation in QCD may only take place in a finite volume. In HIC, the nuclear fireball reaches extremely high temperatures and densities that allow the system to enter the QGP. QCD exhibits a non-trivial and degenerate vacuum structure with different topologically inequivalent sectors. Many authors support the idea that large topological fluctuations connecting different vacua may be possible in a hot medium leading to visible effects via the Chiral Magnetic Effect. In this case, not only peripheral collisions (where CME is present) will show traces of parity breaking. While the system cools down to the hadronic phase, we claim that a new phase characterised by LPB may be crossed. The topological charge resulting from these fluctuations is transmitted to the quark sector inducing a non-trivial axial charge $Q_5^q$, namely an imbalance between the left- and right-handed quarks. In the case of light quarks ($u,d$) and a fireball living for $\tau_\text{fireball}\simeq 5-10$ fm, this axial charge is approximately conserved. In the strange sector, although an axial charge is formed, it will be wiped out due to non-trivial left-right oscillations as a consequence of the large strange quark mass. We introduced an axial chemical potential $\mu_5$ conjugated to $Q_5^q$ in order to investigate the parity-breaking effect. This is the approach we used throughout this thesis.

\medskip

The axial chemical potential is proportional to the time derivative of a spatially homogeneous pseudoscalar axion-like condensate and it may have an isospin structure. In this work we only considered the isosinglet case, which should be more relevant for experiments like RHIC and LHC. However, the isotriplet case can also be implemented as we gave the main indications to work with it. The future experiments FAIR and NICA could be scenarios where the triplet condensate is actually more important.

\medskip
 
In Chapter \ref{ChNJL} we presented the NJL model as a toy model for QCD when parity is violated. This model has received a lot of attention in order to study QCD in extreme conditions. It is not clear that NJL is a good modellisation of QCD. Yet one expects the results of NJL to be qualitatively similar to those in QCD like the mechanism of chiral symmetry breaking through a scalar isosinglet condensation. The study of $\mu_5$ in QCD has been performed in lattice simulations. Unlikely, finite density calculations present serious difficulties both analytically and in the lattice. The simplicity of the NJL model allows to work in a regime where $\mu,\mu_5\neq0$ and its results may help to obtain some realistic insights of QCD.

\medskip

We implemented the NJL model in the presence of both a vector and an axial chemical potentials at zero temperature so as to investigate the different phases of the model. The presence of a non-trivial $\mu_5$ constitutes a source for explicit parity breaking. The interplay of both external drivers in this model was never considered before. We found that the axial chemical potential plays an essential role in the game (unlike $\mu$). First, in the chirally broken phase it leads to a non-trivial dependence of the scalar condensate introducing finite jumps under certain conditions. And second, it may trigger a spontaneous breaking of parity via an isosinglet pseudoscalar condensation $\eta\neq0$ as long as massive quarks are considered and the full $U(2)_L\times U(2)_R$ global symmetry is broken to $SU(2)_L\times SU(2)_R \times U(1)_V$ (i.e. $G_1 \neq G_2$). We considered for simplicity $a_0^0=\langle \bar\psi\tau^3\psi\rangle=0$ thus forbidding flavour violation together with parity conservation. With these considerations, it is remarkable the absence of a phase with a pion-like condensation with a simultaneous violation of flavour and parity symmetries. However, we found an extremely small region in the $\mu-\mu_5$ space of parameters where flavour symmetry may be broken by a non-trivial $a_0^0$ condensate and parity is preserved.

\medskip

We investigated the main features of the chirally broken phase in NJL considering a slightly modified DR scheme to treat UV divergences. We showed the allowed parameter space (including $\mu$ and $\mu_5$) for this phase to be stable. A deep discussion of the jumps in the chiral condensate was presented depending on the values of the external drivers. As it is well known, chiral restoration was found at high values of $\mu$ with a sharper transition as one approaches the limit $m\to0$.

\medskip

The stability of the parity-breaking phase requires a set of conditions for the parameters that make the model very different of QCD. Due to a non-conventional order relation between $G_1$ and $G_2$, the NJL mass spectrum in the chirally broken phase shows that the $\sigma$ and $\pi$ mesons are heavier than $a_0(980)$ and $\eta_q$ respectively. We set the appropriate conditions for both the chirally- and parity-broken phases to be stable in order to analyse the possible phase transition. In this context, $\mu_5$ is determinant (unlike $\mu$) in order to enter the parity-breaking phase. We showed the evolution of the scalar and pseudoscalar condensates with respect the axial chemical potential and also the phase transition line, which was displayed in the $\mu-\mu_5$ plane. For $\mu<M^c$ we have a second order transition while for $\mu>M^c$, it corresponds to a first order one.

\medskip

The addition of an axial chemical potential induces new features in the NJL that had not been observed before. The model itself remains rather simple but all the subtleties arising from a non-trivial $\mu_5$ seem to reflect that QCD may still exhibit many features that we do not understand yet. On the other hand, the fact that an axial chemical potential of the order or a few hundreds of MeV may induce the stability of a parity-violating phase has to be understood as a generic feature of QCD and not only about this specific model. Therefore, we explored how hadronic physics is affected by LPB.

\medskip

In Chapter \ref{ChSmodel} we studied the main effects that parity-violation induces to spin zero mesons. We introduced a generalised sigma model with some of the lightest degrees of freedom in $SU(3)$. Masses and widths of the different states were calculated and a fit to QCD was performed in order to set the parameters of the model. We introduced an axial chemical potential as the time component of a spurious axial-vector field and consequently, new terms mixing states of different parity arose in the Lagrangian. In addition to parity breaking, this mechanism also introduces a Lorentz-symmetry violation that is manifest in the dispersion relation of the different states. Therefore, the properties of in-medium mesons are frame-dependent.

\medskip

First we described the mixing between the scalar and pseudoscalar triplets $a_0$ and $\pi$ and next we solved the system in the isosinglet sector where $\sigma$, $\eta$ and $\eta'$ are all mixed. As expected due to the Lorentz violation, the masses of the new eigenstates are energy-dependent. In both cases, we found that the lightest states corresponding to the distorted $\tilde\pi$ and $\tilde\eta$ become tachyonic for high enough energies. However, that was not a problem since energies are well defined and vacuum remains stable.

\medskip

Due to the breaking of parity, new in-medium \emph{exotic} processes and decays may take place. In the presence of $\mu_5$ the pion gas that commonly forms the HIC fireball may actually be composed of distorted pions. We computed the decay widths of the singlet states into $\tilde\pi$ and observed a non-trivial dependence on the axial chemical potential. The energy-dependent widths show different bumps that have nothing to do with the Lorentz factor but most probably reflect the tachyonic nature of the final state \emph{pions}. The distorted $\eta$ meson acquired a width of order 60 MeV, comparable to the inverse fireball lifetime. Accordingly, we stressed that this state may reach thermal equilibrium within the HIC fireball, fact that could help in understanding the anomalous dilepton production that was studied in Chapter \ref{ChVMD}.

\medskip

We explained that the isosinglet axial-vector meson channel interferes with the axial chemical potential. The condensation of its time component induces a renormalization of the bare $\mu_5$ leading to an effective $\bar\mu_5$. We provided the relation between them and stressed that the effective one should be introduced in the mass-gap equations of the model.

\medskip

In Chapter \ref{ChVMD} we investigated how the violation of parity affects vector mesons. We discussed that the mixing of states of different parities may also take place in the vector/axial-vector sector. However, this coupling is very model dependent since it is not forbidden but it does not appear from a spurious $\mu_5$. Thus the LPB effect was implemented through the Chern-Simmons term. It was shown that LPB affects the dispersion relation of the lightest vector mesons $\rho$ and $\omega$. In the presence of an axial chemical potential their effective masses depend on their polarization and 3-momentum reflecting the breaking of Lorentz and parity symmetries. Accordingly, vector mesons split into three polarizations with different masses. The transversely polarized mesons have their corresponding resonance peak aside the vacuum peak leading to a distorted spectral function that strongly indicates a polarization asymmetry around the vacuum peak.

\medskip

In order to search for possible manifestations of the LPB effect in the vector meson channel, we computed the dilepton production from vector meson decays in central collisions and observed that a non-trivial axial chemical potential leads to an overproduction of dileptons around the $\rho-\omega$ resonance peak. We computed the shape of the $\rho$ spectral function applying the experimental conditions of NA60 and saw that there is a suggesting similarity between our results and the experimental data for $\mu_5\simeq 300$ MeV. However, this comparison was not completely significant as data is not acceptance corrected. We also investigated the case of the PHENIX dilepton excess in the low invariant mass region. The combination of the $\rho$ and $\omega$ contributions compared to the vacuum case showed a clear enhancement of the dielectron production around the resonance peak that strongly depends on the axial chemical potential. We took the theoretical prediction of the PHENIX dilepton yield based 'cocktail' of known hadronic sources and replaced the vacuum $\rho$ and $\omega$ contributions by our in-medium calculations and observed a better agreement with the experimental data around the peak. In addition, the fact that distorted $\eta$ mesons are in thermal equilibrium within the \emph{pion} gas leads to an enhancement of their production of lepton pairs. These considerations together with other many body or in-medium corrections \cite{rapp,renkr,BratkCass,zahed} may help to explain the anomalous results from PHENIX and STAR. And moreover, LPB may also play a non-trivial role in NA60, where current explanations for the in-medium $\rho$ spectral function could not be the only answer due to the large experimental uncertainty in the data.
 
\medskip

A clear signal of LPB would be the verification that dileptons produced for values of the invariant mass above and below the $\rho-\omega$ pole are predominantly of opposite circular polarizations. This requires searching for asymmetries among longitudinal and transverse polarization for different values of the dilepton invariant mass in event-by-event measurements. An angular analysis of the final lepton states could reveal polarization dependences. We presented two angular variables that are not considered in the literature and carry the main information related to the LPB effect. The first one is the angle between the two outgoing leptons produced from the meson decay in the laboratory frame. The second one is defined as the angle between one of the leptons at the laboratory frame and the same lepton in the dilepton rest frame. We stressed that the effects of LPB may be missed if one does not carefully choose the appropriate angular variables. A two-dimensional study of one of the previous angles and the dilepton invariant mass may allow to distinguish (at least) one extra peak aside the resonance pole. We showed the $\rho$ and $\omega$ spectral functions to illustrate this effect and discussed the most efficient ways to search for polarization asymmetry.

\section{Future perspectives}

We would like to emphasize the simplicity of the approach presented here. The features of both scalar and vector mesons only depend on a free parameter $\mu_5$, which is expected to depend on the characteristics of the HIC. In particular, the fits presented in Chapter \ref{ChVMD} use the values (effective temperatures, normalizations, etc.) quoted by the experiments themselves. A method to measure the axial chemical potential was also presented. The work developed about parity breaking in HIC seems interesting by itself. Even more if one considers the consequent phenomenological implications that we found so far and the expected ones. Several appealing investigation lines could be followed in order investigate more details of this phenomenon.

\medskip

A deep study about the origin of a parity-odd medium through the generation of large local topological fluctuations could be helpful to understand the nature of the axial chemical potential and the essence of parity violation in QCD. Some analysis based on the evolution of classical Yang-Mills fields \cite{lappi} reveal the possibility that some \emph{tubular} structure may be generated after a HIC with a characteristic $\vec E\vec B\neq0$. The connections between these domains and topological fluctuations have been briefly analysed in this context. Going into detail about this topic and relate it with the work already done represents one of the more significant analysis that could be immediately performed.

\medskip

Another open line of research is performing some numerical and analytical studies on the lattice, possibly doing strong coupling expansions to tackle the $\mu\neq0$ problem. This work could complement other works in this field. Other possible extensions of the present work could be an analysis of the $\rho$ and $\omega$ line shapes comparing with the NA60 results. Also the inclusion of heavier degrees of freedom, both vector and scalar mesons may also lead to more accurate conclusions. Other objects such as glueballs may also help in a better description of the effective theories we considered. The mixing of states of different parity in the vector sector was not treated due to its model dependency. However, this is an interesting point to analyse in the future as it may result in phenomenological observables that could be eventually detected. On the other hand, a generalisation of the concept of axial chemical potential could be addressed. We assumed this magnitude to be similar to the derivative of a time-dependent axion-like field. However, it is well known that in accelerator collisions, spatial isotropy is a crude approximation. As a matter of fact, the study of a space-dependent axion-like field is already under study \cite{wreflexion}. Finally, a parallel study of an isotriplet axial chemical potential could be performed to examine dense systems and draw possible predictions for future experiments like FAIR and NICA.

\medskip

On the other hand, parity is globally conserved so if a topological fluctuation occurs in a certain domain, another one with different sign must also take place. Hence it is plausible to think that if lepton production from vector mesons is dominated by a sign of the chemical potential (on average); there will be a different reaction that carries the opposite sign, like jets or other phenomena. This is a topic that could spark some interest.

\medskip

Although this thesis does not provide definite answers to the open question of the possibility of local parity breaking in heavy ion collisions, the main conclusions of this thesis should lead us to a deeper understanding of the nature of QCD. And hopefully, our results will be soon contrasted with experiments like ALICE at LHC. In any case, we are living in a fascinating era where theory and experiment may be contrasted very quickly. Remarkable synergies among different sectors of the scientific community are emerging and so in the near future we may witness rapid advances in this area.

\section{List of publications}
\subsection{Publications in journals}
\begin{itemize}
\item A. A. Andrianov, V. A. Andrianov, D. Espriu and X. Planells, \emph{Analysis of dilepton angular distributions in a parity breaking medium}, Phys. Rev. D 90 (2014) 034024, arXiv:1402.2147 [hep-ph].
\item A. A. Andrianov, D. Espriu and X. Planells, \emph{Chemical potentials and parity breaking: the Nambu--Jona-Lasinio model}, Eur. Phys. J. C 74 (2014) 2776, arXiv:1310.4416 [hep-ph].
\item A. A. Andrianov, D. Espriu and  X. Planells, \emph{An effective QCD Lagrangian in the presence of an axial chemical potential}, Eur. Phys. J. C 73 (2013) 2294, arXiv:1210.7712 [hep-ph].
\item A. A. Andrianov, V. A. Andrianov, D. Espriu and X. Planells, \emph{Dilepton excess from local parity breaking in baryon matter}, Phys. Lett. B 710 (2012) 230, arXiv:1201.3485 [hep-ph].
\item A. A. Andrianov, V. A. Andrianov, D. Espriu and X. Planells, \emph{Abnormal enhancement of dilepton yield in central heavy-ion collisions from local parity breaking}, Theor. Math. Phys. 170 (2012) 17.
\end{itemize}

\subsection{Other publications}
\begin{itemize}
\item A. A. Andrianov, D. Espriu and  X. Planells, \emph{An effective theory for QCD with an axial chemical potential}, PoS QFTHEP 2013 (2014) 049, arXiv:1310.4434 [hep-ph].
\item A. A. Andrianov, V. A. Andrianov, D. Espriu and X. Planells, \emph{Implications of local parity breaking in heavy ion collisions}, PoS QFTHEP 2011 (2013) 025, arXiv:1310.4428 [hep-ph].
\item A. A. Andrianov, V. A. Andrianov, D. Espriu and X. Planells, \emph{Abnormal dilepton yield from parity breaking in dense nuclear matter}, AIP Conf. Proc. 1343 (2011) 450, arXiv:1012.0744 [hep-ph].
\item A. A. Andrianov, V. A. Andrianov, D. Espriu and X. Planells, \emph{The problem of anomalous dilepton yield in relativistic heavy-ion collisions}, PoS QFTHEP 2010 (2010) 053.
\end{itemize}

\subsection{Unpublished}
\begin{itemize}
\item A. A. Andrianov, V. A. Andrianov, D. Espriu and X. Planells, \emph{Abnormal dilepton yield from local parity breaking in heavy-ion collisions} (2010), arXiv:1010.4688 [hep-ph].
\end{itemize}
\chapter{Resumen en castellano}\label{ChResum}
\graphicspath{{1-intro/figures/}{figures/}}

La fuerza nuclear fuerte es conocida como la responsable de ligar los n\' ucleos at\' omicos a pesar de la repulsi\' on electromagn\' etica entre los protones. Esta interacci\' on es una de las cuatro fuerzas fundamentales de la naturaleza, al igual que la gravedad y las fuerzas electromagn\' etica y d\' ebil. Las \' ultimas dos interacciones pueden ser tratadas conjuntamente por medio de la teor\' ia unificada de la interacci\' on electro-d\' ebil. Las interacciones electro-d\' ebiles y fuertes, aunque no la gravedad, pueden ser formuladas en t\' erminos de una teor\' ia cu\' antica de campos que es conocida como el Modelo Est\' andar (SM) de las part\' iculas elementales. El comportamiento de las part\' iculas elementales se rige por esta exitosa teor\' ia que ha sido contrastada en numerosas ocasiones con experimentos.

\medskip

Los primeros intentos de tratar con las interacciones fuertes se encuentran en la mitad del siglo pasado. En 1961 Gell-Mann propuso un modelo fenomenol\' ogico \cite{eightfold} para estructurar el gran n\' umero de nuevos hadrones que se estaban encontrando en experimentos con aceleradores en aquella \' epoca. Durante la d\' ecada siguiente la fuerza nuclear fuerte fue formulada a nivel cu\' antico por la Cromodin\' amica Cu\' antica (QCD) \cite{QCD}, una teor\' ia de Yang-Mills no abeliana formulada en $SU(3)$ que describe las interacciones fundamentales entre quarks y gluones \cite{QCDSU3}, llamados partones. Los quarks y gluones fueron introducidos como los componentes b\' asicos de los hadrones y tienen asociado un n\' umero cu\' antico de color debido a la simetr\' ia del grupo de QCD. El lagrangiano de QCD viene dado por
\begin{align}\label{QCDlagesp}
\nonumber \mathcal L_{\text{QCD}}&=-\frac14G^{\mu\nu,a}G_{\mu\nu}^a+\bar q(i\gamma^\mu D_\mu-m)q,\\
D_\mu&=\partial_\mu -igG_\mu^a\lambda^a, \quad G_{\mu\nu}^a=\partial_\mu G_\nu^a-\partial_\nu G_\mu^a+g f^{abc}G_\mu^bG_\nu^c,
\end{align}
donde $q$ es el campo del quark con dos \' indices impl\' icitos en los espacios de color y sabor y el campo dependiente de color del glu\' on es $G_\mu^a$. Tambi\' en se usan la constante de acoplo fuerte $g$ y las matrices de Gell-Mann $\lambda^a$ en la representaci\' on adjunta de $SU(3)$. Existen 3 colores y 6 sabores de quarks. Los quarks son estados fermi\' onicos masivos en la representaci\' on fundamental de la simetr\' ia de grupo de QCD mientras que los gluones son los portadores de la fuerza (sin masa), equivalentemente a los fotones en el electromagnetismo, y pertenecen a la representaci\' on adjunta de $SU(3)$.

\medskip

La atracci\' on fuerte entre pares de quark-antiquark requiere una gran cantidad de energ\' ia para separarlos. Esta energ\' ia excita el estado de vac\' io hasta un punto en que la creaci\' on de otro nuevo par de quarks se vuelve energ\' eticamente favorable y las nuevas part\' iculas r\' apidamente apantallan la carga original. Este es la descripci\' on cualitativa del confinamiento de los quarks, que es responsable del hecho que los estados de color no pueden ser observados de forma aislada en la naturaleza. Este fen\' omeno s\' olo es conocido en QCD y se cree que est\' a ligado a su estructura no abeliana. La falta de una prueba rigurosa del confinamiento a\' un constituye un reto notable para los f\' isicos te\' oricos en la actualidad.

\medskip

Se observ\' o experimentalmente que s\' olo existen dos estructuras principales formadas por quarks. Los hadrones se separan en bariones y mesones. Los bariones son estados de 4 quarks mientras que los mesones est\' an compuestos de una pareja de quark y antiquark. Con tal de preservar el principio de exclusi\' on de Pauli en la formaci\' on de bariones, un nuevo n\' umero cu\' antico se introdujo: la carga de color. Suponiendo que cada quark lleva un color asociado, ambas estructuras pueden ser organizadas de manera que no tengan color. Esta es la raz\' on por la que se dice com\' unmente que s\' olo estados sin color (o blancos) son observables en la naturaleza.

\medskip

Adem\' as de la simetr\' ia gauge, el lagrangiano de QCD \eqref{QCDlagesp} muestra una invariancia bajo tres simetr\' ias globales discretas \cite{PS,coleman}. La paridad $P$ invierte la orientaci\' on del espacio que corresponde a revertir el momento de una part\' icula sin girar su esp\' in. La transformaci\' on de paridad viene dada por
\begin{equation}
q(\vec x,t)\to \eta_P\gamma^0 q(-\vec x,t), \qquad G_\mu(\vec x,t)\to G_\mu(-\vec x,t),
\end{equation}
donde $\eta_P$ es una posible fase gen\' erica que es tomada normalmente como 1 para quarks y leptones sin p\' erdida de generalidad. Para antipart\' iculas, se elige $\eta_P=-1$. La conjugaci\' on de carga $C$ intercambia part\' iculas por antipart\' iculas sin ninguna modificaci\' on de momento o esp\' in a trav\' es de la transformaci\' on
\begin{equation}
q(\vec x,t)\to -i\eta_C C \bar q^T(\vec x,t), \qquad G_\mu(\vec x,t)\to -G_\mu^T(\vec x,t),
\end{equation}
donde $\eta_C$ es otra fase y $C$ es la matriz de conjugaci\' on de carga que obedece las siguientes relaciones
\begin{equation}
C\gamma_\mu C^{-1}=-\gamma_\mu^T, \qquad C\gamma_5 C^{-1}=\gamma_5^T, C^{\dagger}=C^{-1}=-C
\end{equation}
y en la representaci\' on de Dirac viene dada por $C=i\gamma^0\gamma^2$. Finalmente, la reversi\' on temporal $T$ corresponde al intercambio de los conos de luz delantero y trasero, correspondiendo a invertir el momento y el esp\' in de una part\' icula v\' ia
\begin{equation}
q(\vec x,t)\to -i\eta_T\gamma_5C q(\vec x,-t), \qquad G_\mu(\vec x,t)\to G_\mu(\vec x,-t),
\end{equation}
donde $\eta_T$ es de nuevo una fase arbitraria. QCD, como cualquier otra teor\' ia cu\' antica de campos local e invariante Lorentz con un hamiltoniano herm\' itico es tambi\' en invariante bajo la simetr\' ia simult\' anea $CPT$.

\medskip

El espectro de quarks se extiende sobre un amplio rango de masas desde $m_{u,d}\sim 5$ MeV hasta $m_t\sim 175$ GeV, que lleva a una clara jerarqu\' ia de masa/energ\' ia. A una cierta escala de energ\' ia $\Lambda$, los quarks pueden separarse en quarks ligeros con masa $m_q$ y quarks pesados con $m_Q$ de tal manera que se satisface la relaci\' on $m_q\lesssim \Lambda \ll m_Q$. En este contexto a bajas energ\' ias, el efecto de los quarks pesados equivale a redefiniciones de las masas de los quarks ligeros $m_q$ y las constantes de acoplo que involucran los quarks ligeros con correcciones del orden de $\Lambda/m_Q$. Esto es conocido como el teorema del desacoplo \cite{AppelCarazz}. Las teor\' ias de bajas energ\' ias \cite{ChPT} normalmente se formulan para describir la f\' isica de las interacciones fuertes bajo la escalar de ruptura de simetr\' ia quiral $\Lambda\sim 1$ GeV y de ah\' i el n\' umero de sabores de quarks se toma generalmente como $N_f=2,3$ incluyendo siempre los quarks $u$ y $d$, y en ocasiones el quark $s$. Para an\' alisis a escalas de energ\' ias m\' as elevadas, se usan otras t\' ecnicas como la teor\' ia efectiva de quarks pesados \cite{HQET}.

\medskip

En el l\' imite de quarks sin masa, QCD exhibe una simetr\' ia global $U(N_f)_L\times U(N_f)_R$ exacta a nivel cl\' asico, donde $N_f$ es el n\' umero de sabores de quarks considerados. Bajo esta transformaci\' on los campos de quarks se ven modificados de la siguiente manera:
\begin{equation}
q_{L,R}(x)\to\exp\left (\frac i2 \theta^a_{L,R}T^a\right )q_{L,R}(x), \qquad q_{L,R}(x)\to\exp\left (\frac i2 \tilde\theta_{L,R}\right )q_{L,R}(x),
\end{equation}
donde $T^a$ pertenecen a la representaci\' on adjunta de $SU(N_f)$ y $\theta^a_{L,R},\tilde\theta_{L,R}$ son unas constantes arbitrarias que parametrizan la transformaci\' on. Los campos glu\' onicos no se ven afectados por esta simetr\' ia global. Cualquier transformaci\' on $U(N)$ puede descomponerse en $SU(N)\times U(1)$ y por ello, la simetr\' ia global previa puede ser reescrita como $SU(N_f)_L\times SU(N_f)_R\times U(1)_V\times U(1)_A$, donde la parte vectorial $V$ es la suma de los n\' umeros cu\' anticos de quarks derechos e izquierdos y la componente axial $A$ corresponde a la resta de estas cantidades.

\medskip

Las correcciones cu\' anticas revelan que la corriente vector axial singlete asociada a la simetr\' ia bajo $U(1)_A$, que se conserva a nivel cl\' asico, desarrolla una anomal\' ia y por tanto, la verdadera simetr\' ia de la teor\' ia cu\' antica se reduce a $SU(N_f)_L\times SU(N_f)_R\times U(1)_V$. En el Cap\' itulo \ref{Chfase} abordaremos una explicaci\' on m\' as detallada de la anomal\' ia axial. La simetr\' ia $U(1)_V$ se relaciona con la conservaci\' on del n\' umero bari\' onico mientras que el resto se trata de simetr\' ia quiral. Si la simetr\' ia quiral fuese una verdadera simetr\' ia de QCD, uno esperar\' ia un espectro de estados degenerados con paridad opuesta. Sin embargo, esto no se observa en la naturaleza implicando que la simetr\' ia est\' a rota. De hecho, la simetr\' ia se rompe espont\' aneamente ya que los generadores del subgrupo $SU(N_f)_A$ no aniquilan el estado fundamental y por tanto, cada uno de ellos se asocia a un bos\' on de Goldstone sin masa. Los estados f\' isicos correspondientes a estos bosones de Goldstone son los mesones m\' as ligeros $\pi$, $\eta$ y $K$, siendo el \' ultimo importante s\' olo si $N_f=3$. Las masas finitas de estas part\' iculas se entienden como consecuencia de la ruptura expl\' icita de simetr\' ia debido a las masas finitas de los quarks en el lagrangiano de QCD. A pesar de que la simetr\' ia quiral exacta no se satisface realmente en QCD, este tratamiento puede usarse como una aproximaci\' on para trabajar en QCD a bajas energ\' ias desarrollando la llamada Teor\' ia de Perturbaciones Quirales ($\chi$PT), que se discute m\' as abajo.

\medskip

La constante de acoplo de QCD $g$ determina la fuerza de las interacciones entre quarks y gluones y las auto-interacciones de gluones. \' estas \' ultimas son permitidas debido a la naturaleza no abeliana de QCD, al contrario el caso del electromagnetismo donde los fotones s\' olo pueden interaccionar con otras part\' iculas. La renormalizaci\' on de una teor\' ia cu\' antica conduce a la movilidad de su constante de acoplo, que no es realmente constante sino que depende de la escala de energ\' ia $\mu$ a la que un proceso tiene lugar. Una teor\' ia gauge $SU(N_c)$ con $N_f$ sabores de quarks muestra \cite{asymfreed} que la funci\' on $\beta$ a un loop
\begin{equation}
\beta(g)_{\text{1-loop}}=\frac{\partial g}{\partial\log\mu}=-\frac{g^3}{16\pi^2}\left (\frac{11}3N_c-\frac23N_f\right )
\end{equation}
es negativa para $N_f<\frac{11}2N_c$. En particular, el hecho de que QCD con $N_c=3$ y $N_f=6$ quarks se caracteriza por una funci\' on $\beta$ negativa implica una constante de acoplo decreciente con una energ\' ia creciente. A altas energ\' ias, QCD ya no es una teor\' ia que interaccione fuertemente y se acerca al l\' imite libre. Este efecto es conocido como libertad asint\' otica y de alguna manera corresponde al comportamiento opuesto al confinamiento descrito a bajas energ\' ias.

\medskip

Esta visi\' on de QCD permite usar t\' ecnicas perturbativas para c\' alculos de procesos a altas energ\' ias. Las secciones eficaces perturbativas de partones a cortas distancias y las cantidades no perturbativas de largas distancias pueden combinarse gracias a los teoremas de factorizaci\' on de QCD \cite{factorizationQCD}. La contribuci\' on de largas distancias incluye las funciones de distribuci\' on de los partones, funciones de fragmentaci\' on, diferentes tipos de factores de forma, etc. Efectivamente, este m\' etodo se vuelve m\' as preciso as\' i como las energ\' ias aumentan debido a la libertad asint\' otica de QCD.

\medskip

A bajas energ\' ias, los c\' alculos perturbativos no tienen sentido debido al gran acoplo entre quarks y gluones. Los efectos no perturbativos no se pueden evitar en este r\' egimen y por tanto, aparecen muchas dificultades haciendo que QCD sea muy dif\' icil de solucionar anal\' itica o num\' ericamente. Diversas aproximaciones han sido abordadas para superar estos problemas como las simulaciones num\' ericas en el ret\' iculo \cite{wilson}, teor\' ias de campos efectivas \cite{ChPT} o expansiones en $1/N_c$ \cite{1/N}.

\medskip

Una posibilidad de definir rigurosamente una teor\' ia cu\' antica de campos consiste en discretizarla. Los c\' alculos en el ret\' iculo son un mecanismo no perturbativo donde QCD se formula en una cuadr\' icula de puntos en un espacio-tiempo eucl\' ideo. Una separaci\' on fija y finita entre los puntos de la cuadr\' icula representa un regulador UV. Si el tama\~no del ret\' iculo se toma infinitamente grande y la separaci\' on de sus puntos infinitesimalmente peque\~na, se recupera el continuo de QCD \cite{wilson}. La aproximaci\' on de integral de caminos de Feynman se usa para calcular las funciones de correlaci\' on de operadores hadr\' onicos y elementos de matriz de cualquier operador entre estados hadr\' onicos en funci\' on de los grados de libertad fundamentales de quarks y gluones. La masa del pi\' on $m_\pi$ se usa normalmente para examinar la fiabilidad de los c\' alculos en el ret\' iculo ya que \' este es el estado hadr\' onico m\' as ligero y por tanto su longitud de onda correspondiente es la m\' as larga. Si el ret\' iculo no es lo suficientemente grande, la funci\' on de onda del pi\' on no cabe en el ret\' iculo. Por este motivo normalmente se usan grandes masas para los quarks en esta descripci\' on que conducen a un valor no f\' isico para el valor de la masa del pi\' on. Sin embargo, grandes ret\' iculos con espaciado peque\~no corresponden a un gran n\' umero de puntos en la cuadr\' icula y por tanto, simulaciones computacionales que necesitan mucho m\' as tiempo para ser ejecutadas. Adem\' as, las correcciones de efectos de volumen finito tienen que ser calculadas para una comparaci\' on realista con los experimentos.

\medskip

Una aproximaci\' on diferente QCD a bajas energ\' ias consiste en usar teor\' ias de campos efectivas (EFT). Como no son visibles a grandes distancias, los quarks y gluones pueden ser eliminados del modelo mientras que los hadrones se convierten en los grados de libertad relevantes a bajas energ\' ias. Imponiendo las simetr\' ias de QCD a la teor\' ia, cualquier observable puede ser calculado mediante una especie de expansi\' on perturbativa en la que el esquema de contado de potencias determina la importancia de las amplitudes cu\' anticas. De la misma manera que el aumento de la potencia de una constante de acoplo peque\~na se usa en teor\' ia de perturbaciones, EFT normalmente aprovecha el ratio de un par\' ametro peque\~no como masa, momento/energ\' ia. Estas teor\' ias corresponden a la realizaci\' on a bajas energ\' ias de QCD y por consiguiente \' estas son s\' olo v\' alidas hasta una cierta escala m\' axima, haci\' endolas no-renormalizables. $\chi$PT ha demostrado ser una teor\' ia efectiva de mucho \' exito donde se describen las interacciones entre los estados mes\' onicos ligeros (para un an\' alisis detallado ver \cite{ChPT,ChPT2}). Debido a la gran masa del quark $s$, la convergencia de la expansi\' on en el sector $SU(3)$ es en cierto modo m\' as lenta comparada con la versi\' on de $SU(2)$. Es posible generalizar esta teor\' ia mediante la introducci\' on de estados bari\' onicos en la teor\' ia que involucra interacciones bari\' on-bari\' on y bari\' on-mes\' on adicionales. N\' otese que $\chi$PT s\' olo usa como ingredientes las simetr\' ias de QCD y por ello no se trata de un modelo sino de una realizaci\' on a bajas energ\' ias.

\medskip

En esta tesis usaremos dos modelos de bajas energ\' ias: el modelo de Nambu--Jona-Lasinio (NJL) y el modelo sigma. NJL es un modelo microsc\' opico que incluye v\' ertices de 4 fermiones que reemplazan las interacciones de los gluones en QCD. Debido a la presencia de estos t\' erminos, el modelo no es renormalizable pero la ausencia de gluones permite que la teor\' ia sea bosonizada. Despu\' es de este proceso, se obtiene una teor\' ia de mesones con campos escalares y pseudoescalares que puede ser asociada a un modelo sigma con valores concretos de sus par\' ametros. Por otro lado, el modelo sigma corresponde a una teor\' ia particular que deriva de una EFT basada en $\chi$PT asumiendo dominancia escalar en la interacci\' on a bajas energ\' ias. Tanto en NJL como en el modelo sigma, la simetr\' ia quiral se rompe espont\' aneamente debido a la formaci\' on de un condensado din\' amico de quarks quirales. Este condensado transmite una masa constituyente a los quarks. En el sector $u,d$ se espera que \' esta sea del orden de unos 300 MeV (en base a la masa de los nucleones) en contraste con la ligera masa de los quarks desnudos, del orden de 5 MeV.

\medskip

Cuando ninguno de los otros m\' etodos funciona, la \' unica forma de investigar el comportamiento de QCD a bajas energ\' ias consiste en hacer una expansi\' on en t\' erminos del peque\~no par\' ametro $1/N_c$. El l\' imite de gran $N_c$  de una teor\' ia gauge $SU(N_c)$ puede llevar a pensar que este mayor grupo gauge complica la teor\' ia ya que conlleva un aumento del n\' umero de los grados de libertad din\' amicos. Sin embargo, es posible demostrar que este l\' imite es una aproximaci\' on razonable a\' un en el caso de QCD donde $N_c=3$ \cite{1/N}. En el l\' imite de gran $N_c$ el n\' umero de especies de gluones aumenta como $N_c^2$ mientras el n\' umero de quarks lo hace como $N_c$. QCD se convierte en una teor\' ia de un n\' umero infinito de resonancias sin interacci\' on ni anchura de desintegraci\' on \cite{1/N-resonanc}, que son mesones y bolas de gluones ya que las masas de los bariones crece con $N_c$ y pueden desacoplarse de la teor\' ia y ser ignorados sin peligro. La expansi\' on $1/N_c$ normalmente se formula asignando a cada diagrama de Feynman una potencia de $N_c$ \cite{1/N-form}, que es determinada por la topolog\' ia del diagrama. Las contribuciones dominantes son 'planares' y no contienen \emph{loops} de quarks mientras que los t\' erminos subdominantes van suprimidos por un factor $1/N_c$ debido a \emph{loops} de quarks y cada intercambio de gluones no planar es suprimido por $1/N_c^2$. El punto cualitativamente importante de esta aproximaci\' on es el comportamiento de escala en $N_c$ de los diferentes diagramas que permite clasificarlos dependiendo de su relevancia. Dado que $1/N_c$ no es un par\' ametro muy peque\~no, una expansi\' on perturbativa en $1/N_c$ puede requerir la inclusi\' on de un gran n\' umero de t\' erminos de orden superior si se busca un resultado preciso.

\section{Diagrama de fases de QCD}

El hecho de que partones altamente energ\' eticos puedan interaccionar casi libremente sugiere la posibilidad de que la materia confinada pueda sufrir con el tiempo una transici\' on de fase. El sistema es dirigido a un r\' egimen de deconfinamiento donde los grados de libertad relevantes de QCD son los quarks y gluones en lugar de los hadrones y por tanto los estados sin color no existir\' ian m\' as.

\medskip

El estudio del deconfinamiento ha sido abordado por muy diferentes colaboraciones y t\' ecnicas. Los c\' alculos de QCD en el ret\' iculo pueden dar resultados fiables de las propiedades termodin\' amicas de la materia que interacciona fuertemente s\' olo para valores muy peque\~nos del potencial qu\' imico bari\' onico. En este caso se predice una transici\' on de deconfinamiento de tipo h\' ibrido \cite{lattcrossover}. QCD en el ret\' iculo sugiere una transici\' on de fase desde la fase hadr\' onica al QGP a $T_c\sim140-190$ MeV \cite{lattcrit} (para un an\' alisis detallado ver \cite{revlattcrit}).

\medskip

Esta nueva fase, el llamado plasma de quarks y gluones (QGP), podr\' ia ser alcanzado a altas temperaturas y densidades, condiciones que son factibles en colisiones de iones pesados (HIC). De hecho, los recientes resultados del Acelerador Relativista de Iones Pesados (RHIC) muestran una fuerte evidencia de que el QGP ha sido creado en colisiones Au+Au a $\sqrt{s_{NN}}=200$ GeV\cite{QGPatRHIC}.

\medskip

Modificando las condiciones experimentales como la energ\' ia de la colisi\' on o el par\' ametro de impacto es posible explorar un amplio rango de temperaturas y densidades desentra\~nando las diferentes fases de QCD. Mientras el sistema se comprime y calienta, se espera que los hadrones ocupen cada vez m\' as y m\' as espacio, lo que lleva a un solapamiento final que desencadena una 'percolaci\' on' de partones entre hadrones y \' estos acaban siendo liberados. Adem\' as, los quarks ligeros pierden su masa constituyente dentro de los hadrones y adquieren su masa real (desnuda) restaurando la simetr\' ia quiral. Uno de los objetivos cient\' ificos del programa actual de iones pesados de alta energ\' ia es la cuantificaci\' on de las propiedades de la materia del QGP, como por ejemplo su ecuaci\' on de estado.

\medskip

En un experimento de HIC, los n\' ucleos pesados son acelerados a velocidades ultra-relativistas antes de su impacto. Inmediatamente despu\' es de su colisi\' on, los n\' ucleos crean un medio que interacciona fuertemente y alcanza temperaturas extremadamente altas del orden de varios centenares de MeV, y la densidad de la materia nuclear puede ser ampliamente rebasada. Entender el comportamiento de la caliente y densa bola de fuego constituye uno de los retos m\' as notables en QCD. Se cree que estas extremas condiciones termodin\' amicas existen en dos contextos astrof\' isicos: algunos microsegundos despu\' es del 'Big Bang' una etapa transitoria de materia fuertemente interactuante sobrevivi\' o con una temperatura $\sim 200$ MeV (alrededor de $10^{12}$ K) y un peque\~no exceso neto de bariones; en el n\' ucleo de las estrellas de neutrones se cree que las densidades se son varias veces la densidad nuclear. Nuevas regiones del diagrama de fases de QCD pueden ser exploradas mediante los actuales experimentos en aceleradores \cite{Andronic:2009gj,Blaizot:2009ws}. Las primeras colisiones se llevaron a cabo en el Sincrotr\' on de Nivelado Alternado (AGS) en Brookhaven (BNL) con una energ\' ia en el centro de masas de alrededor de $\sqrt s\sim 5$ GeV por nucle\' on y m\' as tarde, el S\' uper Sincrotr\' on de Protones (SPS) del CERN alcanz\' o $\sqrt s\sim$20 GeV por nucle\' on entre muchos otros experimentos de aceleradores. Actualmente RHIC en BNL y LHC en el CERN llegan a energ\' ias de hasta $\sqrt s\sim 200$ GeV y 5 TeV por nucle\' on, respectivamente.

\medskip
\begin{figure}[h!]
\centering
\includegraphics[scale=1.8]{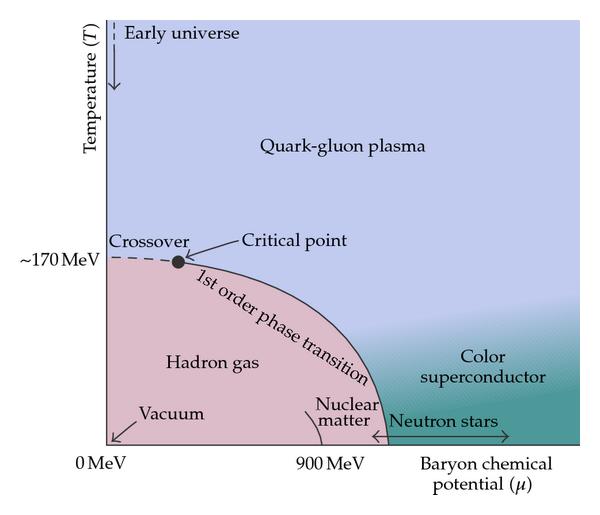}
\caption{Diagrama de fases de QCD dependiendo de la temperatura y densidad extra\' ido de \cite{QCDphasediag}. Se indican las fases reales y las especulativas.}\label{QCDphasediagesp}
\end{figure}

El fen\' omeno del deconfinamiento ha sido investigado tambi\' en en experimentos con aceleradores. La transici\' on h\' ibrida hacia la fase deconfinada que predice QCD en el ret\' iculo parece ser confirmada por recientes experimentos en RHIC \cite{crossovRHIC}. En este r\' egimen de densidades moderadas, diferentes aproximaciones predicen una posible transici\' on de fase de primer orden hacia la fase de deconfinamiento sugiriendo la existencia de un punto cr\' itico que ha sido buscado intensamente en RHIC. Uno de los mayores objetivos del programa de bajas energ\' ias de RHIC es encontrar este punto cr\' itico escaneando diferentes regiones del espacio de fases de los n\' ucleos de oro que se hacen colisionar a diversas energ\' ias. Para encontrar un posible punto cr\' itico se miden momentos superiores de las distribuciones netas de protones en colisiones Au+Au en un amplio rango de potenciales qu\' imicos. Se han observado desviaciones de algunos momentos con respecto a las expectativas te\' oricas a energ\' ias del haz por debajo de $\sqrt{s_{NN}}=27$ GeV por nucle\' on \cite{critpoint}. Sin embargo, se necesita m\' as estad\' istica para obtener conclusiones definitivas. Desde el punto de vista te\' orico, muchas aproximaciones diferentes han planteado e investigado este punto cr\' itico.

\medskip

Tratar con materia acoplada fuertemente complica el entendimiento te\' orico de las propiedades y la din\' amica del medio. En los \' ultimos a\~nos un creciente inter\' es en t\' ecnicas de acoplo fuerte conllev\' o el estudio de los plasmas l\' iquidos que aparecen como fases calientes de deconfinamiento en teor\' ias gauge no abelianas que tienen descripciones hologr\' aficas duales como las teor\' ias gravitacionales en 4+1 dimensiones conteniendo un horizonte de agujero negro. Un ejemplo simple de esta dualidad gauge/gravedad ha sido aplicado en una teor\' ia supersim\' etrica de Yang-Mills $\mathcal N=4$ fuertemente acoplada en el l\' imite de un gran n\' umero de colores. Este tipo de an\' alisis han proporcionado numerosos conocimientos cualitativos de las propiedades del QGP \cite{casalderrey}. En particular, en el l\' imite donde el acoplo de QCD es grande a todas las energ\' ias relevantes, la dualidad gauge/gravedad permiti\' o usar c\' alculos hologr\' aficos para obtener informaci\' on din\' amica detallada de procesos con p\' erdida de energ\' ia.

\medskip

Los estudios te\' oricos de la termodin\' amica del QGP revelaron que \' este se comporta como un l\' iquido de Fermi casi ideal, a pesar de que la investigaci\' on de las caracter\' isticas de flujo a\' un est\' a en marcha. Un importante desarrollo en los \' ultimos a\~nos ha sido la conjetura de que para cualquier fluido t\' ermico puede existir un l\' imite fundamental en el valor de la viscosidad \cite{viscbound}. El principio de indeterminaci\' on requiere una viscosidad diferente de cero. De forma m\' as precisa la conjetura indica que el ratio de la viscosidad sobre entrop\' ia debe satisfacer la relaci\' on $\eta/s\geq 1/4\pi$ y se han desarrollado diferentes investigaciones para medir experimentalmente este ratio \cite{viscos}. La combinaci\' on de diferentes resultados parece concluir que la materia producida est\' a dentro de un factor 2-3 del l\' imite conjeturado y por tanto, \' este deber\' ia corresponder al fluido m\' as ideal que ha sido medido jam\' as. Sin embargo, existen evidencias de que la materia no se comporta como un estado casi ideal de quarks y gluones libres sino m\' as bien como un fluido denso casi perfecto con una absorci\' on casi completa de pruebas de gran momento \cite{zajc}.

\medskip

Adem\' as del QGP y la fase hadr\' onica, otras fases termodin\' amicas han sido discutidas junto con sus correspondientes transiciones de fase. En la Fig. \ref{QCDphasediagesp} se muestra un diagrama de fases de QCD tentativo dependiendo de la temperatura y densidad. Por ejemplo, se cree que la transici\' on l\' iquido-gas de la materia nuclear al gas hadr\' onico ocurre a temperaturas del orden de $T\lesssim 10$ MeV y potenciales qu\' imicos $\mu\simeq 900$ MeV. Otra fase es la llamada fase de color-sabor cerrado (CFL) \cite{CFL}. A densidades muy altas los 3 quarks m\' as ligeros $u,s,d$ pueden formar parejas de Cooper que correlacionan sus propiedades de color y sabor conllevando un comportamiento de superconductividad de color. La fase CFL romper\' ia las simetr\' ias de QCD y exhibir\' ia un espectro y unas propiedades de transporte muy diferentes. Hasta la fecha no se han encontrado evidencias experimentales de esta fase pero se conjetura que esta forma de materia puede existir en los n\' ucleos de las estrellas de neutrones. Con todo, la estructura del diagrama de fases a altas densidades bari\' onicas permanece bastante incierta.

\section{Explorando el diagrama de fases con colisiones de iones pesados}

\begin{figure}[h!]
\centering
\includegraphics[scale=0.4]{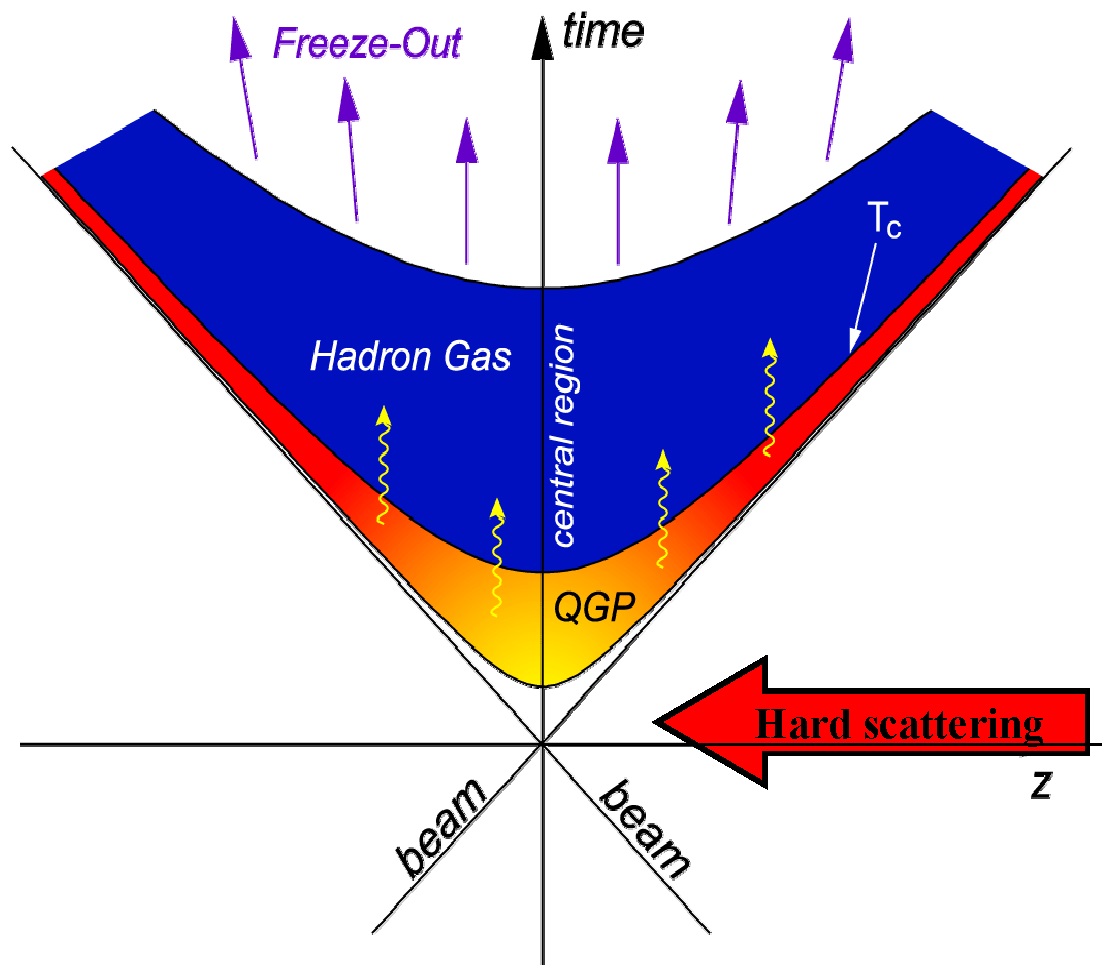}
\caption{Evoluci\' on del medio nuclear creado en HIC dependiendo del tiempo. Despu\' es de la colisi\' on el sistema se calienta y entra en el QGP donde los quarks y gluones quedan deconfinados. A continuaci\' on, un proceso de enfriamiento lento tiene lugar llevando al sistema a la fase hadr\' onica donde los partones se agrupan en bariones y mesones. Finalmente, cuando la temperatura de congelamiento es alcanzada, los hadrones dejan de interactuar.}\label{HICesp}
\end{figure}

La din\' amica de la bola de fuego nuclear creada en HIC no es trivial. En la Fig. \ref{HICesp} presentamos un boceto de una HIC evolucionando en el tiempo y la correspondiente fase donde el sistema se encuentra en cada momento. En las primeras etapas la bola de fuego se comprime por un corto periodo de tiempo entre 0.5 y 1 fm. Durante este tiempo se cree que el sistema alcanza temperaturas de hasta $\sim$600-700 MeV dependiendo del modelo y el experimento (ver \cite{LHClastcall} y las referencias en su interior) y entra el QGP. En la pr\' oxima etapa toda la fenomenolog\' ia asociada a la fase de deconfinamiento tiene lugar mientras el sistema empieza a termalizar debido a las fuertes interacciones entre los componentes del medio. La temperatura cae lentamente hasta los $\sim 200$ MeV, una situaci\' on donde a\' un no est\' a claro si existe una fase mixta. Despu\' es de la expansi\' on que dura entre 5 y 10 fm, los modelos hidrodin\' amicos convencionales y los c\' alculos en el ret\' iculo sugieren que alrededor de $T_c\sim 140-190$ MeV, una transici\' on de fase lleva al sistema a la fase hadr\' onica, donde los quarks y los gluones se reagrupan de nuevo para formar hadrones. Las pruebas experimentales de interacciones hadr\' onicas pueden proporcionar valiosos conocimientos del QGP y de la transici\' on de fase de deconfinamiento. Finalmente, cuando el sistema est\' a lo suficientemente fr\' io, es decir $T\sim 100$ MeV, las colisiones entre hadrones cesan y los componentes del medio quedan congelados, sobreviviendo como part\' iculas en el estado final.

\medskip

A altas energ\' ias, o equivalentemente, a peque\~nos valores de la $x$ de Bjorken, las densidades de gluones son muy elevadas. Bajo estas circunstancias QCD puede ser descrita por una teor\' ia de muchos cuerpos de partones acoplados d\' ebilmente, el llamado Condensado de Vidrio de Color (CGC) \cite{CGC}. En las primeras etapas de una HIC dos n\' ucleos a velocidades relativistas pueden ser considerados como dos tortitas de CGC. Los gluones en las funciones de onda nucleares se encuentran desordenadas y evolucionan muy lentamente en comparaci\' on con la escala de tiempo de una HIC, similar al comportamiento de los vidrios. Su alta densidad es la responsable del t\' ermino condensado. Esta visi\' on de la materia nuclear consiste en un campo cl\' asico coherente que se desintegra en partones cercanos a su capa m\' asica y eventualmente se termaliza para formar el QGP. Esta aproximaci\' on permite el estudio de las condiciones iniciales y el proceso de equilibrio de las HIC. De hecho, el CGC ha de ser considerado como la etapa de no-equilibrio precedente del QGP. Por esta raz\' on, a esta materia se la denomina normalmente Glasma \cite{lappi,Glasma}.

\medskip

\begin{figure}[h!]
\centering
\includegraphics[scale=0.3]{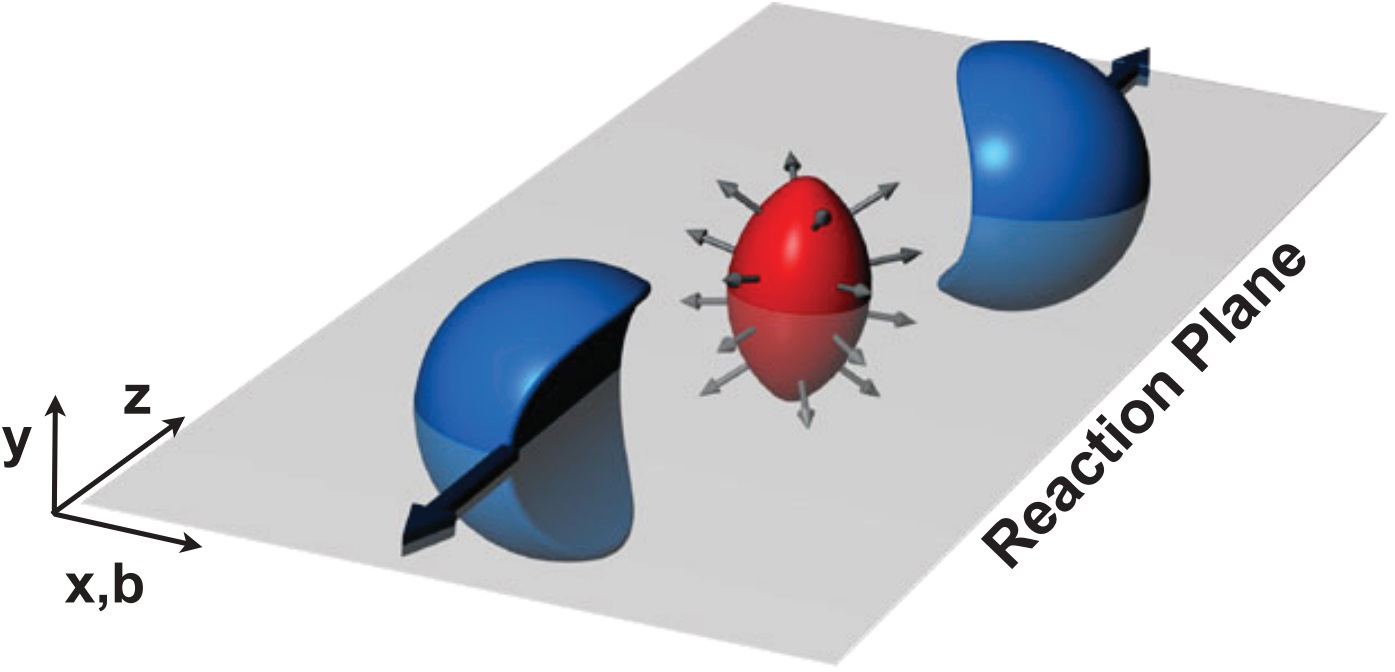}
\caption{En colisiones no centrales, la regi\' on central donde los dos n\' ucleos se solapan presenta una forma parecida a una almendra (gr\' afico extra\' ido de \cite{almond}) que acarrea una anisotrop\' ia inicial con respecto al plano de reacci\' on. Esto se traduce en anisotrop\' ias en las distribuciones de momentos de las part\' iculas.}\label{almondesp}
\end{figure}

Una manera de analizar las propiedades de la materia de QCD en HIC es v\' ia el estudio de la anisotrop\' ia azimutal inducida en colisiones no centrales (para an\' alisis detallados ver \cite{flowreview,almond}). El fluido colectivo anisotr\' opico, es decir, anisotrop\' ias en las distribuciones de momentos de las part\' iculas con respecto al plano de reacci\' on, es altamente sensible a las propiedades del sistema en las primeras etapas de su evoluci\' on (ver Fig. \ref{almondesp}). Esto es debido a que las asimetr\' ias iniciales en la geometr\' ia del sistema que r\' apidamente se desvanecen mientras el sistema se enfr\' ia. La distribuci\' on azimutal de part\' iculas medida con respecto al plano de reacci\' on es convenientemente expandida en una serie de Fourier \cite{flowFourier}. Los diferentes harm\' onicos, generalmente llamados $n$-\' esimo orden del fluido, de esta expansi\' on dependen de la rapidez, del momento transverso y del par\' ametro de impacto de la colisi\' on y han sido medidos en diferentes experimentos \cite{flowexp}. Se han abordado muchos c\' alculos hidrodin\' amicos diferentes para proporcionar descripciones cualitativas de los fluidos colectivos (ver \cite{casalderrey} y las referencias en el interior).

\medskip

Cuando la bola de fuego de una HIC se enfr\' ia y entra en la fase hadr\' onica, los quarks y gluones se reagrupan para formar hadrones. La materia se deja de tener color de nuevo creando un gas de hadrones. Se ha argumentado que tanto en RHIC como en LHC se ha alcanzado el equilibrio t\' ermico a diferencia de otras m\' aquinas previas como SPS. Este medio puede estar formado por muchas part\' iculas diferentes pero la funci\' on de partici\' on viene dominada por los grados de libertad m\' as ligeros. Por tanto, la bola de fuego puede aproximarse por un gas de piones. Otros estados mes\' onicos ligeros pueden estar presentes en peque\~nas cantidades pero los bariones son ciertamente demasiado pesados para jugar un papel significante en este contexto.

\medskip

La evoluci\' on din\' amica del gas de hadrones necesita entenderse correctamente ya que corresponde al paso intermedio entre el QGPP y los estados finales que se miden en los detectores. El mes\' on $\rho$ juega un papel clave en la termalizaci\' on del gas de hadrones. Primero de todo, los mesones $\rho$ interact\' uan fuertemente con los piones y su corto recorrido libre medio permite m\' ultiples procesos de dispersi\' on que lleva a la regeneraci\' on de \' estas part\' iculas a partir de los piones, es decir, el n\' umero efectivo de mesones $\rho$ se ve incrementado con respecto a su producci\' on directa en el proceso de hadronizaci\' on. Y en segundo lugar, parece ser que diversos estudios basados en teor\' ias hadr\' onicas efectivas coinciden en que la funci\' on espectral del mes\' on $\rho$ se ve modificada severamente por efectos del medio nuclear conllevando un fuerte ensanchamiento de la resonancia \cite{rapp,renkr,zahed} con s\' olo un peque\~no desplazamiento de la masa \cite{brown}.

\medskip

Por otra parte, un ensanchamiento de la funci\' on espectral del mes\' on $\rho$ puede resultar de su creaci\' on a partir de los bariones en el denso medio hadr\' onico \cite{rapp}. Un modelo hadr\' onico de muchos cuerpos extendido de la $\rho$ en materia mas\' onica caliente que incluye excitaciones de resonancias y un incremento Bose de piones en la autoenerg\' ia de la $\rho$ implica un ensanchamiento de alrededor de $\Delta\Gamma_\rho\sim 80$ MeV a una temperatura de $T=150$ MeV y una densidad pi\' onica $\varrho_\pi\simeq 0.75\varrho_0$ \cite{RappT}, donde $\varrho_0$ es la densidad nuclear. En materia nuclear fr\' ia otras investigaciones han sido desarrolladas en modelos de resonancia $\rho N$ incluyendo la nube de piones en el medio \cite{Rappmu} mostrando un ensanchamiento a\' un m\' as fuerte de $\sim300$ MeV a $\varrho_N\simeq \varrho_0$ y un desplazamiento de masa comparable de unos $\sim40$ MeV. Por otro lado, el escenario de ca\' ida de masa de los mesones vectoriales en la densa bola de fuego \cite{dropmass} establece un v\' inculo entre las masas hadr\' onicas y el condensado de quarks, el par\' ametro de orden de la restauraci\' on de la simetr\' ia quiral. La masa de la $\rho$ escala \cite{brown} con el condensado de quarks y \' este \' ultimo cae debido a la alta densidad bari\' onica m\' as que por una alta temperatura. N\' otese que ambas aproximaciones se basan en una alta densidad bari\' onica. Los relativos m\' eritos de estas dos diferentes aproximaciones han atra\' ido mucho debate intentando explicar el exceso de dileptones de baja masa invariante que han sido medidos en diferentes experimentos \cite{CERES,NA60,phenix}.

\section{Pruebas duras en colisiones de iones pesados}

Se ha propuesto una gran variedad de se\~nales del medio nuclear y su din\' amica. Las llamadas pruebas duras, es decir procesos que involucran intercambios de grandes momentos, pueden explorar la f\' isica de peque\~nas distancias siempre que los quarks constituyentes sean lo suficientemente pesados, un r\' egimen que puede ser bien descrito por QCD perturbativa que conduce a un buen control te\' orico del asunto.

\medskip

Una prueba dura importante es el conjunto de diferentes estados de quarkonio pesado y sus modificaciones en el medio \cite{MatsuiSatz}. La presencia del QGP apantalla la fuerza atractiva de un par quark-antiquark. La m\' axima distancia que permite la formaci\' on del quarkonio disminuye con la temperatura del medio. La interacci\' on de estos estados con un medio caliente conlleva la modificaci\' on de sus propiedades y su completa disociaci\' on a temperaturas suficientemente altas. La dualidad gauge/gravedad ha mostrado que los mesones de quarkonio pesado permanecen ligados en el plasma y el hecho de que su temperatura de disociaci\' on cae con una creciente velocidad del mes\' on \cite{casalderrey}. Esta ruptura implica la supresi\' on de la producci\' on de quarkonio en colisiones nucleares en comparaci\' on con su equivalente prot\' on-prot\' on. La mayor\' ia de estudios sobre el quarkonio se han centrado en estados de charmonio como por ejemplo el mes\' on $J/\psi$ debido a la peque\~na secci\' on eficaz de producci\' on del quark $b$ a bajas energ\' ias. Sin embargo, los quarks $c$ no son lo suficientemente pesados para una descripci\' on te\' orica fiable y por tanto, sus interacciones con el QGP y su mecanismo de producci\' on en HIC son relativamente inciertos \cite{quarkonium}. Se ha mostrado una supresi\' on fuerte de los mesones $J/\psi$ en HIC \cite{Jpsi,CMSquarkonium} pero el patr\' on de supresi\' on es a\' un confuso y la interpretaci\' on de datos no est\' a totalmente clara ya que el quarkonio tambi\' en queda suprimido en colisiones prot\' on-n\' ucleo \cite{pA}. Recientemente se han publicado datos sobre la supresi\' on del bottomonio a trav\' es del estudio de los mesones $\Upsilon$ en LHC \cite{CMSquarkonium} y RHIC \cite{RHICUpsilon} permitiendo un mejor entendimiento te\' orico de este canal.

\medskip

Los jets de alta energ\' ia producidos en una colisi\' on se ven severamente afectados por un medio que interacciona fuertemente. Cuando un part\' on de gran momento se produce en la primera fase de una colisi\' on n\' ucleo-n\' ucleo, \' este interactuar\' a m\' ultiplemente con los constituyentes del medio antes de su hadronizaci\' on. La energ\' ia de estos partones se ve reducida debido a la p\' erdida de energ\' ia por colisi\' on \cite{perkins} y a la radiaci\' on de gluones inducida por el medio \cite{gross}. La energ\' ia de estos partones se almacena en el QGP y se propaga como una excitaci\' on colectiva o fluido c\' onico \cite{conicflow}. Este efecto conlleva la aparici\' on de picos en correlaciones de part\' iculas que han sido observados experimentalmente \cite{conicflowexp}. El efecto de la extinci\' on de jets en el QGP es la motivaci\' on principal para el estudio de jets al igual que del espectro de part\' iculas de gran momento y correlaciones de part\' iculas en HIC. El estudio de jets altamente energ\' eticos es una de los mejores pruebas duras para investigar el medio nuclear ya que su producci\' on ocurre a escalas de energ\' ia muy altas donde QCD perturbativa garantiza una buena descripci\' on te\' orica. De forma similar, muchas de las propiedades de los jets en el vac\' io tambi\' en est\' an controladas por la f\' isica a altas escalas de energ\' ia y por tanto est\' an bien entendidas desde un punto de vista te\' orico. Las desviaciones observadas en las diferentes propiedades en HIC y en vac\' io han de ser debidas a la interacci\' on de diferentes componentes del jet con el medio nuclear. La precisa reconstrucci\' on de jets permite la medida de las funciones de fragmentaci\' on de los jets y permiten entender mejor las propiedades del caliente y denso QGP. Las primeras evidencias de p\' erdida de energ\' ia en partones fueron observadas en RHIC a partir de la supresi\' on de part\' iculas con alto $p_T$ estudiando el factor de modificaci\' on nuclear \cite{highptsup1} y la supresi\' on de correlaciones de part\' iculas que salen en direcciones opuestas \cite{highptsup2}. El LHC tambi\' en ha observado la extinci\' on de jets \cite{highptsup3} y SPS parcialmente tambi\' en \cite{SPSjet}.

\medskip

Adem\' as de las pruebas duras, otras se\~nales del medio creado en HIC han sido discutidas. Grandes fluctuaciones topol\' ogicas en el vac\' io de QCD pueden inducir la ruptura de paridad en colisiones n\' ucleo-n\' ucleo que llevan al Efecto Magn\' etico Quiral \cite{kharzeev}, que predice una separaci\' on de carga a lo largo del plano de reacci\' on. Adem\' as, los fotones y dileptones (parejas de lept\' on y antilept\' on del mismo sabor) son considerados como potentes pruebas debido a su despreciable interacci\' on con los estados hadr\' onicos finales. Explicaremos estos dos aspectos en detalle en las pr\' oximas secciones.

\section{El Efecto Magn\' etico Quiral}\label{SecCMEesp}

\begin{figure}[h!]
\centering
\includegraphics[scale=0.3]{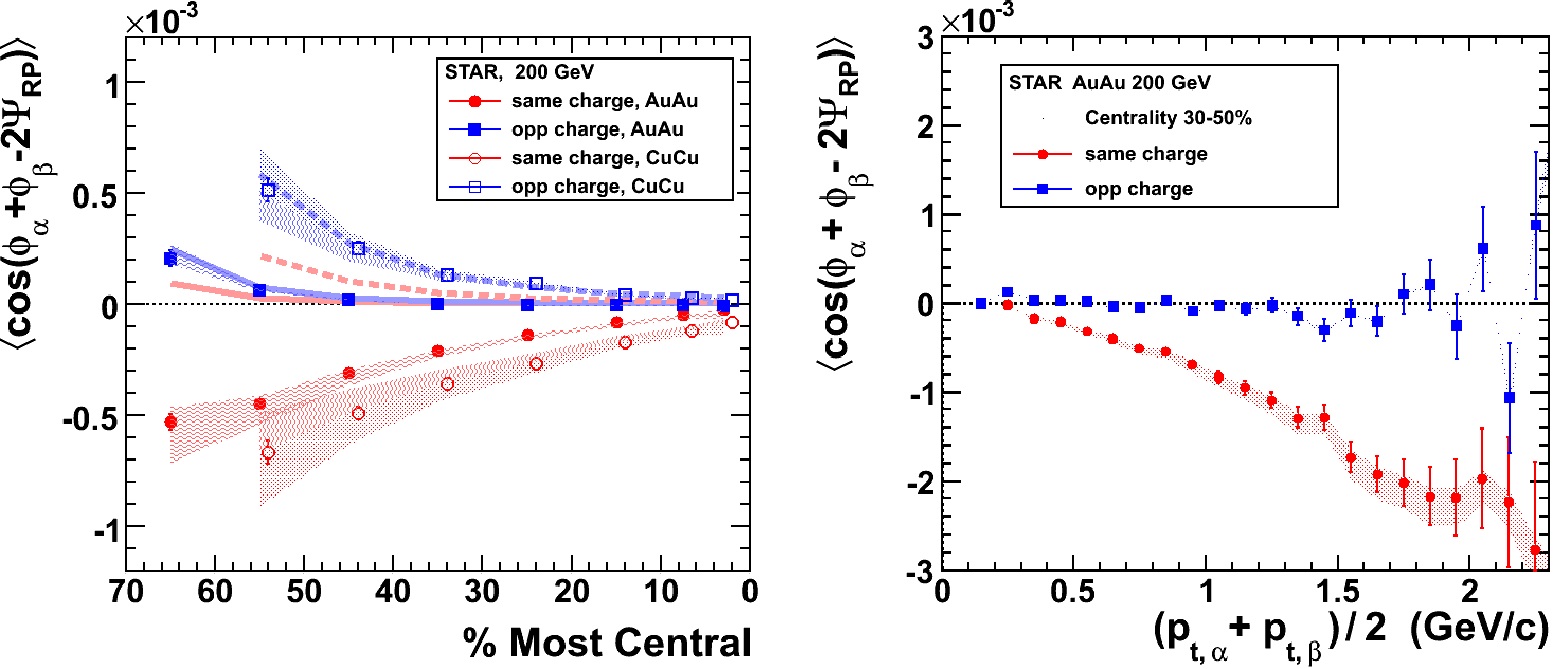}
\caption{El observable que refleja las correlaciones azimutales de igual y opuesta carga se muestra dependiendo de la centralidad de la colisi\' on para choques Au+Au y Cu+Cu (izquierda) y de la semi-suma de los momentos transversos s\' olo para colisiones Au+Au (derecha) extra\' ido de \cite{star}. En el panel izquierdo las l\' ineas continua (Au+Au) y discontinua (Cu+Cu) representan los c\' alculos te\' oricos de correlaciones de 3 part\' iculas.}\label{staresp}
\end{figure}

Un efecto diferente que ha recibido mucha atenci\' on es la separaci\' on de cargas detectada en colisiones n\' ucleo-n\' ucleo en STAR \cite{star}. La medida de un observable no trivial de 3 part\' iculas refleja que las correlaciones azimutales de igual y opuesta carga en cada evento conducen a una peque\~na separaci\' on de cargas sobre el plano de reacci\' on en una HIC. Este observable tiene la ventaja de suprimir las correlaciones de fondo que no est\' an relacionadas con el plano de reacci\' on. Se ha supuesto que la medida de este efecto es la confirmaci\' on del Efecto Magn\' etico Quiral (CME) \cite{kharzeev}, que asume que dominios locales con paridad impar pueden aparecer en HIC. Se espera que las correlaciones de pares con la misma y opuesta carga debido al CME sean similares en magnitud y de signo opuesto. Esto podr\' ia verse modificado por el medio que resultar\' ia en la diluci\' on de las correlaciones entre part\' iculas de signo opuesto \cite{kharzeev}. Parece observarse una diferencia clara en la fuerza de correlaci\' on entre las combinaciones de part\' iculas de carga igual y opuesta (ver Fig. \ref{staresp}). La simulaci\' on de correlaciones de carga opuesta de la correlaci\' on de 3 part\' iculas es similar a la se\~nal medida pero la se\~nal de igual carga es mucho mayor y de signo opuesto a la predicha. Este fen\' omeno tambi\' en ha sido observado en ALICE en LHC \cite{ALICECME}.

La dependencia en la centralidad que se muestra en el panel izquierdo de la Fig. \ref{staresp} coincide con la separaci\' on de carga relativa al plano de reacci\' on. Por otro lado, la magnitud del efecto para part\' iculas de igual carga aumenta con la media de $p_T$ del par tanto en RHIC como en LHC. Esta observaci\' on se encuentra en contradicci\' on con la expectativa inicial \cite{kharzeev}, que predice el dominio de part\' iculas con bajo $p_T$. En cualquier caso, no se pueden exponer conclusiones definitivas debido a la falta de c\' alculos de modelos realistas para las dependencias en la centralidad y el momento del par.

\medskip

Se esperan diferentes din\' amicas de campo magn\' etico en LHC y RHIC, un hecho que deber\' ia ser cuantificado. La se\~nal de ruptura de paridad indica la presencia de una separaci\' on de carga cuya magnitud se encoge con una energ\' ia decreciente del haz, sugiriendo que el volumen del QGP se reduce con colisiones de menor energ\' ia, lo que conlleva una menor separaci\' on de carga. Se ha argumentado que a altas energ\' ias la fuerza del CME deber\' ia decrecer ya que el campo magn\' etico decae mucho m\' as r\' apido \cite{CME-B}.

\medskip

El an\' alisis de diferentes correlaciones azimutales de carga para buscar separaciones de carga en un estado deconfinado de la materia puede ayudar en la posibilidad de probar nuevos efectos en QCD a altas temperaturas y densidades. En particular, la aparici\' on de grandes fluctuaciones topol\' ogicas locales que inducen dominios metaestables con $P$ y $CP$ impar podr\' ian ser detectados junto con la corriente electromagn\' etica inducida por grandes momentos angulares o campos magn\' eticos en HIC no centrales. Adem\' as del estudio presentado, se planea el an\' alisis de correlaciones de harm\' onicos de orden superior, que puede ayudar a proporcionar un mayor entendimiento de las correlaciones dependientes de la carga observadas en RHIC y LHC.

\section{El problema de los dileptones de baja masa}\label{dilepexcessesp}

Se considera que los dileptones son la prueba electromagn\' etica m\' as \' util para desentra\~nar los rasgos principales del medio caliente creado en HIC. \' estos son producidos de forma continua transportando informaci\' on de toda la evoluci\' on del sistema y dada la existencia de una variable adicional, su masa invariante $M_{\ell\ell}$, los dileptones poseen un mejor ratio se\~nal/fondo comparado con los fotones. Dependiendo de la fase de la colisi\' on, los dileptones son emitidos desde diferentes fuentes. Antes de la colisi\' on, los dileptones se producen a trav\' es de una radiaci\' on de frenado coherente debido a la repulsi\' on electromagn\' etica entre los dos n\' ucleos \cite{Bremsstr-dilep} pero esta contribuci\' on es despreciable comparada con otros procesos que le siguen \cite{late-dilep}. Durante el calentamiento del sistema nuclear y antes de que se alcance el equilibrio, la aniquilaci\' on Drell-Yan proporciona la fuente principal de dileptones que puede ser dominante por encima de $M_{\ell\ell}\simeq 3$ GeV \cite{DrellYan}. Cuando el sistema entra en el QGP y termaliza, los dileptones se crean principalmente de la dispersi\' on perturbativa de quarks y antiquarks. A continuaci\' on, el plasma evoluciona a un gas de hadrones donde los dileptones se emiten de la aniquilaci\' on de piones y kaones (entre otros canales) \cite{piK-annihil}. La presencia de resonancias mes\' onicas vectoriales ligeras como $\rho$, $\omega$ y $\phi$ son responsables de un incremento de estos procesos debido a su conexi\' on entre los estados de dos mesones iniciales y los dileptones finales \cite{vectmesons-dilep}. La masa invariante de los dileptones corresponde a la distribuci\' on de masas del mes\' on vectorial en el preciso momento de la desintegraci\' on permitiendo entender mejor el medio a altas temperaturas. En las fases finales de la colisi\' on cuando se congelan las interacciones entre hadrones, s\' olo los estados hadr\' onicos de larga vida (principalmente mesones) como $\pi^0$, $\eta$ y $\omega$ son capaces de producir pares de leptones a trav\' es de su desintegraci\' on Dalitz ya que \' estos ya no interact\' uan entre ellos \cite{KW}. Estos procesos corresponden a un estado inicial que se desintegra en un par de part\' iculas, una de ellas es un fot\' on virtual que posteriormente decae en una pareja de leptones. Debido a la cinem\' atica de estas desintegraciones, todas las fuentes de mesones mencionadas son relevantes para la producci\' on de dileptones con una masa invariante por debajo de 1 GeV.

\medskip

Debido a la relevancia de los dileptones, se han llevado a cabo numerosos estudios para desentra\~nar las principales caracter\' isticas del medio nuclear. Posiblemente uno de los an\' alisis m\' as importantes y antiguos consiste en medir la distribuci\' on de masa invariante de los dileptones, que ha sido analizado por diferentes colaboraciones experimentales durante las \' ultimas d\' ecadas. En colisiones prot\' on-prot\' on la forma de la distribuci\' on detectada puede ser correctamente reproducida por el llamado 'coctel' hadr\' onico, la suma de diversas desintegraciones de hadrones en el estado final con abundancias conocidas. De hecho, los datos $pp$ normalmente se usan para ajustar el peso de los diferentes hadrones para ser usados para otras predicciones. Y mientras los datos prot\' on-n\' ucleo se pueden entender a partir de los resultados $pp$, las colisiones centrales n\' ucleo-n\' ucleo han presentado severas discrepancias con respecto a la contribuci\' on te\' orica esperada \cite{CERES,NA60,phenix}, en particular, un fuerte incremento de dileptones con masa invariante por debajo de 1 GeV. Este desacuerdo refleja que los efectos en el medio han de ser considerados para explicar los resultados anormales en colisiones $AA$.

\medskip

La producci\' on an\' omala de dileptones en colisiones n\' ucleo-n\' ucleo ha sido un problema de larga duraci\' on que ha sido investigado por muchos grupos diferentes, tanto te\' oricos como experimentales (para un an\' alisis detallado ver \cite{reviews}). No se efectuaron medidas de dileptones en el acelerador AGS, donde se esperaban altas densidades bari\' onicas. La colaboraci\' on DLS en BEVALAC lanz\' o los primeros resultados de dielectrones para reacciones Ca+Ca, C+C, He+Ca y Nb+Nb con objetivo fijo a una energ\' ia del haz de 1 GeV por nucle\' on. La primera generaci\' on de datos de DLS \cite{DLSv1} era consistente con los resultados de c\' alculos de modelos de transporte \cite{transpmod} incluyendo las fuentes convencionales de dileptones como la radiaci\' on de frenado $pn$, desintegraciones Dalitz de mesones ligeros, desintegraciones directas de mesones vectoriales y aniquilaci\' on pi\' on-pi\' on, sin incorporar efectos en el medio. Sin embargo, una versi\' on actualizada de las medidas de DLS que inclu\' ia toda la muestra de datos y con un an\' alisis mejorado mostr\' o un incremento de casi un orden de magnitud en la secci\' on eficaz comparado con los resultados previos \cite{DLSv2}. Esta discrepancia ha persistido durante a\~nos y se han abordado numerosos intentos para desenredar los nuevos datos de DLS. De hecho, todos los escenarios en el medio que han explicado con \' exito el incremento de dileptones a m\' as altas energ\' ias en aceleradores m\' as potentes fallan al describir los nuevos datos de dileptones de DLS, un problema que pronto fue conocido como el 'puzle DLS'.

\medskip

El puzle DLS motiv\' o el experimento HADES en SIS, un experimento de nueva generaci\' on que fue dise\~nado para estudiar colisiones C+C, Ni+Ni y Au+Au a unas energ\' ias del haz de hasta 8A GeV con un objetivo fijo. El exceso de dielectrones observado en DLS fue confirmado por los datos de HADES \cite{HADES} haciendo colisionar C+C a 1A y 2A GeV. Recientemente, nuevas aproximaciones han intentado atacar este problema. Por un lado, el efecto fue parcialmente interpretado como si fuese debido a un incremento de la producci\' on del mes\' on $\eta$ en la dispersi\' on prot\' on-neutr\' on \cite{HADES-eta}. Por otro lado, un incremento en la contribuci\' on de radiaci\' on de frenado de acuerdo con c\' alculos de intercambio de un bos\' on condujo a un acuerdo bastante exitoso con los datos publicados por DLS y HADES en colisiones C+C \cite{BratkCass}. Sin embargo, los mismos c\' alculos a\' un subestiman la producci\' on de dileptones en reacciones Ca+Ca y por tanto, el tema a\' un es discutible.

\medskip

En SPS el exceso de dileptones de baja masa en colisiones centrales n\' ucleo-n\' ucleo fue observado de nuevo. Se usaron dos haces diferentes: un haz de azufre a 200 GeV/nucle\' on y uno de plomo a 158 GeV/nucle\' on. La colaboraci\' on CERES present\' o un exceso de dielectrones \cite{CERES} donde se analizaban reacciones S+Au y Pb+Au con objetivo fijo. Adem\' as, mostraron que el incremento, particularmente importante en la regi\' on de masa alrededor de 400-600 MeV, cubr\' ia un amplio rango de momento transverso, pero era mayor a bajo momento transverso. Se observaron resultados anormales similares para dimuones de baja masa en HELIOS/3 \cite{HELIOS} con reacciones S+W. Los resultados de reacciones S+Cu y S+U estudiadas por la colaboraci\' on NA38 junto con las colisiones Pb+Pb en NA50 revelaron un exceso poco significante de dimuones \cite{NA38/50}, un hecho que muy probablemente pudo estar relacionado con un gran corte en $M_T$ aplicado en su an\' alisis. Estas medidas independientes del exceso de dileptones de baja masa en SPS desencadenaron una gran investigaci\' on te\' orica tambi\' en motivada por una posible conexi\' on con la ruptura de simetr\' ia quiral. La importancia de la densidad bari\' onica sugerida por efectos en el mes\' on $\rho$ en el medio animaron a la colaboraci\' on CERES a medir dielectrones de baja masa en colisiones Pb+Au a una energ\' ia baja de 40A GeV donde se esperaba una alta densidad bari\' onica y una menor temperatura \cite{CERES40AGEV}. Se llevaron a cabo numerosas aproximaciones para describir el exceso de dileptones en SPS, donde no est\' a claro que se alcanzase un equilibrio t\' ermico. Finalmente, los datos pudieron ser aparentemente explicados mediante modificaciones en el medio del mes\' on $\rho$ debido a su fuerte acoplo hadr\' onico y otras hip\' otesis menos convencionales \cite{rapp,dropmass,HatsudaLee,brown,BratkCass}.

\medskip

El programa de f\' isica de dileptones de las colaboraciones NA38/NA50, HELIOS/3 y CERES fue continuado por el experimento NA60 tambi\' en en SPS. Se investigaron colisiones n\' ucleo-n\' ucleo v\' ia producci\' on de dimuones en reacciones In+In a 158 GeV/nucle\' on con un objetivo fijo. El experimento exhibi\' o un claro exceso de dimuones a bajas energ\' ias que aumentaba con la centralidad y bajo $p_T$ del par \cite{scomparin-arnaldi2}. Los modelos con m\' as \' exito de la \' epoca fueron comparados con los datos de alta calidad de NA60, que pudieron aislar el canal $\rho$ del resto de fuentes hadr\' onicas. En particular, los escenarios de ca\' ida de masa y de ensanchamiento muestran resultados similares para masas por debajo del pico de resonancia del mes\' on $\omega$ debido a la limitada precisi\' on de los datos, pero entre las resonancias $\omega$ y $\phi$ el caso de ca\' ida de masa parece ser descartado \cite{NA60}. El escenario de ensanchamiento de masa \cite{rapp} se refuerza con los datos de NA60, junto con otros c\' alculos recientes que aseguran reproducir el exceso de dileptones \cite{renkr,zahed}. Con todo, las a\' un considerables incertidumbres experimentales pueden permitir que otros efectos tambi\' en tengan lugar.

\medskip

\begin{figure}[h!]
\centering
\includegraphics[scale=0.3]{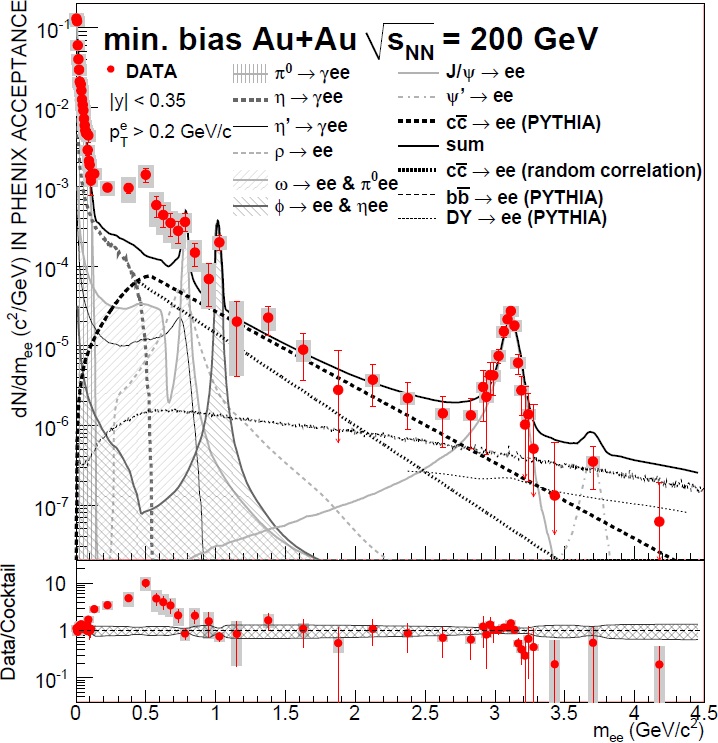}
\caption{Espectro de masa de dielectrones en colisiones Au+Au con m\' inimo sesgo en PHENIX comparado con el coctel de las conocidas fuentes hadr\' onicas extra\' ido de \cite{phenix}. El ratio de los datos sobre el coctel se muestra en el panel inferior. Se observa un claro incremento de casi un orden de magnitud a bajas masas invariantes.}\label{phenixesp}
\end{figure}

RHIC es el primer acelerador con dos haces altamente energ\' eticos impactando a $\sqrt{s_{NN}}=200$ GeV/nucle\' on, donde actualmente se llevan a cabo colisiones Au+Au y son analizadas en los experimentos PHENIX y STAR. PHENIX ha medido un serio exceso de dielectrones a bajas energ\' ias que est\' a presente a todo $p_T$ pero \' este es m\' as pronunciado a bajo $p_T$ del par \cite{phenix}. Se observa una fuerte dependencia en la centralidad para el continuo de baja masa. El incremento aparece en colisiones centrales donde \' este alcanza un factor casi 8 (ver Fig. \ref{phenixesp}). Los resultados de PHENIX son diferentes de los observados en SPS: mientras que en SPS el exceso era pr\' oximo al pico resonante del mes\' on $\rho$, en PHENIX \' este se extiende hacia masas invariantes m\' as bajas. Adem\' as, la producci\' on integrada aumenta m\' as r\' apido con la centralidad de las colisiones que el n\' umero de nucleones participantes. La contribuci\' on de pares correlacionados a partir de desintegraciones semilept\' onicas de mesones con el quark $c$, despreciable a las energ\' ias de SPS, es una fuente importante que ha de ser incluida en el coctel hadr\' onico en las energ\' ias de RHIC. Al contrario que en previos experimentos, los resultados de dileptones de baja masa de PHENIX no han sido reproducidos correctamente por ning\' un modelo con \' exito que haya aparentemente explicado el exceso anteriormente.

\medskip

La colaboraci\' on STAR tambi\' en present\' o un incremento de dielectrones bajo el pico del mes\' on $\omega$ \cite{STARdilept} pero se observ\' o que era significantemente inferior que el medido en PHENIX. Debido a este exceso moderado, los datos de STAR posiblemente se pueden explicar mediante c\' alculos te\' oricos incluyendo un mes\' on $\rho$ ensanchado en el medio. La centralidad y la dependencia en $p_T$ tambi\' en es comparable con las expectativas de c\' alculos a partir de modelos.

\medskip

Finalmente, las colaboraciones de LHC a\' un no se han publicado medidas de dileptones de baja masa invariante.

\medskip

Tambi\' en se ha medido un exceso de dileptones similar con una masa invariante de entre 1 y 3 GeV en SPS y RHIC. Todos los experimentos en SPS presentaron un exceso con respecto a la producci\' on esperada a partir de las dos contribuciones principales en esta regi\' on de masas, las desintegraciones Drell-Yan y semilept\' onicas del quark $c$ \cite{HELIOS,NA38/50-IMR}. Se investigaron diversas explicaciones basadas en hip\' otesis fenomenol\' ogicas de una producci\' on incrementada $c$, como emisiones secundarias Drell-Yan, re-dispersiones en los estados finales que ensanchen la distribuci\' on de $m_T$ de los mesones encantados aumentando la producci\' on, y todas las fuentes de parejas de leptones a partir de interacciones mes\' onicas secundarias. NA60 concluy\' o que el exceso de dimuones en la regi\' on de masas intermedias corresponde a la radiaci\' on t\' ermica emitida en las primeras etapas de la colisi\' on \cite{NA602}. Por el contrario, PHENIX mostr\' o que la producci\' on inclusiva parece estar de acuerdo con la contribuci\' on simulada de mesones encantados \cite{phenix}.

\medskip

Vivimos en una era con numerosos experimentos en curso que pueden obtener informaci\' on relevante sobre la materia de QCD. Muchos modelos han intentado solucionar o explicar diferentes problemas con la adici\' on de nuevos par\' ametros u observables que necesitan ser medidos. Gracias al duro trabajo experimental actual, la medida de estos par\' ametros puede ayudar a confirmar o descartar todos estos modelos. Adem\' as, uno no puede excluir el hecho de que podr\' ian detectarse nuevos fen\' omenos. Este escenario constituye una fuente estimulante de motivaci\' on para la investigaci\' on tanto te\' orica como experimental.

\section{Alcance y visi\' on de conjunto}

En el contexto de HIC, se pueden estudiar nuevos efectos en QCD debido a altas temperaturas y densidades. En particular, en esta tesis exploraremos la posibilidad de que el vac\' io de QCD rompa paridad en un volumen finito. El teorema de Vafa-Witten \cite{vw} proh\' ibe la ruptura espont\' anea de simetr\' ias discretas en teor\' ias gauge. Sin embargo, este teorema no se aplica cuando uno trata con un sistema denso. Las altas temperaturas y densidades alcanzadas en el medio fuertemente interactuante inducido en HIC nos llevan a un escenario realista donde la violaci\' on de paridad en QCD podr\' ia ser perceptible.

\medskip

La violaci\' on de paridad en QCD ya ha sido estudiada por diferentes colaboraciones y aproximaciones \cite{kharzeev,anesp} que discutiremos. Aqu\' i estamos interesados en describir las condiciones termodin\' amicas que podr\' ian conducir a este fen\' omeno en HIC al igual que la fenomenolog\' ia asociada a \' el. Si este efecto tiene lugar, las propiedades de los hadrones en el medio se ver\' ian severamente distorsionadas, lo que provocar\' ia modificaciones inesperadas de las distribuciones de las part\' iculas en el estado final. Mostraremos en detalle c\' omo los mesones escalares y vectoriales se ven influenciados por un medio de paridad impar y adem\' as, buscaremos posibles trazas que puedan ayudar a detectar este efecto en RHIC o LHC.

\medskip

Primeramente, nos centramos en la posible existencia de una nueva fase caracterizada por la violaci\' on de paridad en QCD. Hemos analizado dos aproximaciones diferentes para explorar este efecto. La primera de ellas corresponde al estudio de QCD a densidad finita mediante t\' ecnicas de lagrangianos efectivos acoplados a un potencial qu\' imico bari\' onico no trivial. Un modelo sigma lineal generalizado que retiene los dos multipletes m\' as ligeros de campos escalares y pseudoescalares de $SU(2)$ y que satisface las simetr\' ias de QCD en el vac\' io muestra que para un rango dado de $\mu\neq0$, un condensado pseudoescalar puede aparecer llev\' andonos a una nueva fase de QCD \cite{anesp}. La segunda posibilidad est\' a relacionada con la presencia de burbujas metaestables con $P$ y $CP$ impar en materia caliente debido a grandes fluctuaciones topol\' ogicas en un volumen finito. En este contexto un gran campo magn\' etico o momento angular podr\' ia inducir un campo el\' ectrico perpendicular al plano de reacci\' on y de ah\' i un momento dipolar el\' ectrico \cite{kharzeev}. Este es el Efecto Magn\' etico Quiral que ya ha sido presentado previamente y se dice que ha sido detectado en HIC perif\' ericas en STAR \cite{star}. Siguiendo los argumentos de esta \' ultima aproximaci\' on, investigamos la violaci\' on de paridad en QCD como consecuencia de grandes fluctuaciones topol\' ogicas correspondiendo a un efecto de no-equilibrio que puede ser tratado por un potencial qu\' imico axial externo $\mu_5$.

\medskip

A continuaci\' on implementamos el modelo NJL con un potencial qu\' imico vectorial y axial a temperatura cero para investigar el diagrama de fases del modelo en busca de posibles fases con un condensado pseudoescalar que violen paridad. Vemos que un $\mu_5$ diferente de cero modifica severamente las propiedades conocidas del modelo NJL mientras que la presencia de $\mu$ no afecta demasiado a nuestras consideraciones. Siempre que ciertas relaciones num\' ericas entre los par\' ametros del modelo se satisfagan, se encuentra que la ruptura de paridad espont\' anea es estable. En particular, encontramos una fase con un condensado pseudoescalar singlete de isosp\' in no trivial al igual que un condensado escalar a partir de un valor cr\' itico del potencial qu\' imico axial. Sin embargo, NJL con tales par\' ametros caracter\' isticos no tiene nada que ver con QCD ya que las masas de los conocidos estados mes\' onicos en el vac\' io no son compatibles con la fenomenolog\' ia. A pesar de no poder dar conclusiones definitivas para QCD, los valores de $\mu_5$ que llevan a una fase termodin\' amicamente estable con violaci\' on de paridad sugieren el rango de potenciales qu\' imicos axiales donde se ha de buscar la fase de ruptura de paridad en experimentos de HIC.

\medskip

Despu\' es de investigar los posibles valores de $\mu_5$ que podr\' ian conllevar una violaci\' on de paridad, estudiamos detalladamente las modificaciones de los mesones en el medio. Primero de todo, analizamos los mesones escalares y pseudoescalares en un medio caracterizado por un potencial qu\' imico axial no trivial. Implementamos un modelo sigma generalizado incluyendo los mesones escalares y pseudoescalares m\' as ligeros en la representaci\' on de $SU(3)$ para investigar la mezcla entre los estados de diferente paridad y mismo sabor inducidos por un medio de paridad impar donde de hecho, la distinci\' on entre estados escalares y pseudoescalares no tiene ya sentido. Se introduce el potencial qu\' imico axial como un espurio campo vector axial externo a trav\' es de una derivada covariante. Los mesones en el medio exhiben propiedades dependientes de la energ\' ia como el espectro y las anchuras de desintegraci\' on. Vemos que los grados de libertad m\' as ligeros en cada canal de isosp\' in se convierten en estados taqui\' onicos a altas energ\' ias. Adem\' as, el gas de piones que constituye la bola de fuego en una colisi\' on de HIC una vez alcanza la fase hadr\' onica podr\' ia ser un gas de piones distorsionados debido al efecto de ruptura de paridad. Calculamos las anchuras de desintegraci\' on de todos estos estados a piones distorsionados para buscar posibles resonancias anchas que pudiesen sufrir un proceso de regeneraci\' on como en el caso del mes\' on $\rho$. En particular, el mes\' on $\eta$ distorsionado parece estar en equilibrio t\' ermico dentro del gas de 'piones' debido a su gran anchura, a diferencia de lo que sucede en el vac\' io.

\medskip

En el sector de los mesones vectoriales introducimos el efecto de ruptura de paridad a trav\' es del t\' ermino de Chern-Simons ya que la mezcla de estados de diferente paridad es bastante incierta. Mediante el modelo de Dominancia de Mesones Vectoriales (VMD) introducimos los estados m\' as ligeros en la representaci\' on de $SU(3)$. Sin embargo, el mes\' on $\phi$ se desacopla del resto del sistema y \' este puede ser simplificado a $SU(2)$. Se encuentra una mezcla entre fotones y los mesones $\rho$ y $\omega$ debido al propio mecanismo VMD y al efecto de violaci\' on de paridad. Despu\' es de resolver el sistema mezclado encontramos una relaci\' on de dispersi\' on distorsionada para los mesones vectoriales mientras que los fotones no se ven afectados por los efectos de ruptura de paridad. Como en el caso de los mesones de esp\' in nulo, los mesones vectoriales muestran una masa efectiva dependiente del sistema de referencia y un comportamiento taqui\' onico para una de sus polarizaciones. Las modificaciones en el medio de los mesones vectoriales pueden ser examinadas a trav\' es de sus desintegraciones en dileptones. La forma distorsionada de las funciones espectrales de $\rho$ y $\omega$ predice un incremento de la producci\' on de dileptones alrededor de su pico de resonancia. Vemos que este incremento podr\' ia ayudar a explicar parcialmente el exceso de dileptones de baja masa presentado en la Section \ref{dilepexcessesp} para los experimentos NA60/PHENIX/STAR. Pero esta se\~nal puede ser bastante complicada de detectar. La forma distorsionada de las funciones espectrales de $\rho$ y $\omega$ revelan una asimetr\' ia en sus polarizaciones alrededor del pico en el vac\' io. De ah\' i, describimos c\' omo el efecto de ruptura de paridad puede ser detectado en HIC a partir de la restricci\' on de la cobertura angular en los estados lept\' onicos finales. Se presentan dos variables angulares adecuadas para buscar posibles dependencias en la polarizaci\' on y explicamos c\' omo un an\' alisis combinado de una de estas variables junto con la masa invariante de los dileptones puede ser una traza importante para detectar la ruptura de paridad en HIC.

\section{Lista de publicaciones}
\subsection{Publicaciones en revistas}
\begin{itemize}
\item A. A. Andrianov, V. A. Andrianov, D. Espriu and X. Planells, \emph{Analysis of dilepton angular distributions in a parity breaking medium}, Phys. Rev. D 90 (2014) 034024, arXiv:1402.2147 [hep-ph].
\item A. A. Andrianov, D. Espriu and X. Planells, \emph{Chemical potentials and parity breaking: the Nambu--Jona-Lasinio model}, Eur. Phys. J. C 74 (2014) 2776 [arXiv:1310.4416 [hep-ph]].
\item A. A. Andrianov, D. Espriu and  X. Planells, \emph{An effective QCD Lagrangian in the presence of an axial chemical potential}, Eur. Phys. J. C 73 (2013) 2294 [arXiv:1210.7712 [hep-ph]].
\item A. A. Andrianov, V. A. Andrianov, D. Espriu and X. Planells, \emph{Dilepton excess from local parity breaking in baryon matter}, Phys. Lett. B 710 (2012) 230 [arXiv:1201.3485 [hep-ph]].
\item A. A. Andrianov, V. A. Andrianov, D. Espriu and X. Planells, \emph{Abnormal enhancement of dilepton yield in central heavy-ion collisions from local parity breaking}, Theor. Math. Phys. 170 (2012) 17.
\end{itemize}

\subsection{Otras publicaciones}
\begin{itemize}
\item A. A. Andrianov, D. Espriu and  X. Planells, \emph{An effective theory for QCD with an axial chemical potential}, PoS QFTHEP 2013 (2014) 049 [arXiv:1310.4434 [hep-ph]].
\item A. A. Andrianov, V. A. Andrianov, D. Espriu and X. Planells, \emph{Implications of local parity breaking in heavy ion collisions}, PoS QFTHEP 2011 (2013) 025 [arXiv:1310.4428 [hep-ph]].
\item A. A. Andrianov, V. A. Andrianov, D. Espriu and X. Planells, \emph{Abnormal dilepton yield from parity breaking in dense nuclear matter}, AIP Conf. Proc. 1343 (2011) 450 [arXiv:1012.0744 [hep-ph]].
\item A. A. Andrianov, V. A. Andrianov, D. Espriu and X. Planells, \emph{The problem of anomalous dilepton yield in relativistic heavy-ion collisions}, PoS QFTHEP 2010 (2010) 053.
\end{itemize}

\subsection{Trabajos sin publicar}
\begin{itemize}
\item A. A. Andrianov, V. A. Andrianov, D. Espriu and X. Planells, \emph{Abnormal dilepton yield from local parity breaking in heavy-ion collisions} (2010) [arXiv:1010.4688 [hep-ph]].
\end{itemize}
\appendix
\chapter{Calculation of the fermion determinant in Chapter 3}\label{trick}
In this appendix we address the analysis of the determinant of the fermion operator presented in Eq. \eqref{fermoper}
\[\mathcal M(\mu,\mu_5)=\partial\!\!\!\!\!\!\not\;\;+(M+\vec\tau\vec a)-\mu\gamma_0-\mu_5\gamma_0\gamma_5+i\gamma_5(\vec\tau\vec\pi+\eta).\]
As it has been already stressed in \cite{azcoiti}, the fermion determinant can be proven to be real. The presence of both a vector and an axial chemical potentials does not modify this feature. Invariance under parity and time reversal symmetries also provide some equalities that will be useful for our purposes. Let us introduce some useful formul\ae~ in the Dirac representation where the superscripts $P$ and $T$ stand for parity and time reversal transformations:
\[\text{det}(\mathcal M)=\text{det}(\mathcal M^P)=\text{det}(\mathcal M^T), \quad \Gamma\equiv-i\gamma_0\gamma_5=\Gamma^\dagger, \quad \gamma_2\gamma_\mu\gamma_2^\dagger=\gamma_\mu^*, \quad \Gamma^2=\gamma_2\gamma_2^\dagger=1\]
\[\mathcal M^\dagger(\mu,\mu_5)=\partial\!\!\!\!\!\!\not\;\;+(M+\vec\tau\vec a)-\mu\gamma_0+\mu_5\gamma_0\gamma_5-i\gamma_5(\vec\tau\vec\pi+\eta)\]
\[\gamma_0\partial\!\!\!\!\!\!\not\;\;\gamma_0=\partial\!\!\!\!\!\!\not\;\;^P, \qquad \Gamma\partial\!\!\!\!\!\!\not\;\;\Gamma^\dagger=\partial\!\!\!\!\!\!\not\;\;^T, \qquad \gamma_2\partial\!\!\!\!\!\!\not\;\;\gamma_2^\dagger=\partial\!\!\!\!\!\!\not\;\;^*, \qquad \gamma_0^*=\gamma_0,\qquad \gamma_5^*=\gamma_5\]
Let us recall that external fields are constant and the parity and time reversal transformations do not modify them. Invariance under parity leads to
\begin{align}
\nonumber \text{det}(\mathcal M(\mu,\mu_5))&=\text{det}(\mathcal M^P(\mu,\mu_5))\\
\nonumber &=\text{det}(\partial\!\!\!\!\!\!\not\;\;+(M+\vec\tau\vec a)-\mu\gamma_0+\mu_5\gamma_0\gamma_5-i\gamma_5(\vec\tau\vec\pi+\eta))=\text{det}(\mathcal M^\dagger(\mu,\mu_5)),
\end{align}
where we multiplied the expression by $\gamma_0[\dots]\gamma_0$. This equality shows that the determinant is real independently on the external values. The same process may be done using invariance under time reversal and multiplying by $\Gamma^\dagger[\dots]\Gamma$:
\begin{align}
\nonumber \text{det}(\mathcal M(\mu,\mu_5))&=\text{det}(\mathcal M^T(\mu,\mu_5))\\
\nonumber &=\text{det}(\partial\!\!\!\!\!\!\not\;\;+(M+\vec\tau\vec a)+\mu\gamma_0-\mu_5\gamma_0\gamma_5-i\gamma_5(\vec\tau\vec\pi+\eta))=\text{det}(\mathcal M^\dagger(-\mu,-\mu_5)).
\end{align}
Finally, let us stress that we are only interested in the neutral components of the isotriplet external fields and then we can compute the conjugate of the determinant and multiply at both sides of the expression by $\gamma_2^\dagger[\dots]\gamma_2$ 
\begin{align}
\nonumber [\text{det}(\mathcal M(\mu,\mu_5))]^*&=\text{det}(\mathcal M^*(\mu,\mu_5))\\
\nonumber &=\text{det}(\partial\!\!\!\!\!\!\not\;\;+(M+\vec\tau\vec a)+\mu\gamma_0-\mu_5\gamma_0\gamma_5+i\gamma_5(\vec\tau\vec\pi+\eta))=\text{det}(\mathcal M(-\mu,\mu_5)).
\end{align}
The last two expressions are useful to show
\[\text{det}(\mathcal M(\mu,\mu_5))=\text{det}(\mathcal M(-\mu,-\mu_5))=\text{det}(\mathcal M(-\mu,\mu_5))=\text{det}(\mathcal M(\mu,-\mu_5)).\]
In particular, we will use
\[\text{det}(\mathcal M(\mu,\mu_5))=\text{det}(\mathcal M^\dagger(\mu,\mu_5))=\text{det}(\mathcal M(\mu,-\mu_5))=\text{det}(\mathcal M^\dagger(\mu,-\mu_5)).\]
We shall use $N$ to be even in order the determinant to be positive defined and use the fact that $\det(\mathcal M)^2=\det(\mathcal M^2)$. The development of the product
\begin{align}
\nonumber \mathcal M(\mu,\mu_5)\mathcal M^\dagger(\mu,-\mu_5)=&-\partial^2+M^2+\vec\pi^2+ (\eta^2+\vec a^2)+2 M\vec\tau\vec a+2 \eta\vec\tau\vec\pi\\
\nonumber &+2 \gamma_5(\vec a\times \vec \pi)\vec\tau-\mu^2+\mu_5^2+2\mu\partial_0-2\mu_5\gamma_0\vec \gamma\vec \partial\gamma_5
\end{align}
provides a result which is scalar in flavour except for the term proportional to $\mu_5$. An additional product produces
\begin{align}
\nonumber \mathcal M(\mu,\mu_5)\mathcal M^\dagger(\mu,-\mu_5)\mathcal M(\mu,-\mu_5)\mathcal M^\dagger(\mu,\mu_5)=A'+\vec\tau(\vec \alpha'+\vec{\epsilon'}\gamma_5)
\end{align}
with
\begin{gather}
\nonumber A'=A^2+\vec\alpha^2+\vec\epsilon^2+4\mu_5^2\vec \partial^2, \qquad \vec \alpha'=2A\vec \alpha, \qquad \vec{\epsilon'}=2A\vec{\epsilon}\\
\nonumber A=-\partial^2+M^2+\vec\pi^2+ (\eta^2+\vec a^2)-\mu^2+\mu_5^2+2\mu\partial_0,\\
\nonumber \vec \alpha= 2 (M\vec a+\eta\vec\pi),\\
\nonumber \vec \epsilon= 2 (\vec a\times \vec \pi), \qquad \vec\alpha\vec{\epsilon}=0
\end{gather}
with the property $\vec \alpha'\vec{\epsilon'}=0$. The logarithm of a quantity with such characteristics can be calculated and all the non-diagonal operators in Dirac or flavour space disappear leading to
\begin{equation}
\nonumber \log[A+\vec\tau(\vec \alpha+\vec{\epsilon}\gamma_5)]=\frac12\log[A^2-\vec \alpha^2-\vec{\epsilon}^2].
\end{equation}
The evaluation of the argument leads us to
\begin{align}
\nonumber &A'^2-\vec \alpha'^2-\vec{\epsilon'}^2=\prod_{\pm} \left [-(ik_0+\mu)^2+(|\vec k|\pm \mu_5)^2+M_+^2\right ]\left [-(ik_0+\mu)^2+(|\vec k|\pm \mu_5)^2+M_-^2\right ]
\end{align}
where $M_\pm^2=(M\pm a)^2+(\eta\pm\pi)^2$. Finally the fermion determinant can be written as
\begin{align}
\nonumber \log\text{det}(\mathcal M(\mu,\mu_5))=&\text{Tr}\log\mathcal M(\mu,\mu_5)=\frac18\text{Tr}\log(A'^2-\vec \alpha'^2-\vec{\epsilon'}^2)\\
\nonumber =&\frac18\text{Tr}\sum_{\pm}\Bigg\{\log\left [-(ik_0+\mu)^2+(|\vec k|\pm \mu_5)^2+M_+^2\right ]\\
&+\log\left [-(ik_0+\mu)^2+(|\vec k|\pm \mu_5)^2+M_-^2\right ]\Bigg\},
\end{align}
where the trace operator is given by
\[\text{Tr}(1)=8NT\sum_n\int\frac{d^3\vec k}{(2\pi)^3}[k_0\to \omega^F_n],\]
with $\omega^F_n=(2n+1)\pi/\beta$.

\section{Inclusion of temperature}

A finite temperature can be included with the formalism of the Matsubara frequencies \cite{Matsubara}
\[\int \frac{dk_0}{2\pi}f(k_0)=iT\sum_{n=-\infty}^{\infty}f(i\omega_n)\]
where $\beta T=1$ and
\[\omega_n^B=\frac{2n\pi}\beta,\qquad \omega_n^F=\frac{(2n+1)\pi}\beta\]
are the Matsubara frequencies for fermions or bosons. In our case, we are intersted in integrating Eq. \eqref{trace} which can be written as
\[\sum_n\frac 1{E^2-(i\omega_n^F+\mu)^2}=\frac 1{2E}\sum_n\left [\frac 1{(E+\mu)+i\omega^F_n}-\frac 1{(-E+\mu)+i\omega_n^F}\right ].\]
In order to sum the previous expression, we take the function
\[f(\tau)=n^F(E+\mu)e^{(E+\mu+i\pi T)\tau}+(n^F(E-\mu)-1)e^{(-E+\mu+i\pi T)\tau}\]
and impose periodicity in the interval $\tau\in[0,\beta)$ that is extended to all real values, say $f(\tau+\beta)=f(\tau)$. In the previous expression we used the fermion and boson Boltzmann distributions
\[n^F(E)\equiv \frac 1{\exp(\beta E)+1},\qquad n^B(E)\equiv \frac 1{\exp(\beta E)-1}.\]
The decomposition of $f(\tau)$ in a Fourier expansion
\[f(\tau)=T\sum_{n=-\infty}^\infty c_n e^{-i\omega_n^B\tau}, \qquad c_n=\int_0^\beta f(\tau)e^{i\omega_n^B\tau}d\tau.\]
leads to some coefficients $c_n$ that are directly related to the sum in Eq. \eqref{trace} and thus, one automatically finds that
\[\sum_n \frac 1{E^2-(i\omega_n^F+\mu)^2}=-\frac \beta{2E}\left [n^F(E+\mu)+n^F(E-\mu)-1\right ],\]
which is related to $f(\tau)$.

\medskip

After performing the sum of frequencies, we are left with a term $(|\vec k|\pm\mu_5)^2$ that requires a change of variable $\hat {\vec k}=\vec k(1\pm\mu_5/|\vec k|)$. Then
\begin{equation}\label{chvar}
d^3\vec k=\frac{d^3\hat{\vec k}}{\det\left (\frac{\partial \hat k^i}{\partial k^j}\right )}=d^3\hat{\vec k}\frac 1{\left (1\pm\frac{\mu_5}{|\vec k|}\right )^2}=d^3\hat{\vec k}\left (1\mp\frac{\mu_5}{|\hat{\vec k}|}\right )^2
\end{equation}
For simplicity, let us rescale the 3-momentum as $\vec k=\mu_5\vec x$. Then, the integral we will find is
\[\int d^3\vec k f(|\vec k| \pm \mu_5) \longleftrightarrow \int d^3\vec x f(|\vec x|\pm 1).\]
The previous change of variable allows us to write
\[\int_{R^3} d^3\vec x f(|\vec x|\pm 1)=\int_{\tilde R^3}f(|\vec x'|)d^3\vec x'\left (1\mp\frac 1{|\vec x'|}\right )^2,\]
where $f(|\vec x'|)\equiv f(x')=f(-x')$ due to the structure of Eq. \eqref{trace}. Note that the integration domain changed non-trivially. $\tilde R^3$ corresponds to $|\vec x'_{\pm}|\in [\pm 1,+\infty)$. We can separate the $\pm$ contributions to study them more clearly. The + integral is
\[\int d\Omega\int_{+1}^\infty x'^2f(x')dx'\left (1-\frac 1{x'}\right )^2=\Bigg [\int_{R^3}-\int_{0\leq|\vec x'|\leq 1}\Bigg ]f(|\vec x'|)d^3\vec x'\left (1-\frac 1{|\vec x'|}\right )^2.\]
The - integral is given by
\[\int d\Omega\int_{-1}^\infty x'^2f(x')dx'\left (1+\frac 1{x'}\right )^2=\Bigg [\int_{R^3}+\int_{-1\leq|\vec x'|\leq 0}\Bigg ]f(|\vec x'|)d^3\vec x'\left (1+\frac 1{|\vec x'|}\right )^2,\]
where the last integral can be rewritten as
\begin{align}
\nonumber &\int_{-1\leq|\vec x'|\leq 0}f(|\vec x'|)d^3\vec x'\left (1+\frac 1{|\vec x'|}\right )^2=\int d\Omega\int_{-1}^0f(x')x'^2dx'\left (1+\frac 1{x'}\right )^2\\
\nonumber =&\int d\Omega\int_{1}^0f(-x')(-x')^2(-dx')\left (1+\frac 1{-x'}\right )^2=\int d\Omega\int_{0}^1f(x')x'^2dx'\left (1-\frac 1{x'}\right )^2\\
\nonumber =&\int_{0\leq |\vec x'|\leq 1} f(x')d^3\vec x'\left (1-\frac 1{|\vec x'|}\right )^2
\end{align}
We can generalize the $\pm$ integrals with the following expression:
\begin{align}
\nonumber \int_{\tilde R^3}f(|\vec x'|)d^3\vec x'\left (1\mp\frac 1{|\vec x'|}\right )^2&=\int_{R^3}f(|\vec x'|)d^3\vec x'\left (1\mp\frac 1{|\vec x'|}\right )^2\\
\nonumber &\mp \int_{0\leq |\vec x'|\leq 1} f(x')d^3\vec x'\left (1-\frac 1{|\vec x'|}\right )^2.
\end{align}
In the case one has the sum of both integrals, the second term disappears and one is left with
\[\int d^3\vec x [f(|\vec x|+ 1)+f(|\vec x|- 1)]=2\int f(|\vec x'|)d^3\vec x'\left (1+\frac 1{|\vec x'|^2}\right ),\]
which can be rewritten again in terms of the 3-momentum $\vec k$ as
\[\sum_\pm\int d^3\vec k f(|\vec k|\pm\mu_5)=2\int f(|\vec k|)d^3\vec k\left (1+\frac {\mu_5^2}{|\vec k|^2}\right ).\]

\medskip

With all these ingredients Eq. \eqref{trace} is reduced to the calculation of two different integrals. The first type of integral is finite and reads
\[I_n(\mu)=\sum_\pm\int \frac{d^3\vec k}{(2\pi)^3E_\pm^n}n^F(E_\pm-\mu),\]
where we used the notation $E_\pm^2\equiv (|\vec k|\pm\mu_5)^2+M^2$. In the limit $T\to 0$ the Boltzmann distribution becomes $n^F(E)\to\Theta(-E)$ and then
\begin{align}
\nonumber I_n(\mu)&=2\int \frac{d^3\vec k}{(2\pi)^3E^n}\Theta(\mu-E)\left (1+\frac{\mu_5^2}{|\vec k|^2}\right )\\
\nonumber &=\frac1{\pi^2}\int_0^\infty \frac{k^2dk}{(k^2+M^2)^{n/2}}\Theta(\mu-\sqrt{k^2+M^2})\left (1+\frac{\mu_5^2}{k^2}\right )\\
\nonumber &=\frac1{\pi^2}\Theta(\mu-M)\int_0^{\sqrt{\mu^2-M^2}} \frac{dk}{(k^2+M^2)^{n/2}}\left (k^2+\mu_5^2\right ).
\end{align}
Finally one finds
\begin{equation}
I_1(\mu)=\frac1{2\pi^2}\Theta(\mu-M)\left [\mu\sqrt{\mu^2-M^2}+(2\mu_5^2-M^2)\log\left (\frac {\mu+\sqrt{\mu^ 2-M^2}}M\right )\right ].
\end{equation}
The second type of integral may be divergent and in such case the final result depends on the regularization scheme. In dimensional regularization usual integration techniques lead to
\begin{align}
\nonumber J_n&=\sum_{\pm}\int \frac{d^3\vec k}{(2\pi)^3E_\pm^n}\\
\nonumber &=\frac{M^{3-n}}{4\pi^{3/2}}\left (\frac{M^2}{4\pi\mu_R^2}\right )^{-\epsilon}\frac{\Gamma\left (\frac{n-1}2+\epsilon\right )}{\left (\frac{n-3}2+\epsilon\right )\Gamma\left (\frac n2\right )}\left [1+\frac{2\mu_5^2}{M^2}\left (\frac{n-3}2+(n-2)\epsilon\right )\right ].
\end{align}
In particular we are interested in
\[J_1=-\frac{M^2}{4\pi^2}\left [\left (1-\frac{2\mu_5^2}{M^2}\right )\left (\frac{M^2}{4\pi\mu_R^2}\right )^{-\epsilon}\Gamma(\epsilon)+1-\frac{4\mu_5^2}{M^2}\right ],\]
which can also be calculated by means of a 3-dimensional cut-off leading to
\begin{align}
\nonumber J_1^\Lambda&=\frac1{2\pi^2}\left [\Lambda\sqrt{M^2+\Lambda^2}-(M^2-2\mu_5^2)\log\left (\frac{\Lambda+\sqrt{M^2+\Lambda^2}}M\right )\right]\\
\nonumber &\approx \frac1{2\pi^2}\left [\Lambda^2-(M^2-2\mu_5^2)\log\left (\frac{2\Lambda}M\right )+\frac12 M^2\right].
\end{align}

\medskip

Finally, Eq. \eqref{trace} can be summed leading to
\begin{align}
\nonumber K_1^{\text{DR}}(M,\mu,\mu_5)=&\frac1{2\pi^2}\Bigg [\Theta(\mu-M)\left \{\mu\sqrt{\mu^2-M^2}+(2\mu_5^2-M^2)\log\left (\frac {\mu+\sqrt{\mu^ 2-M^2}}M\right )\right \}\\
&-\frac12 M^2+\frac1{2}(M^2-2\mu_5^2)\left (\frac1{\epsilon}-\gamma_E + 2 -\log\frac{M^2}{4\pi\mu_R^2}\right )\Bigg ],
\end{align}
in dimensional regularization while a 3-dimensional cut-off yields
\begin{align}
\nonumber K_1^\Lambda(M,\mu,\mu_5)=&\frac1{2\pi^2}\Bigg [\Theta(\mu-M)\left \{\mu\sqrt{\mu^2-M^2}+(2\mu_5^2-M^2)\log\left (\frac {\mu+\sqrt{\mu^ 2-M^2}}M\right )\right \}\\
&-\frac12 {M^2}+\frac12 (M^2-2\mu_5^2)\log\frac {4\Lambda^2}{M^2}-\Lambda^2\Bigg ].
\end{align}
\chapter{Dalitz decays in a parity-odd medium}\label{AppDalitz}
\graphicspath{{appendix/figures/}{figures/}}

In the calculation of the dilepton production, the inclusion of the $\rho$ and $\omega$ contributions could seem to be insufficient. In this Appendix we will calculate the Dalitz decays of the main resonances appearing the hadronic 'cocktail' in a parity-odd medium. In fact, the most significant discrepancy in PHENIX data is observed in the region of $M\sim 500$ MeV (see Fig. \ref{phenix}), where one expects an important contribution from all the Dalitz decays. Dalitz decays are one of the main sources of dileptons in the last stages of a HIC after hadron freezout. Although one would not expect in-medium modifications of these processes, their study could help to investigate the possibility that LPB survives at low temperatures.

\medskip

The processes we will consider here are: $\eta\to \gamma e^+e^-$, $\eta'\to \gamma e^+e^-$ and $\omega\to\pi^0e^+e^-$. Let us stress that the Kroll-Wada (KW) formula \cite{landsberg} cannot be applied directly in this context since the masses of both scalar/pseudoscalar and vector mesons depend on the 3-momentum and the calculation has to be redone. For comparison with PHENIX, we will use the same experimental cuts and temperature as discussed in Chapter \ref{ChVMD}. NA60 has also been able to identify the individual contributions from $\eta$ and $\omega$ Dalitz decays and found an enhancement, particularly for the latter\cite{NA60Dalitz}. The behaviour of the $\eta$ decay is very similar as the one predicted by KW. However, NA60 fits the $\omega$ Dalitz decay with a form factor where the $\rho$ meson has zero width. We skip any possible comparison with the latter as the relation of this fit with our results could be misleading.

\section{$\omega$ Dalitz decay}

In this section, we will investigate the Dalitz decay $\omega\to\pi^0e^+e^-$ in order to see if a parity-odd medium could strongly affect the production of dileptons. The $\omega\pi\gamma$ interaction is governed by the Lagrangian
\begin{equation}
 \mathcal L=ie\frac{g_{\omega\pi\gamma}}{m_\omega}\varepsilon^{\mu\nu\rho\sigma} A_\mu\partial_\nu \pi \partial_\rho\omega_\sigma,
\end{equation}
with $g_{\omega\pi\gamma}\simeq g$. The calculation of this process will be performed in two ways. The first one consists in considering the decaying meson as a 'free' state with the only in-medium modification driven by the distorted propagator of the $\rho$ meson that produces the lepton pair. In the second approach we assume that the $\omega$ meson is also modified by the parity-odd medium together with the assumption that it has been created from a previous process that induces a Breit-Wigner distribution for the initial state.

\subsection{Free $\omega$}

In the case that the decaying $\omega$ meson is not affected by LPB, the squared modulus of the amplitude can be written as
\begin{align}
\nonumber \overline{|\mathcal M|^2}=-\frac13 \left (\frac{8\pi g\alpha}{m_\omega M^2}\right )^2 \varepsilon^{\mu\nu\alpha\beta}q_\mu k_\beta \varepsilon^{\mu'\nu'\alpha'\beta'}q_{\mu'} k_{\beta'}g_{\nu\nu'} \sum_{\epsilon,\epsilon'}P_{\alpha\lambda}^\epsilon F^\epsilon P_{\lambda'\alpha'}^{\epsilon'}F^{*\epsilon'}\\
\times\left (p^\lambda p'^{\lambda'}+p^{\lambda'}p'^\lambda-\frac{M^2}2g^{\lambda\lambda'}\right ).
\end{align}
The definition of momenta is the following: $q$ for $\omega$, $k$ for $\rho$, and $p$ and $p'$ for the two leptons. Let us remark that $\epsilon$ and $\epsilon'$ are the polarizations of the $\rho$ meson. The dilepton production is then given by
\begin{align}\label{dilept-omega-dalitz}
\nonumber \frac{dN}{d^4x dM^2}&=\int\frac{d^4q~ d^4p_\pi~ d^4p~ d^4p'}{[(2\pi)^3]^4} \delta(p^2-m_e^2) \delta(p'^2-m_e^2)\delta(q^2-m_\omega^2) \delta(p_\pi^2-m_\pi^2) \\
& \times \int \frac{d^4k}{2\pi}\delta(k^2-M^2)[(2\pi)^4\delta^4(q-k-p_\pi)(2\pi)^4\delta^4(p+p'-k)]\overline{|\mathcal M|} f(q),
\end{align}
where $f(q)$ is a Boltzmann distribution for the $\omega$ meson and the form factors are the same ones as in Eq. \eqref{genericproduct}. An in-medium thermal $\omega$ meson may be disputable but it is necessary to introduce a momentum distribution for it and this could be a reasonable first approach. The results of this approach and the next one are shown together in Fig. \ref{dalitz-omega}.

\subsection{In-medium $\omega$}

The case when the decaying $\omega$ meson is affected by LPB is more involved. First of all, the squared modulus of the amplitude for one of the $\omega$ polarizations can be written as 
\begin{align}\label{m2dalitzw}
\nonumber \overline{|\mathcal M|^2}(\epsilon_\omega)=-\frac13 \left (\frac{8\pi g\alpha}{m_\omega M^2}\right )^2 \varepsilon^{\mu\nu\alpha\beta}q_\mu k_\beta \varepsilon^{\mu'\nu'\alpha'\beta'}q_{\mu'} k_{\beta'}P^{\epsilon_\omega}_{\nu\nu'} \sum_{\epsilon,\epsilon'}P_{\alpha\lambda}^\epsilon F^\epsilon P_{\lambda'\alpha'}^{\epsilon'}F^{*\epsilon'}\\
\times\left (p^\lambda p'^{\lambda'}+p^{\lambda'}p'^\lambda-\frac{M^2}2g^{\lambda\lambda'}\right ).
\end{align}
while $\epsilon_\omega$ corresponds to the $\omega$ meson. The algebraic calculation of this matrix element is rather tricky so we will use a shorthand notation through the definition 
\begin{align}
\overline{|\mathcal M|^2}(\epsilon_\omega)&=\frac23 \left (\frac{4\pi g\alpha}{m_\omega M^2}\right )^2 \sum_{\epsilon,\epsilon'} F^{\epsilon} F^{*\epsilon'}[\epsilon_\omega,\epsilon,\epsilon'],\\
\nonumber [\epsilon_\omega,\epsilon,\epsilon']&=\varepsilon^{\mu\nu\alpha\beta}\varepsilon^{\mu'\nu'\alpha'\beta'}q_\mu q_{\mu'} k_\beta k_{\beta'} P_{\nu\nu'}^{\epsilon_\omega}P_{\alpha\lambda}^\epsilon P_{\lambda'\alpha'}^{\epsilon'}\left (4p^\lambda p^{\lambda'}+M^2g^{\lambda\lambda'}\right ).
\end{align}
We will assume that the $\omega$ meson is created from a previous process that induces a Breit-Wigner distribution for it. In this case the mass-shell condition has to be replaced in Eq. \eqref{dilept-omega-dalitz} by
\begin{equation}
\delta(q^2-m_\omega^2) \quad \Longleftrightarrow \quad \Delta(q^2,\vec q, \epsilon_\omega)=\frac1\pi\frac{\Gamma_\omega m_{\omega,\epsilon_\omega}^2/m_\omega}{(q^2-m_{\omega,\epsilon_\omega}^2)^2+(\Gamma_\omega m_{\omega,\epsilon_\omega}^2/m_\omega)^2},
\end{equation}
and the effective meson mass $m^2_{\omega,\epsilon_\omega}$ is given by the dispersion relation in Eq. \eqref{disp.rel}. The different combinations for the three polarizations can be summarized as follows:
\begin{align}
\nonumber [+\pm\pm]&=[-\mp\mp]=-\frac12 \left [\vec p^2-\frac{(\vec p\vec k)^2}{\vec k^2}-\frac{M^2}2\right ]\left (\frac{E_q}{|\vec q|}\pm\frac{E_k}{|\vec k|}\right )^2 \left (|\vec q||\vec k|\mp\vec q\vec k\right )^2,\\
\nonumber [+\pm\mp]&=[-\pm\mp]=\left (\frac{E_k^2}{\vec k^2}-\frac{E_q^2}{\vec q^2}\right )\left [\frac12\left (\vec p^2-\frac{(\vec p\vec k)^2}{\vec k^2}\right )\left (\vec q^2\vec k^2-(\vec q\vec k)^2\right )-\vec p(\vec q\times\vec k)\right .\\
\nonumber &\left .\times\left (\vec p(\vec q\times\vec k)\mp i|\vec k|\left (\vec q\vec p-\frac{(\vec p\vec k)(\vec q\vec k)}{\vec k^2}\right )\right )\right ],\\
\nonumber [L++]&=[L--]=\left [\vec p^2-\frac{(\vec p\vec k)^2}{\vec k^2}-\frac{M^2}2\right ]\left [(E_qE_k-\vec q\vec k)^2+\frac{(E_q\vec q\vec k-E_k\vec q^2)^2}{\vec q^2}-M^2M_\omega^2\right ],\\
\nonumber [L\pm\mp]&=M_\omega^2\left [(\vec p\vec k)^2-\vec p^2\vec k^2+2\frac{\vec p\vec q}{\vec q^2}(\vec k^2\vec p\vec q-2(\vec p\vec k)(\vec q\vec k))+\frac{(\vec q\vec k)^2}{\vec q^2\vec k^2}((\vec p\vec k)^2+\vec p^2\vec k^2)\right ]\\
\nonumber &-\frac{\lambda(M_\omega^2,M^2,m_\pi^2)}2\left [\vec p^2-\frac{(\vec p\vec k)^2}{\vec k^2}-\frac{M^2}2\right ]\pm 2i\frac{|\vec k|M_\omega^2}{\vec q^2}\vec p(\vec q\times\vec k)\left (\vec p\vec q-\frac{(\vec p\vec k)(\vec q\vec k)}{\vec k^2}\right ),\\
\nonumber [\pm L\pm]&=[\mp\mp L]=\frac1{|\vec k|}\left (E_p-E_k\frac{\vec p\vec k}{\vec k^2}\right )(E_k|\vec q|+E_q|\vec k|)\left (1-\frac{\vec q \vec k}{|\vec q||\vec k|}\right )\\
\nonumber &\times\left (\vec k^2\vec p\vec q-(\vec p\vec k)(\vec q\vec k)\mp i|\vec k|\vec p(\vec q\times\vec k)\right ),\\
\nonumber [\pm L\mp]&=[\mp\pm L]=\frac1{|\vec k|}\left (E_p-E_k\frac{\vec p\vec k}{\vec k^2}\right )(-E_k|\vec q|+E_q|\vec k|)\left (1+\frac{\vec q \vec k}{|\vec q||\vec k|}\right )\\
\nonumber &\times \left (\vec k^2\vec p\vec q-(\vec p\vec k)(\vec q\vec k)\pm i|\vec k|\vec p(\vec q\times\vec k)\right ),
\end{align}
\begin{align}
\nonumber [\pm LL]&=\left [\frac{M^4}2-2\left (E_p-E_k\frac{\vec p\vec k}{\vec k^2}\right )^2\vec k^2\right ]\left (\vec q^2-\frac{(\vec q\vec k)^2}{\vec k^2}\right ),\\
[L\pm L]&=[LL\pm]=[LLL]=0,
\end{align}
being $\lambda(x,y,z)$ the K\"all\' en function and $M_\omega^2$ the squared of the 4-momentum, which is different from the effective mass $m_{\omega,\epsilon}$ that centres the Breit-Wigner distribution when LPB affects the $\omega$ meson.

\begin{figure}[h!]
\centering
\includegraphics[scale=0.22]{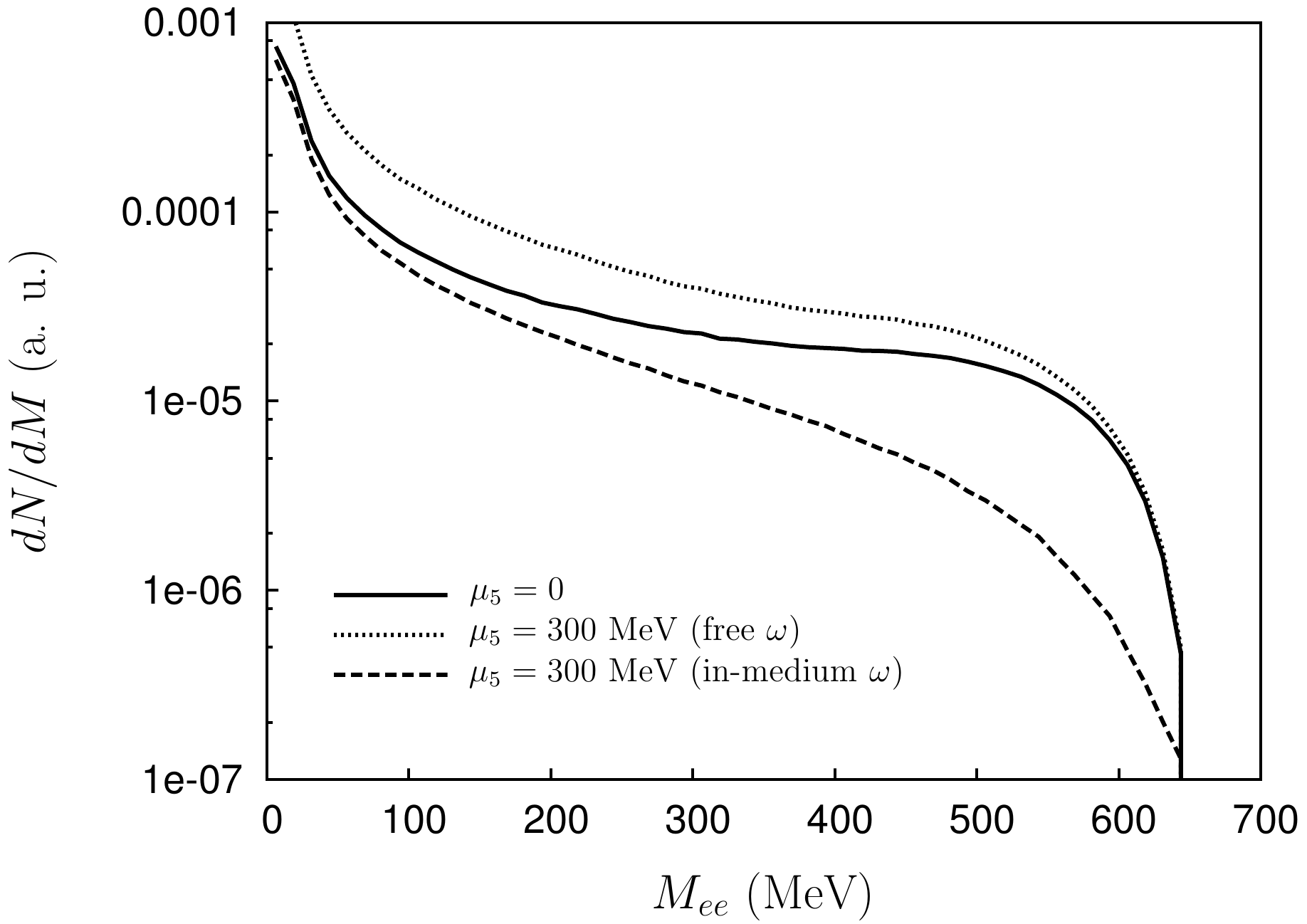}
\caption{The $\omega$ meson Dalitz decay is presented for parity-even matter (solid line) and for LPB with $\mu_5=300$ MeV in two different cases where LPB does and does not affect the $\omega$ meson (in-medium, dashed line; and free, dotted line).}\label{dalitz-omega}
\end{figure}

In Fig. \ref{dalitz-omega} we present the results for the Dalitz decay $\omega\to\pi^0e^+e^-$ using the PHENIX cuts and temperature presented in Chapter \ref{ChVMD}. Note that the curve corresponding to the vacuum process is not implemented through the KW formula since we account for experimental cuts. The cases where LPB affects the Dalitz decay have been enhanced by the same factor we used in Fig. \ref{phcockt} for the $\omega$ contribution in order to fit its resonance peak. The first case where the $\omega$ meson is a free state with an intermediate $\rho$ meson affected by LPB shows a small enhancement in the entire the region. However, this process is several orders of magnitude below the hadronic 'cocktail' and such enhancement does not substantially modify it. The second case where LPB also affects the momentum distribution and dispersion relation of the $\omega$ meson shows a slight suppression. These results suggest that even that LPB could affect the $\omega$ Dalitz decay, this effect would not be visible in the inclusive dilepton production.

\section{$\eta$ and $\eta'$ Dalitz decays}

We will focus now in the Dalitz decays of $\eta$ and $\eta'$. As in the previous section, we separate our analysis in two parts. In the first one we assume that the only LPB effect modifies the $\eta-\eta'$ form factor through the dispersion relation of the $\rho$ and $\omega$ mesons. In the second part we assume that the decaying mesons are also affected by LPB and therefore they are mixed due to a non-trivial $\mu_5$ as described in Chapter \ref{ChSmodel}. Here we will also investigate the coupling of 2 distorted pions to the initial state.

\subsection{Vacuum $\eta$ and $\eta'$}

The chirally-invariant Lagrangian that describes the anomalous decay of light pseudoscalar mesons into a pair of photons is given by \cite{ChPT,PS}
\begin{equation}\label{etagammagamma}
\mathcal L=-\frac{3e^2}{32\pi^2 f}\text{Tr}\left (\Phi Q^2\right )\varepsilon_{\mu\nu\alpha\beta}F^{\mu\nu}F^{\alpha\beta},
\end{equation}
and we use the same notation as in Chapter \ref{ChSmodel}, where we wrote explicitly $\Phi$ in \eqref{Phi}. The matrix of quark charges is $Q=$diag$(\frac23,-\frac13,-\frac13)$. In fact, it is not necessary to compute the whole process. The Lorentz structure of such decay is equivalent to the one we calculated for the process $\omega\to\pi e^+e^-$ in the previous section. In addition the overall constant is unknown and thus, our simulation is the same as the previous one with appropriate changes in the masses of the different states and the $\eta$ and $\eta'$ form factors.

\medskip

The form factors of $\eta$ and $\eta'$ are the same and they can be written as \cite{landsberg}
\begin{align}\label{Ffeta}
F_\eta\equiv F_{\eta\gamma}=F_{\eta'\gamma}=\frac{\frac{m_\rho^2}{m_\rho^2-M^2-i \Gamma_\rho m_\rho}+\frac 19\frac{m_\omega^2}{m_\omega^2-M^2-i \Gamma_\omega m_\omega}}{\frac{m_\rho^2}{m_\rho^2-i \Gamma_\rho m_\rho}+\frac 19\frac{m_\omega^2}{m_\omega^2-i \Gamma_\omega m_\omega}}.
\end{align}
Indeed, one can check that $F_{\eta\gamma}(M=0)=1$ by construction. 

\medskip

\begin{figure}[h!]
\centering
\includegraphics[scale=0.18]{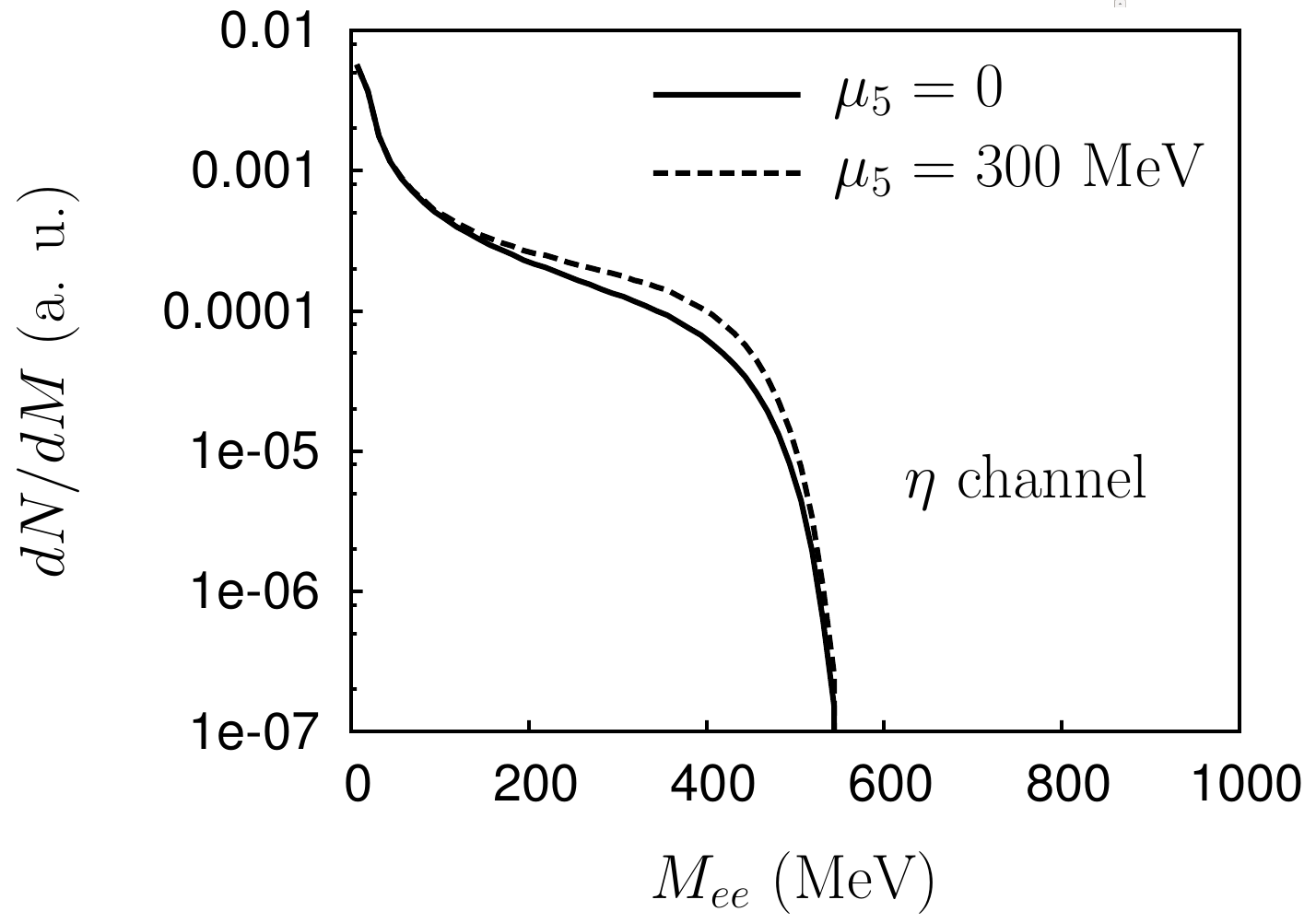} \includegraphics[scale=0.18]{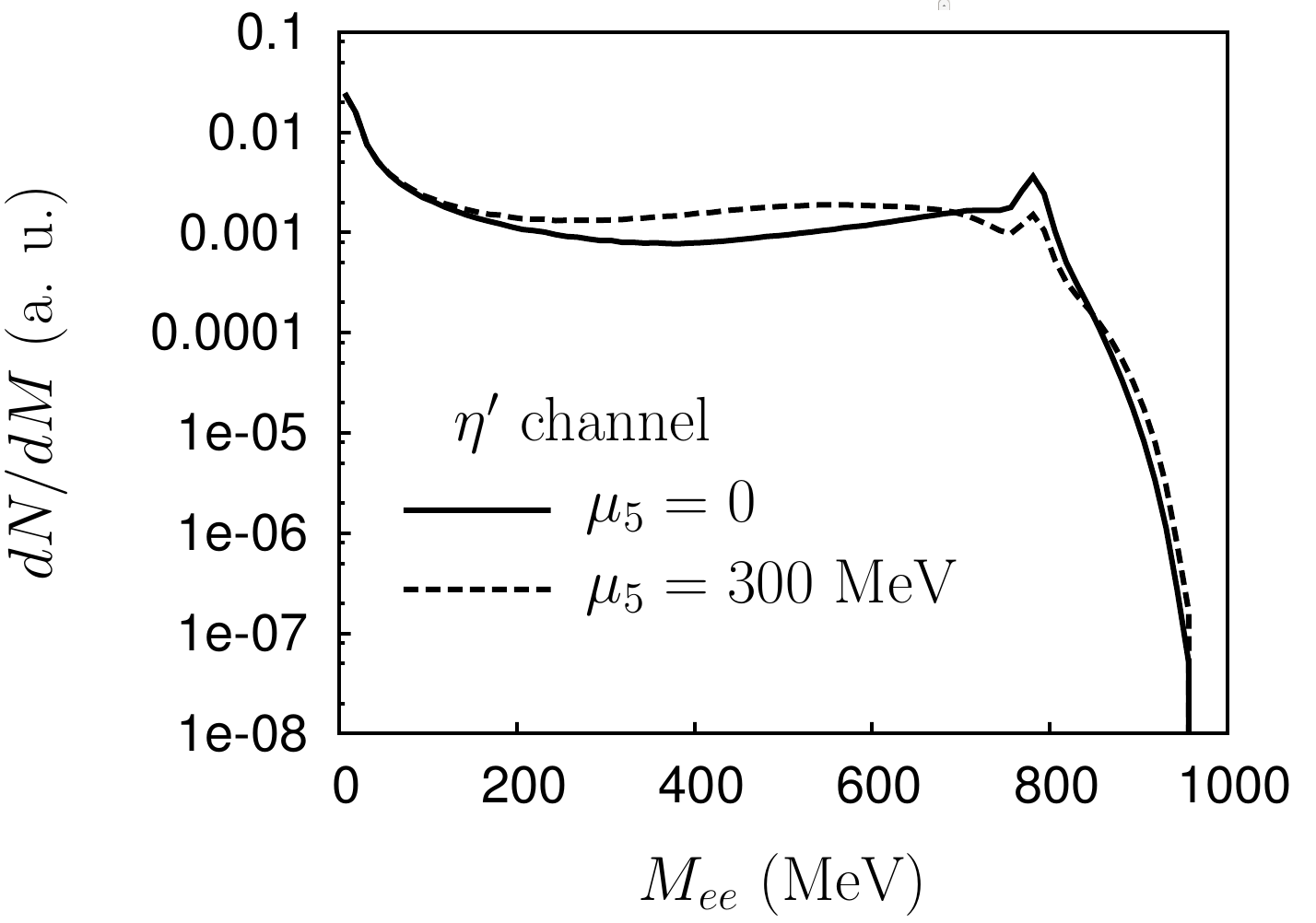}
\caption{The $\eta$ and $\eta'$ Dalitz decays are presented for parity-even matter (solid line) and for LPB with $\mu_5=300$ MeV (dashed line). No clear dilepton excess is found.}\label{dalitz-eta_1}
\end{figure}

In our simulations we include again the PHENIX cuts for the single electron momentum and dielectron rapidity as well as temperature and mass smearing previously discussed. In Fig. \ref{dalitz-eta_1} we show the modification of the dilepton production through this Dalitz process in the $\eta$ and $\eta'$ channels. A parity-odd medium with $\mu_5=300$ MeV slightly distorts the shape of both channels. The in-medium modification of the $\rho$ and $\omega$ propagators only in the form factor (see Eq. \eqref{Ffeta}) seems not to show a clear enhancement of the dielectron production. 

\subsection{Direct decay of in-medium $\eta$ and $\eta'$}

\begin{figure}[h!]
\centering
\includegraphics[scale=0.25]{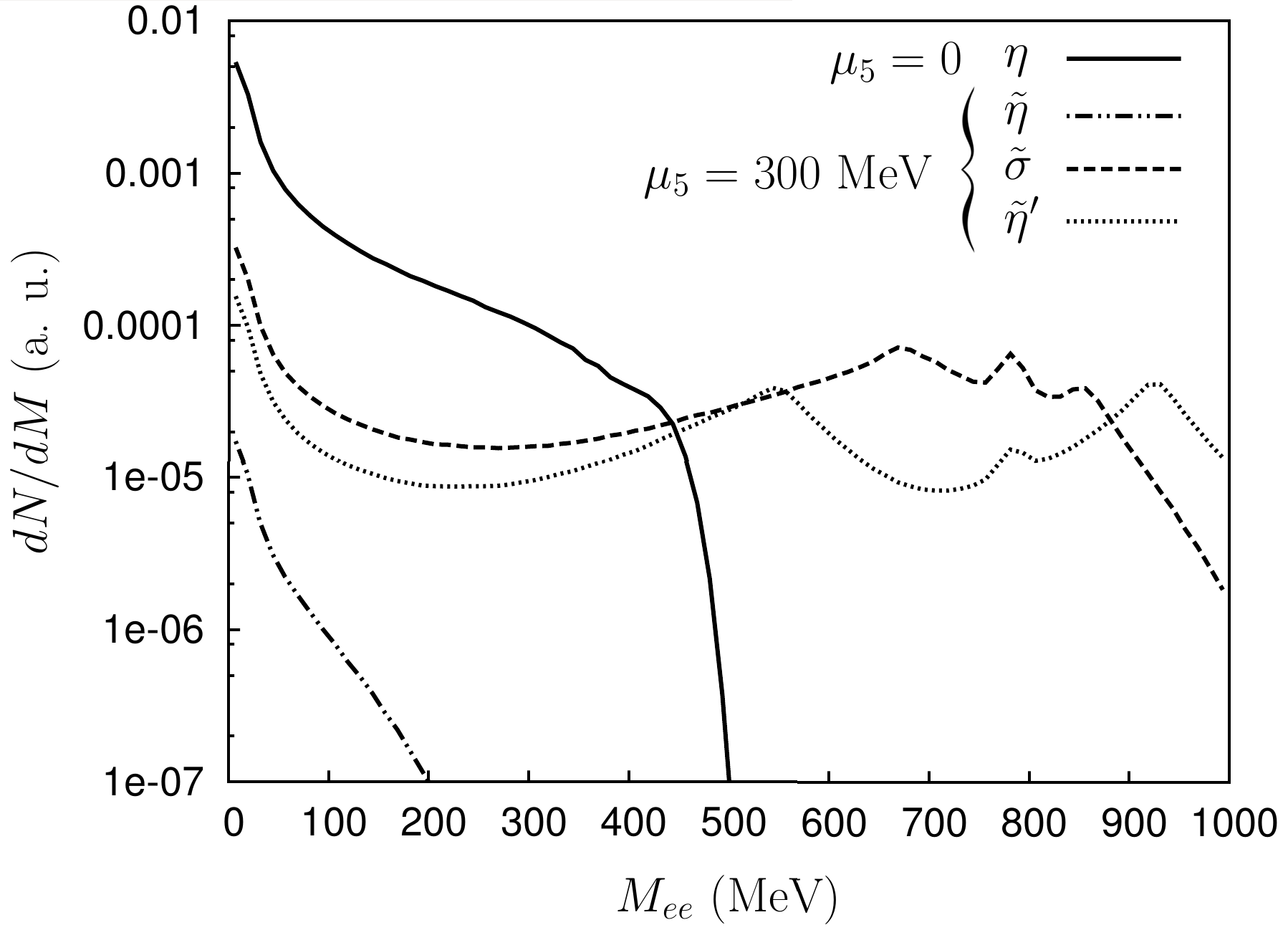}
\caption{The dielectron production from direct $\tilde\eta$ (dashed-dotted line), $\tilde\sigma$ (dashed line) and $\tilde\eta'$ (dotted line) Dalitz decays is shown in a parity-odd medium with $\mu_5=300$ MeV compared to the vacuum contribution of the $\eta$ channel (solid line). The vertical units are taken to coincide with PHENIX experimental data \cite{phenix}, as well as experimental detector cuts and temperature.}\label{dal-dir}
\end{figure}

Here we will consider that $\eta$ and $\eta'$ are affected by LPB and thus, they are mixed with $\sigma$ leading to the eigenstates $\{\tilde\eta,\tilde\sigma,\tilde\eta'\}$ that we generically called $X_i$. After normalizing the kinetic term of the $eta$ fields, the Lagrangian in Eq. \eqref{etagammagamma} reads
\begin{equation}
\mathcal L=-\frac{e^2}{96\pi^2}\left (5\frac{\eta_q}{v_q}+\sqrt 2\frac{\eta_s}{v_s}\right )\varepsilon_{\mu\nu\alpha\beta}F^{\mu\nu}F^{\alpha\beta}
\end{equation}
as a function of the states $\eta_q$ and $\eta_s$. We can rewrite the previous Lagrangian as a function of the proper in-medium eigenstates through Eq. \eqref{eigens}, which yields
\begin{equation}
\mathcal L_i=-\frac{e^2}{96\pi^2}\left (5\frac{C_{\eta_q i}}{v_q}+\sqrt 2\frac{C_{\eta_s i}}{v_s}\right )\varepsilon_{\mu\nu\alpha\beta}F^{\mu\nu}F^{\alpha\beta}X_i.
\end{equation}
With this expression we can compute the decays $X_i\to\gamma e^+e^-$ but we only know the form factors of the parity-defined states, not the ones related to $X_i$. Therefore, it is necessary to express each channel $i$ depending on the QCD states via $X_i=\sum_j b_{ij}\eta_j$ and compute
\begin{equation}
i\mathcal M_{X_i\to \gamma ee}=\sum_j(b_{ij}~i\mathcal M_{\eta_j\to \gamma ee}),
\end{equation}
where $b_{i\sigma}=0$ since $\sigma$ does not couple to 2 photons. Note that $b_{ij}$ is the inverse transformation of $C_{ij}$ with an extra rotation since $b_{ij}$ deals with $\eta$ and $\eta'$ while $C_{ij}$ deals with $\eta_q$ and $\eta_s$. The coefficients $b_{ij}$ read
\begin{equation}
b_{i\eta}=N_i\frac{v_q}{v_0}\left (\cos\psi+\frac{2\sqrt 2cv_s\sin\psi}{m_5^2-m_{\text{eff},i}^2}\right )\; \text{ and } \; b_{i\eta'}=
N_i\frac{v_q}{v_0}\left (\sin\psi-\frac{2\sqrt 2cv_s\cos\psi}{m_5^2-m_{\text{eff},i}^2}\right ).
\end{equation}
With such decomposition, one can readily check that for the Dalitz process we have
\begin{align}
\nonumber \overline{|\mathcal M|^2}_{X_i\to\gamma ee}&=\sum_{j,j'}\frac{\alpha^3 b_{ij}b_{ij'}^*}{9\pi M^4}\left |5\frac{C_{\eta_q i}}{v_q}+\sqrt 2\frac{C_{\eta_s i}}{v_s}\right |^2\left [g_{\lambda\lambda'}(m_{\text{eff},i}^2-M^2)^2+4M^2q_\lambda q_{\lambda'}\right ]\\
&\times\sum_{\epsilon,\epsilon'}F^\epsilon_\eta F^{\epsilon'*}_\eta P_\epsilon^{\lambda\mu}P_{\epsilon'}^{\mu'\lambda'}\left (4p_\mu p_{\mu'}+M^2g_{\mu\mu'}\right ).
\end{align}
We do not show the calculation of the phase space but let us remark the fact that the energy-dependent mass shell of the decaying particle has to be included carefully. The results of this approach are shown in Fig. \ref{dal-dir} where we included the contributions of the in-medium $\tilde\eta$, $\tilde\sigma$ and $\tilde\eta'$ channels compared with the vacuum $\eta$. First of all, note that the $\eta$ contribution in vacuum is not exactly the one shown in Fig. \ref{dalitz-eta_1} since this calculation is performed using the results of the sigma model studied in Chapter \ref{ChSmodel} (recall that we found a discrepancy $\simeq 10\%$ in the $\eta$ mass). The number of expected $\eta$ mesons is approximately the same as $\rho$ mesons both in vacuum and in-medium (see \cite{phenix}). In addition the thermalization of $\rho$ and $\tilde\eta$ in the pion gas are expected to be similar. For these reason we normalized the $\tilde\eta$ channel in the same way as we did with the $\rho$ meson in Chapter \ref{ChVMD}: the vacuum contribution is enhanced by an approximate factor of 1.8 that accounts for multiple regeneration into distorted pions within the fireball. Despite this factor may be controversial, our results and conclusions do not substantially depend on it. As a first approach, the normalization of the $\tilde\eta$, $\tilde\sigma$ and $\tilde\eta'$ channels is assumed to follow a Boltzmann distribution due to the strong coupling to pions. This assumption provides the ratios between the different channels.

The thermalization of the $\tilde\eta$ channel suggests that this state has a dominant contribution from the vacuum $\sigma$ state. As the latter does not couple to electromagnetism, one would expect an important suppression of the dilepton production in the $\tilde\eta$ channel as it can be seen in Fig. \ref{dal-dir}. The $\tilde\sigma$ and $\tilde\eta'$ channels show similar contributions that in general are subdominant compared to the PHENIX 'cocktail'. The only exception is the peak of the $\tilde\eta'$ right wing at $M\simeq925$ MeV. The dilepton yield at this peak is similar to the 'cocktail' contribution but the inclusion of the $\tilde\eta'$ channel would not be visible due to the low resolution of PHENIX data.

\subsection{Indirect decay of in-medium $\eta$ and $\eta'$}

Due to the relatively large width of the decay $X_i\to\tilde\pi\tilde\pi$ and the possible thermal regeneration of $\tilde\eta$ in such 'pion' gas, the indirect Dalitz decays should also be considered. The process $\tilde\pi\tilde\pi\to X_i\to\gamma e^+e^-$ is exactly zero in parity-even matter since the only channel that couples to a pair of pions ($\sigma$) does not couple to electromagnetism and hence it does not decay into $\gamma e^+ e^-$. In order to couple two distorted pions to the state $X_i$, we can use the Feynman rule for the vertex already shown in Eq. \eqref{iM}. The resulting squared element matrix is given by
\begin{align}\label{M2pio}
\nonumber \overline{|\mathcal M|^2}_{\tilde\pi\tilde\pi\to X_i\to\gamma ee}&=\frac{3}{2}\frac{\alpha^3}{9\pi M^4}\left |\sum_i\left [\left (5\frac{C_{\eta_q i}}{v_q}+\sqrt 2\frac{C_{\eta_s i}}{v_s}\right )\left (\sum_{j=\eta,\eta'}b_{ij}\right )\mathcal M_{\tilde\pi\tilde\pi\to X^i} D_{X_i}\right ]\right |^2\\
&\times \left [g_{\lambda\lambda'}(Q^2-M^2)^2+4M^2q_\lambda q_{\lambda'}\right ]\sum_{\epsilon,\epsilon'}F^\epsilon_\eta F^{\epsilon'*}_\eta P_\epsilon^{\lambda\mu}P_{\epsilon'}^{\mu'\lambda'}\left (4p_\mu p_{\mu'}+M^2g_{\mu\mu'}\right ),
\end{align}
where the $X_i$ propagator reads
\begin{equation}
D_{X_i}=\frac{1}{Q^2-m^2_{\text{eff},X^i}-i\frac{\Gamma_{X^i}}{m_{\text{eff},X^i}}Q^2}.
\end{equation}
In order to simplify our numerical calculations, we take the term of the $X_i$ width at rest. In other words, the denominator is simplified as: $Q^2-m^2_{\text{eff},X^i}(q)-i\frac{\Gamma_{X^i}(\vec q=\vec 0)}{m_{\text{eff},X^i}(\vec q=\vec 0)}Q^2$. This approximation should not be very important since the width is dominant only at the very resonance pole. The pole $m^2_{\text{eff},X^i}(q)$ can be equally described by the 3-momentum or the energy. Let us stress that this calculation takes into account all the $X_i$ states and their interferences.

\medskip

\begin{figure}[h!]
\centering
\includegraphics[scale=0.25]{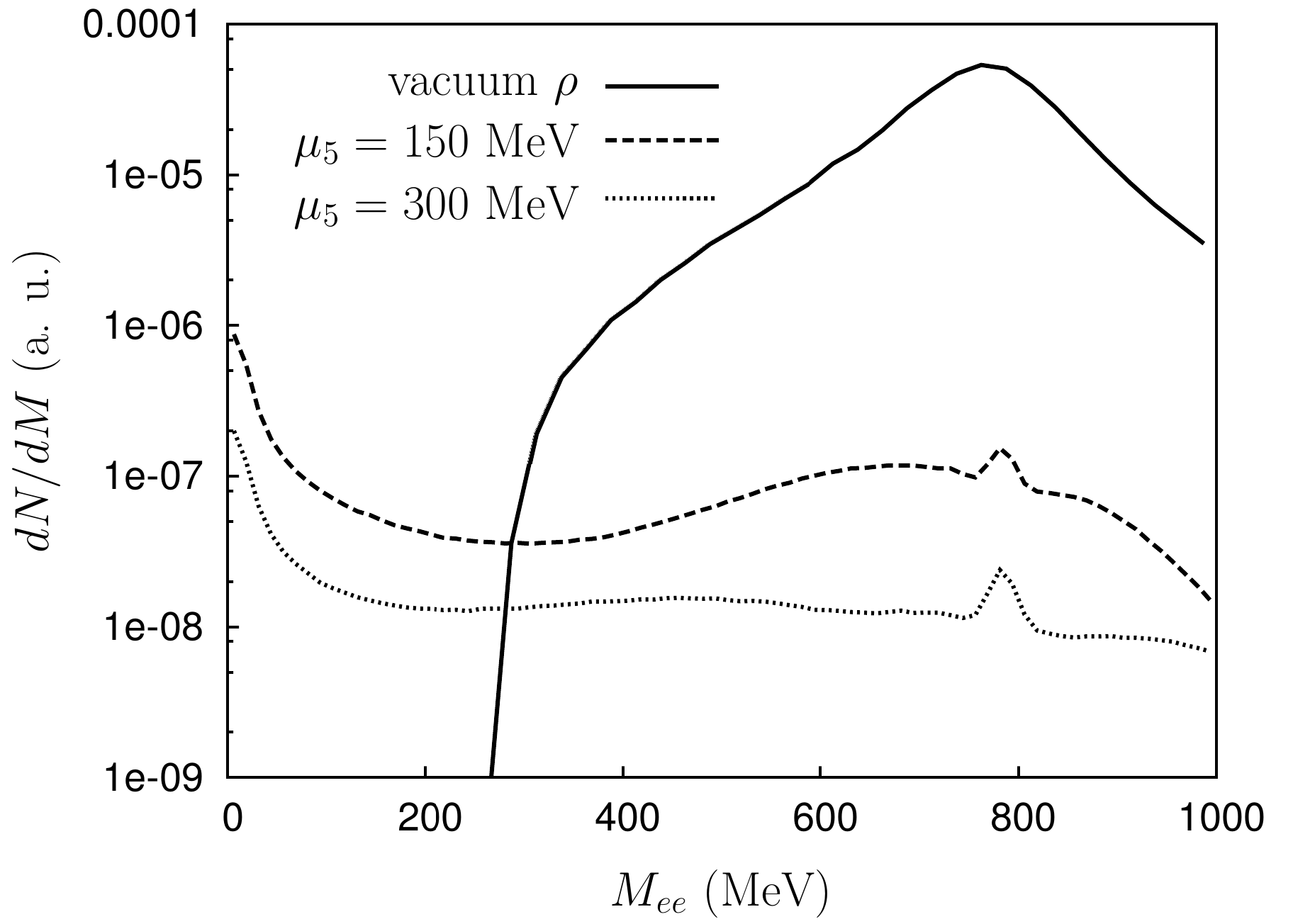}
\caption{The combination of the indirect $\tilde\eta$, $\tilde\sigma$ and $\tilde\eta'$ Dalitz decays is displayed for parity-odd matter with $\mu_5=150$ MeV (dashed line) and 300 MeV (dotted line). For a comparison of the scale, we also show the vacuum $\rho$ contribution (solid line). The vertical units are taken to coincide with PHENIX experimental data \cite{phenix}, as well as experimental detector cuts and temperature.}\label{dal-indir}
\end{figure}

The overall normalization is unknown but as the initial state is the same as in the $\rho$ channel appearing in Chapter \ref{ChVMD}, we used the same constant as a first approximation (see Fig. \ref{phcockt}). In Fig. \ref{dal-indir} we show the combined contribution of the indirect Dalitz decays of in-medium $\tilde\eta$, $\tilde\sigma$ and $\tilde\eta'$. For $\mu_5=150$ MeV, the combined distribution is almost one order of magnitude larger than for $\mu_5=300$ MeV. However, in both cases the dilepton production is completely subdominant. Recall that the vacuum $\rho$ contribution is already one order of magnitude below the 'cocktail' (see Fig. \ref{phenix}). In conclusion, although this new source of dileptons is included in the 'cocktail' of hadronic processes, no visible changes would arise.

\medskip

\section{Summary}

In this Appendix we have calculated the in-medium modifications of the $\omega$, $\eta$ and $\eta'$ Dalitz decays derived from the LPB effect. The $\rho$ and $\omega$ propagators have been changed in the corresponding form factors according to the dispersion relation we found in Chapter \ref{ChVMD}.

\medskip

In the case of the $\omega$ decay, we have analysed two different possibilities. In the first approach we considered that the initial state is unaffected by LPB and the only in-medium dependence is in the propagator of the intermediate $\rho$ meson. In the second situation we assumed that the decaying $\omega$ meson is affected by LPB and the fact that it has been created from a previous process leading to a Breit-Wigner distribution for this state. None of these cases shows a relevant enhancement of the dilepton yield when the in-medium contributions are compared to the vacuum one.

\medskip

Regarding the $\eta$ and $\eta'$ decays, we addressed three different approximations. In the first one, the only LPB effect is in the $\rho$ and $\omega$ propagators that appear in the corresponding form factors. The second approach consists in computing the decays of the distorted states $\tilde\eta$, $\tilde\sigma$ and $\tilde\eta'$ that we discussed in Chapter \ref{ChSmodel}. Finally, in the last approximation we computed the indirect decay of the previous states assuming that they are created from two (distorted) pion scattering in the hadron gas. We saw that LPB does not severely enhance the dilepton production.

\medskip

In summary, the anomalous enhancement of the dielectron production in the region of $M\sim 500$ MeV quoted by PHENIX cannot be explained by the relevant Dalitz decays that form the hadronic cocktail. In other words, although LPB could affect Dalitz decays, it would not be distinguished in the dilepton continuum below the $\rho$ and $\omega$ resonance peak.
\backmatter





\end{document}